\documentclass{aa}%[longauth]{aa}  

\usepackage{graphicx}
\usepackage{natbib}
\usepackage{float}
\usepackage[usenames,dvipsnames]{color} 
\usepackage{morefloats}
\usepackage{longtable}
\usepackage{lscape}
\usepackage{amsmath,upgreek}
\usepackage{txfonts}
\usepackage{textcomp}
\usepackage{gensymb}
\begin{document} 
\newcommand{\dmu}{pc~cm$^{-3}$}
\newcommand{\su}{\sigma^\mathrm{unkn}_{\lg S}}

\title{A LOFAR census of non-recycled
pulsars: average profiles, dispersion measures, flux densities, and spectra}

\titlerunning{A LOFAR census of non-recycled pulsars}

\author{A.~V.~Bilous \inst{\ref{nijmegen},\ref{uva}}
        \and V.~I.~Kondratiev\inst{\ref{astron},\ref{asc}} 
        \and M.~Kramer\inst{\ref{mpifr},\ref{jb}}
        \and E.~F.~Keane\inst{\ref{swinburne},\ref{caastro},\ref{skahq}}
        \and J.~W.~T.~Hessels\inst{\ref{astron},\ref{uva}}
        \and B.~W.~Stappers\inst{\ref{jb}}
        \and V.~M.~Malofeev\inst{\ref{prao}}
        \and C.~Sobey\inst{\ref{astron}}
        \and R.~P.~Breton\inst{\ref{southampton}}
        \and S.~Cooper\inst{\ref{jb}}
        \and H.~Falcke\inst{\ref{nijmegen},\ref{astron}}
        \and A.~Karastergiou\inst{\ref{oxford},\ref{western_cape},\ref{rhodes}}
        \and D.~Michilli\inst{\ref{uva},\ref{astron}}
        \and S.~Os\l{}owski\inst{\ref{bielefeld},\ref{mpifr}}
        \and S.~Sanidas\inst{\ref{uva}}
        \and S.~ter~Veen\inst{\ref{astron}}
        \and J.~van~Leeuwen\inst{\ref{astron},\ref{uva}}
        \and J.~P.~W.~Verbiest\inst{\ref{bielefeld},\ref{mpifr}} 
        \and P.~Weltevrede\inst{\ref{jb}}
        \and P.~Zarka\inst{\ref{lesia},\ref{nancay}}
        \and J.-M.~Grie{\ss}meier\inst{\ref{lpc2e},\ref{nancay}}
        \and M.~Serylak\inst{\ref{western_cape},\ref{nancay}}
        \and M.~E.~Bell\inst{\ref{csiro},\ref{caastro}}
        \and J.~W.~Broderick\inst{\ref{oxford}}
        \and J.~Eisl\"offel\inst{\ref{tautenburg}}
        \and S.~Markoff\inst{\ref{uva}}
        \and A.~Rowlinson\inst{\ref{uva},\ref{astron}}
        }

\institute{
Department of Astrophysics/IMAPP, Radboud University Nijmegen, P.O. Box 9010, 6500 GL Nijmegen, The Netherlands\label{nijmegen}
\and
Anton Pannekoek Institute for Astronomy, University of Amsterdam, Science Park 904, 1098 XH Amsterdam, The Netherlands\\ \email{A.Bilous@uva.nl}\label{uva}
\and
ASTRON, the Netherlands Institute for Radio Astronomy, Postbus
2, 7990 AA Dwingeloo, The Netherlands\label{astron} 
\and
Astro Space Centre, Lebedev Physical Institute, Russian Academy of Sciences, Profsoyuznaya Str. 84/32, Moscow 117997, Russia\label{asc} 
\and
Max-Planck-Institut f\"ur Radioastronomie, Auf dem H\"ugel 69, 53121 Bonn, Germany\label{mpifr} 
\and
Jodrell Bank Centre for Astrophysics, School of Physics and Astronomy, University of Manchester, Manchester M13 9PL, UK\label{jb}
\and
Centre for Astrophysics and Supercomputing, Swinburne University of Technology, Mail H30, PO Box 218, VIC 3122, Australia\label{swinburne}
\and
ARC Centre of Excellence for All-sky Astrophysics (CAASTRO), The University of Sydney, 44 Rosehill Street, Redfern, NSW 2016, Australia\label{caastro}
\and
SKA Organisation, Jodrell Bank Observatory, Lower Withington, Macclesfield, Cheshire, SK11 9DL, UK\label{skahq}
\and
Pushchino Radio Astronomy Observatory, 142290, Pushchino, Moscow region, Russia\label{prao}
\and
School of Physics  and  Astronomy,  University  of  Southampton, SO17 1BJ, UK\label{southampton}
\and
Oxford Astrophysics, Denys Wilkinson Building, Keble Road, Oxford OX1 3RH, UK\label{oxford}
\and
Department of Physics \& Astronomy, University of the Western Cape, Private Bag X17, Bellville 7535, South Africa\label{western_cape}
\and
Department of Physics and Electronics, Rhodes University, PO Box 94, Grahamstown 6140, South Africa\label{rhodes}
\and
Fakult\"at f\"ur Physik, Universit\"at Bielefeld, Postfach 100131, 33501 Bielefeld, Germany\label{bielefeld} 
\and
LESIA, Observatoire de Paris, CNRS, UPMC, Universit\'e Paris-Diderot, 5 place Jules Janssen, 92195 Meudon, France\label{lesia}
\and
Station de Radioastronomie de Nan\c{c}ay, Observatoire de Paris, PSL Research University, CNRS, Univ. Orl\'{e}ans, OSUC, 18330 Nan\c{c}ay, France\label{nancay}
\and
LPC2E - Universit\'{e} d'Orl\'{e}ans / CNRS, 45071 Orl\'{e}ans cedex 2, France\label{lpc2e}
\and
CSIRO Astronomy and Space Science, PO Box 76, Epping, NSW 1710, Australia\label{csiro}
\and
Th\"uringer Landessternwarte, Sternwarte 5, D-07778 Tautenburg, Germany\label{tautenburg}
}

\date{Received November 2015 / Accepted 2015}

  \abstract{      
We present first results from a LOFAR census of non-recycled pulsars.
The census includes almost all such pulsars known (194 sources) at
declinations ${\rm Dec}> 8\degree$ and Galactic latitudes $|{\rm Gb}| > 3\degree$, 
regardless of their expected flux densities and
scattering times. Each pulsar was observed for $\geq 20$\,minutes 
in the contiguous frequency range of 110--188\,MHz. Full-Stokes data were recorded.  
 We present the dispersion measures, flux
densities, and calibrated total intensity profiles for the 158 pulsars
detected in the sample.  The median uncertainty in census dispersion
measures ($1.5 \times 10^{-3}$\,pc\,cm$^{-3}$) is ten times smaller, on average,
than in the ATNF pulsar catalogue.  We combined census flux
densities with those in the literature and fitted the resulting
broadband spectra with single or broken power-law functions. For 48
census pulsars such fits are being published for the first
time. Typically, the choice between single and broken power-laws, as
well as the location of the spectral break, were highly influenced by
the spectral coverage of the available flux density measurements. In
particular, the inclusion of measurements below 100\,MHz appears
essential for investigating the low-frequency turnover in the spectra
for most of the census pulsars. For several pulsars, we compared
the spectral indices from different works and found the typical spread of
values to be within 0.5--1.5, suggesting a prevailing underestimation
of spectral index errors in the literature.  The census observations
yielded some unexpected individual source results, as we describe in
the paper.  Lastly, we will provide this unique sample of wide-band,
low-frequency pulse profiles via the European Pulsar Network Database. 
}


\maketitle
%
%________________________________________________________________

\section{Introduction}
\label{sec:intro}

\begin{table*}
\caption{Criteria used to select the sources for the LOFAR census of non-recycled pulsars.}              % title of Table
\label{table:criteria}      
\centering                                      
\begin{tabular}{|l|l|l|}          
\hline\hline 
  Parameter  & Criteria & \parbox[t][][t]{10cm}{Reasoning } \\ 
\hline
 Declination (Dec) &  $\mathrm{Dec}>8\degree$ & 
 \parbox[t][][t]{11cm}{Maximise the telescope sensitivity (which degrades with zenith angle, ZA, as $\sim\cos^2\mathrm{ZA}$).}\\
 
 Galactic latitude ($\mathrm{Gb}$) & $|\mathrm{Gb}|>3\degree$ &
 Avoid higher sky background temperatures in the Galactic plane.\\
 
 Surface magnetic field ($B_\mathrm{surf}$) & $B_\mathrm{surf}>10^{10}$\,G & 
 \parbox[t][][t]{11cm}{LOFAR observations of recycled pulsars were part of a separate project \citep{Kondratiev2015}.}\\
 
 Position error ($\epsilon_\mathrm{RA}$, $\epsilon_\mathrm{Dec}$) & 
 \parbox[t][][t]{2cm}{$\epsilon_\mathrm{RA}<130\arcsec$ or $\epsilon_\mathrm{Dec}<130\arcsec$} & 
 \parbox[t][][t]{11cm}{In order to avoid non-detections or biased flux density measurements due to mispointings, 
 the source position error was required to be less than the full-width at half-maximum of LOFAR’s HBA full-core tied-array beam, 
 pointed towards zenith, at the shortest wavelength observed   \citep[$130\arcsec$, ][]{vanHaarlem2013}.}\\
 
 Association & field pulsar & 
 \parbox[t][][t]{11cm}{Excluding globular cluster pulsars aimed at simplifying time budget calculation 
 (``one pulsar per pointing'') for the initial version of census proposal and persisted by accident. 
 Thus, four otherwise suitable pulsars in M15 and M53 are  missing from the census sample.}\\
\hline                                         
\end{tabular}
\end{table*}

Since their discovery almost 50 years ago \citep{Hewish1968}, the pulsations from pulsars -- rapidly rotating, 
highly magnetised neutron stars -- have been successfully detected over the entire electromagnetic spectrum, 
from the low radio frequencies at the edge of the ionospheric transparency window \citep[10\,MHz, ][]{Hassall2012} 
up to the very high-energy photons \citep[1.5\,TeV, ][]{MagicCol2015}. 
It is currently accepted that radiation processes at the various wavelengths of the electromagnetic spectrum are governed 
by several distinct emission mechanisms, with emission coming from different regions within the pulsar magnetosphere 
or from the star's surface \citep{Lyne2012}. 

The radio component of pulsar spectra is undoubtedly generated by coherent processes in the relativistic plasma of the pulsar 
magnetosphere \citep{Lyne2012}, but, despite decades of study, the exact mechanisms of the pulsar radio emission 
still remain unclear. Solving this problem would not only contribute to our knowledge of plasma physics under 
these extreme conditions, but also improve the understanding of pulsars as astrophysical objects and as probes 
of the interstellar medium (ISM).  

The lowest radio frequencies (below 200\,MHz) can provide valuable information for tackling this problem, because 
at these very low frequencies pulsar emission undergoes several interesting transformations, 
e.g. spectral turnover \citep{Sieber1973} or rapid profile evolution
\citep{Phillips1992}. %or puzzling low-frequency single-pulse behaviour \citep{Ulyanov2012,Hassall2013}. 
However, observing below 200\,MHz is challenging because the diffuse Galactic radio continuum emission, with its 
strong frequency dependence \citep[e.g. ][]{Lawson1987}, significantly contributes to the system temperature at these 
lower frequencies. At the same time, the deleterious effects of propagation in the ISM become ever more powerful 
\citep{Stappers2011}, sometimes making it difficult to disentangle intrinsic pulsar signal properties from effects 
imparted by the ISM. Nevertheless, various properties of low-frequency pulsar radio emission have been previously 
investigated with the help of a number of telescopes around the globe\footnote{In particular, average pulse profiles 
and flux density measurements have been obtained with the DKR-1000 and LPA telescopes 
of Pushchino Radio Observatory \citep{Izvekova1981,Malofeev2000}, Ukrainian T-shaped
Radio telescope \citep[UTR-2,][]{Bruk1978}, Arecibo telescope \citep{Rankin1970}, Gauribidanur T-array \citep{Deshpande1992}, 
Cambridge 3.6 hectare array \citep{Shrauner1998}, and the Westerbork Synthesis Radio Telescope \citep{Karuppusamy2011}.}.

The last decade was marked by rapid development of both hardware and computing capabilities, which made 
 wide-band pulsar observing at low frequencies possible. A major receiver upgrade has been done on UTR-2 \citep{Ryabov2010} and there are
three new telescopes operating below 200 MHz: LOFAR \citep[LOw-Frequency ARray, the Netherlands;][]{vanHaarlem2013},
MWA  \citep[Murchinson Widefield Array, Australia;][]{Tingay2013} and LWA \citep[Long Wavelength Array, USA;][]{Taylor2012}.

LOFAR has already been used for exploring low-frequency pulsar emission, e.g. wide-band average profiles \citep{Pilia2015}, 
polarisation properties 
\citep{Noutsos2015}, average profiles and flux densities of millisecond pulsars \citep{Kondratiev2015}, drifting 
subpulses from PSR B0809+74 \citep{Hassall2013}, nulling and mode switching in PSR B0823+26 \citep{Sobey2015}, 
as well as mode switching in PSR B0943+10 \citep{Hermsen2013,Bilous2014}. 

The first LOFAR pulsar census of non-recycled pulsars is a logical extension of these studies. 
Within the census project, we have performed single-epoch observations of a large sample of 
sources in certain regions of the sky, without any preliminary selections based on estimated peak flux density of the 
average profile or expected scattering time. Census observations resulted in  full-Stokes datasets spanning the 
frequency ranges of LOFAR's high-band antennas (HBA, 110--188\,MHz) and, for a subset of pulsars, the low-band antennas 
(LBA, 30--90\,MHz). The information recorded can be used for investigating the emission properties of about 150 pulsars, 
with the possibility of both average and single-pulse analyses. Based on census data, the properties of the ISM can 
also be explored using dispersion measures (DMs), scattering times, and rotation measures (RMs).

This paper presents the first results of the high-band part of the census project (analysis 
of the LBA data is deferred to subsequent work).
Sect.~\ref{sec:obs} describes the sample selection, observing setup and 
the initial data processing for the HBA data. In Sect.~\ref{sec:DM} we discuss the source detectability 
versus scattering time and DM, and discuss the DM variation rates obtained by comparing 
census DMs to the values from the literature. The flux calibration 
procedure is explained in Sect.~\ref{sec:calib}. In Sect.~\ref{sec:spec} we combine the HBA flux density measurements 
with previously published values and analyse the broadband pulsar spectra. A summary is given in Sect.~\ref{sec:summary}.

The results of the census measurements (flux densities, DMs and total intensity pulse profiles) will be soon 
made available through the European Pulsar Network (EPN) Database for Pulsar Profiles\footnote{\url{http://www.epta.eu.org/epndb}}, 
as well as via a dedicated LOFAR web-page\footnote{\url{http://www.astron.nl/psrcensus/}}.

\begin{figure*}
   \centering
 \label{fig:sources}   
 \includegraphics[scale=1.0]{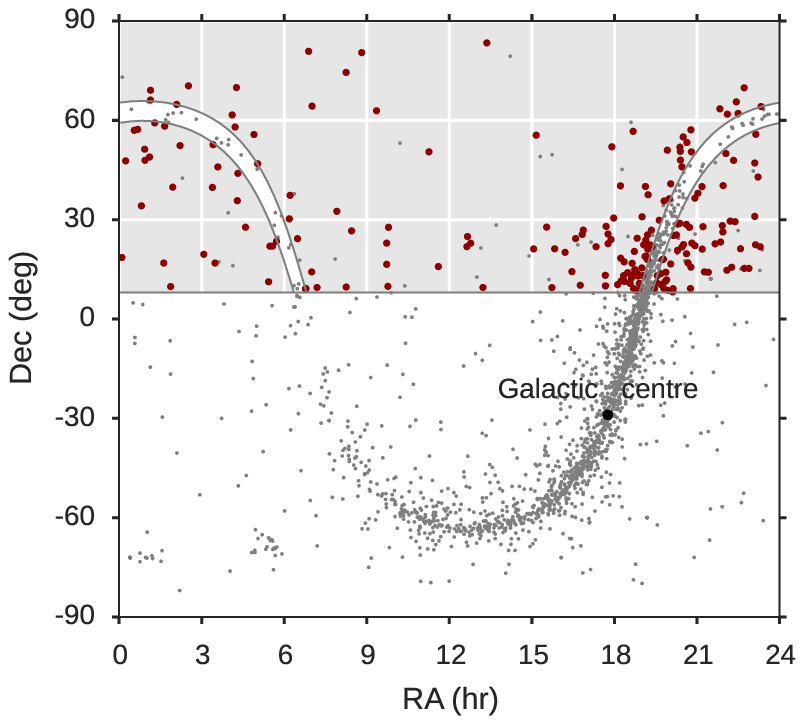}\includegraphics[scale=1.0]{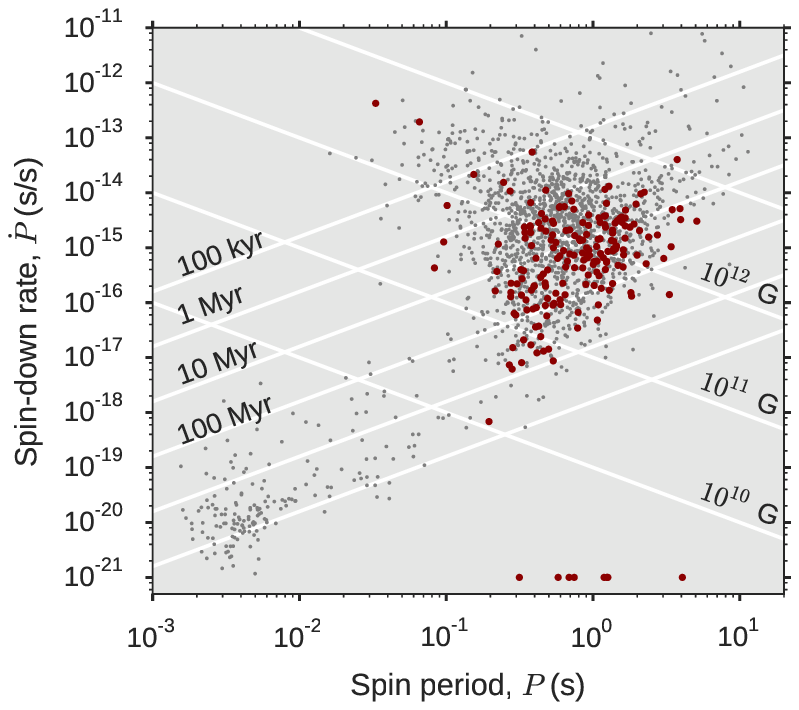}
  \caption{\textit{Left:} Distribution of all known pulsars  from the ATNF pulsar catalogue (grey dots)
  and the LOFAR census pulsars (red circles)  on the sky in equatorial coordinates. 
 The cuts in declination and Galactic latitude made for the LOFAR census sample are shown as grey lines 
 (\(\mathrm{Dec} > 8^{\circ}\) and \(|\mathrm{Gb}|>3^{\circ}\), respectively). For the full list of selection
 criteria see Table~\ref{table:criteria}. 
 \emph{Right:} Distribution of all known pulsars (grey dots) and the LOFAR census pulsars (red circles) 
 on the period--period derivative, \(P-\dot{P}\), diagram. Pulsars with an unknown \(\dot{P}\) are shown at 
 \(\dot{P}\mathrm{=10^{-21}\,s\,s^{-1}}\) in the diagram.
 }
\end{figure*}

\section{Observations and data reduction}
\label{sec:obs}

\subsection{Sample selection}

We selected the sample of known radio pulsars from version 1.51 of the ATNF Pulsar 
Catalogue\footnote{\texttt{http://www.atnf.csiro.au/people/pulsar/psrcat/}} 
(\citealt{Manchester2005}; hereafter ``pulsar catalogue'') that met the criteria summarised in Table~\ref{table:criteria}.
For some of the pulsars that did not satisfy the positional accuracy we were able to find ephemerides with
better positions based on timing observations with the Lovell telescope at Jodrell Bank and the 100-m Robert C. Byrd Green Bank 
Telescope. Such pulsars were included in the census sample.

In total, 194 pulsars were observed. Figure~\ref{fig:sources} shows the distribution of census sources on 
the sky and on the standard period--period derivative ($P-\dot{P}$) diagram.

\subsection{Observations}

Observations were conducted in February--May 2014 using the HBAs of the LOFAR core stations in the frequency 
range of 110--188\,MHz (project code LC1\_003). 
Complex-voltage data from the stations were coherently summed. The total observing band was split 
into 400 sub-bands, 195\,kHz each. Each sub-band was additionally split into 32$-$256 channels and 
full-Stokes samples were recorded in PSRFITS format \citep{Hotan2004}, with time resolution of 163.84$-$1310.72\,$\upmu$s, 
depending on the number of channels in one sub-band. Larger number of channels was chosen for pulsars with
higher DMs in order to mitigate the intra-channel dispersive smearing. For a more detailed description of 
LOFAR and its pulsar observing modes, we refer a reader to  \citet{vanHaarlem2013} and \citet{Stappers2011}.

Each pulsar was observed during one session for either 1000 spin periods, or at least 20\,min. The PSRFITS 
data were subsequently stored in the LOFAR Long-Term Archive\footnote{\url{http://lofar.target.rug.nl/}}.
Observations were pre-processed with the standard LOFAR pulsar pipeline \citep{Stappers2011}, which uses the 
PSRCHIVE software package \citep{Hotan2004,vanStraten2010}. The data were dedispersed and folded with ephemerides either from 
the pulsar catalogue or the timing observations with the Lovell or Green Bank telescopes. For the Crab pulsar,
we used the Jodrell Bank Crab monthly ephemeris\footnote{\url{http://www.jb.man.ac.uk/pulsar/crab.html}} \citep{Lyne2015}.
For folding the data, we chose the number of phase bins to be equal to the power of two 
that matched the original time resolution most closely (but not exceeding 1024). Sometimes, in order to increase
the signal-to-noise (S/N) ratio, the number of bins was reduced by a factor of two, four or eight. For all pulsars, the smearing in one channel 
due to incoherent dedispersion was less than one profile bin at the centre of the band and less than 2.5 bins for the 
lowest frequency channel. Folding produced 1-min sub-integrations and the archives were averaged in frequency to 400 
channels. In this paper we focus only on total intensity data.

\begin{figure*}
   \centering
 \includegraphics[scale=0.9]{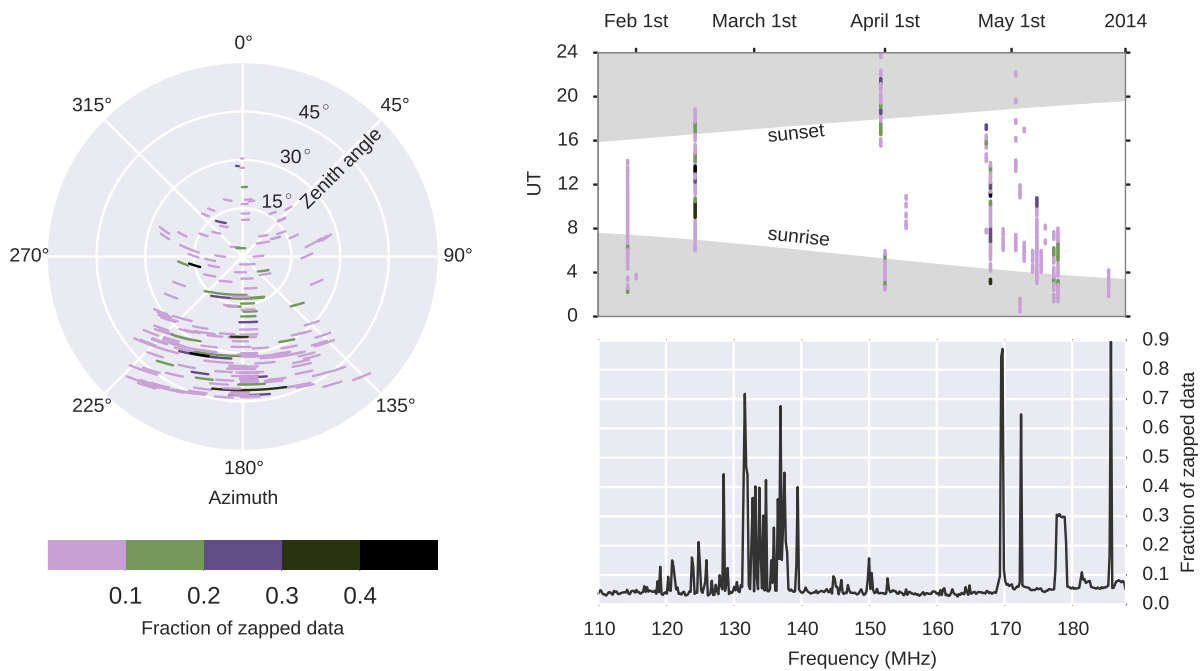}
 \caption{\textit{Left:} Fraction of zero-weighted data calculated according to Eq.~\ref{eq:rfi} as a function of zenith 
 angle  and azimuth of a source. Each stripe corresponds  to a single session. Pulsars were observed close to transit 
 and the North celestial pole at the LOFAR core has a zenith angle of approximately $48\degree$. Low-altitude observations 
 are not necessarily more corrupted by RFI and the interference does not come from any preferable azimuth, though
 the statistics here are limited. 
 \textit{Right, top:} RFI fraction as a function of Universal Time (UT) and date. Colour coding is the same as on the 
 left subplot. The RFI situation changes rapidly from one observation to another, likely due to the beamed nature of terrestrial signals. 
 \textit{Right, bottom:} Fraction of zero-weighted data versus observing frequency in each of 400 sub-bands.}
 \label{fig:rfi}
\end{figure*}

\subsection{RFI excision}

The 77 hours of census observations sampled radio signals from various on-sky directions during both day and night. 
This makes these data suitable for exploring RFI (radio frequency interference: any kind of unwanted signals of 
non-astrophysical origin) environment on the site of the LOFAR core stations. 

A selective analysis of small random subsets of data, performed with the \texttt{rfifind} program from the 
PRESTO\footnote{\url{http://www.cv.nrao.edu/~sransom/presto/}} software package \citep{Ransom2001}, showed that the 
majority of RFI was shorter than one minute in duration and/or narrower in frequency than a 195-kHz sub-band 
\citep[see also][]{Offringa2013}. 
The real-time excision of RFI, however, was not possible on the full time- and frequency-resolution data due to 
limited computing power, and thus was performed on folded archives with 1-min sub-integrations and 195-kHz sub-bands. 
To clean the data we used the {\tt clean.py} tool from the {\tt CoastGuard} package\footnote{\url{https://github.com/plazar/coast_guard}} 
\citep{Lazarus2016}.

In general, the observations were not severely affected by RFI. The median fraction of data, zero-weighted due to RFI, was 
only 5\%. This was calculated using:
\begin{equation}
\label{eq:rfi}
\mathrm{\frac{N_{RFI}}{N_{\text{sub-int}}\times N_{\text{sub-band}}},}
\end{equation}
where \(\mathrm{N_{RFI}}\) is the number of the zero-weighted [sub-integration, sub-band] 
cells, and \(\mathrm{N_{\text{sub-int}}}\) and \(\mathrm{N_{\text{sub-band}}}\) are the total numbers of sub-integrations 
and sub-bands for each observation.
Twelve observing sessions had more than 20\% of the data zero-weighted, with the maximum RFI fraction equal to 46\%. 
The fraction of zero-weighted data may vary dramatically between two consecutive sessions. Most census observations were 
conducted during the daytime, however we did not notice any improvement in the RFI situation during the night.
RFI did not appear to come from any specific altitude or azimuth (Fig.~\ref{fig:rfi}).

Figure~\ref{fig:rfi} (bottom right) shows also the fraction of zero-weighted data versus observing frequency. Individual noisy 
sub-bands in this frequency range are likely to be affected by air-traffic control systems, the Dutch emergency 
paging system C2000, satellite signals and digital audio broadcasting \citep[for the specific list of frequencies, 
see Table~1 in][]{Offringa2013}.

\subsection{Detection and ephemerides update}

For most of our pulsars the epoch of observation lay outside the validity range of the available ephemerides. Thus, 
we expected the observed pulsar period, $P$, and DM to be somewhat different from the values predicted by the ephemerides.
We performed initial adjustment of $P$ and DM with the PSRCHIVE program \texttt{pdmp}, which maximises integrated S/N 
 of the frequency- and time- integrated average profile over the set of trial values of $P$ 
and DM. The output plots from \texttt{pdmp} (namely, maps of integrated S/N values versus trial parameters together with 
time-integrated spectra and frequency-integrated waterfall plots) were visually inspected for a pulsar-like signal. 
Out of 194 census pulsars, 158 were detected in such a manner, all with integrated S/N greater than 8. For detected
pulsars, \texttt{pdmp} DMs were used to make a template profile and subsequently improve $P$ and DM estimates with the
\texttt{tempo2} timing software\footnote{\url{https://bitbucket.org/psrsoft/tempo2}} \citep{Hobbs2006} 
in \texttt{tempo1} emulation mode. 
In most cases the new values of $P$ found by \texttt{tempo2} were very similar to the initial periods. The difference 
between the new and initial values of $P$, $\delta P$, was in most cases smaller than 5 times the new period error 
($\epsilon_P$, reported by \texttt{tempo2} from the least-squares fit). For three moderately bright pulsars, with integrated S/N between 14 and 50, 
 $\delta P$ ranged from seven to $20\epsilon_P$. For the bright (integrated S/N of about 500) binary pulsar 
 PSR B0655$+$64 $\delta P$ was as large as $290\epsilon_P$.

\section{Dispersion measures}
\label{sec:DM}

Owing to the relatively low observing frequencies and the large fractional bandwidth, 
for most of the detected pulsars (except for a few faint ones with broad profiles) 
we were able to measure DMs much more precisely than previous measurements in the pulsar catalogue:
our median DM error (provided by \texttt{tempo2}) is 
0.0015\,\dmu, whereas for the same pulsars the median DM uncertainty in the pulsar catalogue is 0.025\,\dmu~(see 
Table~\ref{table:main} for both measured and catalogue DMs). 
The median relative difference between census DMs and the ones from the pulsar catalogue was 
$|\delta \mathrm{DM}|/\mathrm{DM} = 0.18\%$. 
However, for a few pulsars the relative DM correction was  
$\gtrsim 0.1$, and, in the most extreme case of \object{PSR J1503+2111}, the pulsar was detected at a DM 3.6 times lower than 
the previously published value (see Sect.\,\ref{ssubsec:DM_outl}).

We note that at our level of DM measurement precision a Doppler shift of the observed radio frequency, caused by 
Earth's orbital motion should be taken into account. This effect, if not corrected for, would cause an apparent sinusoidal annual 
DM variation with a relative amplitude of $|\delta \mathrm{DM}/\mathrm{DM}|\lesssim |\varv_\mathrm{orb}/c| \approx 0.01\%$,
where $\varv_\mathrm{orb}$ is the Earth's orbital velocity.
DM measurements can also be biased by profile evolution (intrinsic or caused by scattering, see also discussion in 
Sect.~\ref{subsec:DM_disc}), but accounting for profile evolution is beyond the scope of this work.

\begin{figure}
   \centering 
 \includegraphics[scale=1.0]{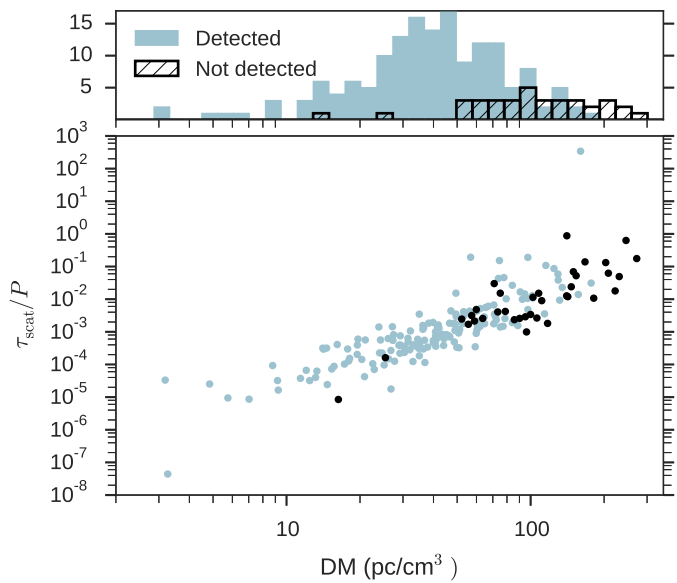}
 \caption{ Detected pulsars (light blue dots) and non-detected ones (black dots) versus DM and the scattering time
 at 150\,MHz divided by each pulsar's period. Scattering time at 1\,GHz was taken from the pulsar catalogue (if available) or 
 from the NE2001 model and scaled to 150\,MHz with  Kolmogorov index of $-4.4$. See the text for the discussion  of outliers.}
 \label{fig:detect_DM_scat}   
\end{figure}

\begin{figure*}
   \centering
 \includegraphics[scale=1.0]{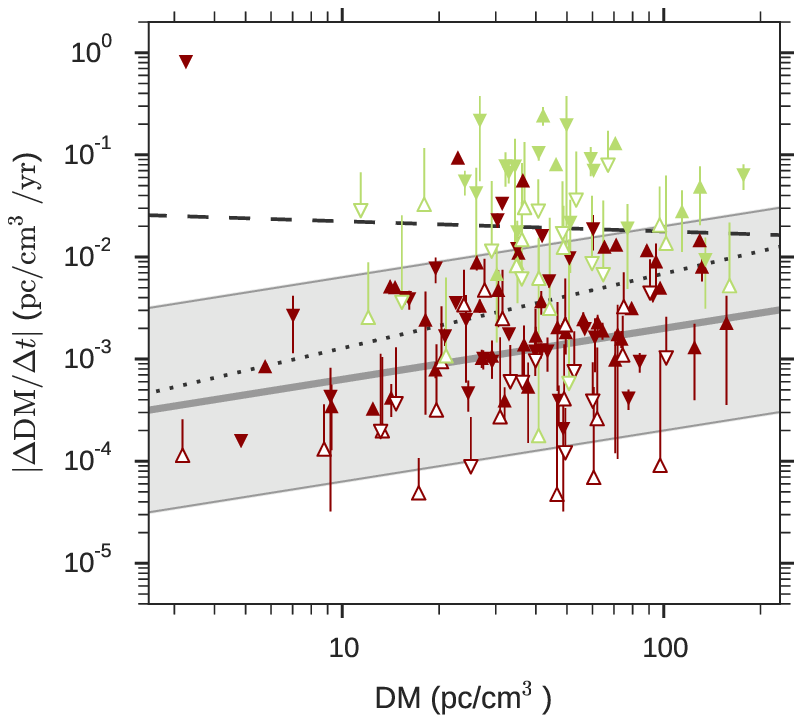}\includegraphics[scale=1.0]{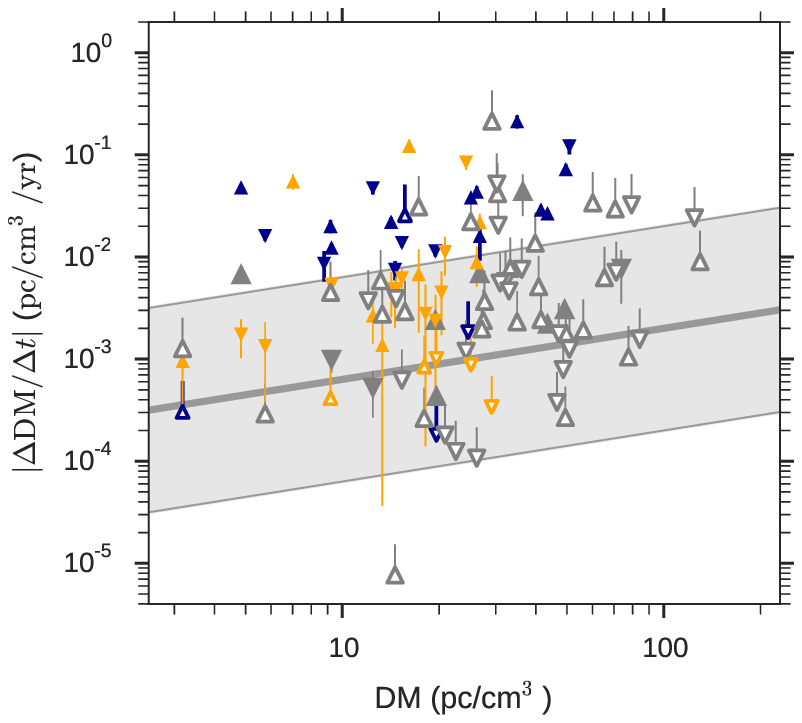}
 \caption{DM variation rates versus census DMs. On both panels the upward/downward triangles indicate DM values 
 increasing/decreasing with time. Unfilled triangles mark pulsars without significant DM variation rate (i. e. with $|\Delta\mathrm{DM}/\Delta t|$ 
 smaller than its uncertainty). For such pulsars the lower parts of the errorbars are not shown. 
  \textit{Left:} DM variation rates, obtained by comparing census DM measurements to the 
 pulsar catalogue values.  Lighter (green) marks
 show pulsars which have relatively large DM errors ($>0.1$\,\dmu), in either the census measurements 
 or the pulsar catalogue. The dotted and dashed lines show unweighted and weighted linear fit to the data in log-log space, respectively, 
 excluding  the outlier in the top left corner.
 The relation of \citet[][grey line]{Hobbs2004b} is overplotted together with one order-of-magnitude scatter reported by the authors (grey shade). 
 The outlier at $\mathrm{DM}\approx3$\,\dmu, PSR J1503$+$2111 is discussed in Sect.~\ref{ssubsec:DM_outl}.
 \textit{Right:} Rate of DM variations calculated by comparing census measurements to the recent low-frequency
 observations of \citet[][larger grey markers]{Pilia2015}, \citet[][dark blue markers]{Stovall2015} and
 \citet[][light orange markers]{Zakharenko2013}. See text for discussion.}
 \label{fig:DM_grad}   
\end{figure*}

It is instructive to plot the detections/non-detections versus DM and the expected scattering time over the pulsar 
period (Fig.~\ref{fig:detect_DM_scat}). The scattering time $\tau_\mathrm{scat}$ was scaled to 150\,MHz from the 
values at 1\,GHz \citep[obtained
from the pulsar catalogue or NE2001 Galactic free electron density model;][]{Cordes2002} with Kolmogorov 
index\footnote{Note that 
the scattering time is only a rough estimate, as its value changes by a factor of eight between the 
edges of HBA band (assuming a spectral index of $-4.4$). Also, the spectral index itself can deviate from the Kolmogorov value \citep{Lewandowski2015}.} of 
$-4.4$. As was expected, non-detected pulsars lay mostly at higher DM and $\tau_\mathrm{scat}/P$. Still, LOFAR HBAs 
are capable of detecting non-recycled pulsars up to at least $\mathrm{DM}=180$\,\dmu~(PSR B1930+13). 
Two non-detected pulsars, PSR J2015+2524 and PSR J2151+2315, have small DMs of $<30$\,\dmu. They  
are faint sources and have not been detected at lower frequencies \citep{Camilo1995,Han2009,Lewandowski2004,Zakharenko2013}.

One of the pulsars, \object{PSR B2036+53}, was detected despite the discouraging predictions of the NE2001 model. This pulsar, 
located at Galactic coordinates $\mathrm{Gl} = 90\fdg37$, $\mathrm{Gb} = 7\fdg31$ with a DM of about 160\,\dmu, 
has a predicted $\tau_\mathrm{scat}(150\,\mathrm{MHz})=486$\,s. The pulsar appeared to show little scattering 
in the upper half of the HBA band and to have $\tau_{\mathrm{scat}}\approx 0.4$\,s at 129\,MHz. Moreover, the pulsar has 
been previously detected with the LPA telescope (Pushchino, Russia) at frequencies close to 100\,MHz \citep{Malov2010}. In NE2001 a region of intense 
scattering has been explicitly modelled in the direction towards PSR B2036+53. This decision was based on higher-frequency  
scattering measurements for this pulsar, although the authors neither quote them directly, nor point to the profile data. 
The average profile at 1408\,MHz from \citet{Gould1998}, available via the EPN, seems to show a small scattering 
tail, with $\tau_\mathrm{scat}$ approximately in agreement with NE2001. However, both the S/N and the time resolution of 
the 1408\,MHz profile are not high. Being extrapolated down to 700\,MHz with the Kolmogorov index, $\tau_\mathrm{scat}$ 
from NE2001 would have caused an order of magnitude larger profile broadening than was observed by \citet{Gould1998} and 
\citet{Han2009}. It is, therefore, possible that modelling the region of intense scattering towards this pulsar 
is not necessary. For comparison, the scattering time from the
\citet{Taylor1993} Galactic electron density model is equal to 40\,ms at 129\,MHz (scaled from 1\,GHz with Kolmogorov index).
This model does not include the region of intense scattering towards PSR B2036+53 and the predicted scattering time is much closer
to the measured value.

\subsection{DM variations}
\label{subsec:DM_disc}

Because of the relative motion of the pulsar/ISM with respect to an Earth-based observer, the DM along any given line of 
sight (LOS) will gradually change with time. Systematic monitoring of DMs reveals that, in general, $\mathrm{DM}(t)$ series
consist of slowly varying (i.e. approximately linear) components superposed with stochastic or periodic variations 
\citep{Keith2013,Coles2015}. The interpretation of these variations can cast light on the turbulence in the ionised electron 
clouds in the interstellar plasma \citep{Armstrong1995}, though the interpretation may be more complex than usually assumed 
\citep{Lam2016}. 

Due to the high precision achievable for each single DM measurement, low-frequency DM monitoring can be particularly useful 
for investigating small, short-term DM variations. Within the census project, however, the DM along any particular LOS was 
measured only once. Nevertheless, some crude estimates on the DM variation rates can be obtained by comparing census DMs 
with previously published values. 

We calculated the rate of DM variation for the census pulsars by comparing their DMs to those obtained from the pulsar 
catalogue:
\begin{equation}
\label{eq:DMvar}
|\Delta\mathrm{DM}/\Delta t| = \left|\frac{\mathrm{DM_{cat}}-\mathrm{DM_{cen}}}{(\mathrm{DMepoch_{cat}}-\mathrm{DMepoch_{cen}})/365.25}\right|,
\end{equation}
where the epochs of DM measurements were expressed in MJD.  The errorbars were set by the error of DM determination: 
\begin{equation}
\epsilon_{|\Delta\mathrm{DM}/\Delta t|} = \frac{\sqrt{\epsilon^2_\mathrm{DM_{cat}}+\epsilon^2_{\mathrm{DM_{cen}}}}}{\left|\mathrm{DMepoch_{cat}}-\mathrm{DMepoch_{cen}}\right|/365.25},
\end{equation} 
and the pulsars without records of DM uncertainties or DM epochs in the pulsar catalogue\footnote{For the Crab pulsar 
we used the oldest entry from the Jodrell Bank monthly ephemerides, namely $\mathrm{DM}=56.834\pm0.005$\,\dmu~at 
$\mathrm{DMepoch}=45015$.} were excluded from the sample. This resulted in a sample of 146 pulsars, with DMs between 
3 and 180\,\dmu, and 5--30 years between the catalogue and census measurement epochs.

We have compared our results to a similar study in \citet{Hobbs2004b}, who measured the rate of DM variations for about 
100 pulsars with DMs between 3 and 600\,\dmu. Their analysis was based on 6--34 years of timing data, taken mostly at 
400$-$1600\,MHz. Figure~\ref{fig:DM_grad} (left) shows $|\Delta\mathrm{DM}/\Delta t|$ versus DM for census observations 
together with the approximate spread of $|\Delta\mathrm{DM}/\Delta t|$ values from \citet{Hobbs2004b}. Most of our pulsars 
have rates of DM variation similar to the ones from \citet{Hobbs2004b}. However, there are some deviations:
the nearby PSR J1503+2111 exhibits unusually large $|\Delta\mathrm{DM}/\Delta t|=0.8$\,\dmu\,$\mathrm{yr^{-1}}$
(see Sect.~\ref{ssubsec:DM_outl}) and there is an excess of larger DM variations for pulsars with $\mathrm{DM}>10$\,\dmu. 
It is interesting to note that pulsars with larger DM variation rates have relatively large reported DM errors: 
$\epsilon_\mathrm{DM}>0.1$\,\dmu, mostly for the pulsar catalogue DMs. Although DM uncertainties are explicitly included 
in the error bars in Fig.~\ref{fig:DM_grad}, it is possible that some of the DM measurements have unaccounted systematic 
errors, with a probability of such error underestimation being larger for pulsars with larger quoted $\epsilon_{\mathrm{DM}}$ 
(e.g. because of low S/N of the profile). 

In addition, we must note that census DM measurements were obtained under the simplifying assumption of the absence of 
profile evolution within the HBA band. It is currently unclear how different profile evolution models would affect the 
measured DM values. As a very approximate estimate, allowing the fiducial point to drift by 0.01 in spin phase 
(10\% of the typical width of an average pulse) across the HBA band would lead to a median DM change of 0.03\,\dmu, 
20 times larger than the median DM precision. This would alter the  observed $|\Delta\mathrm{DM}/\Delta t|$ typically 
by about 0.002\,\dmu\,$\mathrm{yr^{-1}}$,  but for some pulsars the change could be as large as 0.01\,\dmu\,$\mathrm{yr^{-1}}$.

Investigating the dependence of DM variation rate on DM can provide basic information for simple models of interstellar 
plasma fluctuations. \citet{Backer1993}, based on a sample of 13 pulsars with DMs between 2 and 200\,\dmu, 
have found that $|\Delta\mathrm{DM}/\Delta t| \sim \sqrt{\mathrm{DM}}$, which motivated the authors to propose 
a wedge model of electron column density gradients in the ISM.
\citet{Hobbs2004b} found a similar dependence: 
\begin{equation}
\label{eq:hobbs}
|\Delta\mathrm{DM}/\Delta t|\approx0.0002\times\mathrm{DM}^{0.57\pm0.09}_\mathrm{pc\,cm^{-3}} \quad \mathrm{pc\,cm^{-3}\,yr^{-1}}.   
\end{equation}
Both \citet{Backer1993} and \citet{Hobbs2004b} note a large (order of magnitude) scatter of data points around the fitted 
relation. At least partially, this scatter may be due to the dispersion in pulsar transverse velocities, since 
$|\Delta\mathrm{DM}/\Delta t|$ depends also on the transverse velocity of a pulsar.

For the census data\footnote{Excluding the outlier PSR J1503+2111.}, the unweighted fit of the following function:
\begin{equation}
\label{eq:DMvarfit}
\lg |\Delta\mathrm{DM}/\Delta t|_{\mathrm{pc\,cm^{-3}\,yr^{-1}}} = \lg A + B\lg \mathrm{DM_{pc\,cm^{-3}}},
\end{equation}
resulted in a relation which was close to \citet{Hobbs2004b}, with
$B=0.7\pm0.2$ and $A\approx0.0002$\, ($\lg A = -3.6 \pm 0.4$). Assigning each $|\Delta\mathrm{DM}/\Delta t|$ 
a weight inversely proportional to the measurement uncertainty 
yielded a fit with $B=-0.1\pm0.1$ and $A\approx0.03$ ($\lg A = -1.5 \pm 0.2$), however the usefulness of this approach is limited since 
the contribution from the transverse velocities and possible measurement bias due to profile evolution are not taken into account.
 Excluding the insignificant (value smaller than the error) $|\Delta\mathrm{DM}/\Delta t|$ or the ones with 
 $\epsilon_\mathrm{DM}>0.1$\,\dmu~ did not affect the fitting results substantially, except for when PSR J1503+2111 was included.

Overall, it is hard to make any definitive conclusion about the relation between $|\Delta\mathrm{DM}/\Delta t|$ and DM 
based on census data. Future improvements may result from extending the sample to larger DMs, including more pulsars
with $\mathrm{DM}<10$\,\dmu, making DM($t$) measurements with the same profile model, 
better quantification of DM gradients (e.g. separating piecewise linear segments and removing contributions of  
stochastic or periodic variations), and accounting for the contribution from transverse velocities.

Besides the pulsar catalogue, we compared the census DMs to recently published DM measurements taken within the last 
five years at frequencies less than or equal to HBA frequencies \citep{Pilia2015,Stovall2015,Zakharenko2013}.
\citet{Pilia2015} observed 100 pulsars with the LOFAR HBA antennas approximately two years before
the census observations presented here. At this time of data acquisition, just over half of the current HBA band and 
fewer core stations were available in tied-array mode, resulting in lower (by a factor of a few) S/N in the average profiles and
larger errors in DM determination.
The other two works report DMs measured at frequencies below 100\,MHz. \citet{Zakharenko2013} observed nearby 
($\mathrm{DM}<30$\,\dmu) pulsars in the frequency range of 16.5--33\,MHz using the UTR-2 telescope. Observations were taken during 
three sessions in 2010--2011. The authors do not specify the exact epoch of DM measurements, so the inferred uncertainty of 
DMepoch is included in errors on $|\Delta\mathrm{DM}/\Delta t|$ in Fig.~\ref{fig:DM_grad}. \citet{Stovall2015} report the results of 
broadband (35--80\,MHz) pulsar observations with the LWA telescope. Their DM measurements were taken close in time to the census 
ones: the maximum offset between the DM epochs was about $\pm1$\,yr. 
Some of the census pulsars were observed in more than one of these three works, thus 
they have multiple  $|\Delta\mathrm{DM}/\Delta t|$ plotted in Fig.~\ref{fig:DM_grad} (right). 
The calculated DM variation rates for such pulsars could differ from each other by one-two orders of magnitude.

In general, DMs from at least two aforementioned works\footnote{Except for \citet{Pilia2015}, but 
the DMs there have large uncertainties.} also suggest the excess of larger DM variation rates as comparing
to \citet{Hobbs2004b}. This trend is the most obvious for the shortest timespan $|\Delta\mathrm{DM}/\Delta t|$ 
based on DMs from \citet{Stovall2015}. 
The fact that DM variation rates are larger on smaller timescales can be explained by the larger relative 
influence of shorter-term stochastic variations. However, the bias in $|\Delta\mathrm{DM}/\Delta t|$ 
introduced by DM offsets due to the differences in modelling the frequency-dependent profile evolution and scattering
will also be relatively larger because of the shorter time span in the denominator of Eq.~\ref{eq:DMvar}. 

Making DM measurements in the presence of profile evolution and scattering is a complex task \citep{Hassall2012,Pennucci2014,Liu2014}.
Nevertheless, this provides valuable information about both the ISM (electron content and turbulence parameters) and  
pulsar magnetospheres (such as modelling the location of a fiducial phase point and the evolution of components around it; 
e.g. \citealt{Hassall2012}). 
Scattering has a steep dependence on frequency, and profile evolution is usually more rapid at lower frequencies. Thus, 
combining census observations with lower frequencies  (or even with higher-frequency data for pulsars with large scattering 
times) can serve as a good data sample for broadband profile modelling, bias-free DM measurements, and scattering time estimates. 
Finally, it would be interesting to investigate frequency dependence of DM values, arising from different sampling of the ISM,
due to frequency-dependent scattering \citep{Cordes2016}. We will defer such analysis to a subsequent work.

\subsection{PSR J1503$+$2111}
\label{ssubsec:DM_outl}
PSR J1503$+$2111 exhibited an unusually large DM variation rate of about $-0.8$\,\dmu$\mathrm{yr^{-1}}$. 
It has $\mathrm{DM_{cat}}= 11.75 \pm 0.06$\,\dmu~\citep{ChampionA2005} and $\mathrm{DM_{cen}} =  3.260 \pm 0.004$\,\dmu, 
with measurements taken 11\,yr apart. In our observations, folded with $\mathrm{DM_{cat}}$, the pulsar is 
clearly visible across the entire HBA band and its profile exhibits a characteristic quadratic sweep with a net delay 
between the edges of the band equal to 0.6 of spin phase. 
Thus, we are confident that at the epoch of census observation the DM of PSR J1503$+$2111 was substantially 
different from $\mathrm{DM_{cat}}$.

PSR J1503$+$2111 was discovered only a decade ago and is relatively poorly studied. The DM value in the pulsar catalogue
comes from the discovery paper of \citet{ChampionA2005}. The authors obtained initial ephemerides (including DM) based 
on 430-MHz timing data taken with the Arecibo telescope. They subsequently refined the DM using observations at four 
frequencies between 320 and 430\,MHz, while keeping other ephemeris parameters fixed. If the real DM value was close 
to 3\,\dmu~at the time of observations, then the time delay from the highest to the lowest frequencies in their setup 
would be around $-150$\,ms (0.04 of spin phase), much larger than the reported residual time-of-arrival rms of 0.9\,ms. However, such DM error 
could pass unnoticed while folding observations within one band ($-7$\,ms delay, about 10\% of pulse width reported). The 
pulsar was subsequently observed by \citet{Han2009}, in a frequency band spanning from 726 to 822\,MHz. At these 
frequencies the profile smearing due to incorrect DM would be only 0.004 of spin phase, much smaller than the profile 
width (0.03 of spin phase) presented in their work. \citet{Zakharenko2013} failed to detect PSR J1503$+$2111 at 16--33 MHz,
searching for a pulsar signal with a set of trial DM values within 10\% of $\mathrm{DM_{cat}}$. 
If the DM at the epoch of \citet{Zakharenko2013} observations was close to 3.26\,\dmu, the delay due to the DM offset at their 
frequencies would be equal to 30 phase wraps, completely smearing the profile.

To summarise, it is possible that the DM of this pulsar was incorrectly estimated in \citet{ChampionA2005} and went 
unnoticed since then. However, only future observations of this pulsar can show whether the DM along this LOS 
exhibits an anomalously large variation rate.

\begin{figure*}
 \centering
 \includegraphics[scale=0.9]{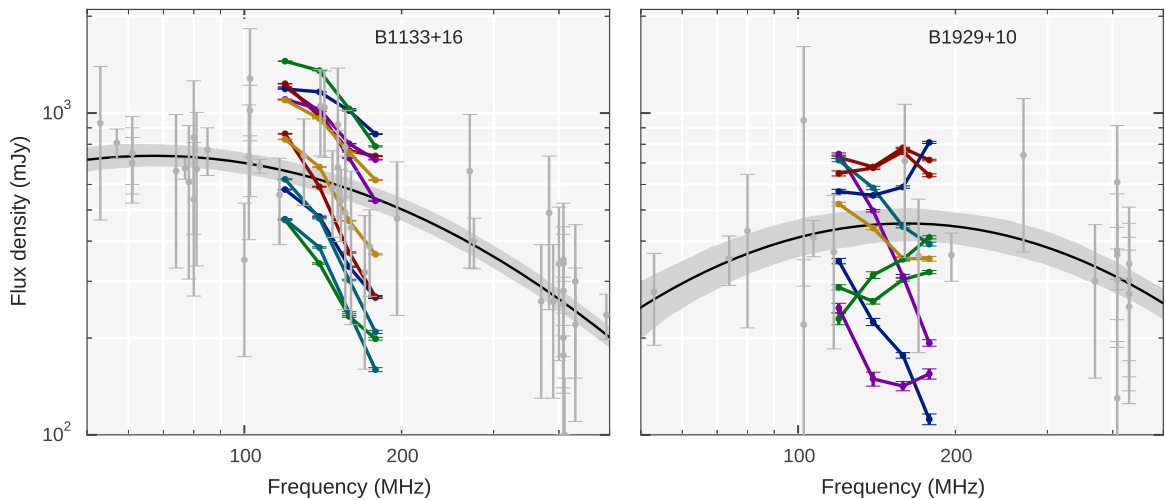}
  \caption{Flux density measurements from individual timing sessions ($S_\mathrm{meas}$, coloured connected dots) for two 
  out of the ten pulsars used for flux density uncertainty estimates. Separate grey dots are flux density values from the 
  literature, with the errors estimated according to the procedure described in Sect.~\ref{subsec:fit}. The black line 
  indicates $S_\mathrm{lit}$, the parabola fit for literature flux density points within 10$-$1000\,MHz. The shaded 
  region marks 68\% uncertainty on $S_\mathrm{lit}$.} 
 \label{fig:timing_spectra}
\end{figure*}

\section{Flux density calibration}
\label{sec:calib}

\subsection{Overview and error estimate}

The flux density scale for a given [sub-integration, sub-band] cell was calibrated with the radiometer equation \citep{Dicke1946}:

\begin{equation}
 S_\mathrm{mean} = \dfrac{T_\mathrm{sys}}{G\sqrt{n_p \Delta f t_\mathrm{obs} n^{-1}_\mathrm{bin} }} \times \langle\mathrm{S/N}\rangle,
\end{equation}
where $T_\mathrm{sys}=T_\mathrm{A}+T_\mathrm{sky}$ is the total system temperature (antenna plus sky background), $G$ is the telescope gain, 
$n_p$ is the number of polarisations summed (2 for census data), $\Delta f$ is sub-band width, $t_\mathrm{obs}$ is 
the length of sub-integration, $n_\mathrm{bin}$ is the number of spin phase bins, and $\langle\mathrm{S/N}\rangle$ is 
the mean signal-to-noise ratio of the pulse profile in a given [sub-integration, sub-band] cell. For LOFAR, both 
antenna temperature and gain have a strong dependence on frequency, with the latter also varying with the elevation and 
azimuth of the source observed. In this work we used the latest version of the LOFAR pulsar flux calibration software, 
described in \citet{Kondratiev2015}. This software uses the Hamaker beam model \citep{Hamaker2006} 
and \texttt{mscorpol}\footnote{\url{https://github.com/2baOrNot2ba/mscorpol}} package by Tobia Carozzi to calculate 
Jones matrices of the antenna response for a given HBA station, frequency and sky direction. The antenna gain is 
further scaled with the actual number of stations used in a given observation. The HBA antenna temperature, $T_\mathrm{A}$, 
is approximated as a frequency-dependent polynomial derived from the measurements of \citet{Wijnholds2011}. The background sky 
temperature, $T_\mathrm{sky}$, is calculated using 408-MHz maps of \citet{Haslam1982}, scaled to HBA frequencies as $\nu^{-2.55}$ \citep{Lawson1987}. 
The mean S/N of the pulse profile is calculated by averaging the normalised signal over the pulse period, 
with the normalisation performed using the mean and standard deviation of the data in the manually selected off-pulse 
window (the region in pulse phase without visible emission).
Zero-weighted sub-bands and/or sub-integrations are ignored and  do not contribute to the overall flux density calculation.
For a more detailed review of the calibration technique we refer the reader to \citet{Kondratiev2015}.

The nominal error on the flux density estimation in \citet{Kondratiev2015}, $\epsilon_{S\mathrm{nom}}$, is set by the standard 
deviation of the data in the off-pulse window. The real uncertainty of 
a single flux density measurement is much larger, being augmented by many factors, including (but not limited to) 
the intrinsic variability of the source, scintillation in the ISM, imperfect knowledge of the system parameters, and some 
other possible effects that are unaccounted for, e.g. uncalibrated phase delays introduced by the ionosphere
or the presence of strong sources in the sidelobes. 

To provide a more realistic uncertainty, we took advantage of regular LOFAR pulsar timing observations.
Within this project, a number of both millisecond 
and normal pulsars were observed with HBA core stations on a monthly basis. From this sample of pulsars we selected ten 
bright non-recycled pulsars with relatively well-known spectra available from the literature\footnote{Namely, PSRs~B0809+74, 
B0823+26, B1133+16, B1237+25, B1508+55, B1919+21, B1929+10, B2016+28, B2020+28, and B2217+47. PSRs B0823+26 and B1237+25 undergo 
mode switches, but their flux densities did not show larger variance in comparison to the other eight sources.}.
These pulsars were observed for 5--20 min at different elevations and azimuths in December 2013--November 2014, 
approximately in the same time span as the census observations. The distribution of LOS directions 
approximately coincides with that for the census pulsars. In particular, both timing and census sources were observed 
only at relatively high elevations (EL), $\mathrm{EL>40}^{\circ}$.

\begin{table*}
\caption{Percentiles of $S_\mathrm{lit}/S_\mathrm{meas}$ distribution for ten bright pulsars with HBA timing observations.}              % title of Table
\label{table:flux_err}      % is used to refer this table in the text
\centering                                      % used for centering table
{\newcommand{\mc}[3]{\multicolumn{#1}{#2}{#3}}
\begin{tabular}{l|cccc|cc|c}          % centered columns (4 columns)
\hline\hline                        % inserts double horizontal lines
& \mc{4}{c|}{4 sub-bands} & \mc{2}{c|}{2 sub-bands} & band-integrated\\
& 120\,MHz & 139\,MHz & 159\,MHz & 178\,MHz & 130\,MHz & 168\,MHz & 149\,MHz\\ 
\hline
 median & 0.8 & 1.0 & 1.0 & 1.2 & 0.9 & 1.1 & 1.0\\
16th percentile & 0.5 & 0.6 & 0.6 & 0.6 & 0.6 & 0.6 & 0.6\\
84th percentile & 1.5 & 1.6 & 1.7 & 2.2 & 1.5 & 1.9 & 1.6\\
\hline                                         
\end{tabular}}
\tablefoot{$S_\mathrm{lit}$ is obtained from a fit through literature flux density values and $S_\mathrm{meas}$ is the HBA flux density.
The 16th and 84th percentiles were used for estimating the uncertainty of a single flux density measurement (see text for 
details).}

\end{table*}

We processed selected timing observations and measured the pulsar flux densities in the same manner as for census sources. 
Examination of the flux density values obtained revealed two to four times larger fluctuations than would have been expected by 
scintillation alone, with a flux density rms on the order of 50\% for the band- and session-integrated flux densities. 
This is at least partly due to the imperfect model of telescope gain, since we record a dependence of the measured flux 
density on the source elevation. The number of flux density measurements, however, is too small to construct a robust 
additional gain correction. Thus, we leave it for future work. 

Figure~\ref{fig:timing_spectra} shows measured spectra with respect to literature points for two pulsars from the timing 
sample. In order to estimate the error of a single flux density measurement (and check for any systematic offsets between 
LOFAR and literature flux densities), we constructed a ``reference'' flux density curve $S_\mathrm{lit}$ by fitting a 
parabola to the literature flux density values in logarithmic space (within the range 10$-$1000\,MHz). The fit was performed using a Markov 
chain Monte-Carlo (MCMC) algorithm\footnote{\url{https://github.com/pymc-devs/pymc}}  and the region between 16th
and 84th percentiles of $S_\mathrm{lit}$ values (obtained from posterior distributions of the fitted parabola parameters) 
is shown with a grey shade. This formal uncertainty in $S_\mathrm{lit}$ should be treated as an approximation of 
the actual uncertainty, since the fitted 
curve can shift by an amount larger than the grey-shaded area if new flux density measurements are added\footnote{This suggests 
a frequent underestimation of the flux density errors quoted in the literature. See also Sects.~\ref{subsec:fit} and 
\ref{subsec:fit_res}.}. We then analysed the distribution of $S_\mathrm{lit}/S_\mathrm{meas}$ values\footnote{Instead of 
using a single value of $S_\mathrm{lit}$ for a given frequency bin, we used a distribution of values calculated from the posterior 
distribution of the parabola fit parameters. In such a way we interpreted the uncertainty in $S_\mathrm{lit}$.},
where the flux density obtained from timing observations, $S_\mathrm{meas}$, is in the denominator. For band-integrated flux 
densities, the median value of $S_\mathrm{lit}/S_\mathrm{meas}$ was 1.0 and $0.6<S_\mathrm{lit}/S_\mathrm{meas}<1.6$
with 68\% probability.

Thus, for the sample of ten timing pulsars the combined influence of all uncertainties, other than 
$\epsilon_{S\mathrm{nom}}$\footnote{The average profiles of timing pulsars had large $\langle\mathrm{S/N}\rangle$, and thus
$\epsilon_{S\mathrm{nom}}\ll S_\mathrm{meas}$.} caused $S_\mathrm{meas}$ to spread around $S_\mathrm{lit}$ with a
magnitude of the spread equal to about $0.5S_\mathrm{meas}$. Assuming the uncertainties to be similar for all census pulsars, 
we adopted a total error on the single flux density measurement $\epsilon_S = \sqrt{(0.5S)^2+\epsilon_{S\mathrm{nom}}^2}$.
This assumption is reasonable, since we expect two major contributors to the flux density uncertainty, namely, our 
imperfect knowledge of the gain and interstellar scintillation, to influence both timing and census observations to a similar 
extent. This is justified because the distribution of source elevations and expected modulation indices due to scintillation (see 
Appendix~\ref{app:scint} for the latter) were similar for both timing and census sources.

We also examined the error distribution for flux densities measured in halves and quarters of the HBA band. We discovered 
that, in general, the slopes of the timing spectra are inconsistent with the literature, with lower-frequency 
$S_\mathrm{meas}$ being consistently overestimated and higher-frequency $S_\mathrm{meas}$
underestimated (see Table~\ref{table:flux_err}). 

Two alternative models of antenna gain were also tested. The first model was based on full electromagnetic simulations 
of an ideal 24-tile HBA sub-station including edge effects and grating lobes \citep{Arts2013}. The second model used a simple 
$\sim\sin^{1.39}(\mathrm{EL})$ scaling of the theoretical frequency-dependent value of the antenna effective area 
\citep{Noutsos2015}. For both models the telescope gain at zenith was similar to the Hamaker-Carozzi model, but 
they predict two to three times larger gains for the lowest census elevations of $40\degree$. 
This meant that pulsar flux densities could be two to three times smaller than those calculated using the
Hamaker-Carozzi model, which made them less consistent with $S_\mathrm{lit}$ values.

For several bright pulsars we also estimated the preliminary
140-MHz flux densities using images from the Multifrequency Snapshot Sky Survey 
 \citep[G. Heald, private communication; see also][]{Heald2015}. The same technique was applied to 
compare the imaging flux density values, $S_\mathrm{MSSS}$, and $S_\mathrm{lit}$. We found that 16th--84th percentiles of $S_\mathrm{MSSS}$ 
agree with $S_\mathrm{lit}$ within 40\%, similarly to the flux densities from the timing campaign.

\subsection{Application to census pulsars}

The flux density calibration was performed on folded, 195-kHz-wide, 1-min sub-integrations. In addition to the synchrotron 
background from \citet{Haslam1982}, we have checked for any bright sources in the primary beam. In all cases the Sun, 
the Moon, and the planets were far away ($>4\fdg7$) from a pulsar position.
A review of the 3C and 3CR catalogues \citep{Edge1959,Bennett1962} did not reveal any nearby extended sources, except in the case 
of the Crab pulsar (3C144) and PSR J0205$+$6449 (3C58). For the Crab pulsar, the contribution from the nebula was estimated 
with the relation $S_\mathrm{Jy} \approx 955 \nu_\mathrm{GHz}^{-0.27}$ \citep{Bietenholz1997,Cordes2004}. Similar flux density 
values were quoted in the 3C catalogue. At 75\,MHz, the solid angle occupied by the nebula 
\citep[radius of $4\arcmin$,][]{Bietenholz1997} is larger than the full-width at the half-maximum
of the LOFAR HBA beam \citep[$2\arcmin$, for the 
2-km baseline, according to the table B.2 in][]{vanHaarlem2013}, thus only approximately one quarter of the nebula 
is contributing to the system temperature. For PSR J0205$+$6449, 
the supernova remnant does not significantly add to the system temperature (5–10\%, depending on the observing frequency), 
so its contribution was neglected. 

For all pulsars, except for the Crab, we were able to find an off-pulse region in each sub-band/sub-integration, 
although in some cases the off-pulse region was small (about 10\% of pulse phase). 
Sometimes the band- and time-integrated average profile exhibited faint pulsed emission in the selected off-pulse region, 
resulting in somewhat underestimated flux densities. However, because of the large number of channels and sub-integrations in 
our observing setup (400 and $\geqslant20$, respectively), we expect this underestimation to be well within the quoted errors. 
For the Crab pulsar, the standard deviation of the noise was calculated after subtracting a polynomial fit to the profile.

The band-integrated flux density values, together with the adopted uncertainties $\epsilon_S$ are quoted in Table~\ref{table:main}. 
For non-detected pulsars we give $3\epsilon_{S\mathrm{nom}}$ as an upper limit. We note that such upper limits must be 
taken with caution, since non-detections can occur for reasons unrelated to the intrinsic pulsar flux density (for 
example, scattering or unknown error in the pulsar position).

\section{Spectra}
\label{sec:spec}

\subsection{Introduction}

The mean (averaged over period) flux density $S_\nu$ of a pulsar observed at a frequency $\nu$ is one of the main 
observables of pulsar emission. Flux density measurements provide constraints on the pulsar emission mechanism 
\citep{Malofeev1980,Ochelkov1984}. They are crucial for deriving the pulsar luminosity function \citep[which is further 
used to study the birth rate and initial spin period distribution of the Galactic population of radio pulsars, e.g. 
][]{Lorimer1993,FaucherGiguere2006}, and for planning the optimal frequency coverage of future pulsar surveys. 

At present, even the most well-studied pulsar radio spectra consist of flux density measurements obtained from 
observations performed under disparate conditions and with different observing setups. The situation is further 
complicated by interstellar scintillation and intrinsic pulsar variability \citep{Sieber1973,Malofeev1980}. As a result, 
flux densities measured at the same frequency by different authors may disagree by up to an order of magnitude.

Despite these difficulties, it has been established that in a wide frequency range of approximately 0.1--10\,GHz,
pulsar radio spectra are usually well-described by a simple power-law relation:
\begin{equation}
S_\nu = S_0\left(\nu/\nu_0\right)^\alpha, 
\end{equation}
where $S_0$ is the flux density at the reference frequency $\nu_0$, and $\alpha$ is the spectral index.
At the edges of this frequency range some spectra start deviating from a single power-law, exhibiting a so-called 
low-frequency turnover at $\lesssim100$\,MHz \citep{Malofeev1993}, or high-frequency flattening around 30\,GHz 
\citep{Kramer1996}. Some pulsars show evidence of a spectral break even in the centimetre wavelength range 
\citep{Maron2000} and there is a subclass of pulsars with distinct spectral turnover around 1\,GHz \citep{Kijak2011}. 

Based on the extensive number of published flux density measurements (see Table~\ref{table:main} for the full list of 
references), we constructed radio spectra for 182 census pulsars (Figs.~\ref{fig:spectra_nondet}--\ref{fig:prof_sp_16}). 
Among those spectra, 24 consisted of the literature points only, since the corresponding pulsars were not detected in census observations. 
Twelve remaining pulsars were not detected in the census observations and had no previously published flux density values.

\subsection{Fitting method}
\label{subsec:fit}

It is customary to fit pulsar spectra with a single or broken power-law (PL), although this approach 
may be only an approximation to the true pulsar spectrum \citep{Maron2000,Loehmer2008}. Still, this parametrisation is 
useful for cases with a limited number of measurements and allows direct comparison to previous work. 

The fit was performed in $\lg S$--$\lg \nu$ space. We used a Bayesian approach, making a statistical model of the data and 
using an MCMC fitting algorithm to find the posterior distributions of the fitted parameters. In general, we modelled each 
$\lg S$ as a normally distributed random variable with the mean $\lg S_{\mathrm{PL}}$ defined by the PL dependence and a standard 
deviation $\sigma_{\lg S}$ reflecting any kind of flux density measurement uncertainty:
\begin{equation}
 \lg S \sim \mathrm{Normal}(\lg S_\mathrm{PL}, \sigma_{\lg S} ). 
\end{equation}
A normal distribution was chosen for the sake of simplicity and for the lack of a better knowledge of the real uncertainty 
distribution.

Depending on the number of flux density measurements and their frequency coverage, $\lg S_\mathrm{PL}$ was approximated 
either as a single PL (hereafter ``1PL''):
 \begin{equation}
  \lg S_\mathrm{1PL} = \alpha \lg (\nu/\nu_0) + \lg S_0,
 \end{equation}
 a broken PL with one break (2PL):
 \begin{equation}
\lg S_\mathrm{2PL} = \begin{cases}
          \alpha_\mathrm{lo}\lg(\nu/\nu_0) + \lg S_0,  & \nu<\nu_\mathrm{br}\\
        \alpha_\mathrm{hi}\lg(\nu/\nu_\mathrm{br}) + \alpha_\mathrm{lo}\lg(\nu_\mathrm{br}/\nu_0) + \lg S_0, & \nu>\nu_\mathrm{br},
        \end{cases}
 \end{equation}
or a broken PL with two breaks (3PL):
\begin{equation}
\lg S_\mathrm{3PL} = \begin{cases}
          \alpha_\mathrm{lo}\lg(\nu/\nu_0) + \lg S_0,  & \nu<\nu^\mathrm{lo}_\mathrm{br}\\
        \alpha_\mathrm{mid}\lg(\nu/\nu^\mathrm{lo}_\mathrm{br}) + \alpha_\mathrm{lo}\lg(\nu^\mathrm{lo}_\mathrm{br}/\nu_0) + \lg S_0, & \nu^\mathrm{lo}_\mathrm{br}<\nu<\nu^\mathrm{hi}_\mathrm{br}\\
        \alpha_\mathrm{hi}\lg(\nu/\nu^\mathrm{hi}_\mathrm{br}) + \alpha_\mathrm{mid}\lg(\nu^\mathrm{hi}_\mathrm{br}/\nu_0) + \lg S_0, & \nu>\nu^\mathrm{hi}_\mathrm{br}.
        \end{cases}
\end{equation}
For all PL models, the reference frequency $\nu_0$ was taken to be the geometric average of the minimum and maximum frequencies 
in the spectrum, rounded to hundreds of MHz. 

If the number of spectral data points was small (two to four, with measurements within 10\% in frequency treated as a single 
group), we fixed $\sigma_{\lg S}$ at the known level, defined by the reported errors:
$\sigma_{\lg S}\equiv\sigma^\mathrm{kn}_{\lg S}=0.5[\lg(S+\epsilon^\mathrm{up}_S)-\lg(S-\epsilon^\mathrm{lo}_S)]$. For 
census measurements the errors were taken from Table~\ref{table:flux_err} and added in quadrature to 
$\epsilon_{S\mathrm{nom}}$\footnote{If the number of literature flux density measurements was small or if the pulsar had 
a low S/N in the census observation, then we included only one, band-integrated census flux density measurement to the 
spectrum. For brighter pulsars with better-known spectra, we used census flux densities measured in halves or quarters of 
the band.}. The errors on the literature flux densities were assigned following the essence of the procedure described in 
\citet{Sieber1973}\footnote{Namely, flux density measurements based on many ($\gtrsim5$) sessions, spread over more than a 
year, with errors given as the standard deviation, were considered reliable and we quoted the original error 
reported by the authors. For the flux density measurements based on a smaller number of sessions, or spread over a smaller 
time span, we adopted an error of 30\%, unless the quoted error was larger. 
%(note that the median flux error in the years-long multi-frequency
%study of \citet{Lorimer1995} is 15\%). 
In case of flux densities based on one session or with an uncertain observing setup, we adopted an error of 50\% 
unless the quoted errors were larger.}.

When the number of data points was larger (more than six or, sometimes, five groups), we introduced an additional fit parameter, 
the unknown error $\sigma^\mathrm{unkn}_{\lg S}$. This error represents any additional flux density uncertainty, not
reflected by $\sigma^\mathrm{kn}_{\lg S}$, for example intrinsic variability, or any kind of unaccounted propagation or 
instrumental error. The total flux density uncertainty of any measurement was then taken as the known and unknown errors 
added in quadrature. We fit a single $\sigma^\mathrm{unkn}_{\lg S}$ per source, although, strictly speaking, unknown 
errors may be different for each separate measurement.

In the presence of a fitted $\sigma^\mathrm{unkn}_{\lg S}$ all three PL models will provide a good fit to the data, since
any systematic deviation between the model and the data points will be absorbed by $\sigma^\mathrm{unkn}_{\lg S}$. 
Thus, in order to discriminate between models, we examined the posterior distribution of $\sigma^\mathrm{unkn}_{\lg S}$.
We took 1PL as a null hypothesis and rejected it in favour of 2PL or 3PL if the latter gave statistically smaller
$\sigma^\mathrm{unkn}_{\lg S}$: the difference between the mean values of the posterior distributions of $\sigma^\mathrm{unkn}_{\lg S}$ 
was larger than the standard deviations of those distributions added in quadrature.

For the sparsely-sampled spectra, where no $\sigma^\mathrm{unkn}_{\lg S}$ was fitted, we adopted 1PL as the single model. 
In a few cases, when the data showed a hint of a spectral break, we fitted 2PL with break frequency fixed at the frequency 
of the largest flux density measurement. For such pulsars we give both 1PL and 2PL values of the fitted parameters.
 
Some flux density measurements were excluded from the fit.
Since most of the flux density measurements were performed for the pulsed emission,
we did not take into account the continuum flux density values for the Crab pulsar at 10--80\,MHz 
from \citet{Bridle1970}. For PSR B1133+16 we excluded the measurements from \citet{Stovall2015},
since they were an order-of-magnitude larger than numerous previous measurements in the same frequency range.
Judging from the visual examination of well-measured spectra, sometimes the upper limit on flux  densities could be 
an order-of-magnitude smaller than actual measurements in the same frequency range. Thus, we considered
both census and literature flux density upper limits to be approximate at best and did not attempt to fit for 
the lower limits on spectral index.

\subsection{Results}
\label{subsec:fit_res}

Out of the 194 census pulsars, 165 had at least two flux density measurements (census or literature), making them suitable 
for a spectral fit. The majority of pulsars, 124 sources, were well-described with the 1PL model (Table~\ref{table:1pl}), 
although the choice of the model was greatly influenced by the small number of data points available. Four 1PL pulsars 
(namely PSRs J1238$+$21, J1741$+$2758, B1910$+$20, and J2139$+$2242) show signs of spectral break, but the number of flux 
density measurements was too small to fit for a break frequency. For these pulsars we provide the values of the 2PL parameters 
with a break frequency fixed at the frequency of maximum flux density (Table~\ref{table:2pl}). The remaining 41 sources 
showed preference for a broken power-law with a single break (36 pulsars, Table~\ref{table:2pl}), or two breaks (five pulsars, 
Table~\ref{table:3pl}). Pulsars best described with a 2PL or 3PL model usually have a larger number of flux density measurements.

\subsection{Discussion}

\subsubsection{Flux variability}

It is interesting to examine the distribution of $\su$ among the pulsars for which this parameter was fitted.
The 46 of these sources were best described by the 1PL model. About $60\%$ of them showed moderate 
flux density scatter with $\su\lesssim0.15$, indicating roughly $\pm$30\% variation in flux density
measurements at a single frequency. Six 1PL pulsars had $\su>0.3$, with flux density varying by 
a factor of two to six. Among those, four pulsars (PSRs B0053$+$47, B0450$+$55, B1753$+$52, and J2307$+$2225) had 
separate outliers in their spectra, hinting at a deviation from the 1PL model, or simply indicating a large 
unaccounted error in a single flux density measurement. PSRs B0643$+$80 and J1740$+$1000 had a large spread of 
flux density measurements across the entire spectrum. Interestingly, PSR B0643$+$80 was shown to have burst-like emission in 
one of its profile components \citep{Malofeev1998}. The other pulsar, PSR J1740$+$1000, was proposed as a candidate 
for a gigahertz-peaked spectrum pulsar by \citet{Kijak2011}, although the authors note contradicting flux density 
measurements at 1.4\,GHz. These contradicting points were excluded from the spectral analysis conducted by 
\citet{Dembska2014} and \citet{Rajwade2015}, who approximated the spectrum of PSR J17400$+$1000 with a parabola or fitted the 
PL with a free-free absorption model directly. Our measurements show that the flux density of PSR J1740$+$1000 does not 
decrease at HBA frequencies, exceeding the extrapolation of the fits by \citet{Dembska2014} and \citet{Rajwade2015} by a 
factor of 10--20. We suggest this pulsar does not have a gigahertz-peaked spectrum, but a regular PL with potentially
large flux density variability. 

For both 2PL and 3PL pulsars, the median value of $\su$ was 0.1. Two 2PL pulsars, PSR B0943+10 and 
PSR B1112+50, had $\su>0.3$. The former is a well-known mode-switching pulsar \citep{Suleymanova1984} 
and the latter has a 20-cm profile that is unstable on the time scale of our observing session \citep{Wright1986}.
This can explain an order-of-magnitude difference between the flux densities obtained from the census measurements 
and the ones reported by \citet{Karuppusamy2011}, which were performed in exactly the same frequency range. 

\begin{figure}
   \centering
  \includegraphics[scale=0.8]{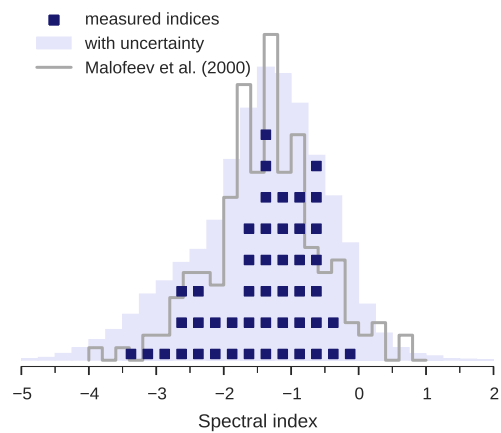}
 \caption{Distribution of spectral indices for 48 pulsars without previously published spectral fits. 
 Each of the 48 dark squares marks the mean of the posterior distribution of the spectral index $\alpha$. The lighter histogram
 was constructed using the whole posterior distribution of $\alpha$ for all pulsars, and thus reflects the uncertainty in the
 spectral index determination.  The grey line marks the distribution of spectral indices for 175 non-recycled pulsars in 
 the similar  frequency range (100--400\,MHz) from \citet{Malofeev2000}.}
 \label{fig:alpha_distrib}    
\end{figure}

\begin{figure}
   \centering
 \includegraphics[scale=0.75]{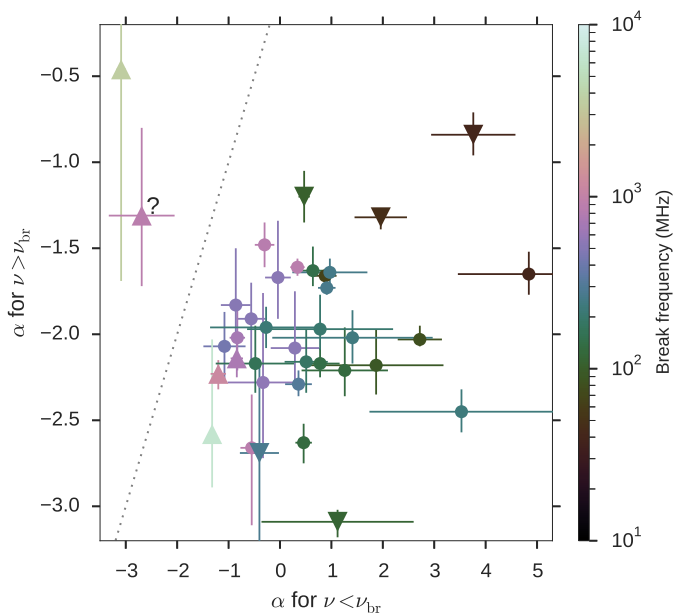}
 \caption{Spectral indices below and above the spectral break 
 for 32 pulsars with relatively well-measured spectra (see text for details). 
 Pulsars with a single spectral break are marked  with circles. 
 For pulsars with two spectral breaks, the lower-frequency one is marked with the downward triangles
 and the higher-frequency one with the upward triangles. The colour indicates the frequency of the break and the
 dotted line corresponds to no change in the spectral index. The question mark indicates PSR B2303$+$30, for which the value 
 of the high-frequency spectral  index was greatly influenced by a single flux density measurement. 
 Note that for $\nu_\mathrm{br}\gtrsim 500$\,MHz the change in spectral index is relatively moderate, 
 whereas for $\nu_\mathrm{br}\lesssim 300$\,MHz the spectral index below the break takes (sometimes large)
 positive values. This corresponds to the previously known ``high-frequency cut-off'' and ``low-frequency turnover'' 
 in pulsar spectra. }
 \label{fig:ind_break}
\end{figure}

\subsubsection{New spectral indices}
In total, 48 pulsars from the census sample did not have previously published spectral fits. The spectra of these pulsars 
typically consisted of a small ($\lesssim5$) number of points, usually limited to the frequency range 100 -- 400/800\,MHz.
Among these 48 pulsars, only PSR J2139+2242 showed signs of a spectral break, however the paucity of available
measurements should be kept in mind.

The distribution of new spectral indices (Fig.~\ref{fig:alpha_distrib}) is comparable in shape to the spectral 
index distribution constructed from a larger sample of 175 non-recycled pulsars in a similar frequency range (between 
102.5 and 408\,MHz) by \citet{Malofeev2000}. Both in \citet{Malofeev2000} and in our work the mean value of the spectral 
index, $\bar{\alpha}=-1.4$, is flatter than the mean spectral index measured at frequencies above 400\,MHz
\citep[e.g. $\bar{\alpha}=-1.8$ in][]{Maron2000}. This can be interpreted as a sign of low-frequency flattening or 
turnover \citep{Malofeev2000}. However, as has been noted by \citet{Bates2013}, the shapes of observed spectral index 
distributions may be greatly affected by the selection effects connected to the frequency-dependent sensitivity
of pulsar surveys. Thus, a comparison of spectral index distributions obtained in the different frequency ranges should 
be done with caution unless the distributions are based on the same sample of sources.

\begin{figure}
   \centering
  \includegraphics[scale=1.0]{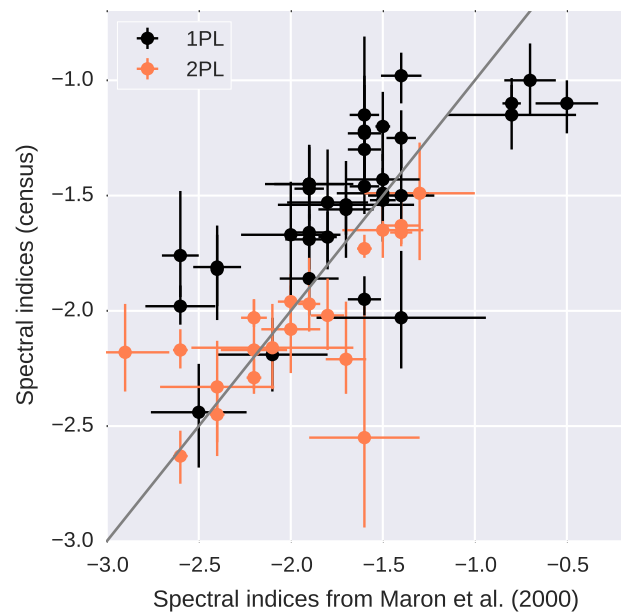}
 \caption{Comparison between spectral indices from  \citet{Maron2000} and this work. Black circles indicate 35 pulsars 
 with spectra best described with a single PL by \citeauthor{Maron2000} (frequency range of 400\,MHz--1.6/5\,GHz)
 and a single PL in our work (frequency range of typically 100\,MHz--5\,GHz). The spectral indices in our work tend to be more 
 flat, indicating a possible low-frequency turnover somewhere close to 100\,MHz. For comparison, spectral indices for 21 
 pulsars with clearly identified low-frequency turnovers below the frequency range of \citet{Maron2000} are shown with
orange circles. }
 \label{fig:flat}    
\end{figure}

\subsubsection{Spectral breaks}

For both 2PL and 3PL pulsars the fitted break frequencies often had asymmetric posterior distributions, with the shape 
of a distribution substantially influenced by the gaps in the frequency coverage of $S_\nu$ measurements.
Sometimes the shape of a spectrum at lower frequencies was clearly affected by scattering (e.g. for the Crab pulsar,
PSRs J1937$+$2950, B1946$+$35, and some others). Large scattering ($\tau_\mathrm{scat}\approx P$) smears the pulse profile, 
effectively reducing the amount of observed pulsed emission. This results in flatter negative, or even large positive 
values of $\alpha$.

\begin{figure*}
   \centering 
 \includegraphics[scale=0.9]{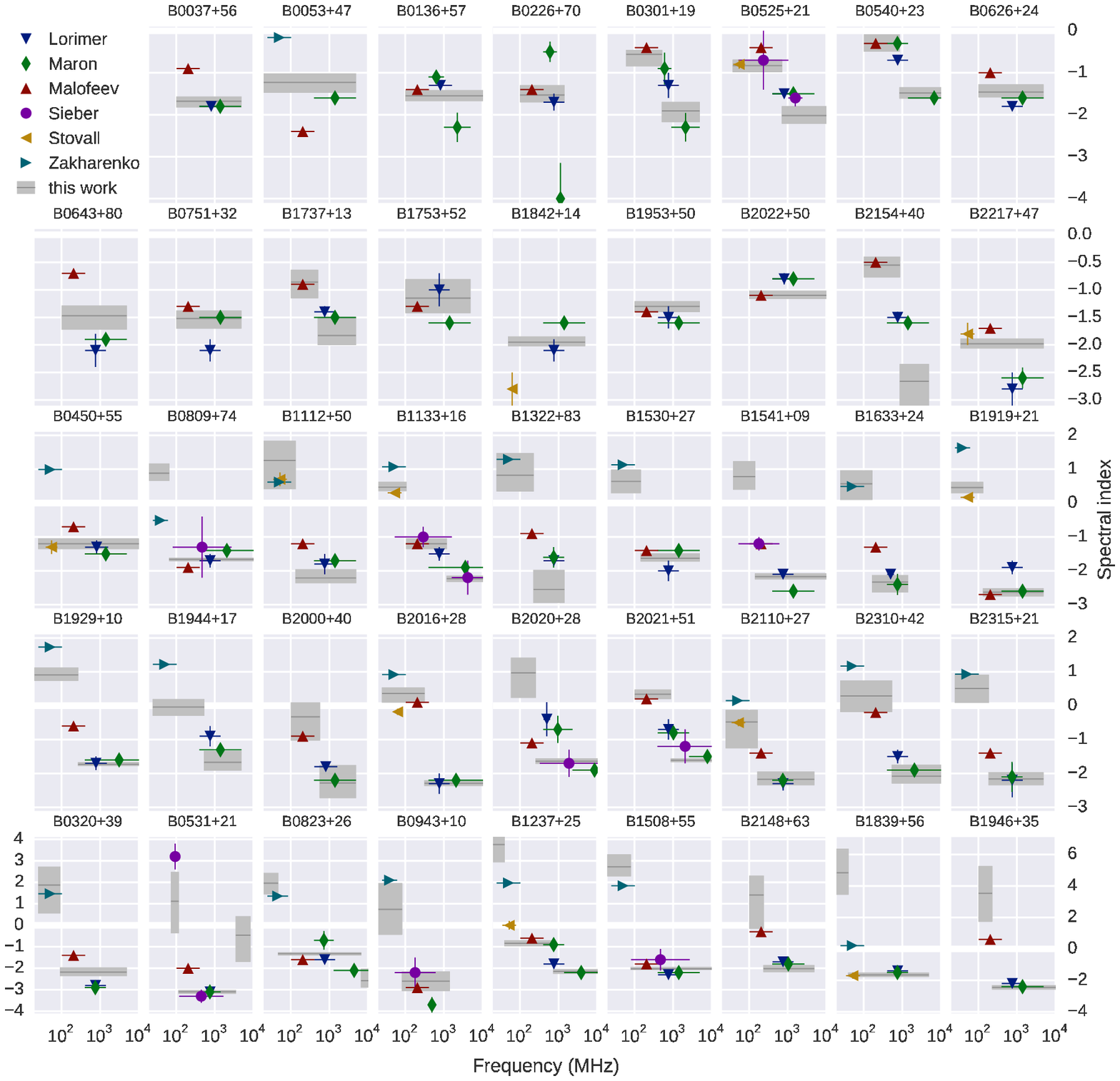}
 \caption{ Comparison of spectral indices from this work (grey-shaded rectangles) to the ones reported by  \citet{Sieber1973}, 
 \citet{Lorimer1995}, \citet{Maron2000}, \citet{Malofeev2000}, \citet{Zakharenko2013}, and \citet{Stovall2015} (coloured 
 points with error bars). The horizontal extent of grey rectangles and horizontal error bars on literature values represent the frequency 
 range over which the index was measured. The vertical extent/error bar marks the reported uncertainty (for \citealt{Malofeev2000} 
 and \citealt{Zakharenko2013} no errors were given). Spectral fits in this work include data points from all these works together 
 with other published values and our own measurements. The pulsars are grouped by the spectral index plotting limits and ordered 
 by right ascension within each group. }
\label{fig:ind_compar}   
\end{figure*}

For the negative spectral indices, the $\alpha$ at frequencies above the break is generally steeper than at frequencies 
below the break, with the exception of PSRs B0531$+$21, B0114$+$58, and B2303$+$30, for which the spectra flatten at higher 
frequencies. It must be noted that for the latter two sources the flattening is based on a single flux density measurement 
and more data are needed to confirm the observed behaviour. In case of the Crab pulsar, the flux density measurements at 
5 and 8\,GHz \citep{Moffett1996} suggest that spectral index may flatten somewhere between 2 and 5\,GHz. 
Such flattening coincides with a dramatic profile transformation happening in the same frequency range \citep{Hankins2015} 
and is also suggested by spectral index measurements for the individual profile components \citep{Moffett1999}. 

Figure~\ref{fig:ind_break} shows the spectral indices below and above the break frequency for the 32 census pulsars with a relatively 
well-known spectra with $\nu_\mathrm{min}<200$\,MHz and $\nu_\mathrm{max}>4$\,GHz, consisting of at least 10 flux density 
measurements. The data confirm a previously noticed tendency \citep{Sieber1973}: in most cases, if the break happens at rather 
higher frequencies ($\nu_\mathrm{br}\gtrsim 500$\,MHz), the change in slope is relatively moderate, with $\alpha_\mathrm{lo} - 
\alpha_\mathrm{hi} \approx 1$ or 2 (the so-called ``high-frequency cut-off''). For the breaks at lower frequencies 
($\nu_\mathrm{br}\lesssim 300$\,MHz), the change is more dramatic, with low-frequency $\alpha$ close to or larger than nought, 
the so-called ``low-frequency turnover''\footnote{Several pulsars outside the census sample are known to have spectra turning 
over at higher frequencies of $\sim 1$\,GHz \citep{Kijak2011}. Such high turnover frequency is tentatively explained 
by thermal free-free absorption in a pulsar surroundings.}.
Possible statistical relationships between the turnover frequency\footnote{The works of Pushchino group 
\citep[e. g.][]{Malov1981} consider
``maximum frequency'', which coincides with turnover frequency if $\alpha_\mathrm{lo}>0$.}, cut-off frequency and pulsar period
had been previously investigated in several works \citep[e.g.][]{Malofeev1980, Izvekova1981}, and a number of theoretical 
explanations was proposed \citep{Malofeev1980, Ochelkov1984, Malov1991, Petrova2002, Kontorovich2013}.

Because of the typically large spread of the same-frequency $S_\nu$ measurements, the reliable identification of the break frequencies 
is feasible only for the well-known spectra, composed of multiple, densely spaced flux density measurements, obtained in a 
wide frequency range.
The census data alone appears to be insufficient for making a substantial contribution to the spectral break identification.  
The HBA band appears to be situated close to or within the frequency range where a spectral turnover is likely to happen for 
the majority of non-recycled pulsars (at least in the census sample), and, apart from a few dozens of the previously 
well-studied sources, the literature spectra of census pulsars were scarcely sampled or did not extend below the HBA band 
(i.e. below 100\,MHz). The studies of low-frequency turnover would considerably benefit from the future flux density 
measurements at frequencies $<100$\,MHz, which could be obtained with the currently operating 
LWA, UTR-2, LOFAR LBA, and the future standalone NenuFAR LOFAR Super Station in Nan\c{c}ay \citep{Zarka2012}.

The census data still provide an indirect indication of the low-frequency turnover in some relatively poorly sampled spectra. 
This evidence comes from the comparison of the spectral indices for pulsars best fitted with 1PL in the work of \citet{Maron2000} 
and in our work. The census sample contains 35 such pulsars, with the indices mostly based on the data from 
100\,MHz--5\,GHz. The corresponding indices from \citet{Maron2000} are based on flux measurements between 400\,MHz and 
1.6/5\,GHz. %, but  occasionally extending to as low as 20\,MHz and as high as 10/30\,GHz). 
There is a statistical preference for a spectral index to be flatter in our broader frequency range (Fig.~\ref{fig:flat}), 
which could indicate a turnover happening somewhere close to 100\,MHz\footnote{Scattering could, in principle, cause the observed 
flattening of the low-frequency part of the spectrum. However, only two out of 35 sources had visibly scattered profiles in the 
HBA frequency range.}. As a comparison, we plot the spectral indices for 21 pulsars with clearly identified low-frequency 
turnover happening below the frequency range of \citet{Maron2000}. Such pulsars generally exhibit better spectral index agreement.

\subsubsection{Comparison to the indices from the literature}
\label{subsec:broadband}

For some of the census pulsars several spectral index estimates have already been made by various authors. 
Our spectra generally include the flux densities reported in those works, together with the other published 
flux densities and our own measurements. Thus, it is interesting to compare our spectral indices to the ones
available from the literature. Such a comparison can serve as an estimate of how much the spectral index values 
could change with addition of new data points in the same frequency range or with expansion of the frequency coverage.

We have compiled the spectral index measurements from several works with single or broken PL fits, 
namely from \citet{Sieber1973}; \citet{Lorimer1995}; \citet{Maron2000}, and \citet{Malofeev2000}. 
A recent set of spectral index measurements 
for frequencies below 100\,MHz from \citet{Zakharenko2013} and \citet{Stovall2015} were added as well. The collection 
of spectral indices versus the respective frequency ranges for the pulsars with at least three such literature 
measurements is shown in Fig.~\ref{fig:ind_compar}. 

Two features can be gleaned from this plot. Firstly, even if 
all authors approximate the spectrum with a single PL in the frequency range available to them, sometimes the
spectral indices in more narrow frequency sub-ranges may differ from each other and from the spectral index in a 
broader frequency range. This spread can reach a magnitude of $\delta \alpha$ of about 0.5 or even 1.5 (e.g. for PSRs 
B0053$+$47, B1929$+$21, and others). This can be explained by a systematic error in individual flux density measurements 
that bias the spectral index estimate, if the number of data points is small or the frequency range is narrow.
 
Secondly, different authors do not always agree on the location of the spectral breaks (e.g. for PSRs B0136$+$57, 
B2020$+$51, and others). Sometimes the narrowband spectral index shows clear gradual evolution with frequency across the
radio band (e.g. for PSRs B1133$+$16 and B1237$+$25). To a certain degree, this gradual evolution can be influenced by the scatter of 
flux density measurements, which ``smooths'' the breaks. However, one can also question whether a collection of PLs is a good 
representation of broadband spectral shape in general.

More complex models of broadband pulsar spectra (usually including a PL and absorption components) 
have been proposed by several authors (e.g. \citealt{Sieber1973}; \citealt{Malofeev1980} for the low-frequency turnover, and 
\citealt{Lewandowski2015}; \citealt{Rajwade2015} for the gigahertz-peaked spectra). 
However, in general,  progress has been impeded by the lack of accepted emission theories and the poor sampling of the 
available spectra. The empirical parabola fit in $\log\nu$--$\log S$ space was proposed by \citet{Kuzmin2001} and 
subsequently used for approximating the gigahertz-peaked spectra \citep[e.g.][]{Dembska2014}.
Recently, a simple three-parameter functional form for pulsar spectra was suggested by \citet{Loehmer2008}.
This form is based on a flicker noise model, which assumes the observed pulsar emission to be a collection of separate 
nanosecond-scale shots. However, the flicker noise model predicts flattening of the spectrum at lower frequencies 
($\alpha\approx0$), not the turnover ($\alpha>0$), and thus requires more development.

\section{Summary}
\label{sec:summary}

We have observed 194 non-recycled northern pulsars with the LOFAR high-band antennas in the frequency range of 
110--188\,MHz. This is a complete (as of September 2013) census of known non-recycled radio pulsars at $\mathrm{Dec}>8\degree$ 
and $|\mathrm{Gb}|>3\degree$, excluding globular cluster pulsars and those with poorly defined positions. 
Each pulsar was observed contiguously for 20\,min or at least 1000 spin periods.

We have detected 158 pulsars, collecting one of the largest 
samples of low-frequency wide-band data. These observations provided a wealth of information for the ongoing investigation 
of the low-frequency properties of pulsar radio emission, including the time-averaged and single-pulse full-Stokes analyses. 
Precise measurements of the ISM parameters (such as DM, RM, and scattering) along many lines of sight are being obtained 
as well. 

In this work we present the DMs, flux densities and flux-calibrated profiles for 158 detected pulsars. 
The average profiles, DM  and flux density measurements will be made publicly available via the 
EPN Database of Pulsar Profiles and on a dedicated LOFAR web-page\footnote{\url{http://www.astron.nl/psrcensus/}}. 
The LBA (30--90\,MHz) extension of the census has also been observed and
its results will be presented in subsequent works.

Owing to the large fractional bandwidth, we were able to measure DMs with typically ten times better 
precision than in the pulsar catalogue (with the median error of census DMs equal to $1.5\times10^{-3}$\,\dmu).
For our DM measurements we assumed the absence of profile evolution across the HBA band. The value of DM could
change by an amount that is an order of magnitude larger than the measurement error under different profile evolution assumptions. 
We computed the rate of secular DM variations by comparing census and catalogue DMs.
The spread of the rates was similar to that found by \citet{Hobbs2004b}, however, 
we also find an excess of larger DM variation rates for the pulsars with larger reported DM errors, possibly 
indicating that some DM errors were unaccounted for. 
We also compared our DMs to similar recent measurements at $\nu<100$\,MHz and found generally more preference
for the larger DM variation rates, which may be due to the relative influence of shorter-timescale stochastic variation in 
the ISM, but also may be caused by offsets in DM introduced by profile evolution and scattering. 

For two of the detected pulsars the ISM properties differ from what was expected. PSR J1503+2111 was
detected at a DM of 3.260\,\dmu,~in contrast to the pulsar catalogue value of 11.75\,\dmu~\citep{ChampionA2005}. 
We suggest that the catalogue value may be incorrect. Another pulsar, PSR B2036$+$53, appeared to have a scattering time 
three orders of magnitude smaller than predicted by the NE2001 model. 

\object{PSR J1740+1000}, previously considered to have a gigahertz-peaked spectrum, was detected at the LOFAR HBA frequencies
with 20 times larger flux density than predicted by the fits in \citet{Dembska2014} and \citet{Rajwade2015}. 
We argue that this pulsar has a normal power-law spectrum with an unusually large amount of
flux density variability.

We have constructed spectra for census pulsars by combining our measurements with those from the literature.
Spectra were fitted with a single or a broken PL (with one or two breaks). 
Out of 165 spectra (some spectra consisted only of the literature points when the pulsars were not detected in the census), 
124 were best fitted using a single PL, although those were also spectra with the smallest number of flux density measurements.
We also include spectral fits for 48 pulsars that do not have a previously published value. 
The distribution of spectral indices for these pulsars agrees with that found in a similar 
frequency range  by \citet{Malofeev2000} and is generally flatter
($\bar{\alpha}=-1.4$) than the distribution at higher frequencies \citep[$\bar{\alpha}=-1.8$,][]{Maron2000}.

The census data alone appear to be insufficient for making a considerable contribution 
to low-frequency turnover studies, since the HBA frequency range is located close to or within the frequency range 
where spectral turnover is likely to happen for the majority of non-recycled pulsars (at least for the census sample). 
Since the lower-frequency data (below 100\,MHz) are normally absent or scarce, for many sources the turnover can be 
observed only as a slight flattening of the spectrum at the low-frequency edge.
Studies of the low-frequency turnover will considerably benefit from 
future flux measurements below 100\,MHz. 

For a sub-sample of pulsars with relatively well-measured spectra, 
we have compared the obtained broadband spectral indices to the ones reported in the literature. It appears
that sometimes  spectral indices in more narrow sub-ranges may differ considerably ($\delta \alpha$ of $\approx0.5-1.5$)
from each other and from the spectral index in a broader frequency range.
This can be explained by errors that are unaccounted for in individual flux density measurements, which
bias the spectral index estimate if the number of data points is small or the frequency range is narrow.
 Also, different authors do not always agree on the location of the spectral breaks and 
sometimes the narrowband spectral index  shows clear gradual evolution with frequency across the 
radio band. To some degree this gradual evolution can be influenced by
the scatter of flux density measurements. However, this may also be an indication that
a collection of PLs is not a good representation of the broadband spectral shape of pulsar radio emission.

\begin{acknowledgements}
This work makes extensive use of \texttt{matplotlib}\footnote{\url{http://matplotlib.org/}} \citep{Hunter2007}, 
\texttt{seaborn}\footnote{\url{http://stanford.edu/~mwaskom/software/seaborn/}} Python plotting libraries, and NASA 
Astrophysics Data System. 
We acknowledge the use of the European Pulsar Network (EPN) database at the Max-Planck-Institut f\"ur Radioastronomie,
and its successor, the new EPN Database of Pulsar Profiles, developed by Michael Keith and
maintained by the University of Manchester.

AVB thanks Olaf Maron for kindly provided flux density measurements, D\'{a}vid Cseh for LOFAR calibration discussions, 
Tim Pennucci for his contribution to the Bayesian analysis of flux calibration uncertainty, Menno Norden for an RFI discussion,
and Tobia Carozzi for development of the \texttt{mscorpol} package.  
The research leading to these results has received funding from the European Research Council under the European
Union's Seventh Framework Programme (FP7/2007-2013) / ERC grant agreement nr. 337062 (DRAGNET; PI Hessels).
SO is supported by the Alexander von Humboldt Foundation. 

The LOFAR facilities in the Netherlands and other countries, under different ownership,
are operated through the International LOFAR Telescope foundation (ILT) as an international 
observatory open to the global astronomical community under a joint scientific
policy. In the Netherlands, LOFAR is funded through the BSIK program for interdisciplinary 
research and improvement of the knowledge infrastructure. Additional funding is
provided through the European Regional Development Fund (EFRO) and the innovation
program EZ/KOMPAS of the Collaboration of the Northern Provinces (SNN). ASTRON
is part of the Netherlands Organisation for Scientific Research (NWO).

\end{acknowledgements}

\bibliographystyle{aa} 
\bibliography{census_bibliography} 

%\newpage
\appendix

\section{Scintillation}
\label{app:scint}

Inhomogeneities in the ISM introduce phase modulations to the propagating pulsar radio emission and cause the observed
flux density to fluctuate both in time and radio frequency. To estimate the influence of scintillation on census flux densities,
we used predictions of a simple thin-screen Kolmogorov model, summarised in \citet{Lorimer2005}. The scintillation 
bandwidth was determined as $\Delta f = 1.16/(2\pi \tau_\mathrm{scat})\times($150\,MHz/1\,GHz$)^{4.4}$, where 
$\tau_\mathrm{scat}$ is scattering time at 1\,GHz from the NE2001 model \citep{Cordes2002}. For PSR B2036+53 we changed the obviously 
incorrect NE2001 prediction, $\tau_\mathrm{scat}($150 MHz$) = 486$\,s, to a more reasonable $\tau_\mathrm{scat}($150\,MHz$) 
\approx 200$\,ms, estimated from the census data at the lower edge of HBA band (note that Table~\ref{table:main} lists NE2001
value of $\tau_\mathrm{scat}$ for this pulsar, for consistency with other sources).

For all census pulsars the scintillation bandwidth $\Delta f$ was of the order of 0.1\,kHz$-$1\,MHz, indicating that the strong
scintillation regime dominates ($\sqrt{f/\Delta f}>1$). To calculate the diffractive scintillation (DISS) time scales, we used
distances and transverse velocities from the pulsar catalogue. The velocities were taken to be 200\,km~s$^{-1}$ if no direct measurements 
were available. DISS time scales ranged from a few seconds to a few minutes. Thus, for all pulsars band- and session-integrated 
flux density measurements were averaged over many scintles both in the time and frequency domain, and thus were not expected to 
be influenced by DISS: the modulation index $m_\mathrm{DISS}$ (rms of the flux density divided by its mean value) was $\approx 0.001$. 
The refractive scintillation (RISS), however, was much more prominent, with typical $m_\mathrm{RISS} \approx 0.1$. The expected 
values of total modulation index are given in Table\,\ref{table:main}.

\section{Tables}
\label{app:fit}

Table~\ref{table:main} contains observation summary.  The columns indicate: pulsar name; spin period; 
observing epoch; observation length; peak S/N of the average profile; DM from the pulsar catalogue; 
measured census DM; expected NE2001 scattering time at 150\,MHz divided by pulsar period; expected
modulation index due to scintillation in the ISM; 
mean flux density within the HBA band (upper limit for the non-detected pulsars), 
and the literature references to previous flux density measurements. The values 
in brackets indicate the errors on the last one or two significant digits.

Tables \ref{table:1pl}--\ref{table:3pl} contain fitted parameters for the pulsars with the spectra
modelled with a single PL, a PL with one break and a PL with two breaks, respectively. The columns 
include pulsar name; spectral frequency span; number of data points in spectrum, $N_p$; 
the reference frequency, $\nu_0$; flux density at the reference frequency, $S_0$; 
spectral index, $\alpha$ (or indices in case of broken PLs), and fitted flux density scatter, $\su$ 
(if applicable, see Sect.~\ref{subsec:fit}).
Tables \ref{table:2pl} and \ref{table:3pl} also include break frequency, $\nu_\mathrm{br}$, together with its 68\% 
uncertainty range.

\onecolumn
\setlength\LTcapwidth{7.1in}%\textwidth} % default: 4in (rather less than \textwidth...)
\setlength{\tabcolsep}{2pt}
\renewcommand{\arraystretch}{1.05}
\begin{center}
\footnotesize%\tiny
\begin{longtable}{|l|c|c|c|c|l|l|c|c|l|l|}
\caption{Observation summary, DM and flux density measurements. \label{table:main}}\\
\hline
\parbox[t][][t]{1.7cm}{\centering PSR} & 
\parbox[t][][t]{0.9cm}{\centering Period $P$\\ (s) } & 
\parbox[t][][t]{1.4cm}{\centering Observing \\ epoch (MJD)} &
\parbox[t][][t]{1.0cm}{\centering Obs. time \\ (min)} & 
\parbox[t][][t]{0.8cm}{\centering Peak S/N} & 
\parbox[t][][t]{1.5cm}{\centering $\mathrm{DM}_\mathrm{cat}$ \\ (pc cm$^{-3}$)} & 
\parbox[t][][t]{1.6cm}{\centering $\mathrm{DM_\mathrm{cen}}$ \\ (pc cm$^{-3}$)} & 
\parbox[t][][t]{1.0cm}{\centering Expec-\\ted $\tau_\mathrm{scat}/P$\\ } &
\parbox[t][][t]{1.0cm}{\centering Exp. mod. index } &
\parbox[t][][t]{1.5cm}{\centering Mean HBA \\ flux \\(mJy) } &
\parbox[t][][t]{3.4cm}{\centering Literature flux references } %&
%\parbox[t][][t]{1.5cm}{\center Notes } 
\\  
\hline
\endfirsthead
\multicolumn{6}{c}%
{\tablename\ \thetable\ -- \textit{Continued from previous page}} \\
\hline
\parbox[t][][t]{1.7cm}{\centering PSR} & 
\parbox[t][][t]{0.9cm}{\centering Period $P$\\ (s) } & 
\parbox[t][][t]{1.4cm}{\centering Observing \\ epoch (MJD)} &
\parbox[t][][t]{1.0cm}{\centering Obs. time \\ (min)} & 
\parbox[t][][t]{0.8cm}{\centering Peak S/N} & 
\parbox[t][][t]{1.5cm}{\centering $\mathrm{DM}_\mathrm{cat}$ \\ (pc cm$^{-3}$)} & 
\parbox[t][][t]{1.6cm}{\centering $\mathrm{DM_\mathrm{cen}}$ \\ (pc cm$^{-3}$)} & 
\parbox[t][][t]{1.0cm}{\centering Expec-\\ted $\tau_\mathrm{scat}/P$\\ } &
\parbox[t][][t]{1.0cm}{\centering Exp. mod. index } &
\parbox[t][][t]{1.5cm}{\centering Mean HBA \\ flux \\(mJy) } &
\parbox[t][][t]{3.4cm}{\centering Literature flux references } %&
%\parbox[t][][t]{1.5cm}{\center Notes } 
\\  
\hline
\endhead
\hline \multicolumn{6}{r}{\textit{Continued on next page}} \\
\endfoot 
\multicolumn{11}{l}{\parbox[t][][t]{7.1in}{\vspace{-4mm}{
\tablebib{ [1] \citet{Bridle1970}; [2] \citet{Barr2013}; [3] \citet{Burgay2013}; 
[4] \citet{Boyles2013}; [5] \citet{Bartel1978}; [6] \citet{ChampionA2005}; [7] \citet{ChampionB2005}; [8] \citet{Camilo1995}; 
[9] \citet{CamiloB1996}; [10] \citet{CamiloA1996}; [11] \citet{Camilo2002}; [12] \citet{Downs1979}; [13] \citet{Dembska2014}; 
[14] \citet{Davies1977}; [15] \citet{Deshpande1992}; [16] \citet{Downs1973}; [17] \citet{Damashek1978}; [18] \citet{Dewey1985}; 
[19] \citet{Fomalont1992}; [20] \citet{Han2009}; [21] \citet{HobbsA2004}; [22] \citet{Hulse1975}; [23] \citet{Izvekova1979}; 
[24] \citet{Izvekova1981}; [25] \citet{Janssen2009}; [26] \citet{Kijak2007}; [27] \citet{Kramer1997}; [28] \citet{Kijak1998}; 
[29] \citet{Kijak2011}; [30] \citet{Kuzmin1986}; [31] \citet{Krzeszowski2014}; [32] \citet{Kuzmin1978}; [33] \citet{Karuppusamy2011}; 
[34] \citet{Kramer1996}; [35] \citet{Lane2014}; [36] \citet{Lorimer2002}; [37] \citet{Lorimer2006}; [38] \citet{Loehmer2008}; 
[39] \citet{Lewandowski2004}; [40] \citet{LorimerA2005}; [41] \citet{Lorimer1995}; [42] \citet{Lommen2000}; [43] \citet{Manchester1971}; 
[44] \citet{Malofeev1993}; [45] \citet{Malofeev1999}; [46] \citet{McLaughlin2002}; [47] \citet{Morris1981}; [48] \citet{Moffett1999}; 
[49] \citet{Maron2000}; [50] \citet{Manchester1978}; [51] \citet{Malofeev1980}; [52] \citet{Malofeev2000}; [53] \citet{Manchester1981}; 
[54] \citet{Navarro2003}; [55] \citet{Pletsch2012}; [56] \citet{Rankin1981}; [57] \citet{Rankin1970}; [58] \citet{Ray1996}; 
[59] \citet{Slee1986}; [60] \citet{Stinebring1990}; [61] \citet{Seiradakis1995}; [62] \citet{Sayer1997}; [63] \citet{Stovall2015}; 
[64] \citet{Stokes1986}; [65] \citet{Shrauner1998}; [66] \citet{Sieber1987}; [67] \citet{Thorsett1993}; [68] \citet{Taylor1995}; 
[69] \citet{Vivekanand1983}; [70] \citet{Wielebinski1993}; [71] \citet{Zakharenko2013}.}}}}\\
\multicolumn{11}{l}{\parbox[t][][t]{7.0in}{\textbf{Notes.}
\tablefoottext{$\sharp$} {For PSR B2036+53, NE2001 predicts $\tau(150\,\mathrm{MHz})\approx486$\,s, whereas census observations suggest
$\tau(150\,\mathrm{MHz})\approx0.2$\,s. For this pulsar, the expected modulation index was based 
on census data estimate of the scattering time. See Sect.~\ref{sec:DM} for discussion.}}}
\endlastfoot
\hline 
% Some of the rows in the table were shuffled to bring the order to numerical by pulsar name (excluding B/J)
J0006+1834 & 0.694 & 56703.58 & 20 & \hphantom{000}6 & \hphantom{0}12.0\,(6) & \hphantom{0}11.41\,(55) & 4e-05 & 0.20 & \hphantom{00}6.5 $\pm$ 3.3 & \parbox[t][][t]{3.4cm}{\tiny  8, 52, 71} \\ 
B0011+47 & 1.241 & 56773.46 & 21 & \hphantom{00}36 & \hphantom{0}30.85\,(7) & \hphantom{0}30.405\,(13) & 2e-04 & 0.13 & \hphantom{00.}35 $\pm$ 18 & \parbox[t][][t]{3.4cm}{\tiny  17, 20, 24, 41, 45, 49, 61} \\ 
J0033+57 & 0.315 & 56773.48 & 20 & \hphantom{0}$\dots$ & \hphantom{0}76 & \hphantom{0}$\dots$ & 2e-02 & 0.08 & \hphantom{00000}<2.0 & \parbox[t][][t]{3.4cm}{\tiny  \dots} \\ 
B0037+56 & 1.118 & 56753.42 & 20 & \hphantom{00}38 & \hphantom{0}92.595\,(9) & \hphantom{0}92.5146\,(25) & 1e-02 & 0.07 & \hphantom{00.}33 $\pm$ 17 & \parbox[t][][t]{3.4cm}{\tiny  18, 28, 41, 52, 61} \\ 
B0045+33 & 1.217 & 56703.61 & 21 & \hphantom{00}98 & \hphantom{0}39.94\,(4) & \hphantom{0}39.922\,(15) & 5e-04 & 0.11 & \hphantom{00.}13 $\pm$ 6 & \parbox[t][][t]{3.4cm}{\tiny  18, 41, 49, 52, 61, 68} \\ 
B0052+51 & 2.115 & 56776.32 & 36 & \hphantom{00}35 & \hphantom{0}44.125\,(15) & \hphantom{0}44.0127\,(24) & 4e-04 & 0.11 & \hphantom{00}9.4 $\pm$ 4.7 & \parbox[t][][t]{3.4cm}{\tiny  18, 20, 41, 49, 52} \\ 
B0053+47 & 0.472 & 56703.59 & 20 & \hphantom{00}19 & \hphantom{0}18.09\,(4) & \hphantom{0}18.1354\,(13) & 2e-04 & 0.16 & \hphantom{00}7.5 $\pm$ 3.8 & \parbox[t][][t]{3.4cm}{\tiny  18, 28, 41, 52, 68, 71} \\ 
B0105+65 & 1.284 & 56784.40 & 22 & \hphantom{00}88 & \hphantom{0}30.46\,(5) & \hphantom{0}30.5482\,(14) & 3e-04 & 0.12 & \hphantom{00.}41 $\pm$ 20 & \parbox[t][][t]{3.4cm}{\tiny  14, 24, 41, 45, 52, 53, 61} \\
\hline                                   
\end{longtable}
\end{center}
\newpage

\tiny
\setlength{\tabcolsep}{2pt}
\renewcommand{\arraystretch}{1.01}
\begin{center}
\begin{longtable}{|l|l|c|c|c|c|c||l|l|c|c|c|c|c|}
\caption{Fit results for pulsars with a single PL spectrum.  
} \label{table:1pl} \\ 

\hline 
\parbox[t][][t]{1.2cm}{\centering PSR } & 
\parbox[t][][t]{1.4cm}{\centering Frequency \\ span \\ (MHz)} & 
\parbox[t][][t]{0.9cm}{\centering \# of points,\\ $N_p$ } & 
\parbox[t][][t]{0.8cm}{\centering Ref. \\ freq., \\ $\nu_0$ (MHz)} & 
\parbox[t][][t]{0.8cm}{\centering Ref. \\ flux, \\ $S_0$ (mJy)} & 
\parbox[t][][t]{1.5cm}{\centering Spectral \\ index, \\ $\alpha$ } & 
\parbox[t][][t]{0.8cm}{\centering Fitted flux \\ scatter, \\ $\su$ } &
\parbox[t][][t]{1.5cm}{\centering PSR } & 
\parbox[t][][t]{1.4cm}{\centering Frequency \\ span \\ (MHz)} & 
\parbox[t][][t]{0.9cm}{\centering \# of points, \\ $N_p$} & 
\parbox[t][][t]{0.8cm}{\centering Ref. \\ freq., \\ $\nu_0$ (MHz)} & 
\parbox[t][][t]{0.8cm}{\centering Ref. \\ flux, \\ $S_0$ (mJy)} & 
\parbox[t][][t]{1.5cm}{\centering Spectral \\ index, \\ $\alpha$ } & 
\parbox[t][][t]{0.8cm}{\centering Fitted flux \\ scatter, \\ $\su$ } \\ \hline 
\endfirsthead
\hline \hline
\multicolumn{14}{l}{\parbox[t][][t]{17.5cm}{\textbf{Notes.}
\tablefoottext{$\sharp$} {These pulsars have also broken PL fit, with break frequency fixed at the frequency of the largest
 measured flux density. See Table~\ref{table:2pl} for the values of fitted parameters.}}}
\endlastfoot
% Some of the rows in the table were shuffled to bring the order to numerical by pulsar name (excluding B/J)
J0006+1834 & 149 -- 430\hphantom{000} & 2\hphantom{0} & \hphantom{0}300 & \hphantom{0}0.6 & $-3.3 \pm 0.9$ & \dots & B1848+12 & 102 -- 1606\hphantom{00} & 9\hphantom{0} & \hphantom{0}400 & \hphantom{0}5.8 & $-1.9 \pm 0.2$ & 0.09\\ 
B0011+47 & 102 -- 4850\hphantom{00} & 17 & \hphantom{0}700 & \hphantom{0}7.8 & $-1.0 \pm 0.1$ & 0.10 & B1848+13 & 102 -- 4850\hphantom{00} & 9\hphantom{0} & \hphantom{0}700 & \hphantom{0}2.8 & $-1.5 \pm 0.2$ & 0.17\\ 
B0037+56 & 102 -- 4850\hphantom{00} & 10 & \hphantom{0}700 & \hphantom{0}2.2 & $-1.7 \pm 0.1$ & 0.11 & B1852+10 & 430 -- 1400\hphantom{00} & 2\hphantom{0} & \hphantom{0}800 & \hphantom{0}3.0 & $-2.0 \pm 0.8$ & \dots\\ 
B0045+33 & 102 -- 1400\hphantom{00} & 8\hphantom{0} & \hphantom{0}400 & \hphantom{0}2.4 & $-2.4 \pm 0.2$ & 0.15 & J1859+1526 & 149 -- 774\hphantom{000} & 3\hphantom{0} & \hphantom{0}300 & \hphantom{0}2.3 & $-1.6 \pm 0.6$ & \dots\\ 
B0052+51 & 102 -- 4850\hphantom{00} & 11 & \hphantom{0}700 & \hphantom{0}2.3 & $-1.0 \pm 0.2$ & 0.09 & J1900+30 & 149 -- 430\hphantom{000} & 2\hphantom{0} & \hphantom{0}300 & \hphantom{0}2.1 & $-1.5 \pm 0.9$ & \dots\\ 
B0053+47 & \hphantom{0}20 -- 4850\hphantom{00} & 8\hphantom{0} & \hphantom{0}300 & \hphantom{0}4.3 & $-1.2 \pm 0.2$ & 0.45 & J1901+1306 & 149 -- 430\hphantom{000} & 2\hphantom{0} & \hphantom{0}300 & \hphantom{0}1.4 & $-2.2 \pm 0.9$ & \dots\\ 
B0105+65 & 102 -- 1420\hphantom{00} & 11 & \hphantom{0}400 & \hphantom{0}9.0 & $-1.4 \pm 0.2$ & 0.16 & J1903+2225 & 430 -- 774\hphantom{000} & 2\hphantom{0} & \hphantom{0}600 & \hphantom{0}0.9 & \hphantom{0}$0.5 \pm 1.4$ & \dots\\ 
B0105+68 & 102 -- 1408\hphantom{00} & 6\hphantom{0} & \hphantom{0}400 & \hphantom{0}3.3 & $-1.5 \pm 0.3$ & 0.10 & J1906+1854 & 149 -- 774\hphantom{000} & 3\hphantom{0} & \hphantom{0}300 & \hphantom{0}7.7 & $-1.2 \pm 0.4$ & \dots\\
\end{longtable}
\end{center}

\tiny
\setlength{\tabcolsep}{2pt}
\renewcommand{\arraystretch}{1.05}
\begin{center}
\begin{longtable}{|l|l|c|c|c|c|c|c|c|c|}
\caption{Fit results for pulsars where the spectrum was modelled with a broken PL.} \label{table:2pl} \\ 

\hline 
\parbox[t][][t]{1.2cm}{\centering PSR } & 
\parbox[t][][t]{1.5cm}{\centering Frequency \\ span \\ (MHz)} & 
\parbox[t][][t]{1.0cm}{\centering \# of points, \\ $N_p$} & 
\parbox[t][][t]{1.0cm}{\centering Ref. \\ freq., \\ $\nu_0$ (MHz)} & 
\parbox[t][][t]{1.0cm}{\centering Ref. \\ flux, \\ $S_0$ (mJy)} & 
\parbox[t][][t]{1.5cm}{\centering Spectral \\ index, \\ $\alpha_\mathrm{lo}$ } & 
\parbox[t][][t]{1.0cm}{\centering Break freq., \\ $\nu_\mathrm{br}$ \\ (MHz)} &
\parbox[t][][t]{2.0cm}{\centering Uncertainty \\ range for \\ $\nu_\mathrm{br}$ \\ (MHz)  } & 
\parbox[t][][t]{1.5cm}{\centering Spectral \\ index, \\ $\alpha_\mathrm{hi}$ } & 
\parbox[t][][t]{1.0cm}{\centering Fitted flux \\ scatter, \\ $\su$ }

\\ \hline 
\endfirsthead

\hline \hline
\endlastfoot
% Some of the rows in the table were shuffled to bring the order to numerical by pulsar name (excluding B/J)
B0114+58 & 102 -- 32000 & \hphantom{0}9 & 1800 & \hphantom{00}0.5 & $-2.2 \pm 0.3$ & 1087 & 894 -- 1407 & $-0.7 \pm 0.3$ &  0.18 \\ 
B0301+19 & \hphantom{0}61 -- 4850\hphantom{00} & 34 & \hphantom{0}500 & \hphantom{0}24.0 & $-0.6 \pm 0.3$ & \hphantom{0}514 & 407 -- 821\hphantom{0} & $-1.9 \pm 0.3$ &  0.10 \\ 
B0320+39 & \hphantom{0}25 -- 4850\hphantom{00} & 23 & \hphantom{0}300 & \hphantom{0}39.0 & \hphantom{$-$}$1.9 \pm 1.3$ & \hphantom{00}92 & \hphantom{0}64 -- 129\hphantom{0} & $-2.2 \pm 0.2$ &  0.12 \\ 
B0458+46 & 102 -- 4850\hphantom{00} & 10 & \hphantom{0}700 & \hphantom{00}8.7 & $-0.6 \pm 0.3$ & \hphantom{0}754 & 618 -- 1084 & $-1.9 \pm 0.5$ &  0.06 \\ 
B0523+11 & \hphantom{0}61 -- 4750\hphantom{00} & 20 & \hphantom{0}500 & \hphantom{0}14.0 & $-1.1 \pm 0.4$ & \hphantom{0}435 & 232 -- 658\hphantom{0} & $-2.1 \pm 0.2$ &  0.05 \\ 
B0525+21 & \hphantom{0}38 -- 14800 & 44 & \hphantom{0}700 & \hphantom{0}33.0 & $-0.8 \pm 0.2$ & \hphantom{0}709 & 524 -- 1100 & $-2.0 \pm 0.2$ &  0.14 \\ 
B0540+23 & 102 -- 32000 & 30 & 1800 & \hphantom{00}8.2 & $-0.3 \pm 0.2$ & \hphantom{0}845 & 678 -- 1212 & $-1.5 \pm 0.1$ &  0.12 \\ 
B0809+74 & \hphantom{0}13 -- 14800 & 65 & \hphantom{0}400 & 110.0 & \hphantom{$-$}$0.9 \pm 0.3$ & \hphantom{00}66 & \hphantom{0}58 -- 72\hphantom{0}\hphantom{0} & $-1.66 \pm 0.05$ &  0.15 \\
\end{longtable}
\end{center}

\tiny
\setlength{\tabcolsep}{2pt}
\renewcommand{\arraystretch}{1.10}
\begin{center}
\begin{longtable}{|l|l|c|c|c|c|c|c|c|c|c|c|c|}
\caption{Fit results for pulsars where the spectrum was modelled by a PL with two breaks. 
} \label{table:3pl} \\ 

\hline 
\parbox[t][][t]{1.2cm}{\centering PSR } & 
\parbox[t][][t]{0.8cm}{\centering Frequency \\ span \\ (MHz)} & 
\parbox[t][][t]{0.8cm}{\centering \# of points, \\ $N_p$} & 
\parbox[t][][t]{0.8cm}{\centering Ref. \\ freq., \\ $\nu_0$ (MHz)} & 
\parbox[t][][t]{0.9cm}{\centering Ref. \\ flux, \\ $S_0$ (mJy)} & 
\parbox[t][][t]{1.4cm}{\centering Spectral \\ index, \\ $\alpha_\mathrm{lo}$ } & 
\parbox[t][][t]{1.2cm}{\centering Lower \\ break freq., \\ $\nu^\mathrm{lo}_\mathrm{br}$ \\ (MHz)} &
\parbox[t][][t]{1.3cm}{\centering Uncertainty \\ range for \\ $\nu^\mathrm{lo}_\mathrm{br}$ \\ (MHz)  } & 
\parbox[t][][t]{1.4cm}{\centering Spectral \\ index \\ $\alpha_\mathrm{mid}$ } & 
\parbox[t][][t]{1.2cm}{\centering Higher \\ break freq., \\ $\nu^\mathrm{hi}_\mathrm{br}$ \\ (MHz)} &
\parbox[t][][t]{1.5cm}{\centering Uncertainty \\ range for \\ $\nu^\mathrm{hi}_\mathrm{br}$ \\ (MHz)  } &
\parbox[t][][t]{1.7cm}{\centering Spectral \\ index, \\ $\alpha_\mathrm{hi}$ } &
\parbox[t][][t]{1.0cm}{\centering Fitted flux \\ scatter, \\ $\su$ }
\\ \hline 
\endfirsthead
\hline \hline
\endlastfoot
B0531+21 & \hphantom{0}73 -- 8400\hphantom{00} & 48 & \hphantom{0}800 & \hphantom{0}73.0 & \hphantom{$-$}$1.1 \pm 1.5$ & \hphantom{0}132 & 116 -- 141\hphantom{0} & $-3.09 \pm 0.09$ &  3506 & 2626 -- 4822 & $-0.5 \pm 1.2$ & 0.04 \\ 
B0823+26 & \hphantom{0}20 -- 32000 & 63 & \hphantom{0}800 & \hphantom{0}32.0 & \hphantom{$-$}$2.0 \pm 0.5$ & \hphantom{00}55 & \hphantom{0}47 -- 62\hphantom{0}\hphantom{0} & $-1.32 \pm 0.07$ &  6643 & 3291 -- 9216 & $-2.6 \pm 0.6$ & 0.05 \\ 
B1133+16 & \hphantom{0}17 -- 32000 & 122 & \hphantom{0}700 & 120.0 & \hphantom{$-$}$0.5 \pm 0.2$ & \hphantom{0}125 & 106 -- 145\hphantom{0} & $-1.2 \pm 0.2$ &  1184 & 981 -- 1609 & $-2.23 \pm 0.09$ & 0.10 \\ 
B1237+25 & \hphantom{0}20 -- 24620 & 75 & \hphantom{0}700 & \hphantom{0}54.0 & \hphantom{$-$}$3.8 \pm 1.2$ & \hphantom{00}45 & \hphantom{0}40 -- 52\hphantom{0}\hphantom{0} & $-0.8 \pm 0.1$ &  \hphantom{0}734 & 582 -- 869\hphantom{0} & $-2.1 \pm 0.1$ & 0.12 \\ 
B2303+30 & \hphantom{0}50 -- 4850\hphantom{00} & 18 & \hphantom{0}500 & \hphantom{0}13.0 & $-0.4 \pm 0.4$ & \hphantom{0}311 & 274 -- 392\hphantom{0} & $-2.7 \pm 0.6$ &  \hphantom{0}876 & 642 -- 1221 & $-1.3 \pm 0.5$ & 0.07 \\
\end{longtable}
\end{center}
\begingroup
\renewcommand{\cleardoublepage}{}
\renewcommand{\clearpage}{}
\section{Spectra}
\label{app:prof_spec}
\begin{figure*}
\includegraphics[scale=0.48]{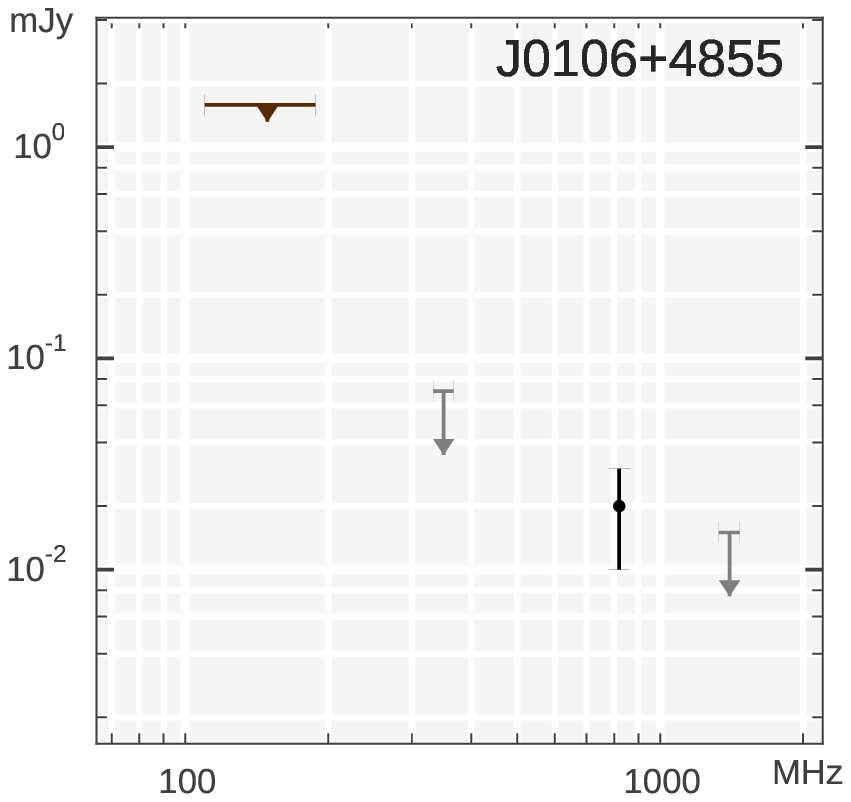}\includegraphics[scale=0.48]{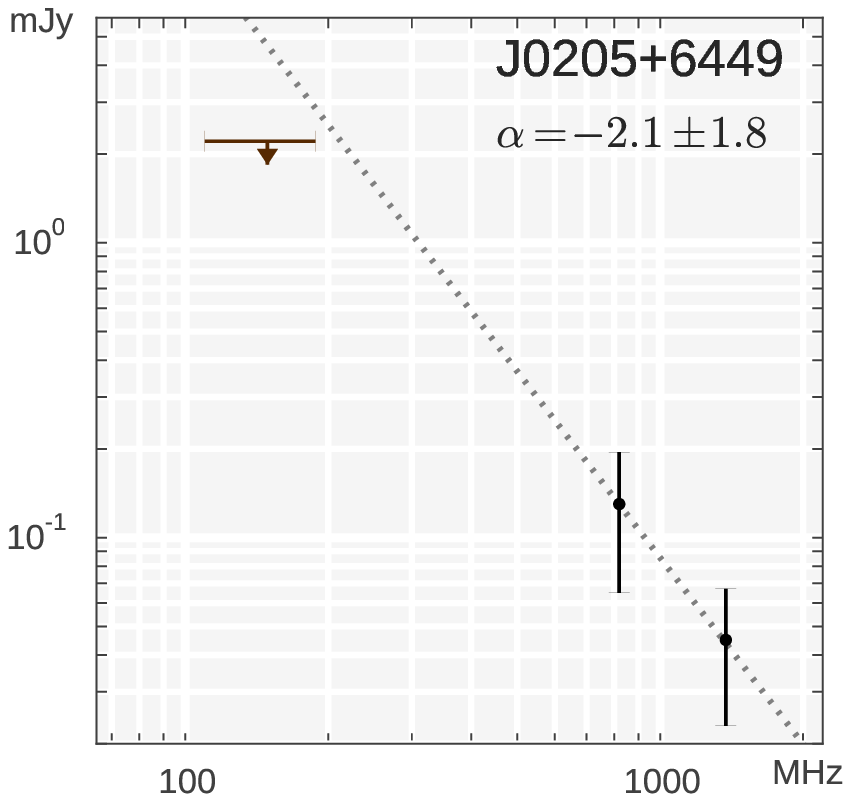}\includegraphics[scale=0.48]{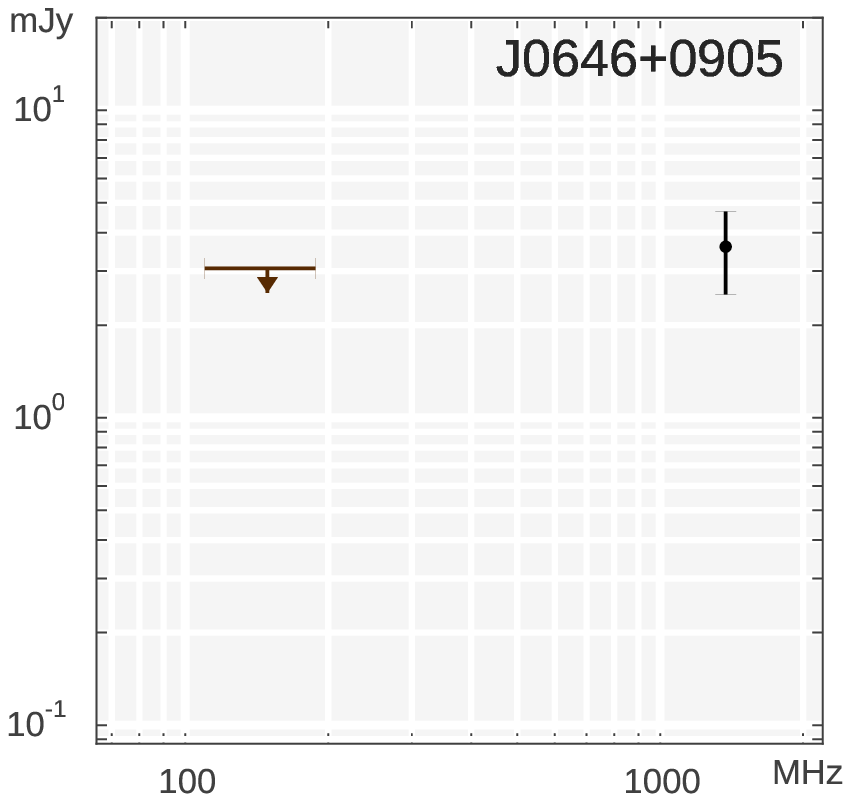}\includegraphics[scale=0.48]{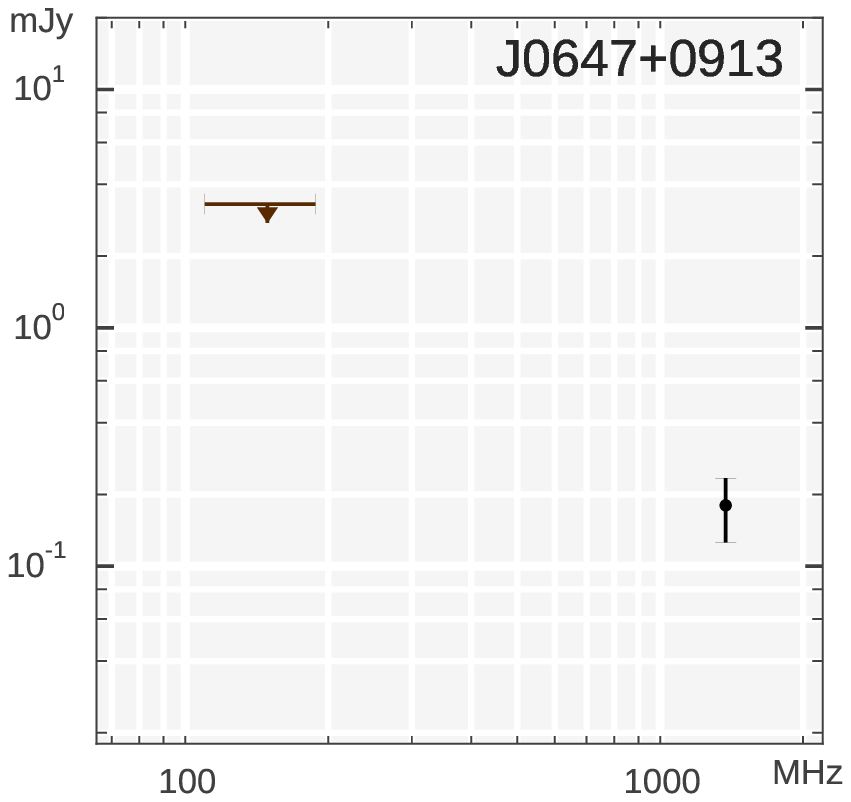}
\includegraphics[scale=0.48]{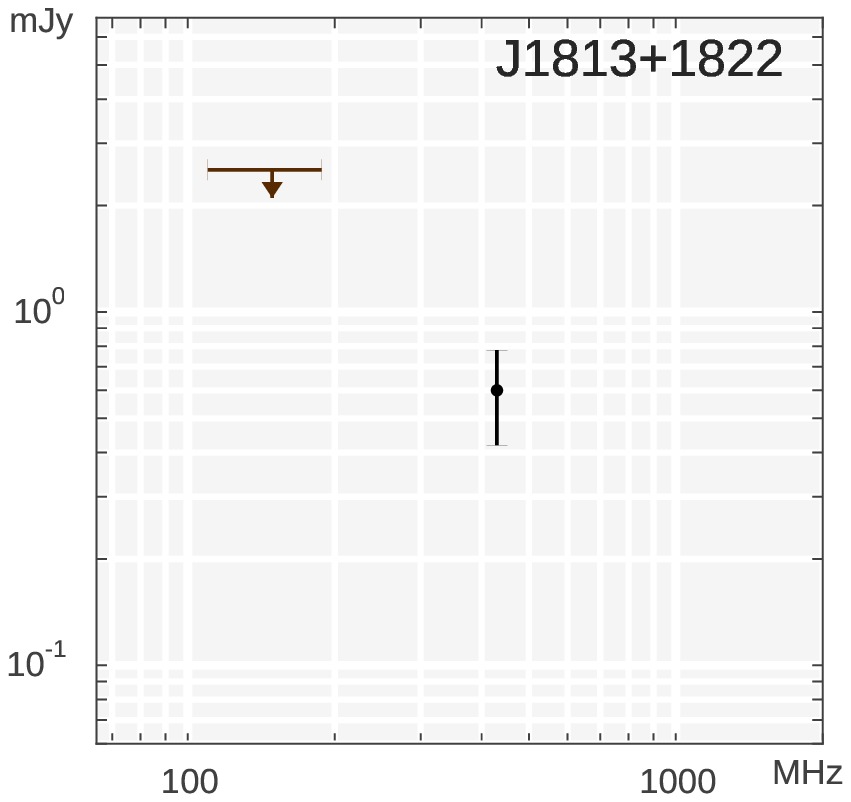}\includegraphics[scale=0.48]{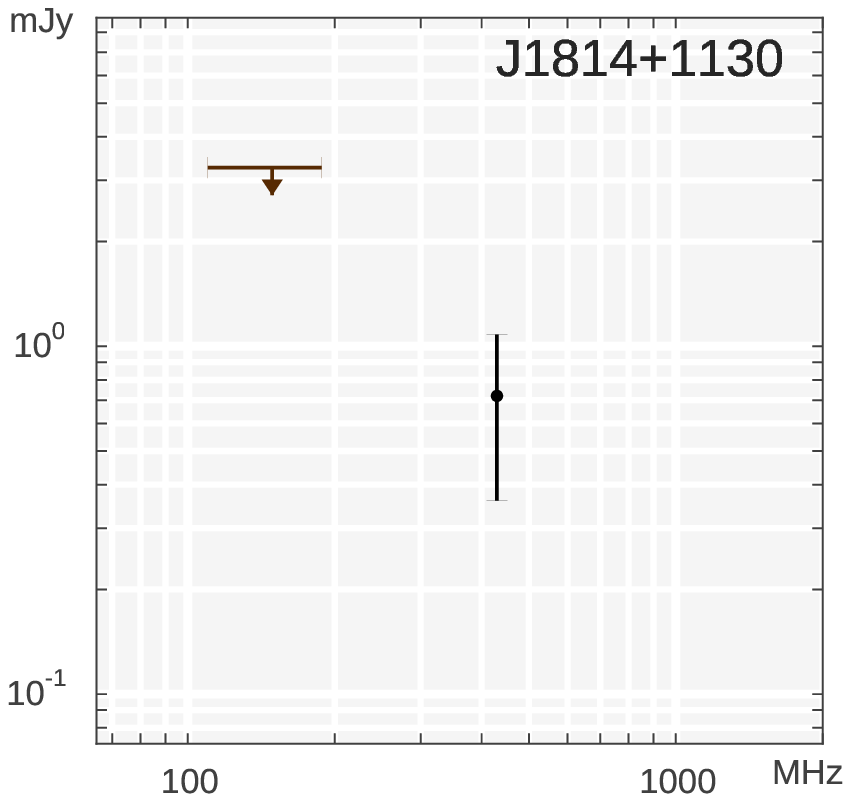}\includegraphics[scale=0.48]{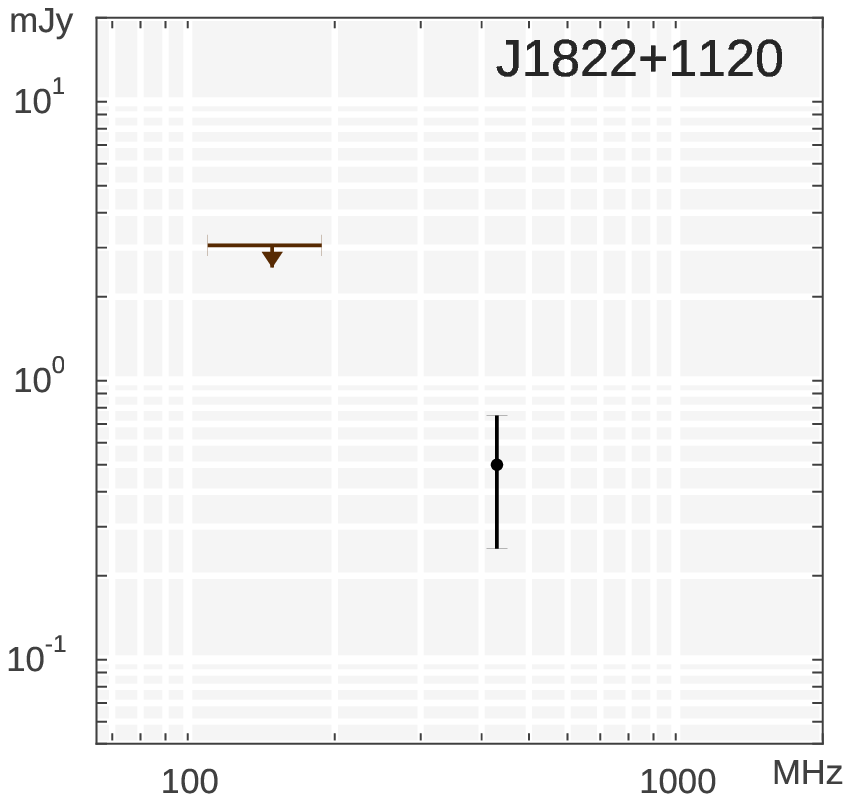}\includegraphics[scale=0.48]{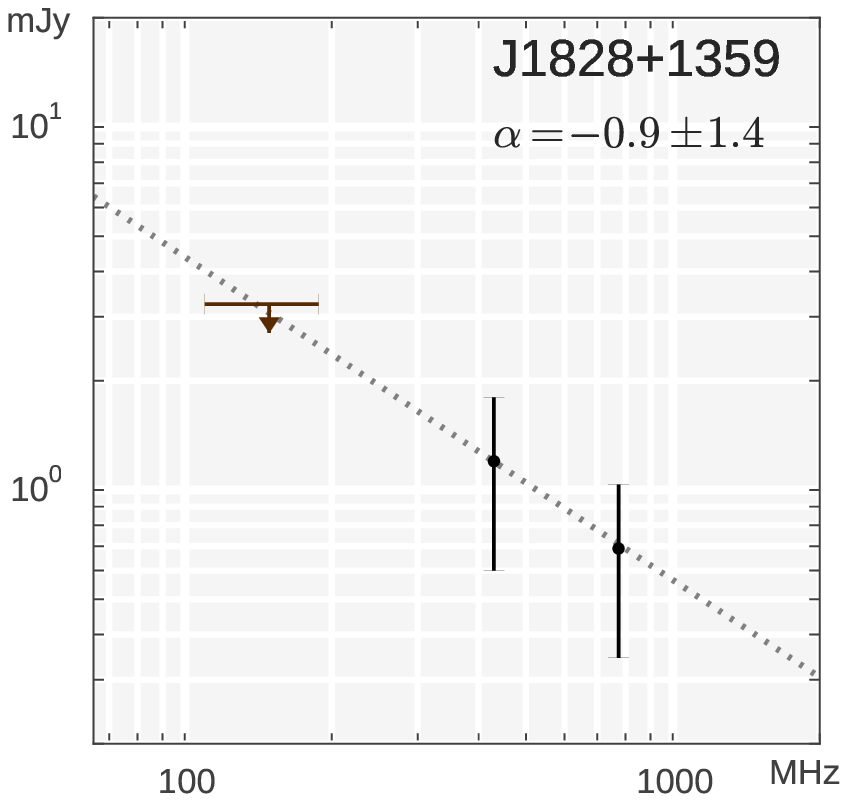}
\includegraphics[scale=0.48]{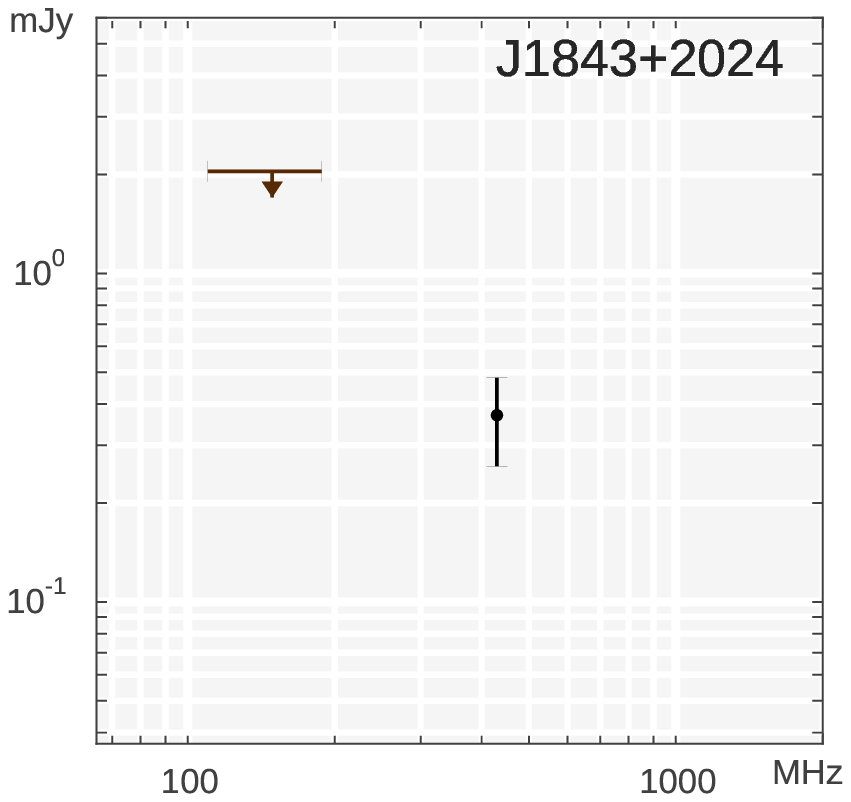}\includegraphics[scale=0.48]{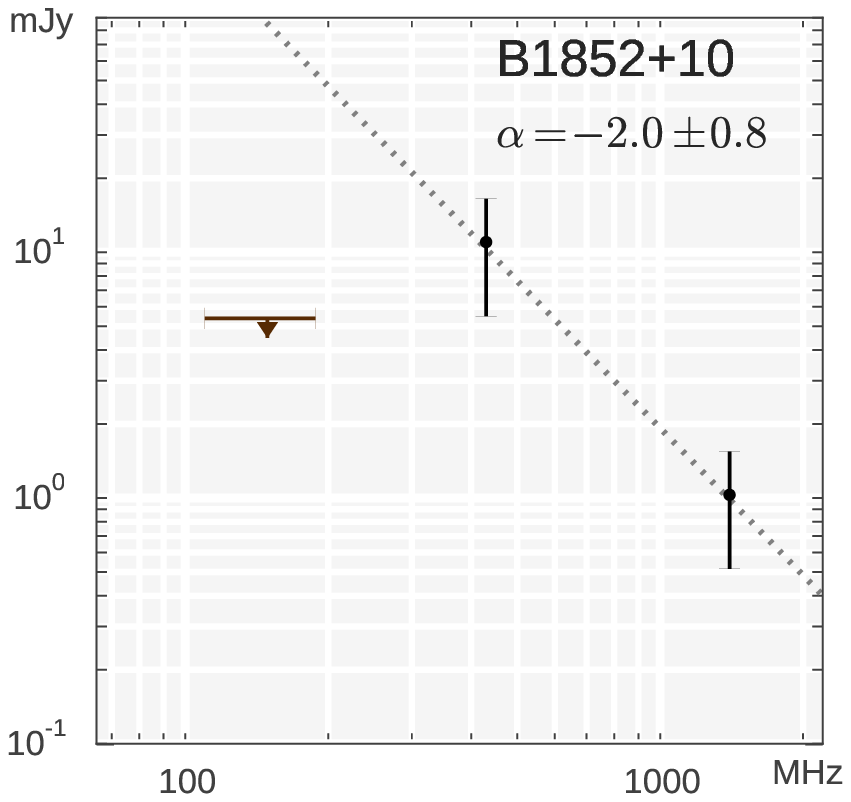}\includegraphics[scale=0.48]{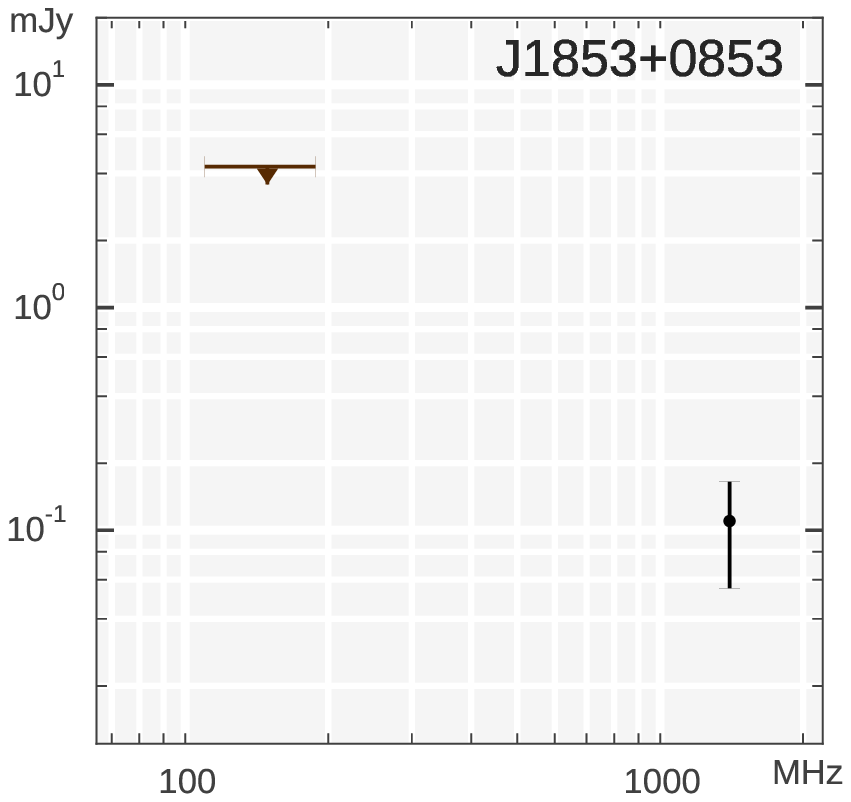}\includegraphics[scale=0.48]{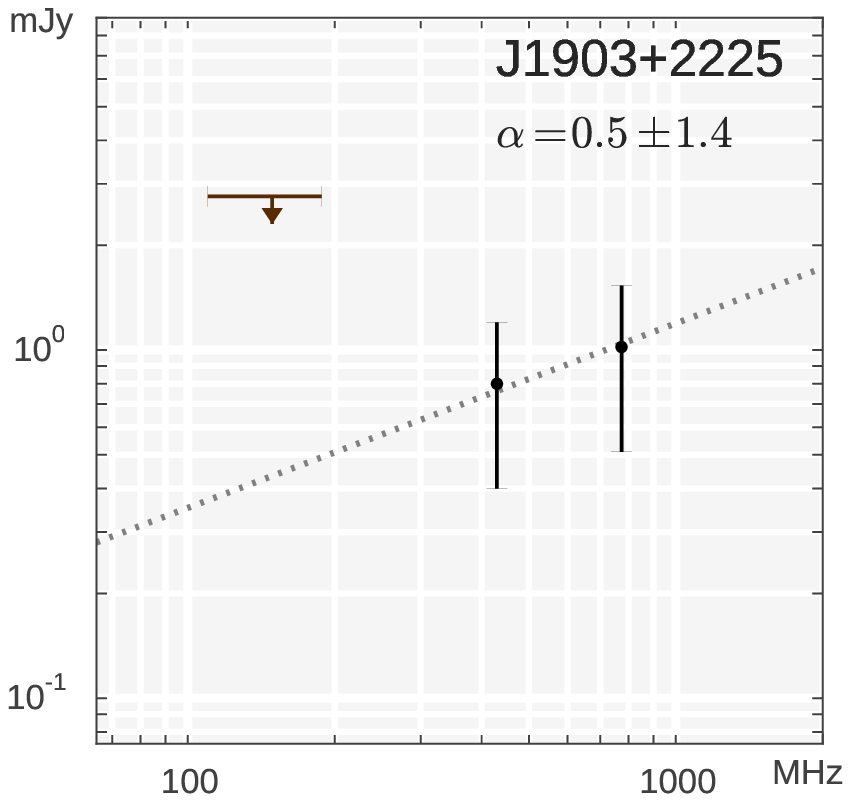}
\includegraphics[scale=0.48]{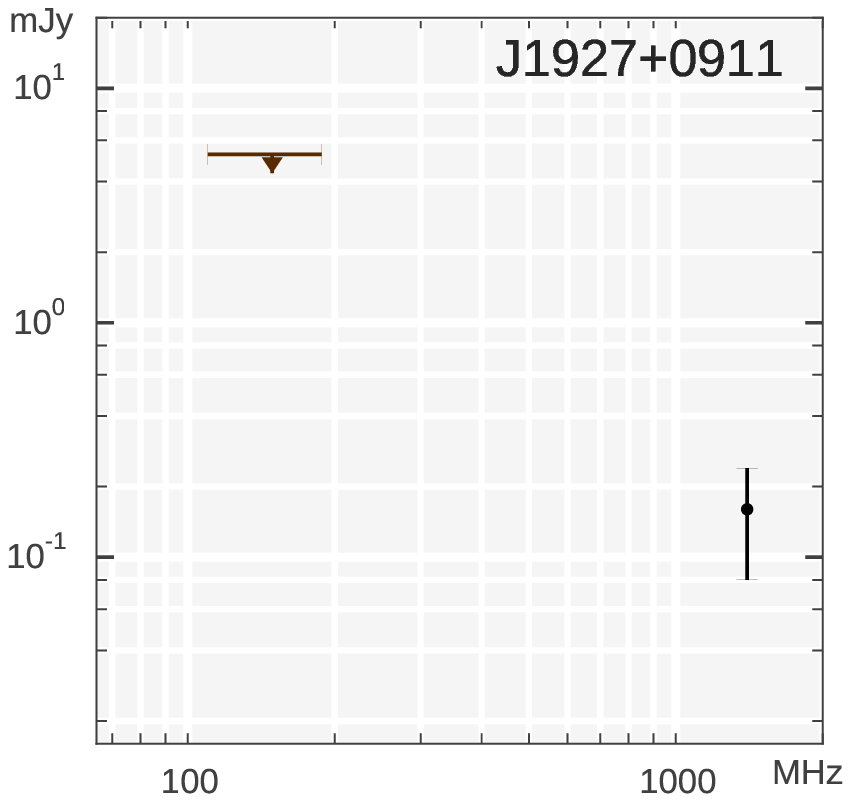}\includegraphics[scale=0.48]{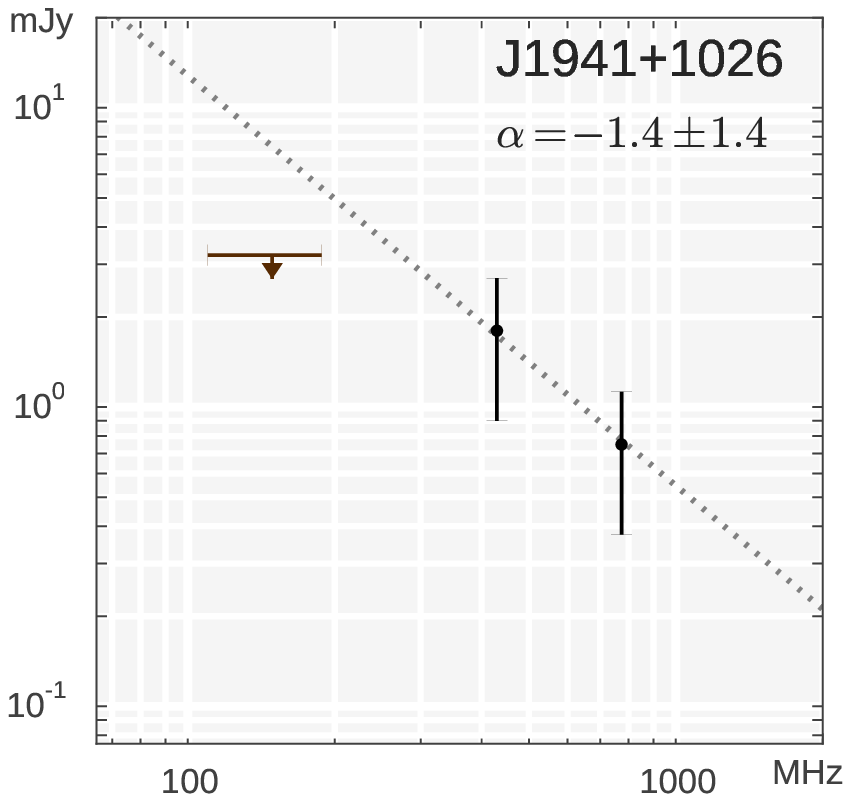}\includegraphics[scale=0.48]{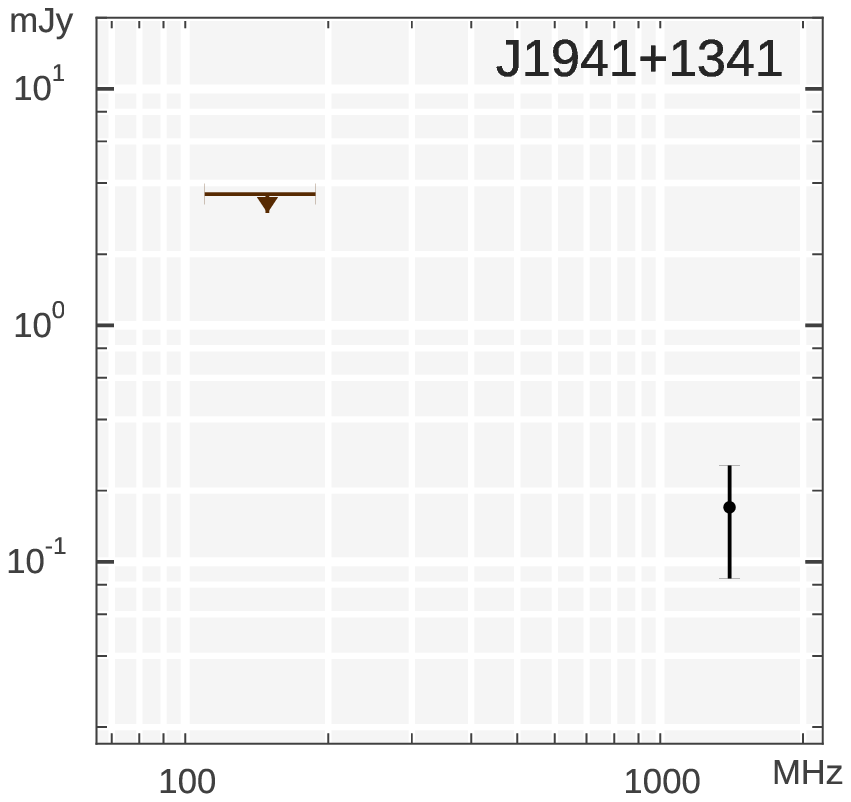}\includegraphics[scale=0.48]{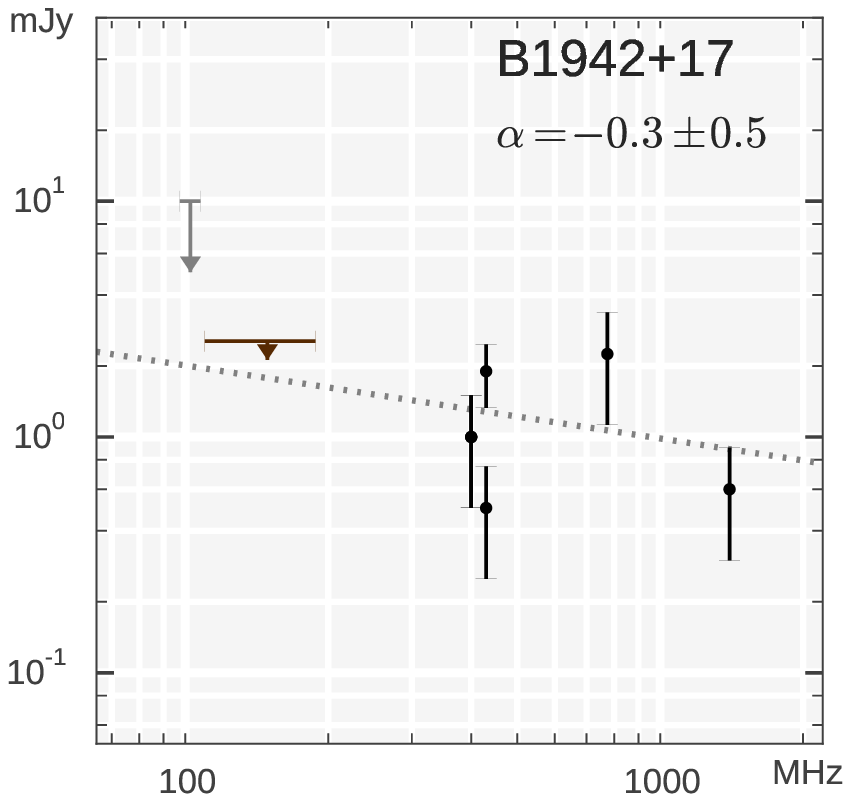}
\includegraphics[scale=0.48]{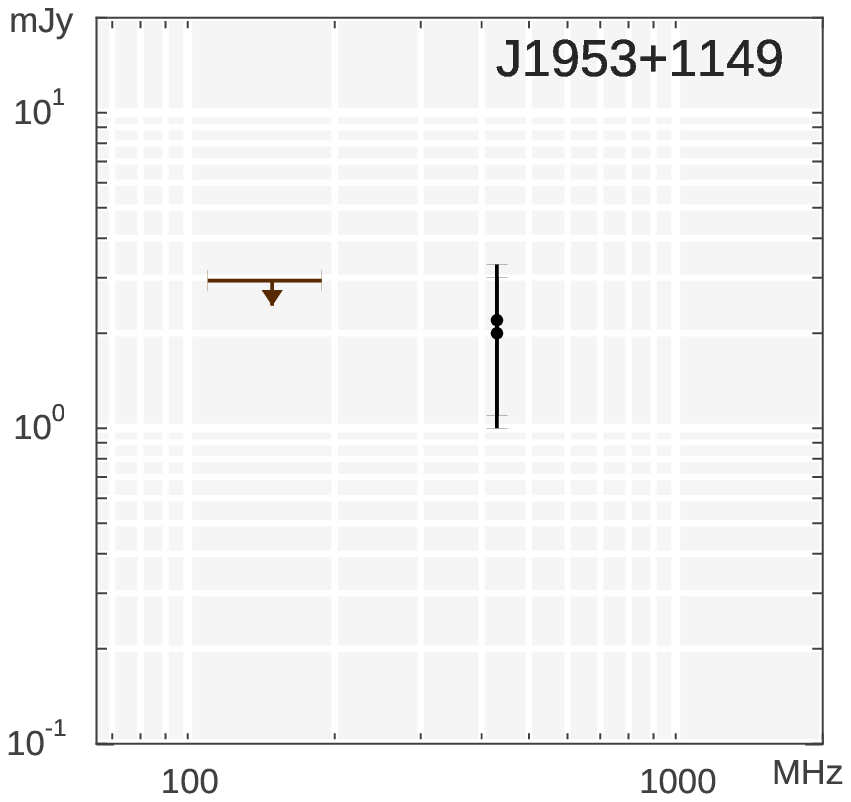}\includegraphics[scale=0.48]{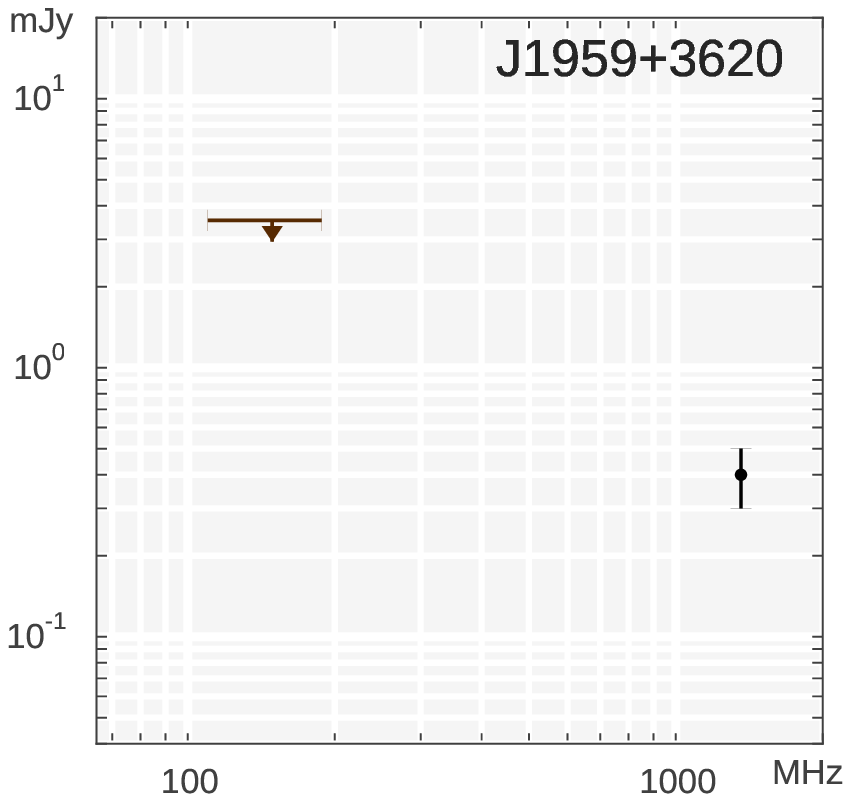}\includegraphics[scale=0.48]{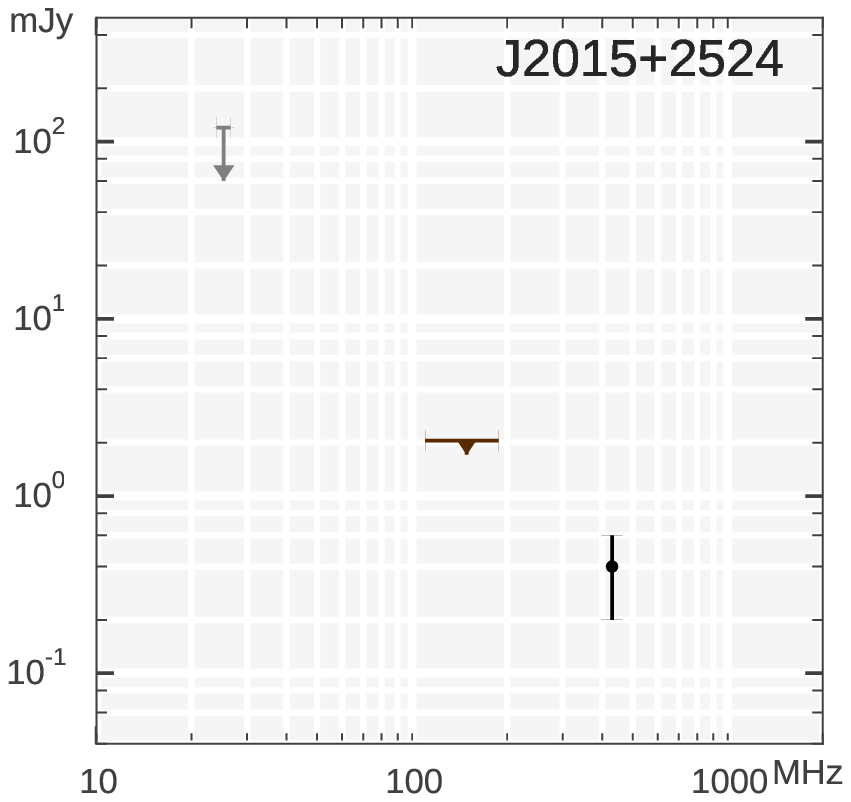}\includegraphics[scale=0.48]{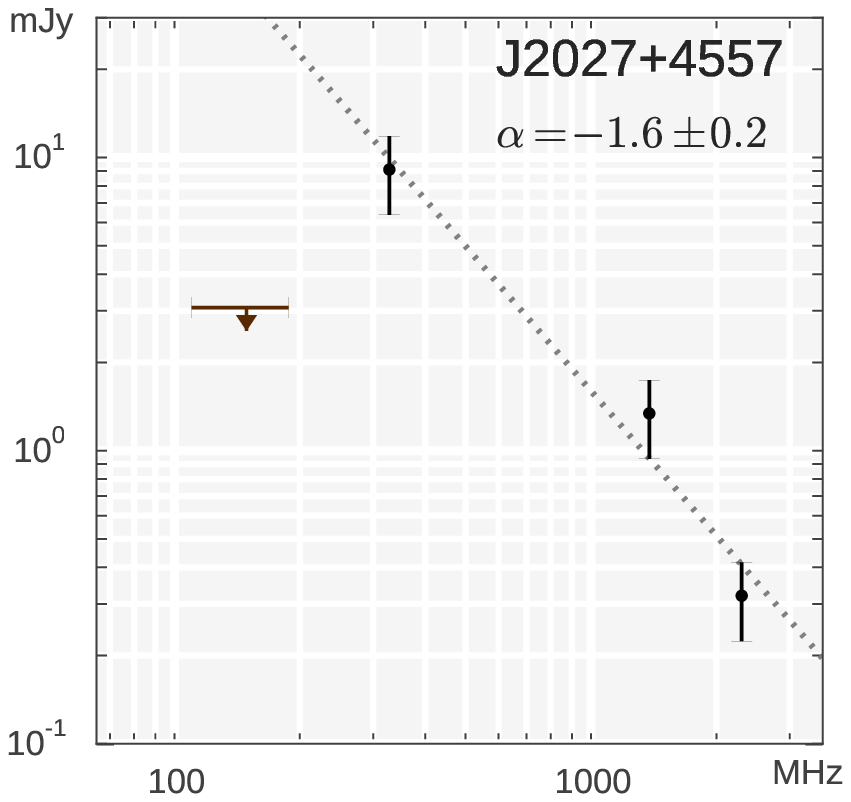}
\includegraphics[scale=0.48]{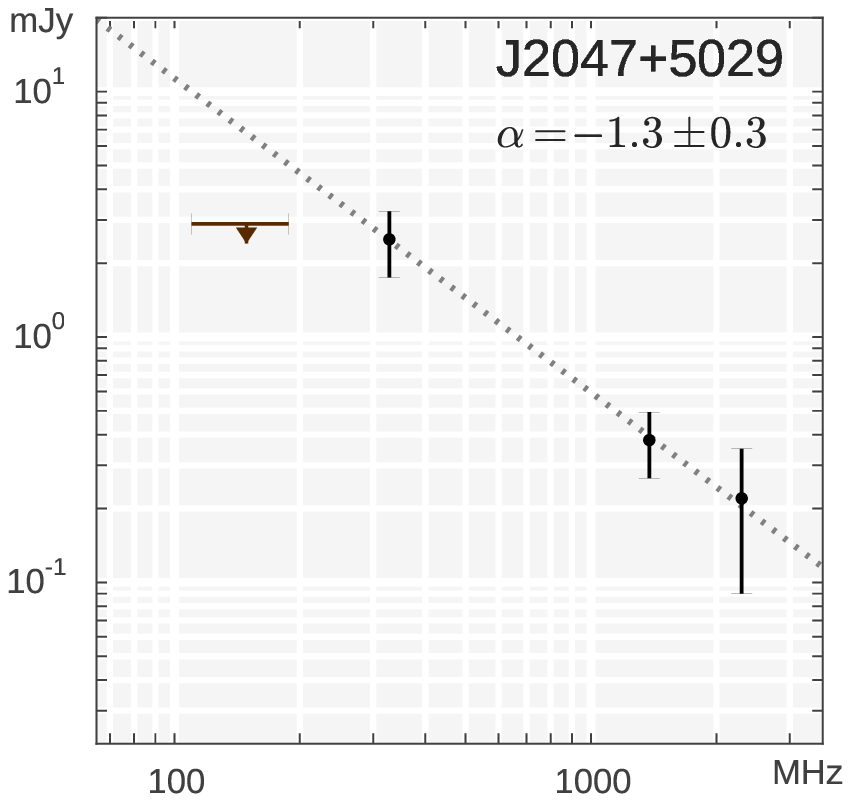}\includegraphics[scale=0.48]{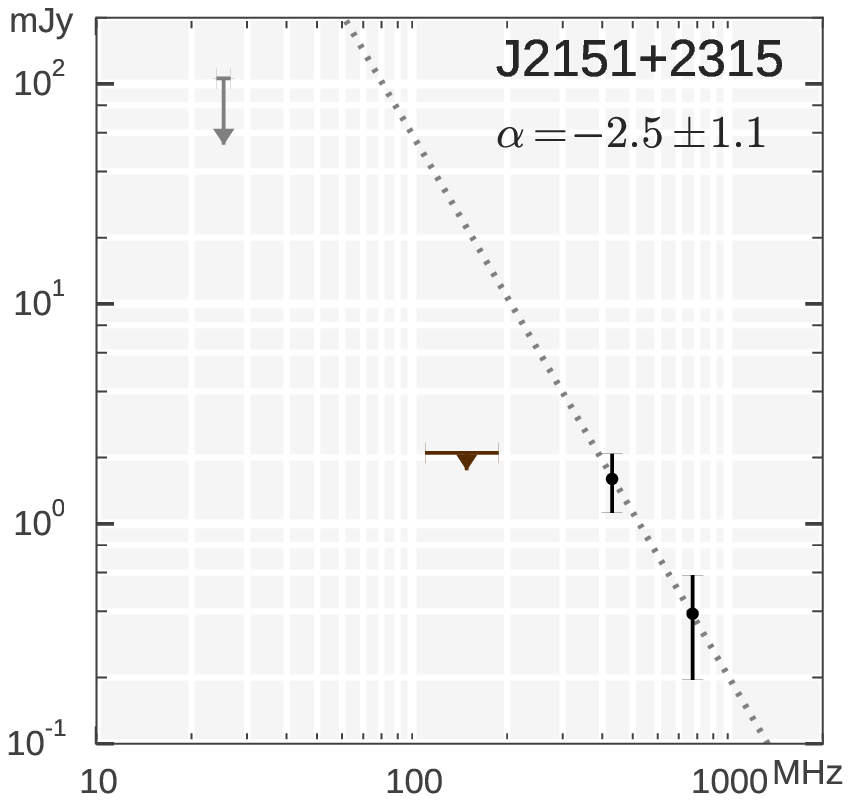}\includegraphics[scale=0.48]{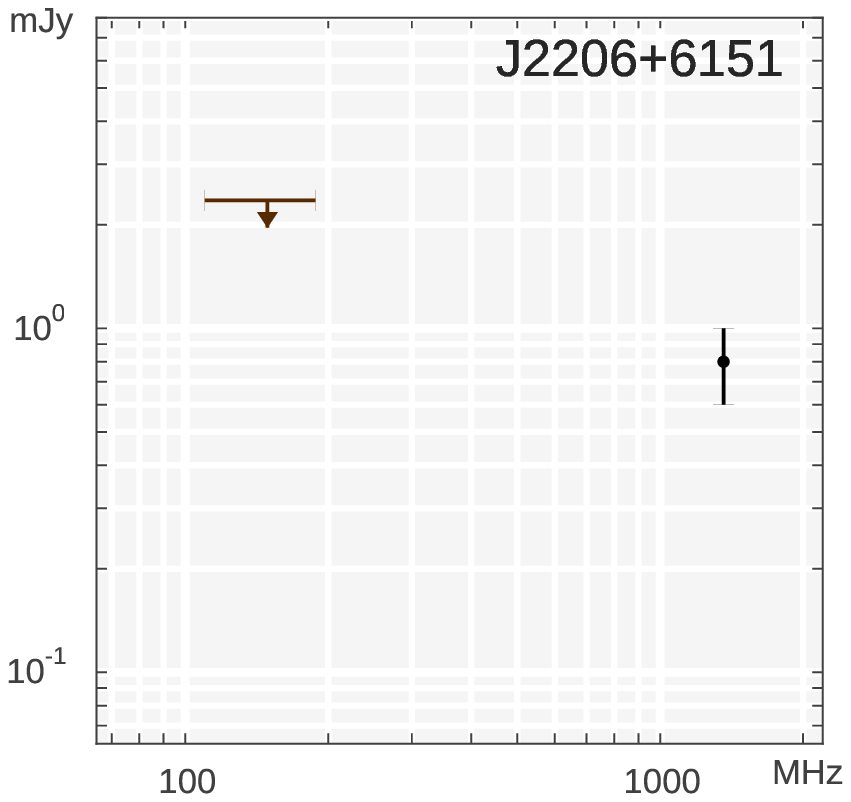}\includegraphics[scale=0.48]{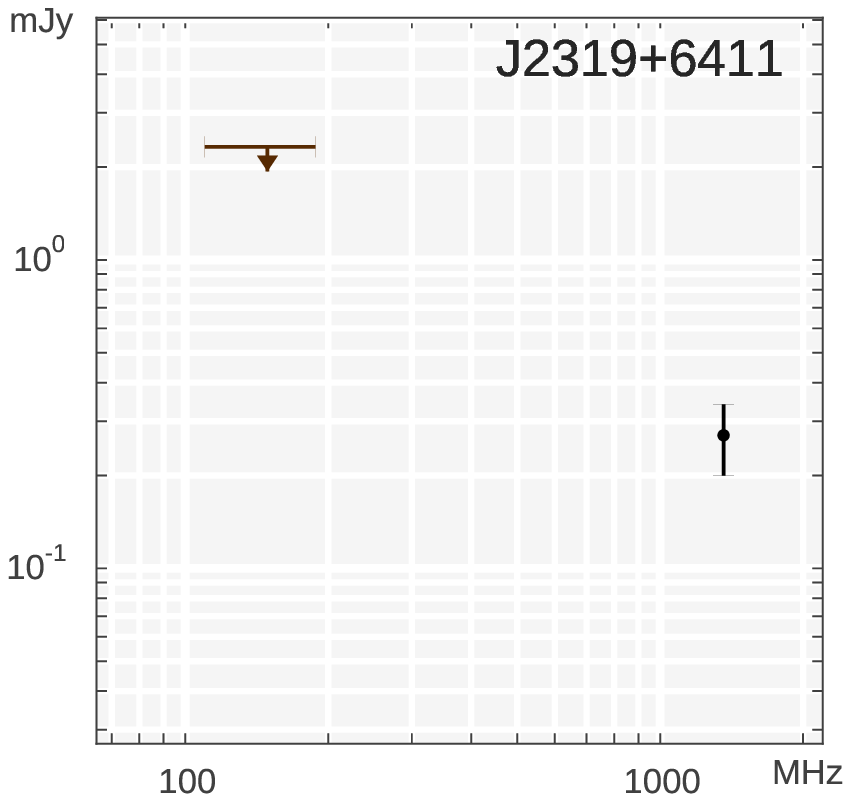}
\caption{Spectra of the 24 pulsars which were not detected in the census observations and have at least one previously reported flux density measurement. Some of these pulsars
had enough literature data points to make a spectral fit. The upper limits on flux densities were not included in such fits. }
\label{fig:spectra_nondet}
\end{figure*}

\begin{figure*}
\centering
\includegraphics[scale=0.440]{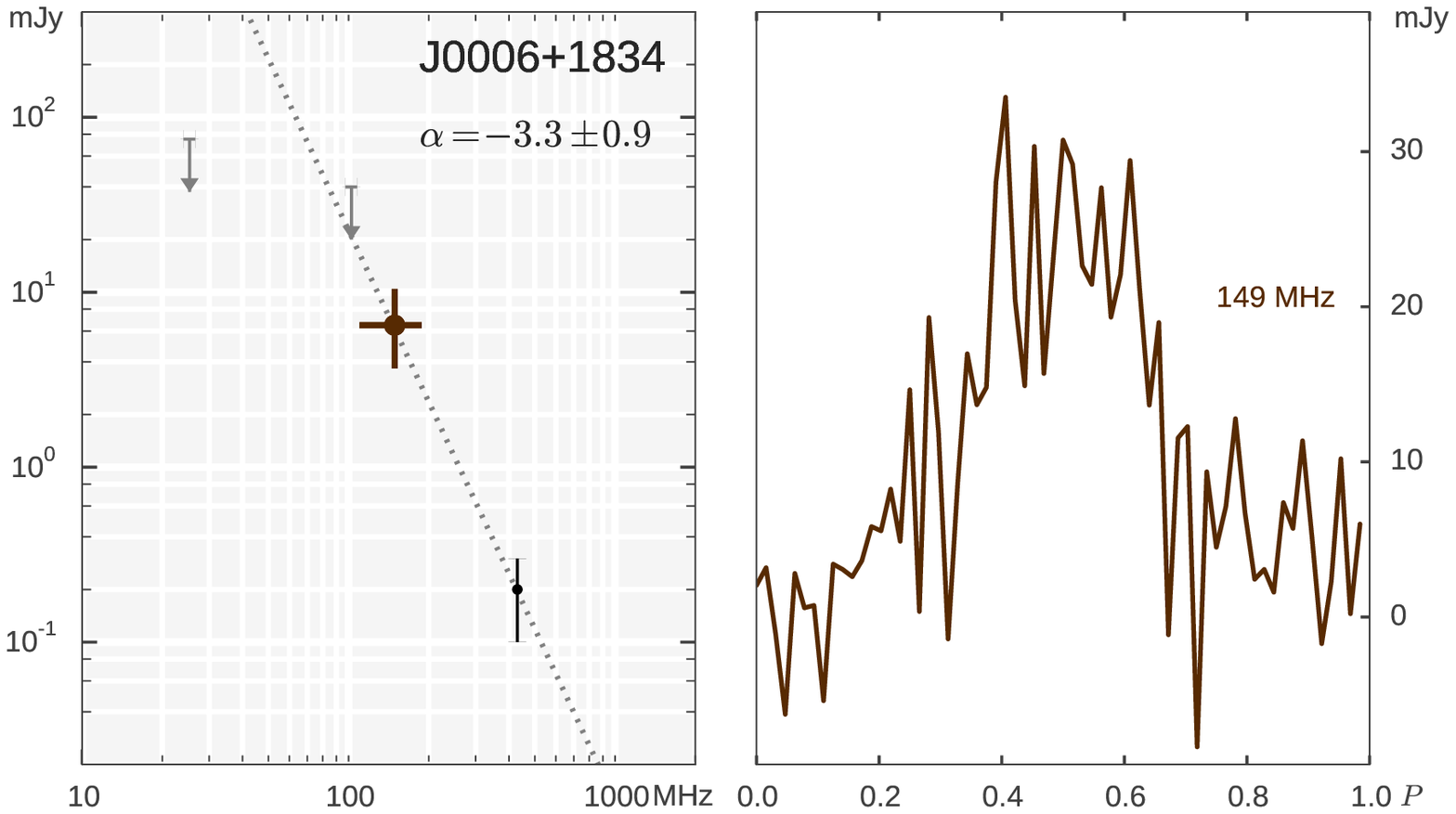}\includegraphics[scale=0.440]{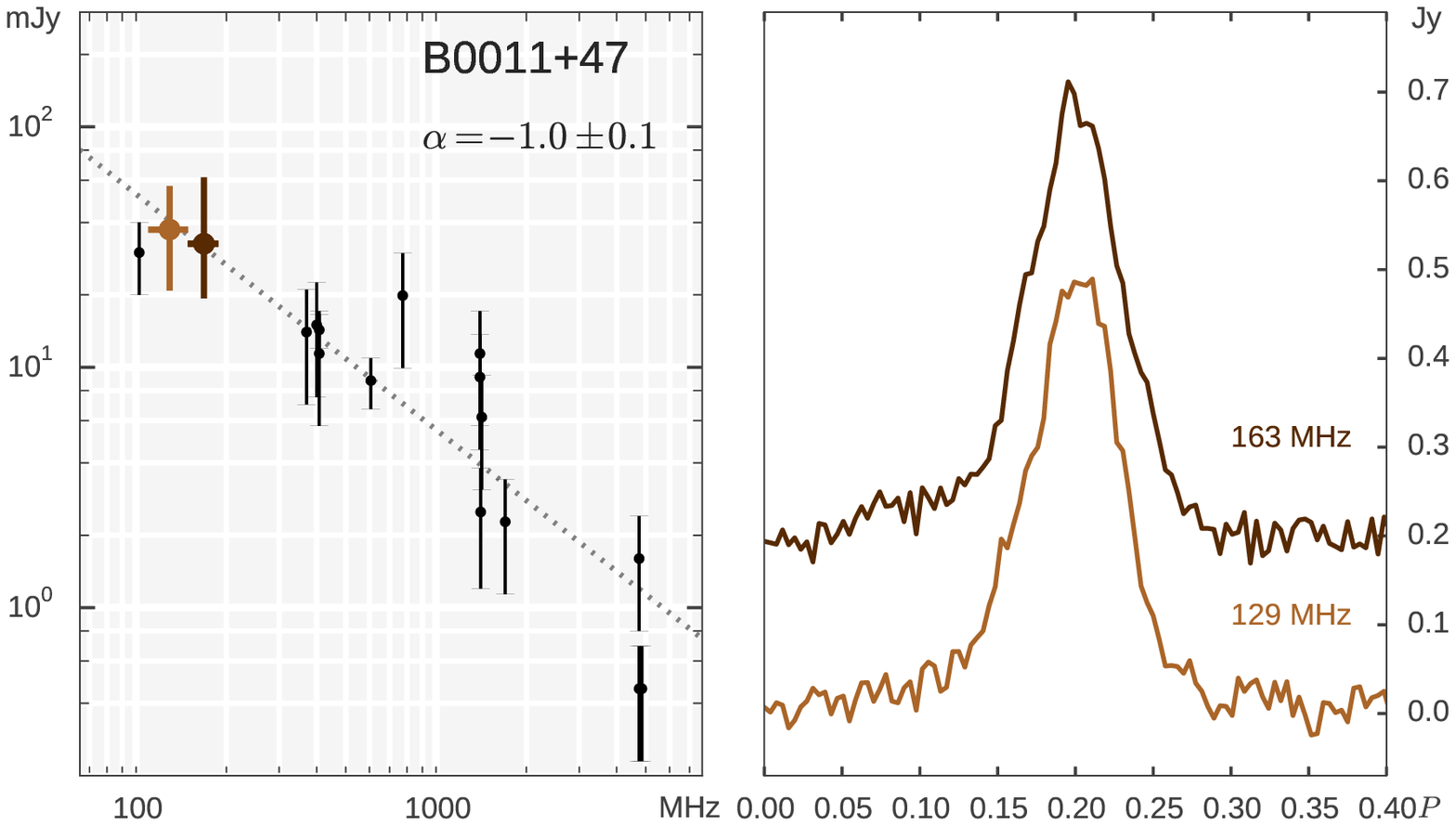}
\includegraphics[scale=0.440]{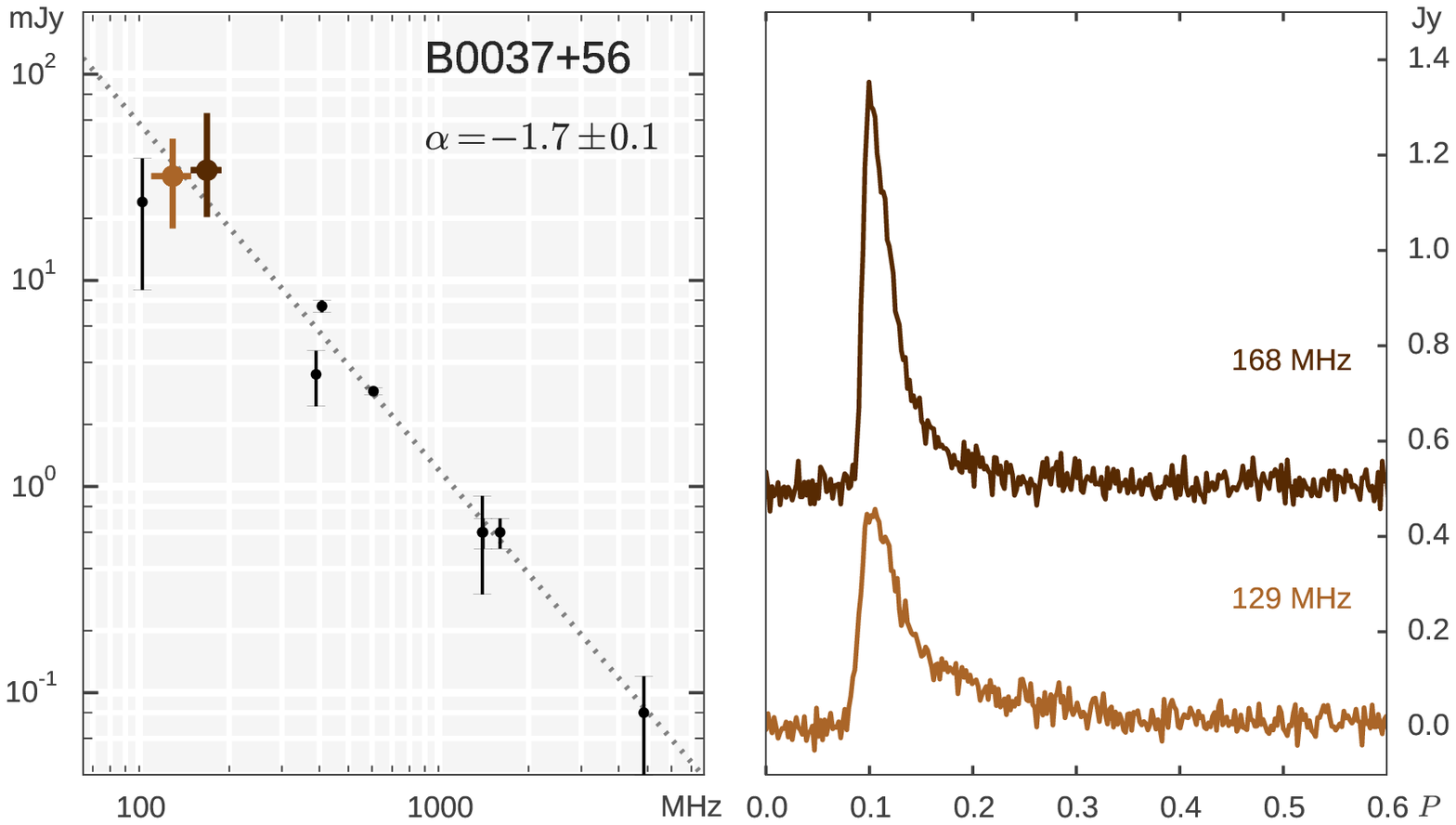}\includegraphics[scale=0.440]{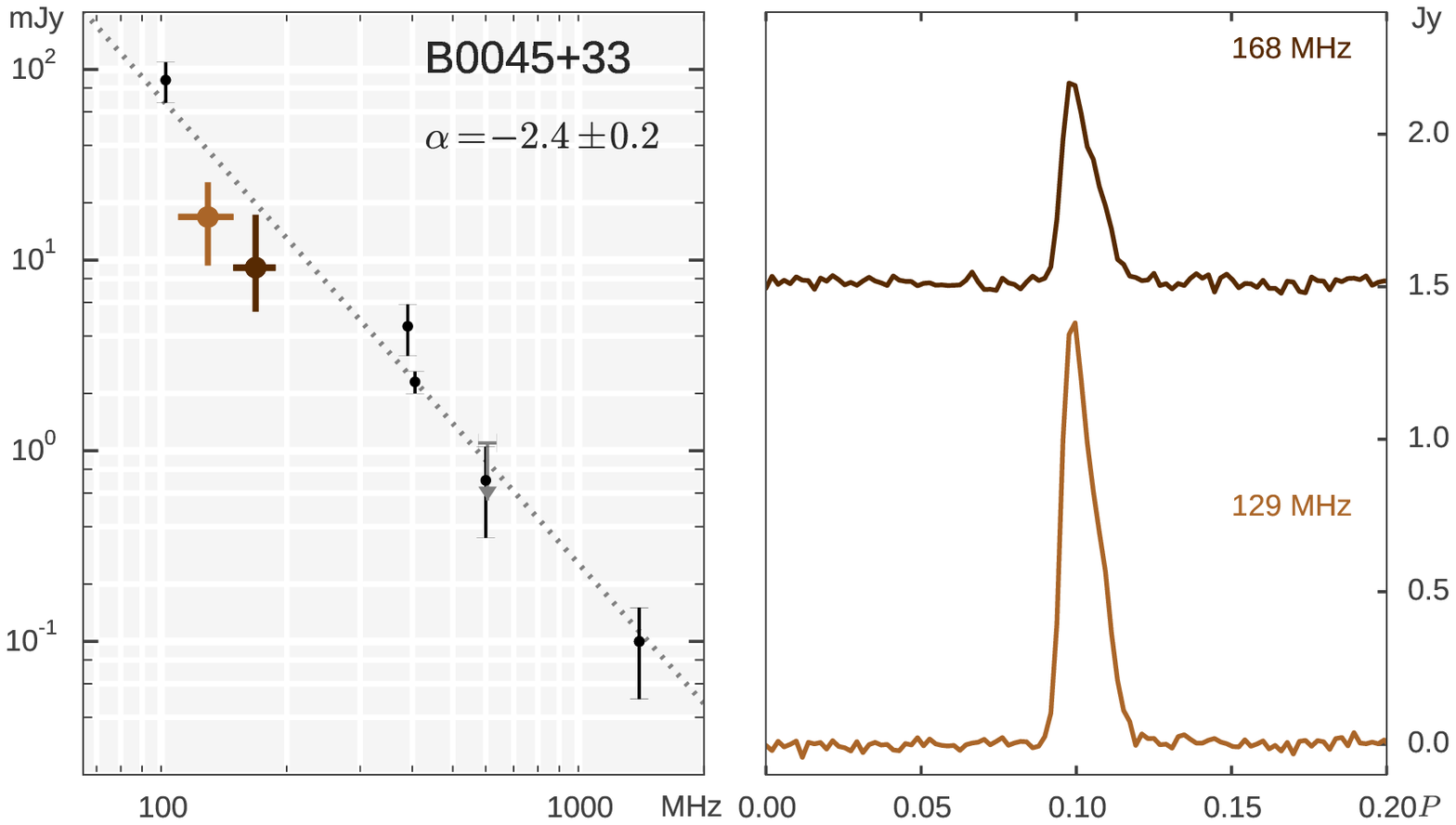}
\includegraphics[scale=0.440]{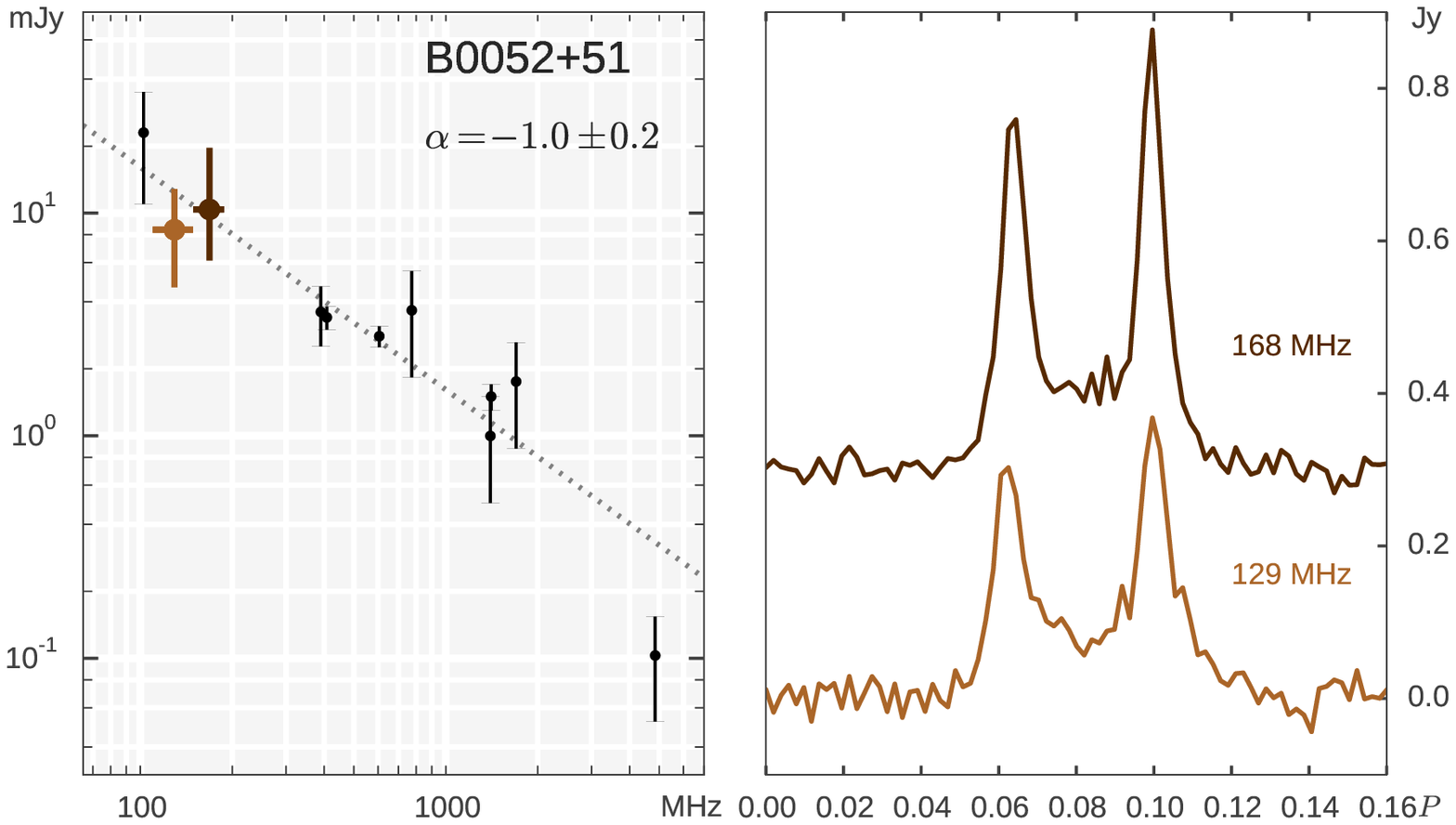}\includegraphics[scale=0.440]{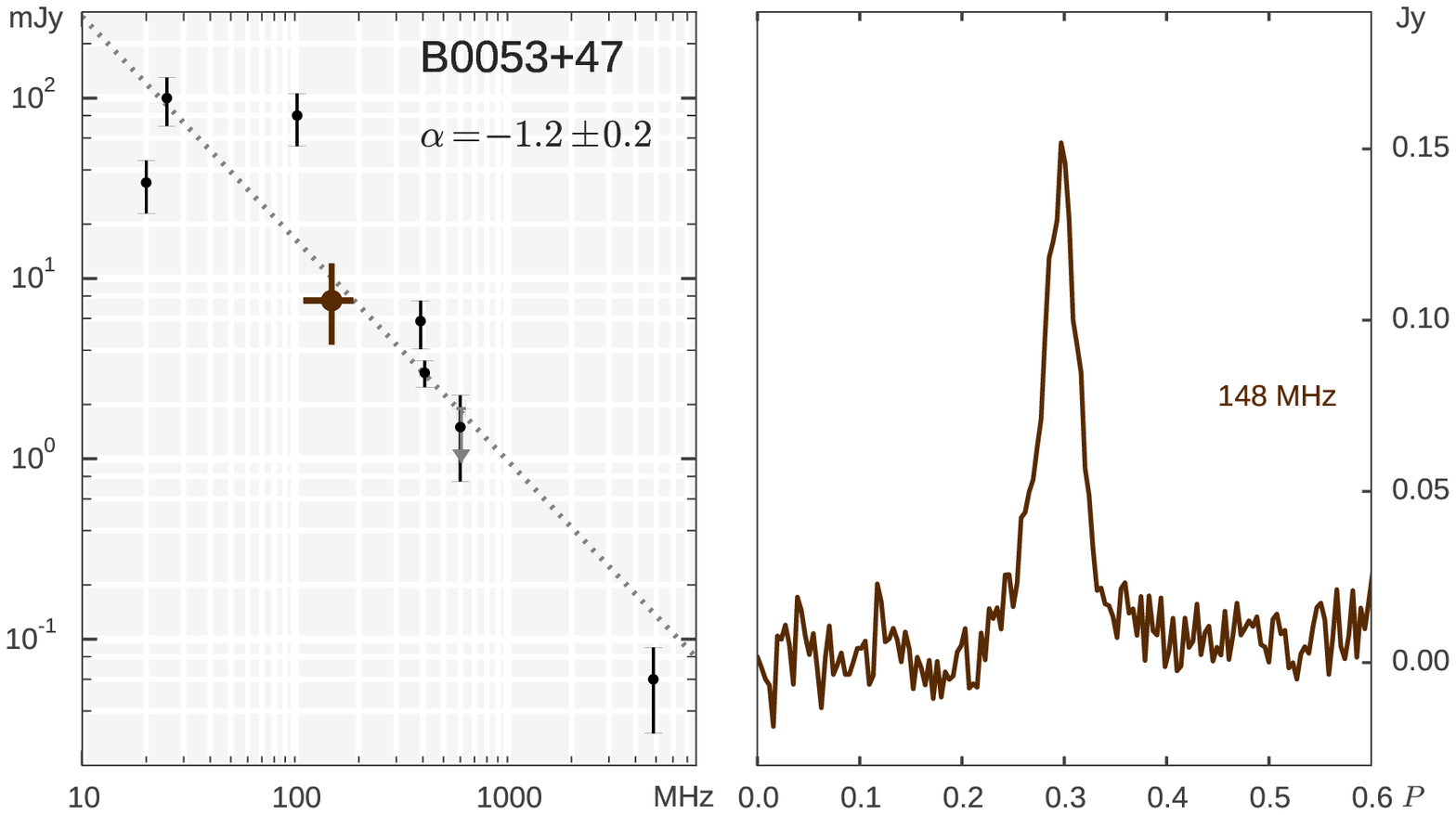}
\includegraphics[scale=0.440]{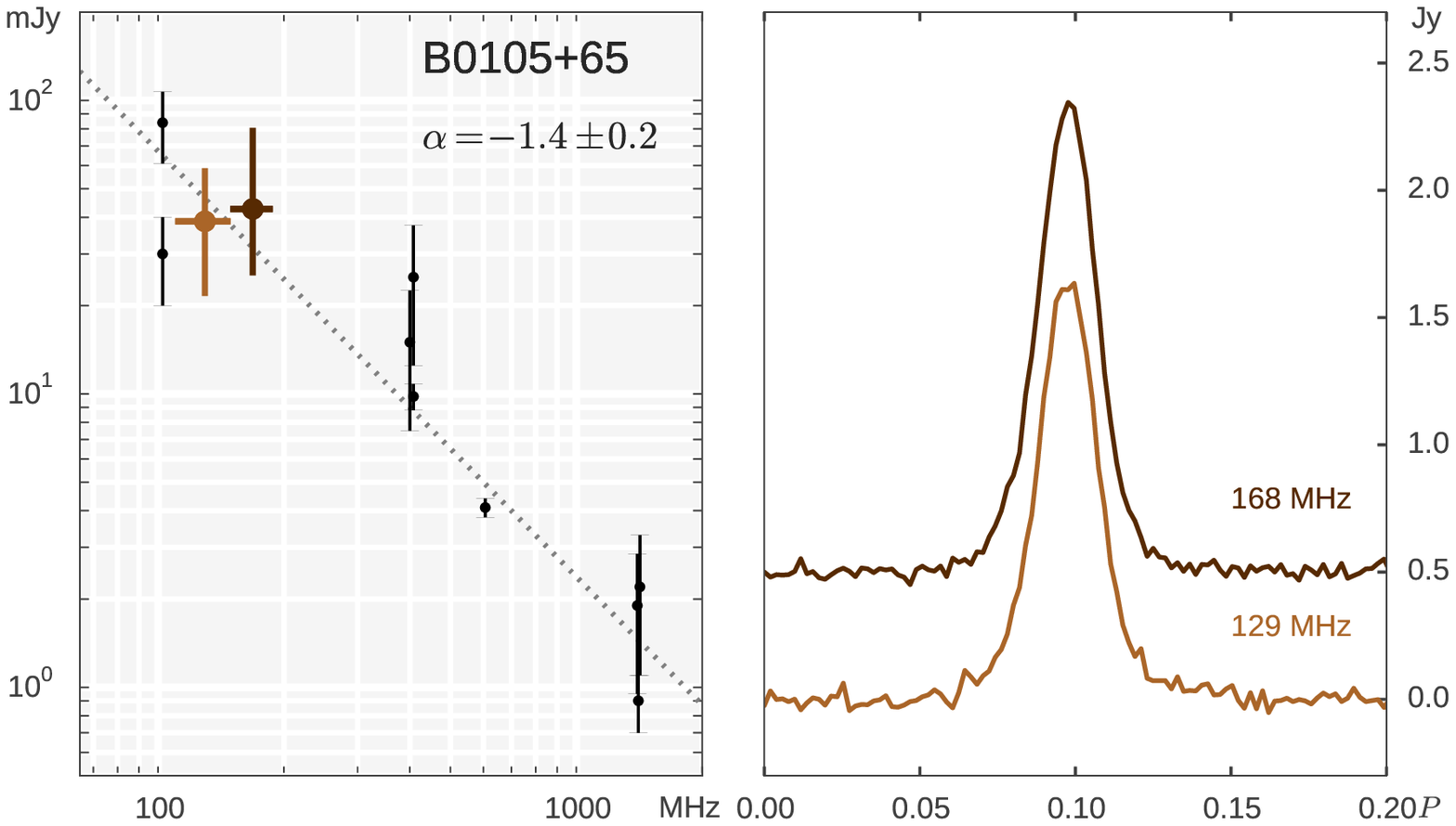}\includegraphics[scale=0.440]{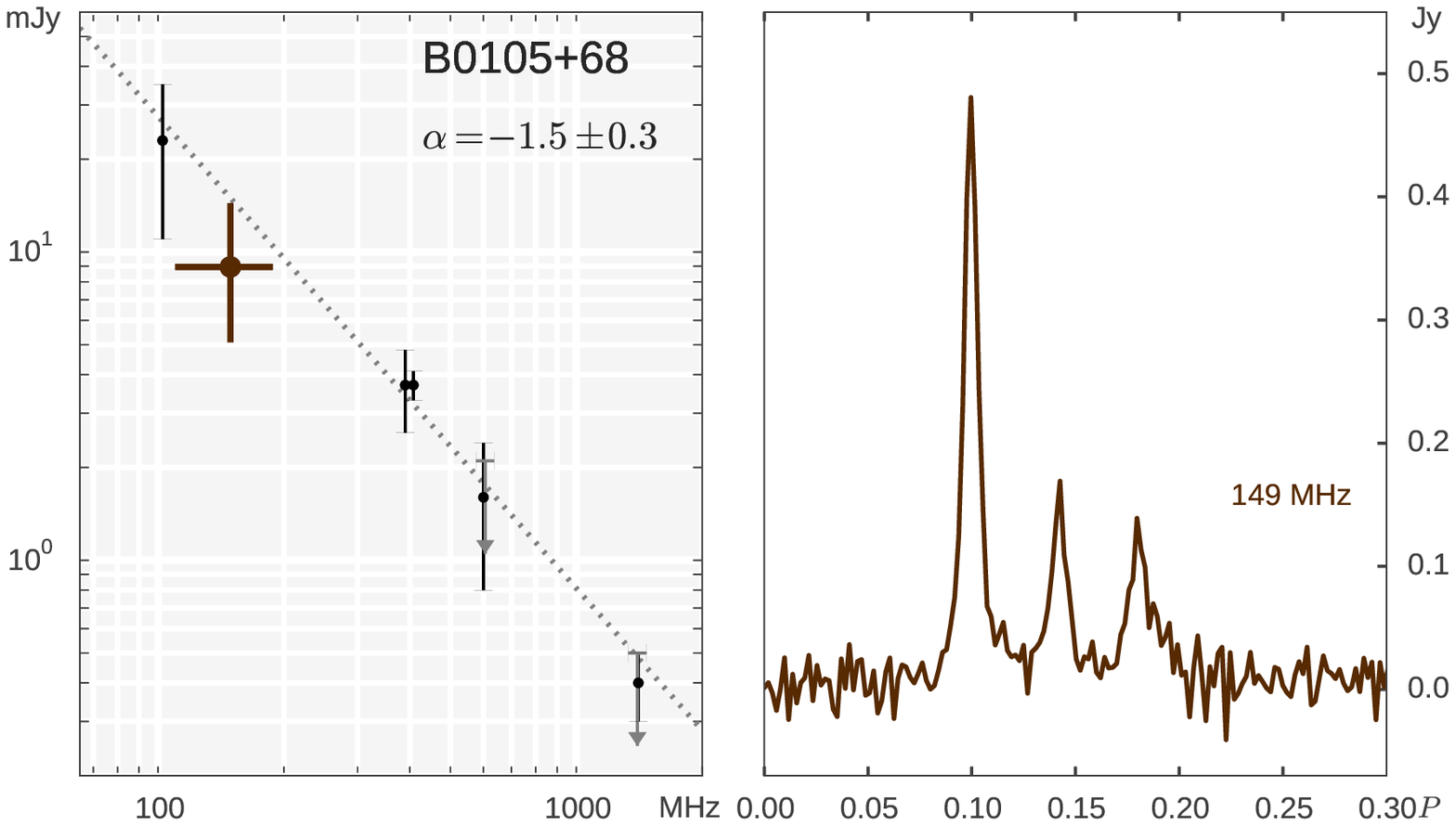}
\includegraphics[scale=0.440]{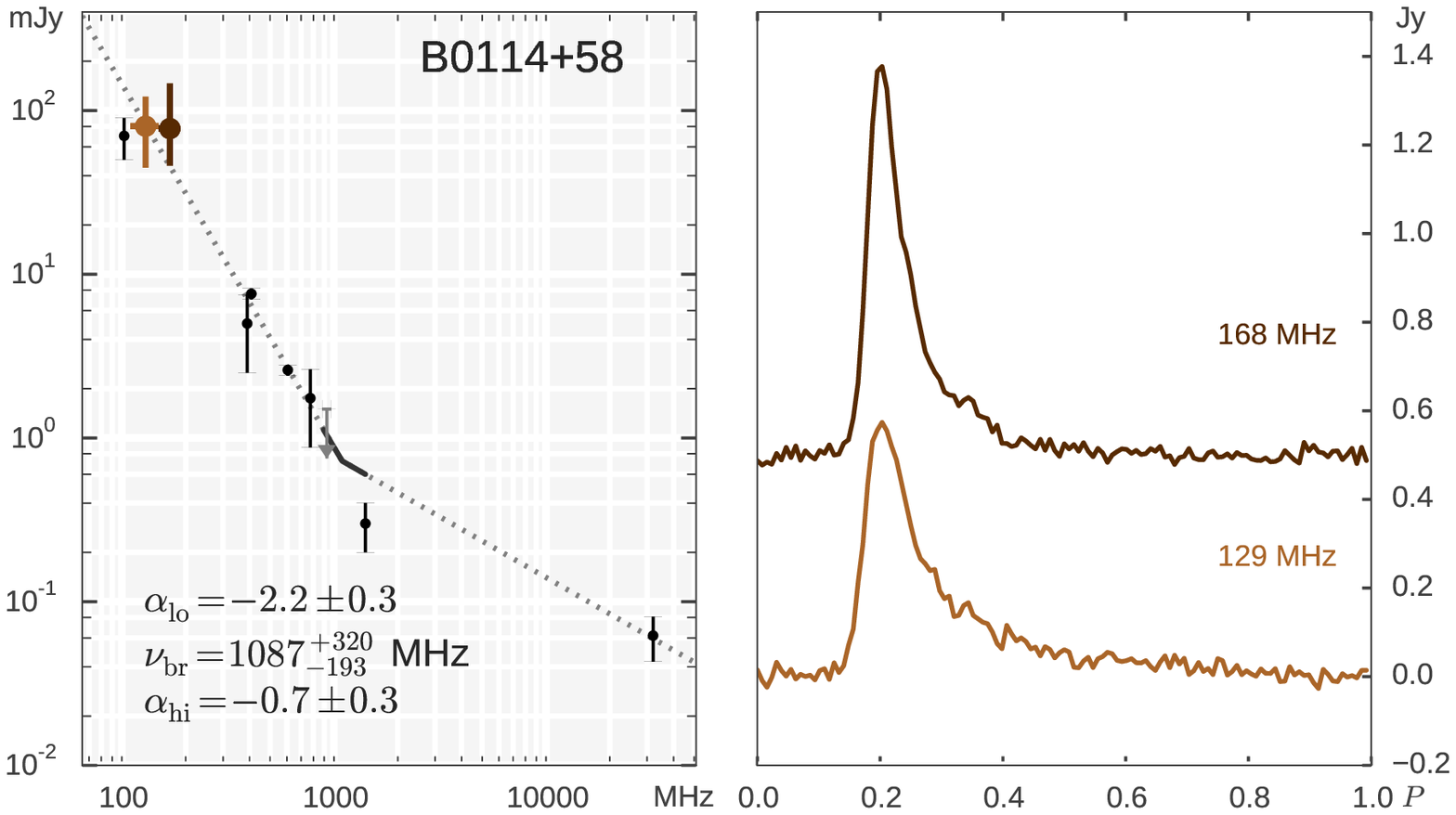}\includegraphics[scale=0.440]{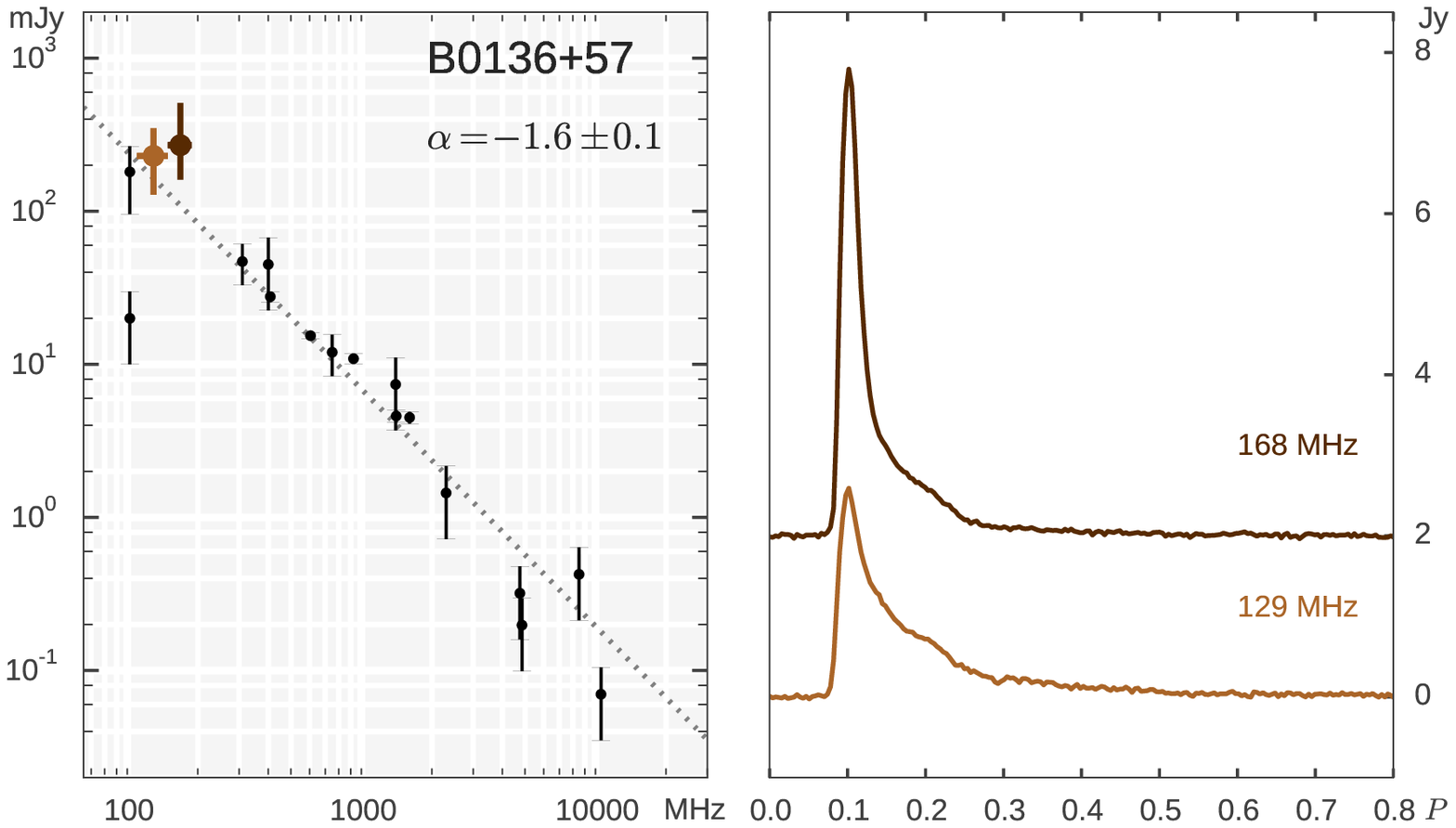}
\caption{For each pair of plots, \textit{Left:} Spectra of radio emission for pulsars detected in the census. Smaller black error bars 
mark literature flux densities, the larger coloured dots indicate the census measurements at various frequencies (with 
the horizontal errorbars indicating the frequency span of a given census measurement). See text for both 
census and literature flux density errors and upper limit discussion. In the case of a multiple-PL fit, the uncertainty on 
break frequency is marked with a broken black line. \textit{Right:} Flux-calibrated average profiles for census observations 
(only the manually selected on-pulse region is shown). Multiple profiles per band are shown with a constant flux offset between separate sub-bands.
The choice of the number of sub-bands was influenced by the peak S/N ratio of the average profile, the presence of 
profile evolution within the observing band and the number of previously published flux density values.}
\label{fig:prof_sp_1}
\end{figure*}

\begin{figure*}
\includegraphics[scale=0.48]{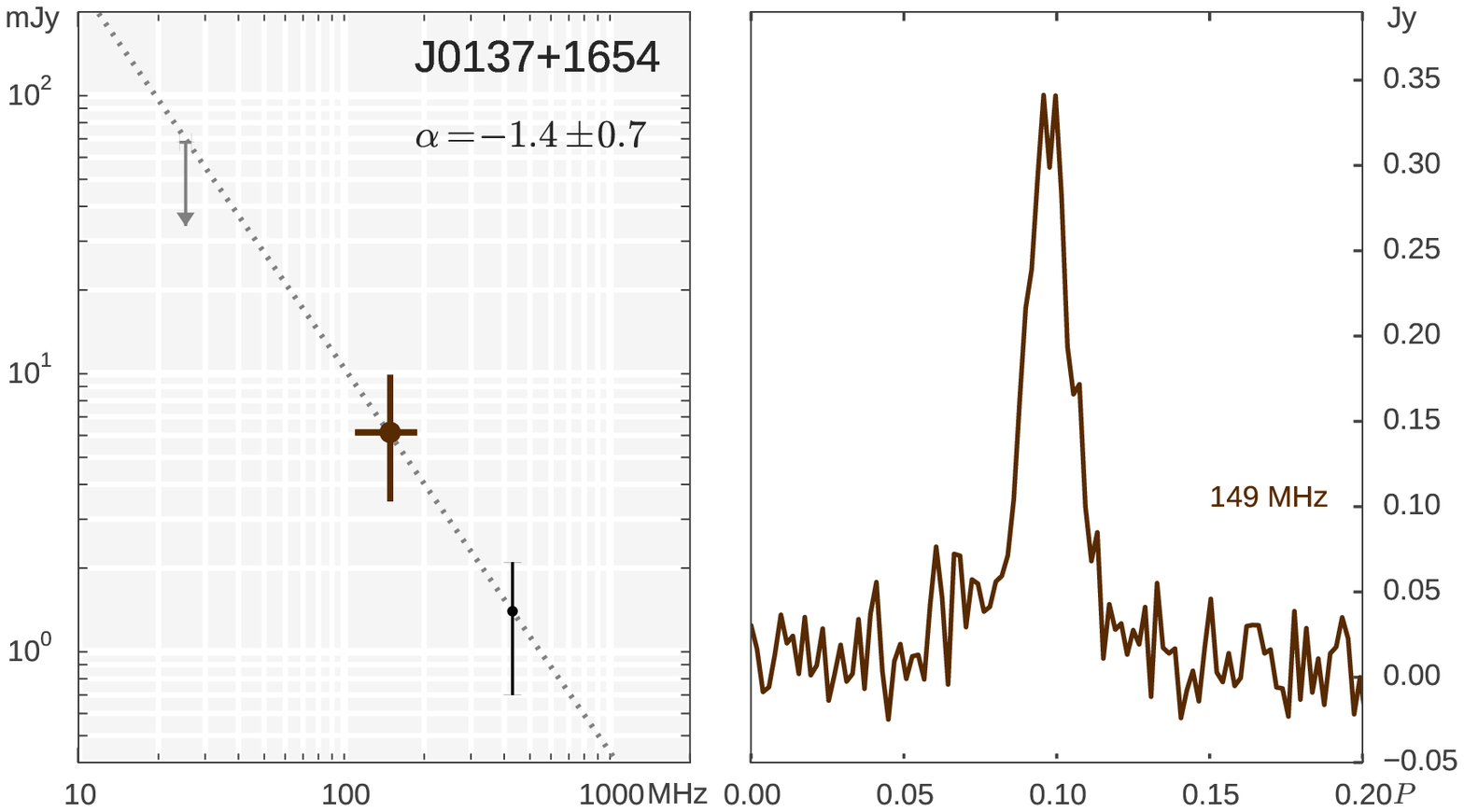}\includegraphics[scale=0.48]{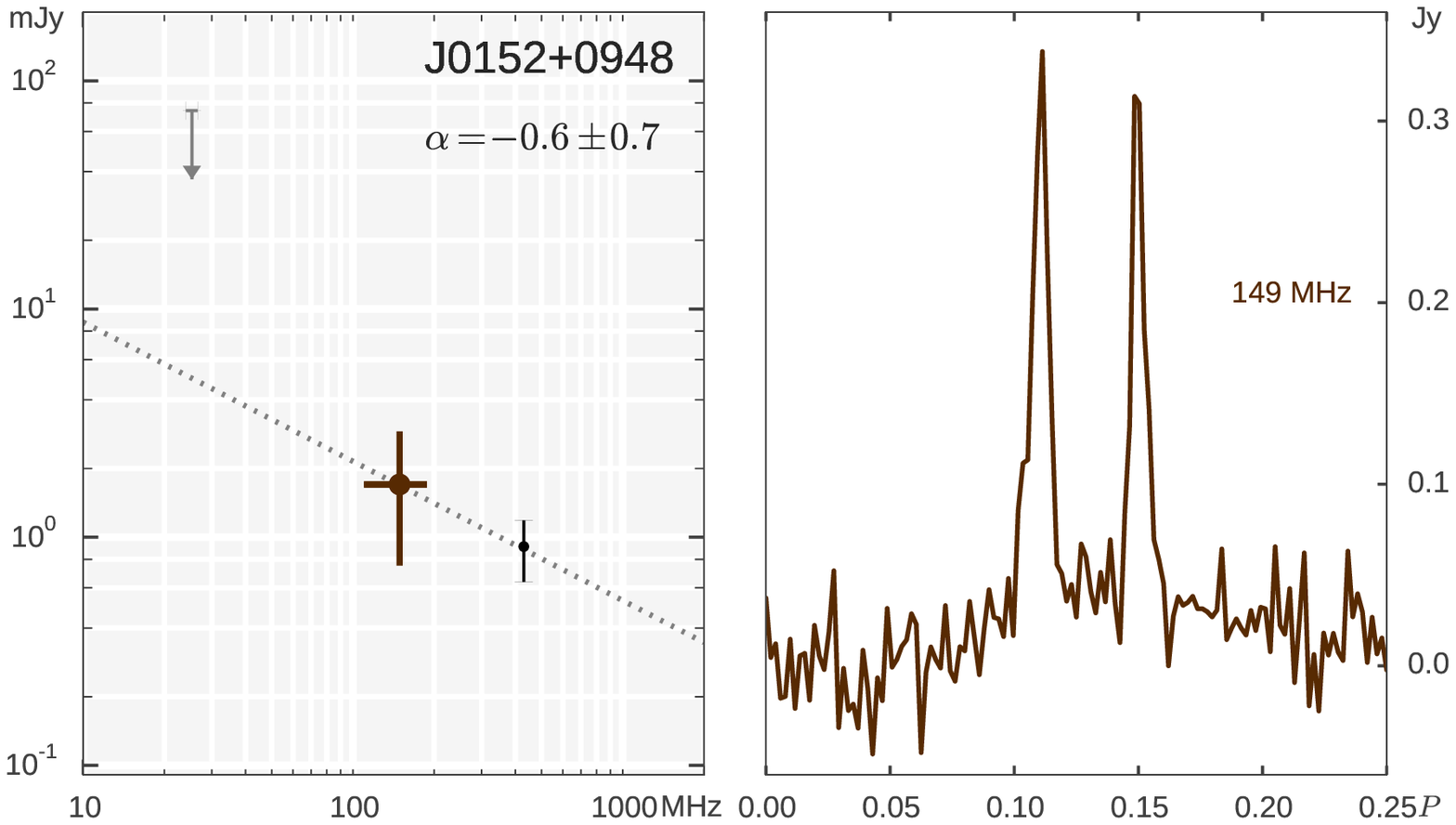}
\includegraphics[scale=0.48]{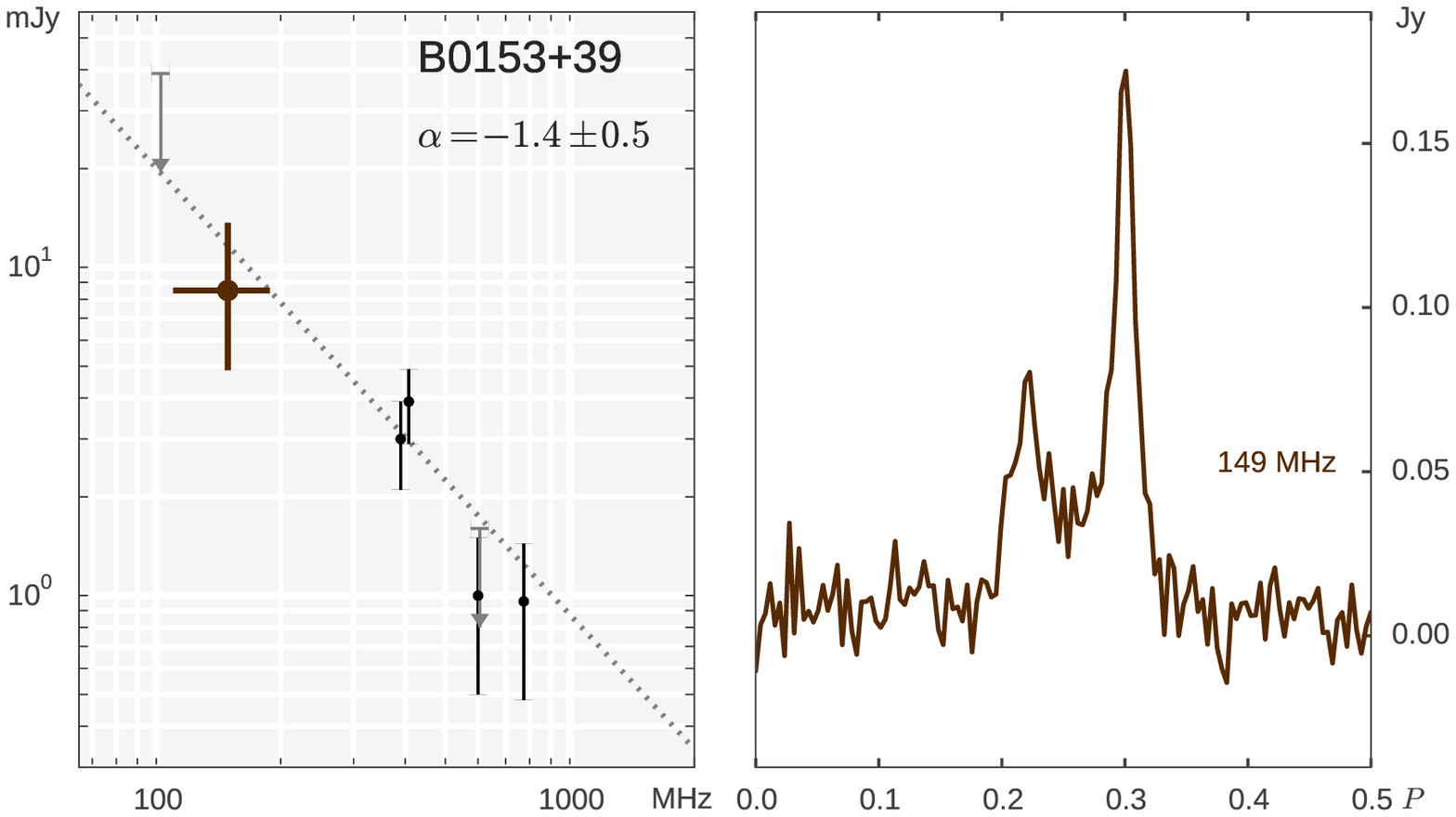}\includegraphics[scale=0.48]{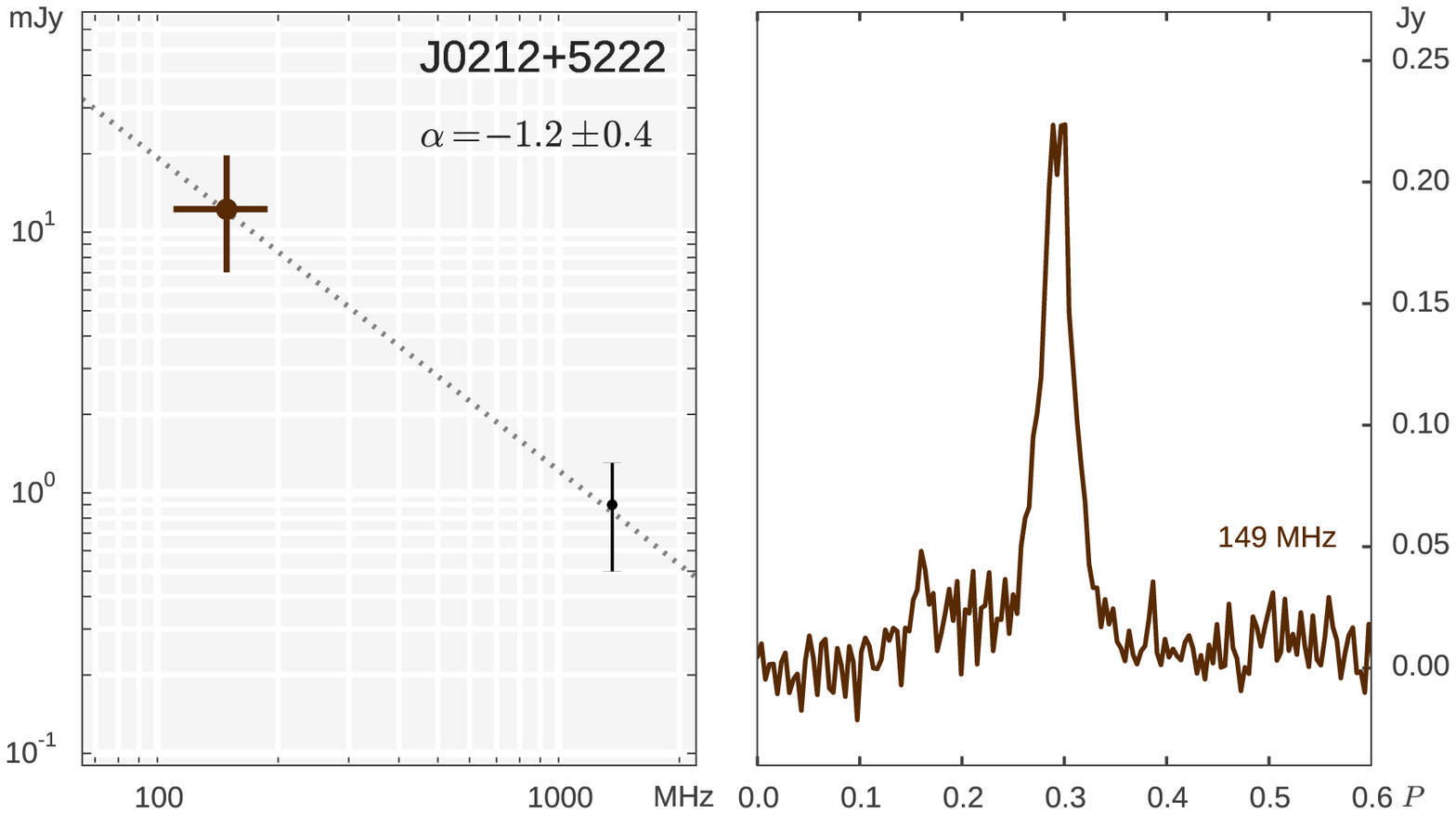}
\includegraphics[scale=0.48]{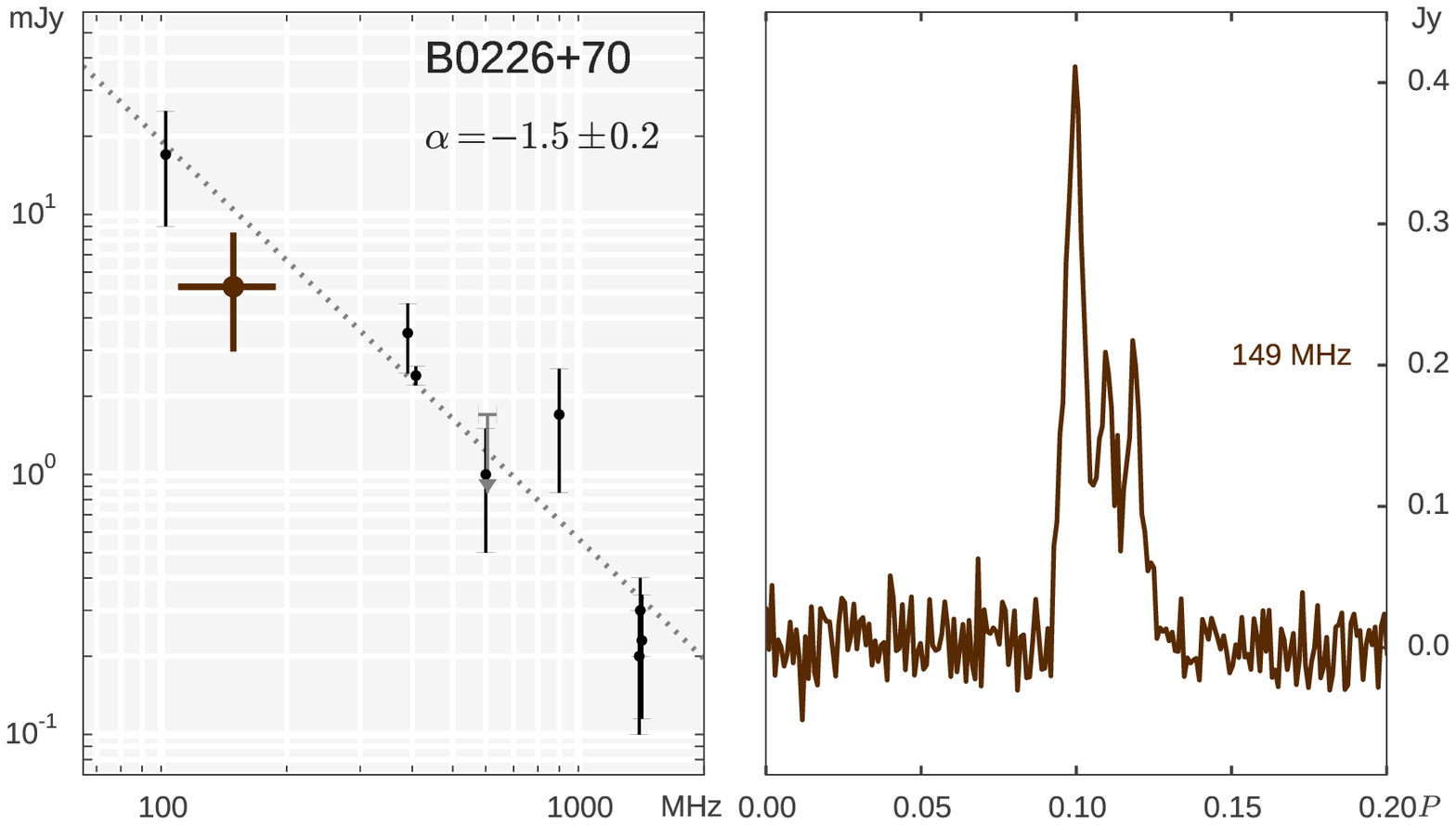}\includegraphics[scale=0.48]{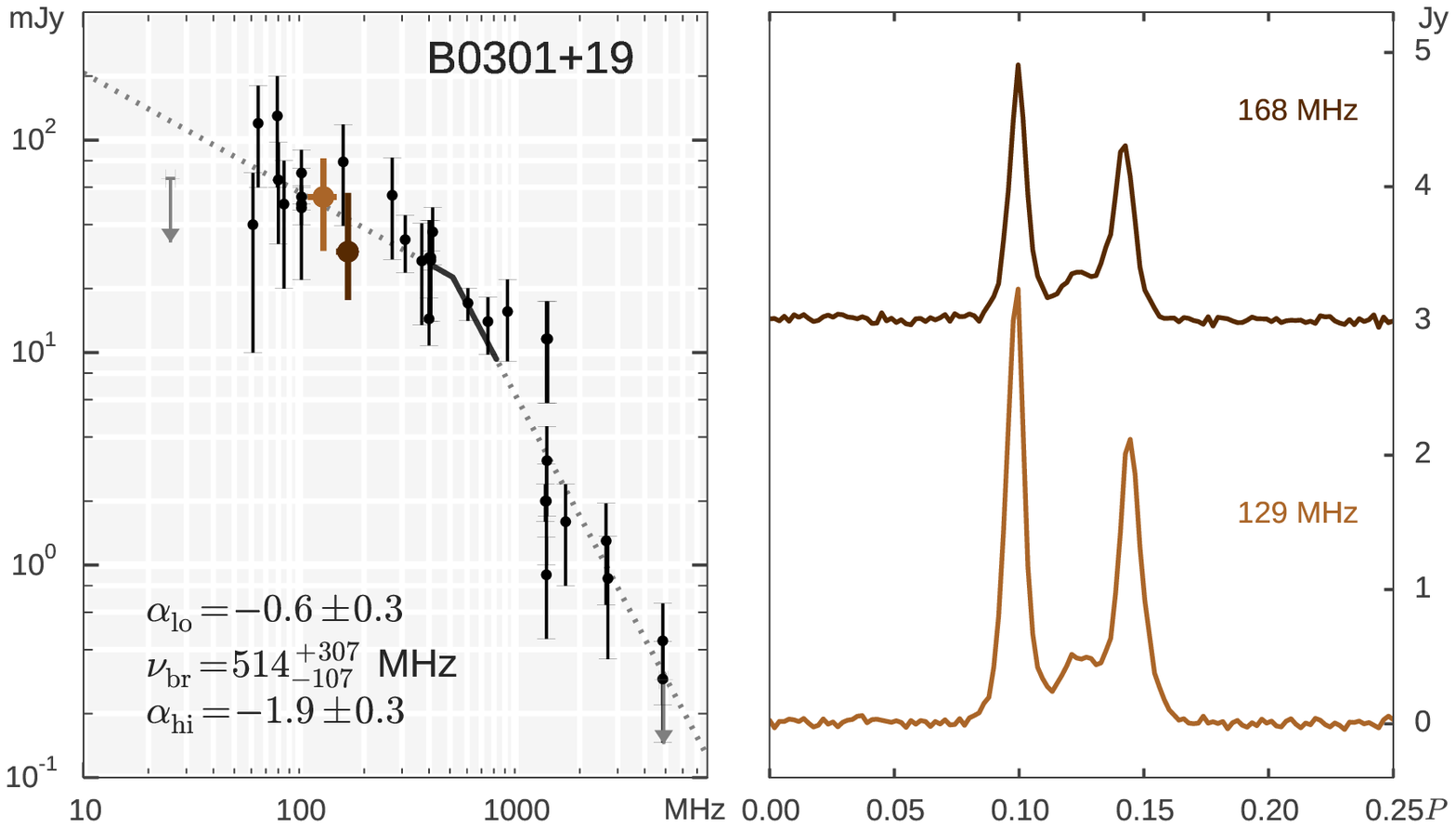}
\includegraphics[scale=0.48]{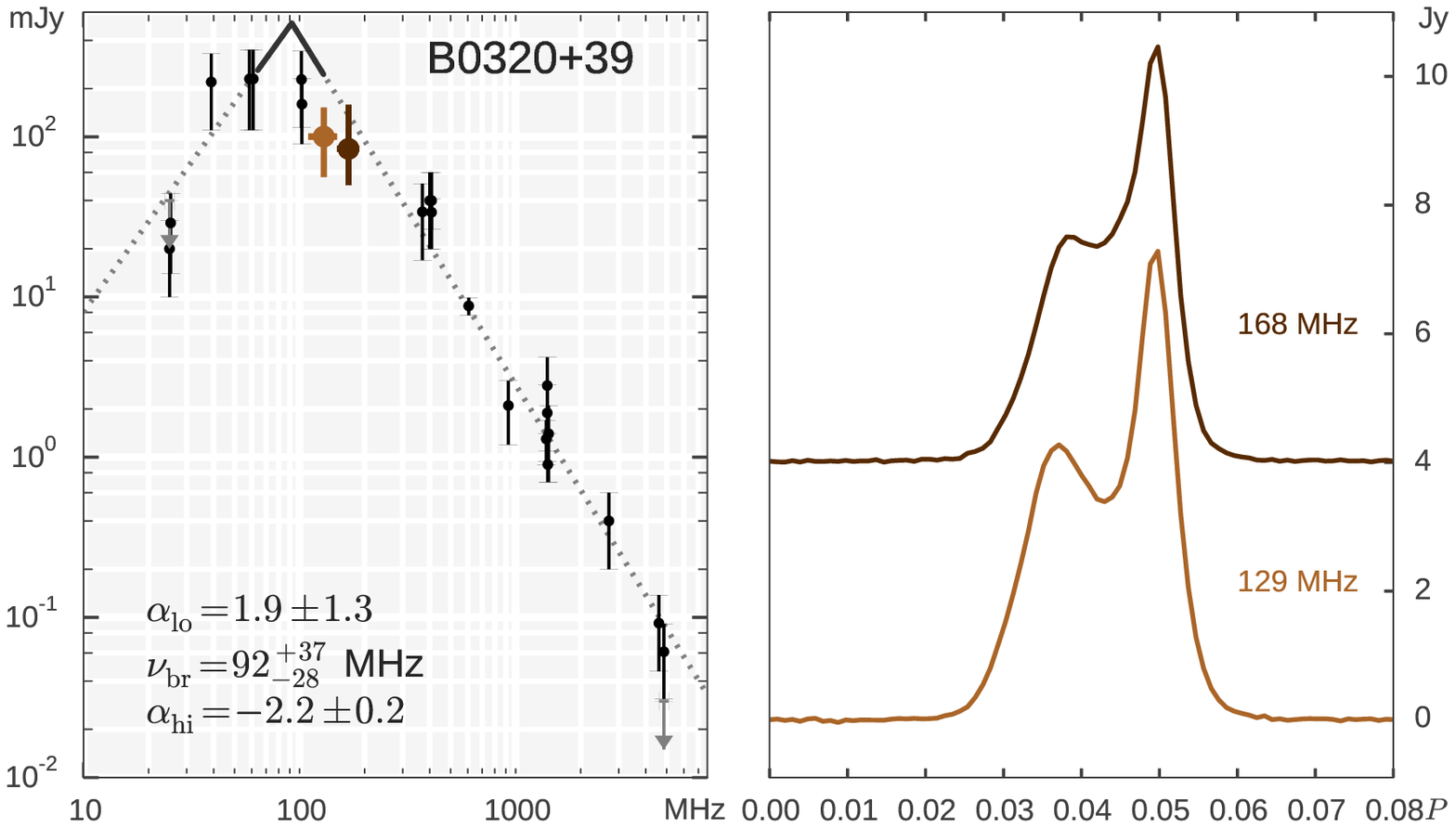}\includegraphics[scale=0.48]{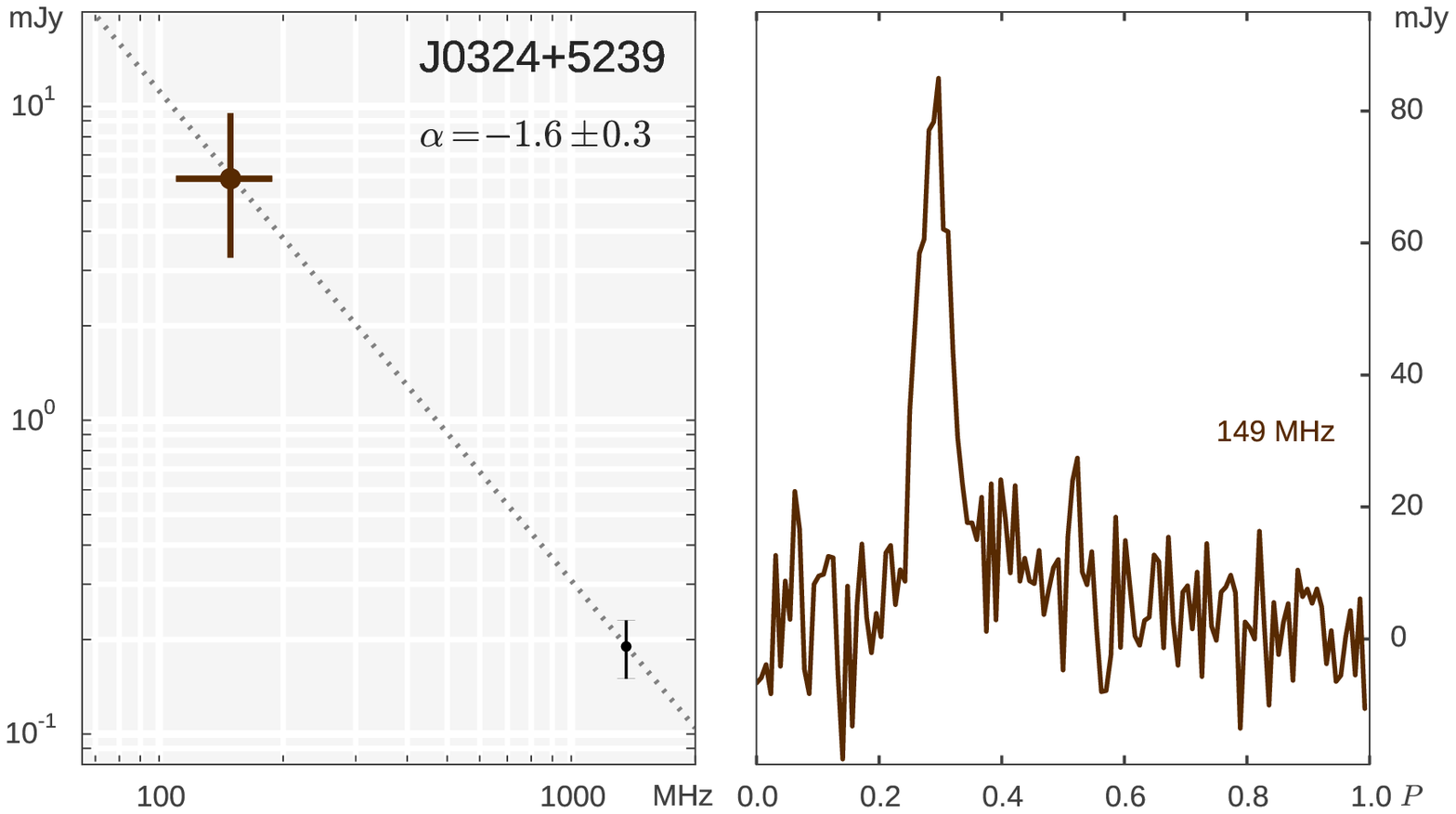}
\includegraphics[scale=0.48]{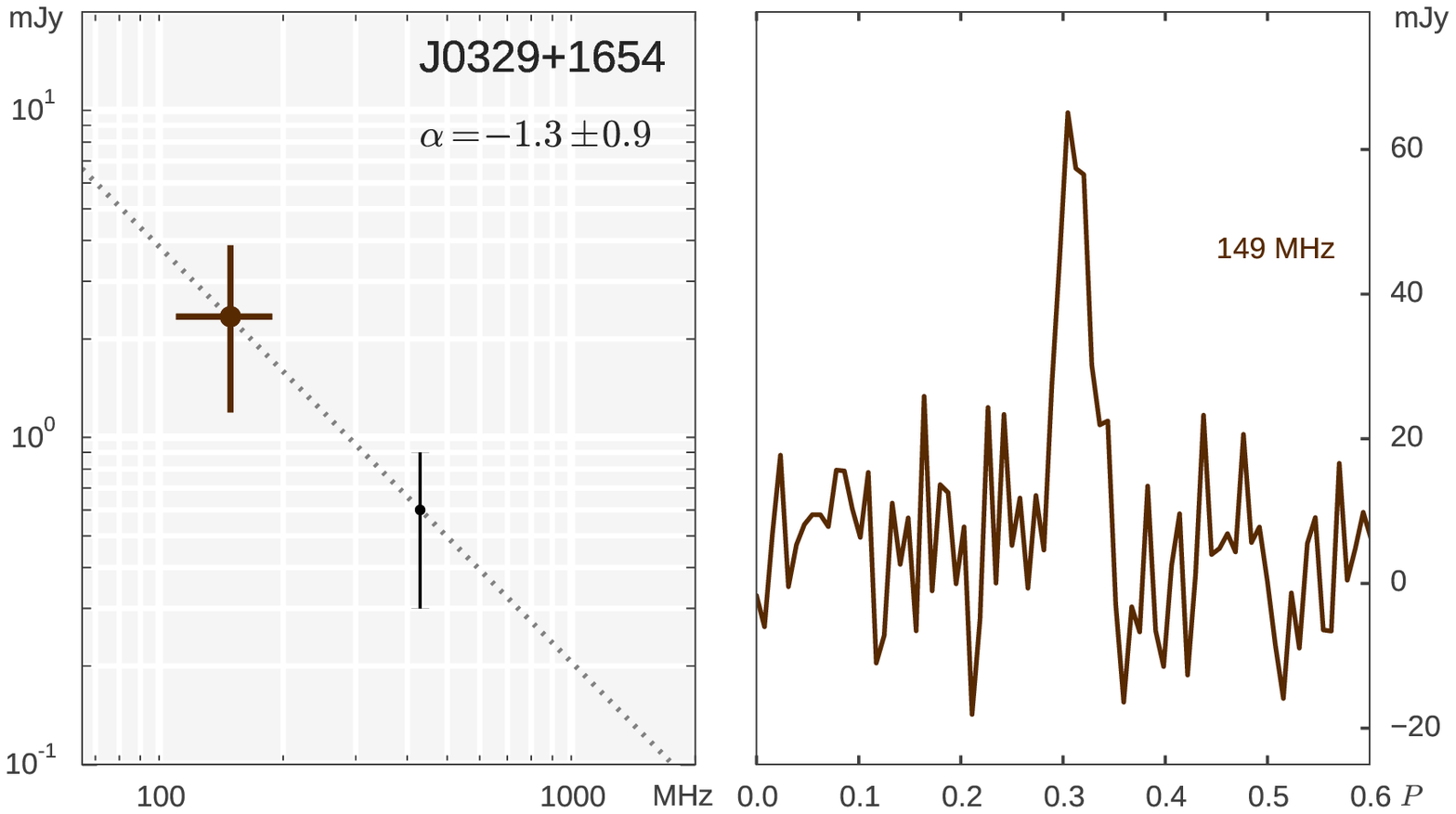}\includegraphics[scale=0.48]{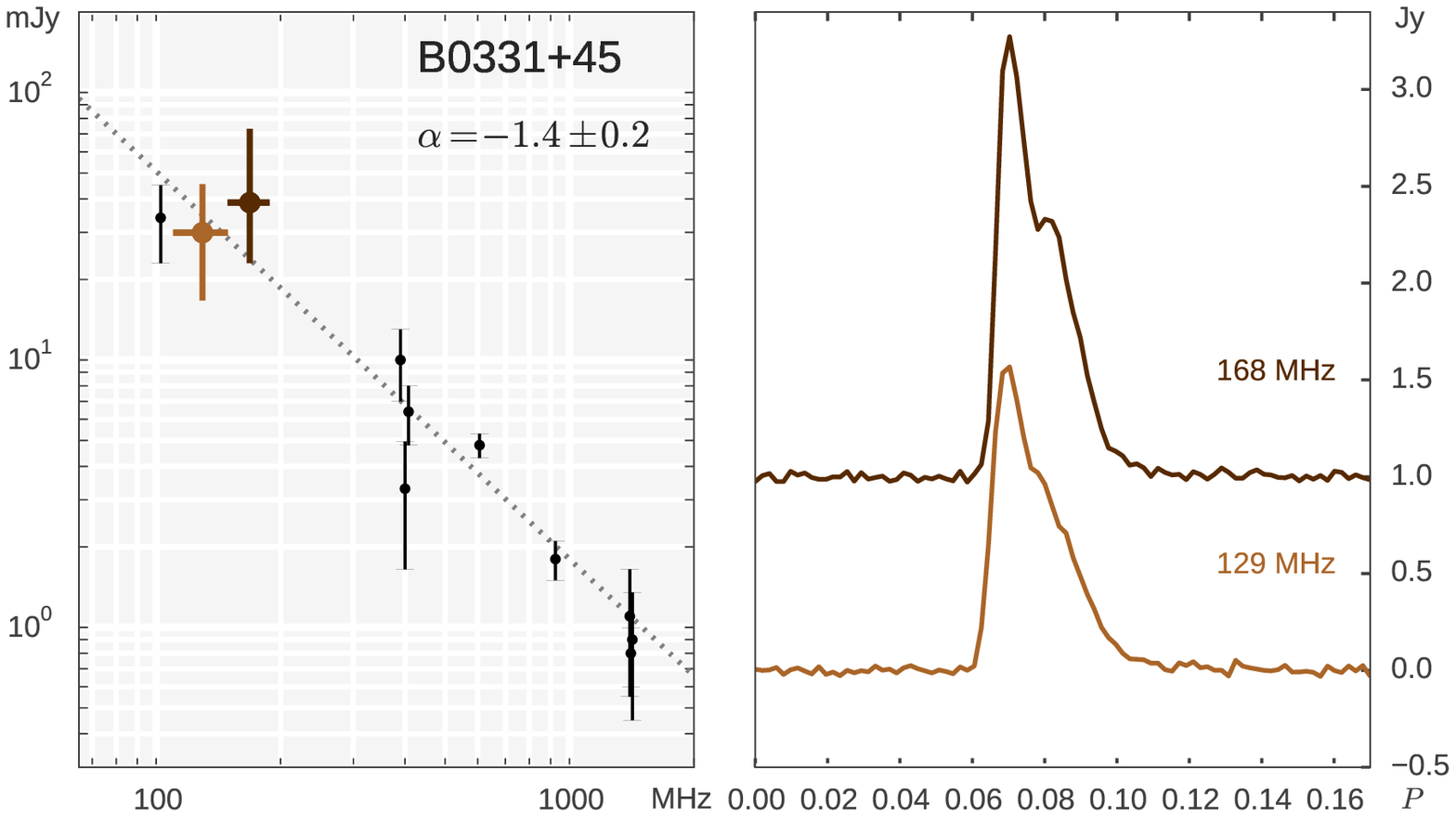}
\caption{See Figure~\ref{fig:prof_sp_1}.}
\label{fig:prof_sp_2}
\end{figure*}

\begin{figure*}
\includegraphics[scale=0.475]{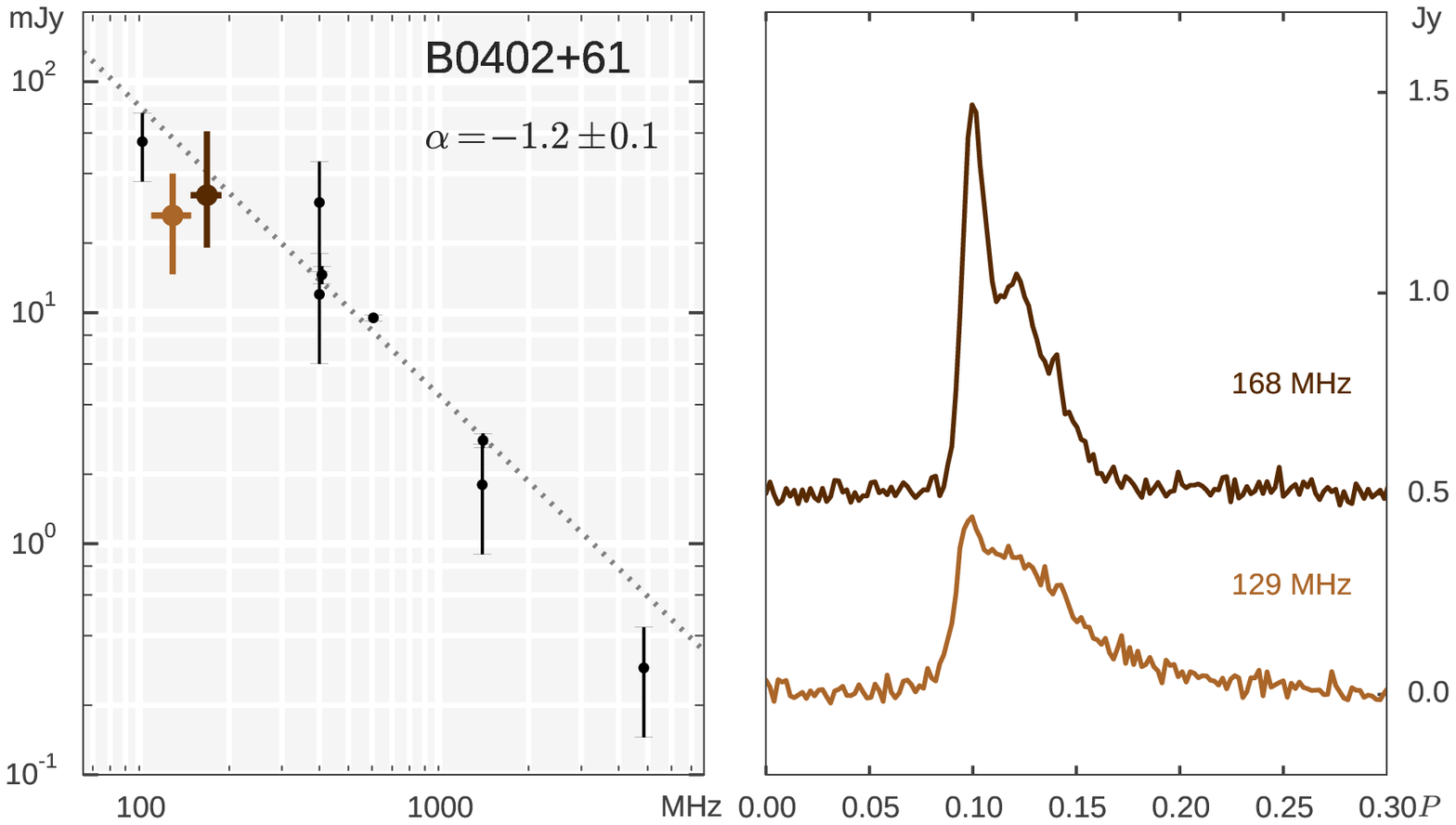}\includegraphics[scale=0.475]{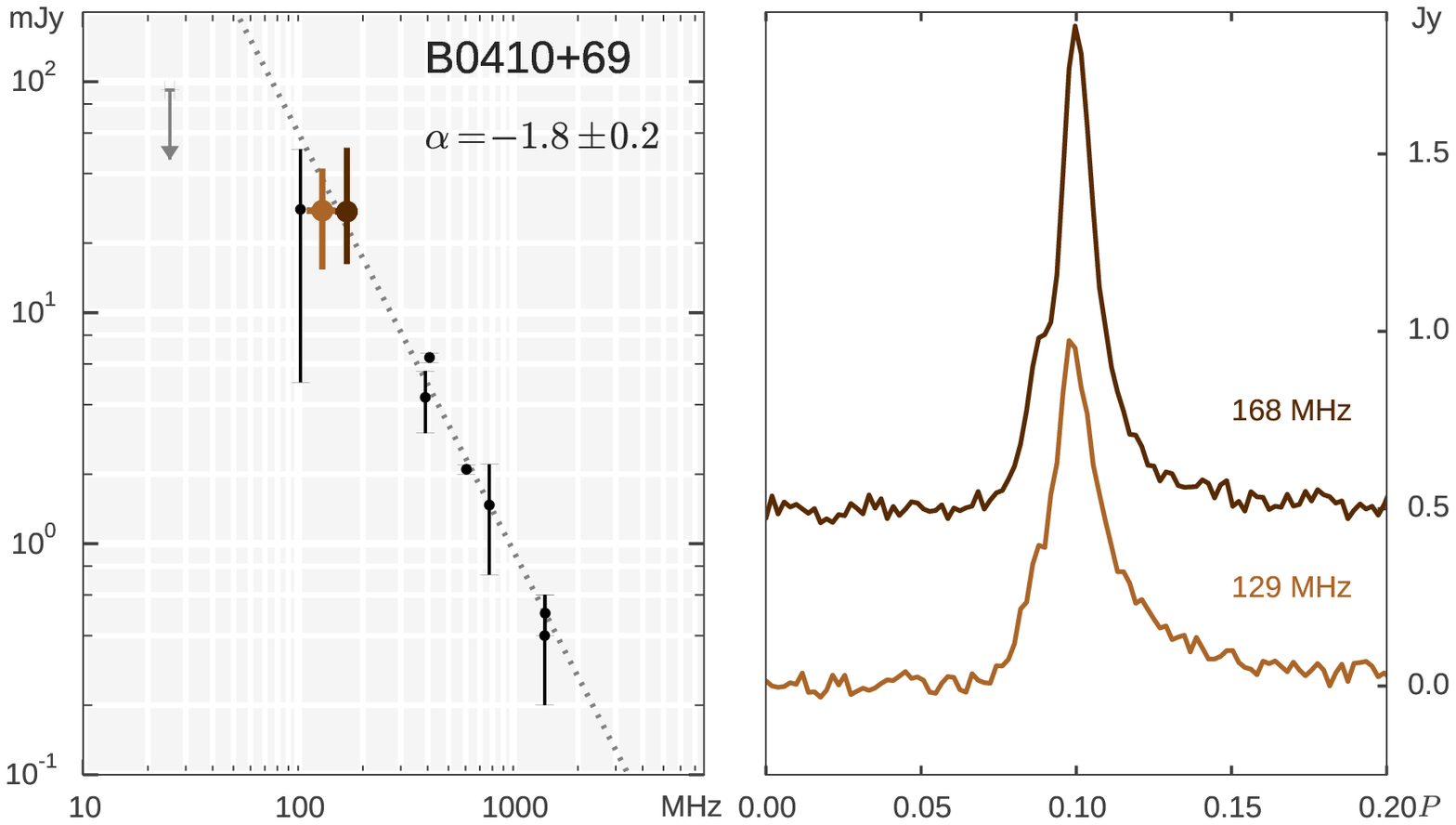}
\includegraphics[scale=0.475]{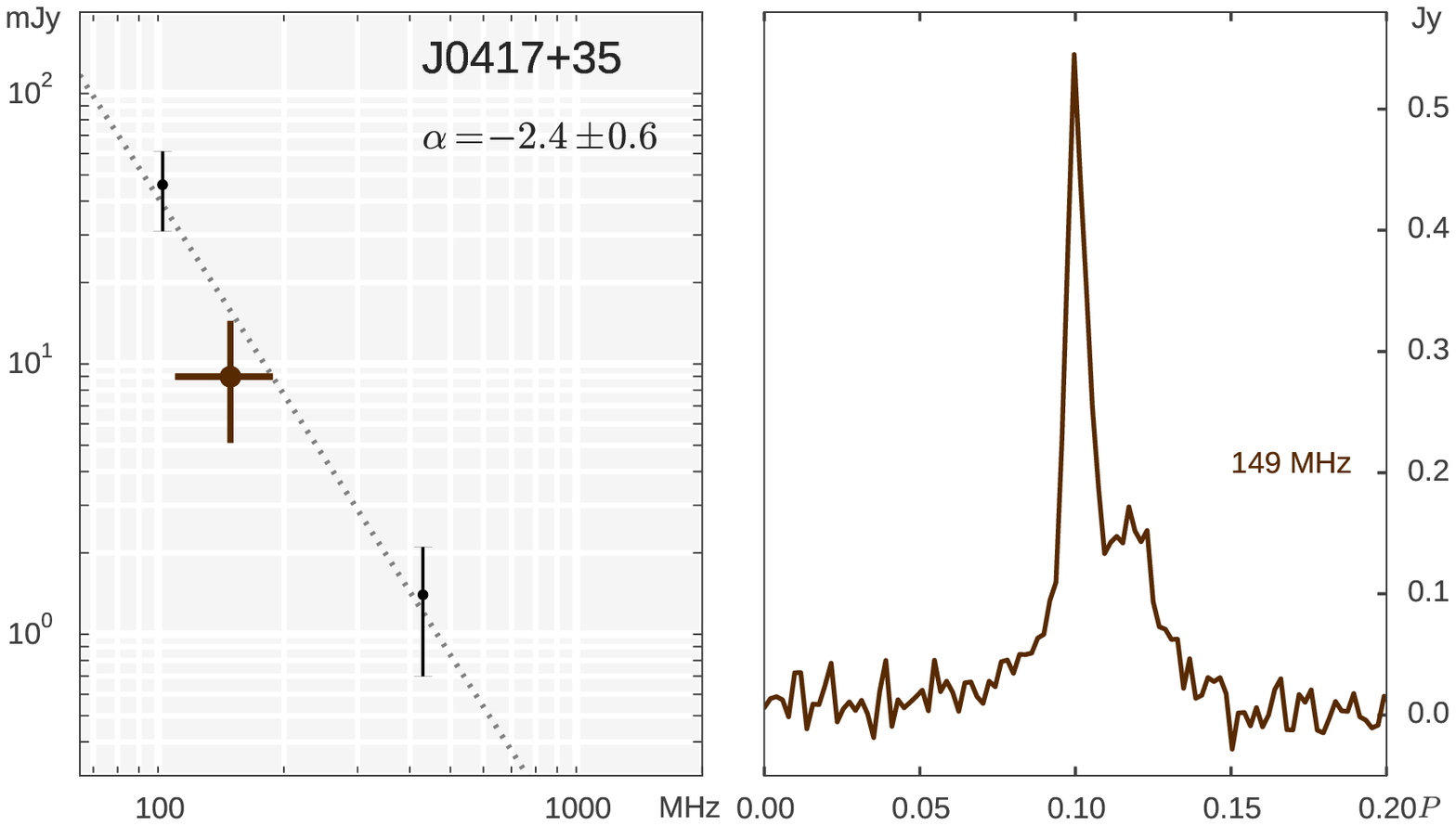}\includegraphics[scale=0.475]{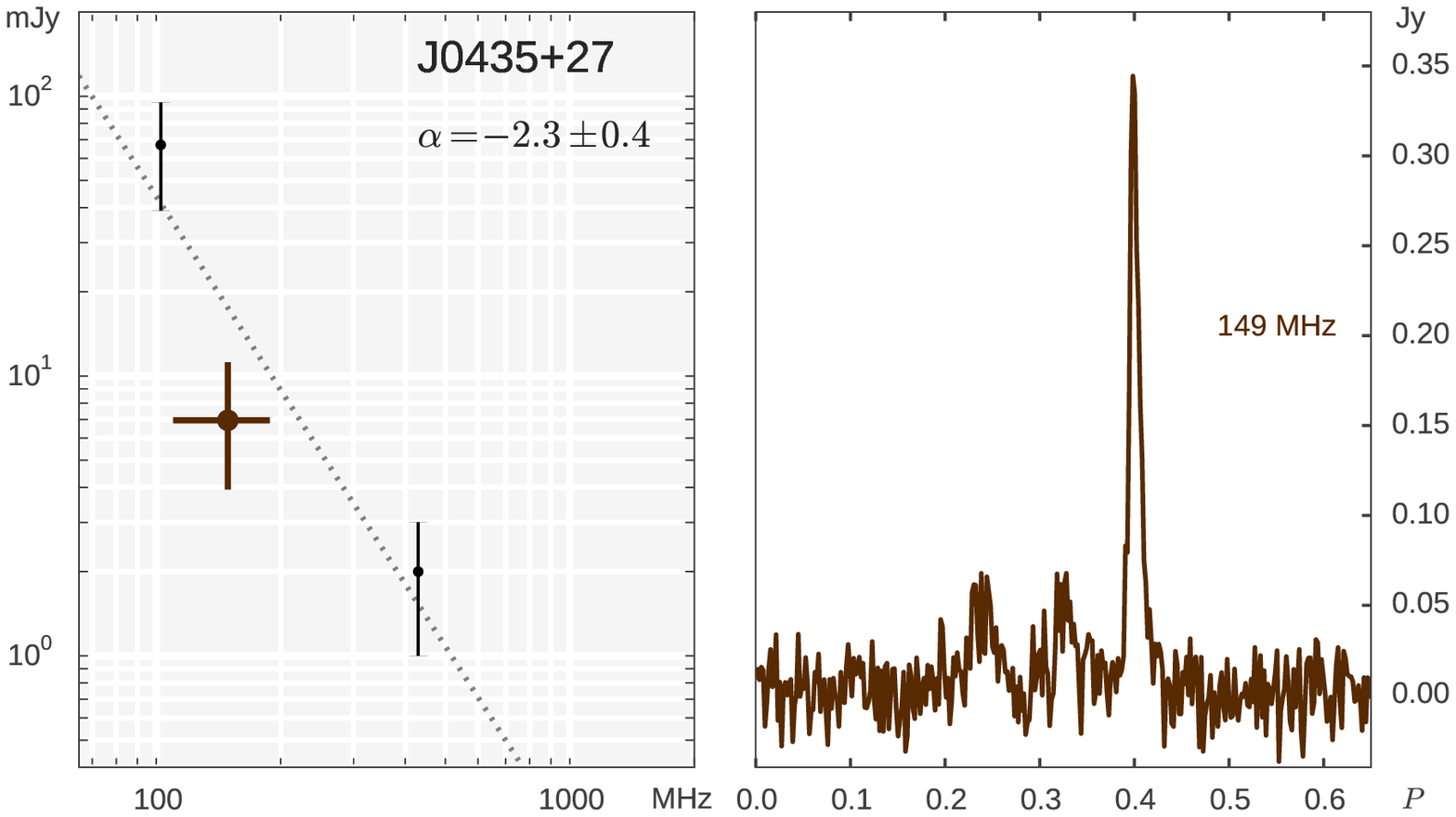}
\includegraphics[scale=0.475]{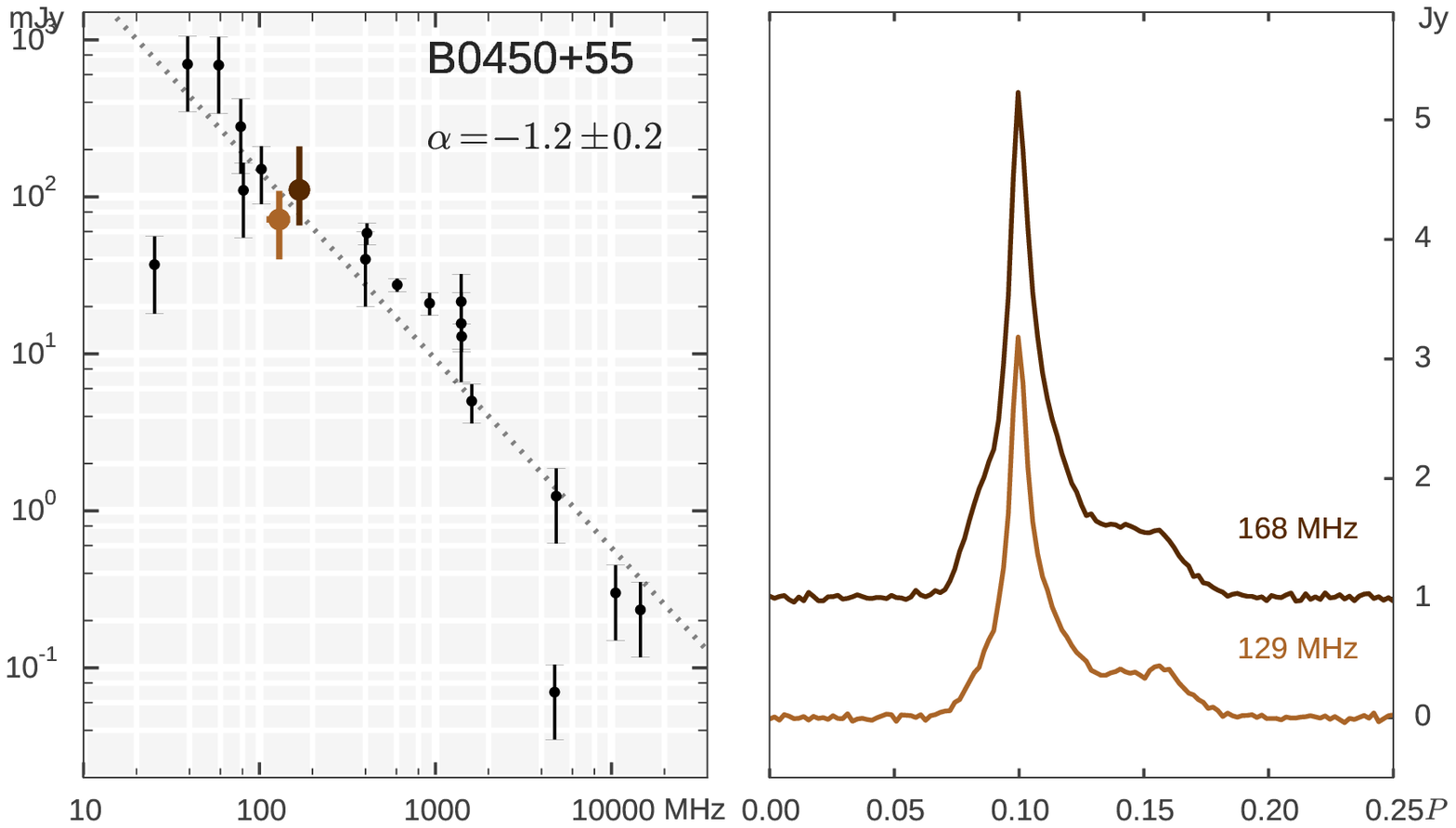}\includegraphics[scale=0.475]{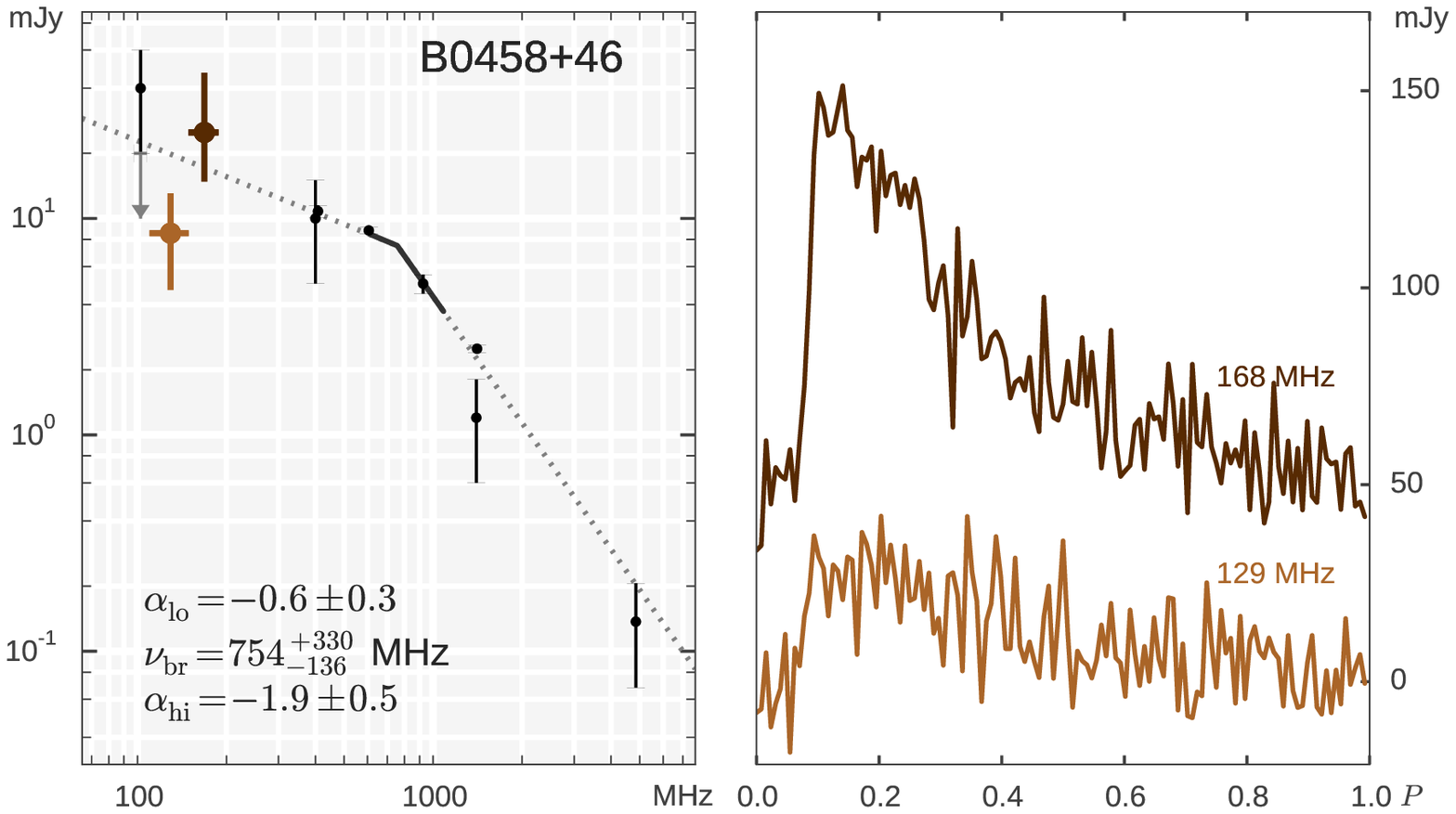}
\includegraphics[scale=0.475]{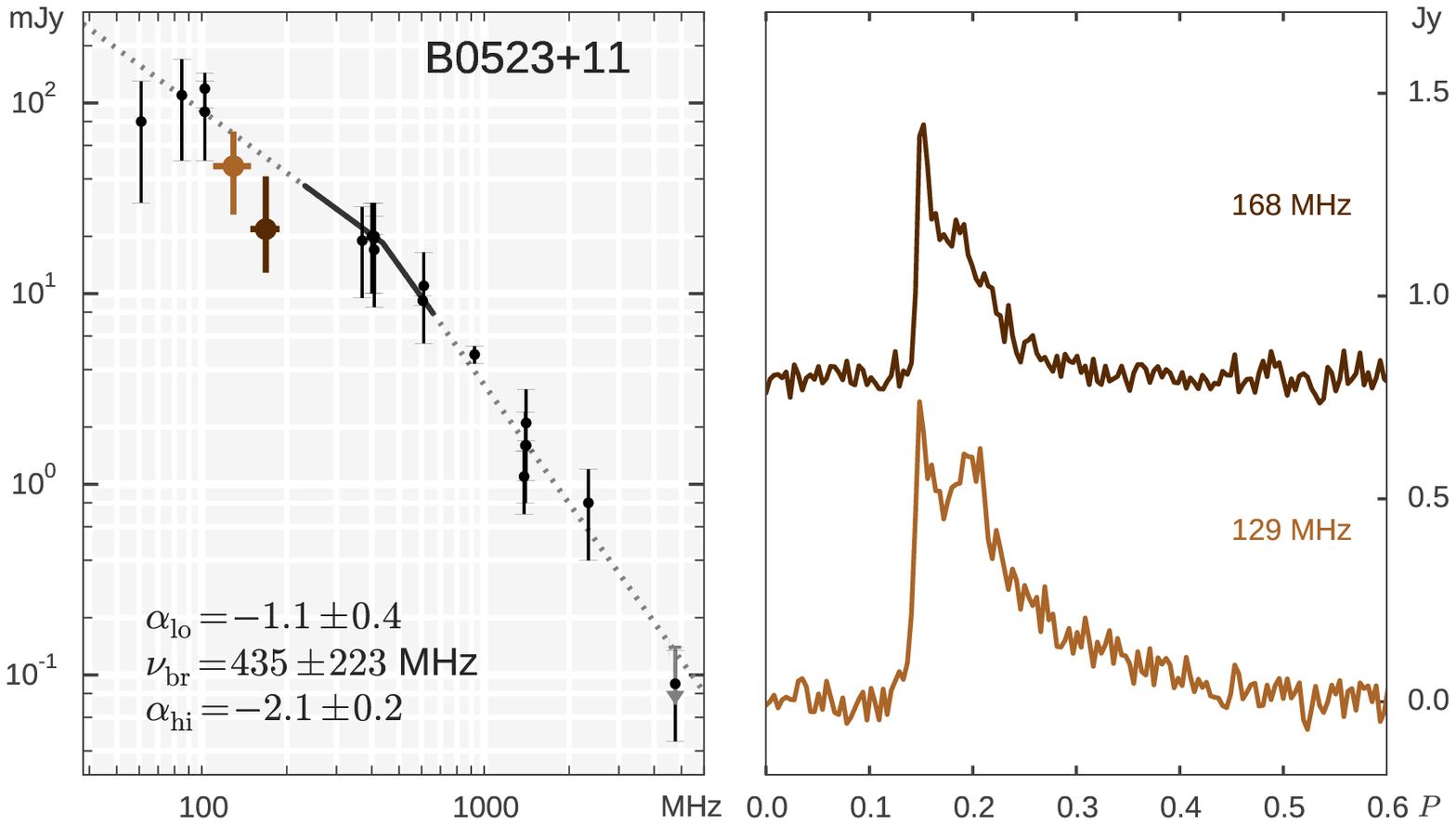}\includegraphics[scale=0.475]{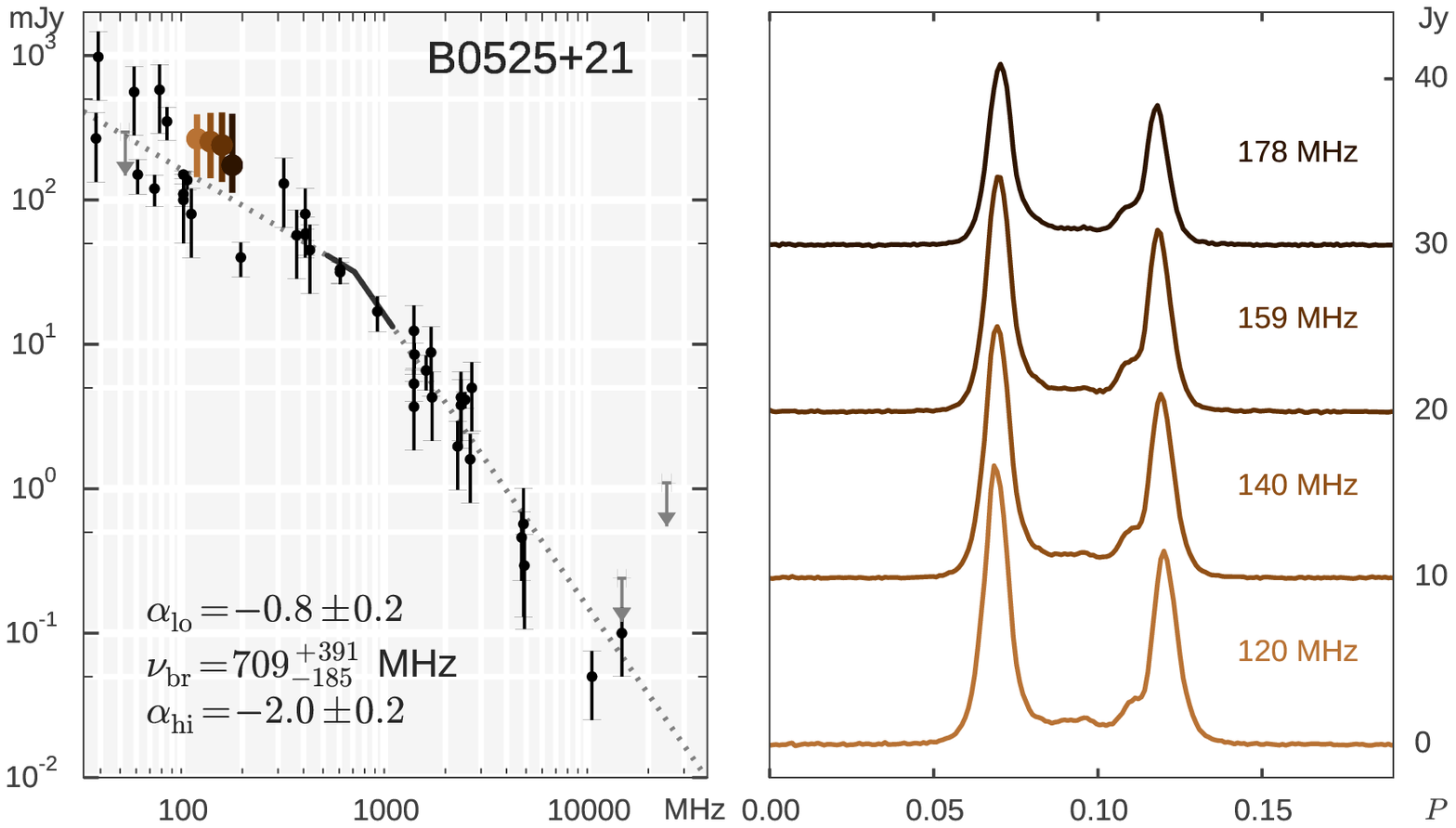}
\includegraphics[scale=0.475]{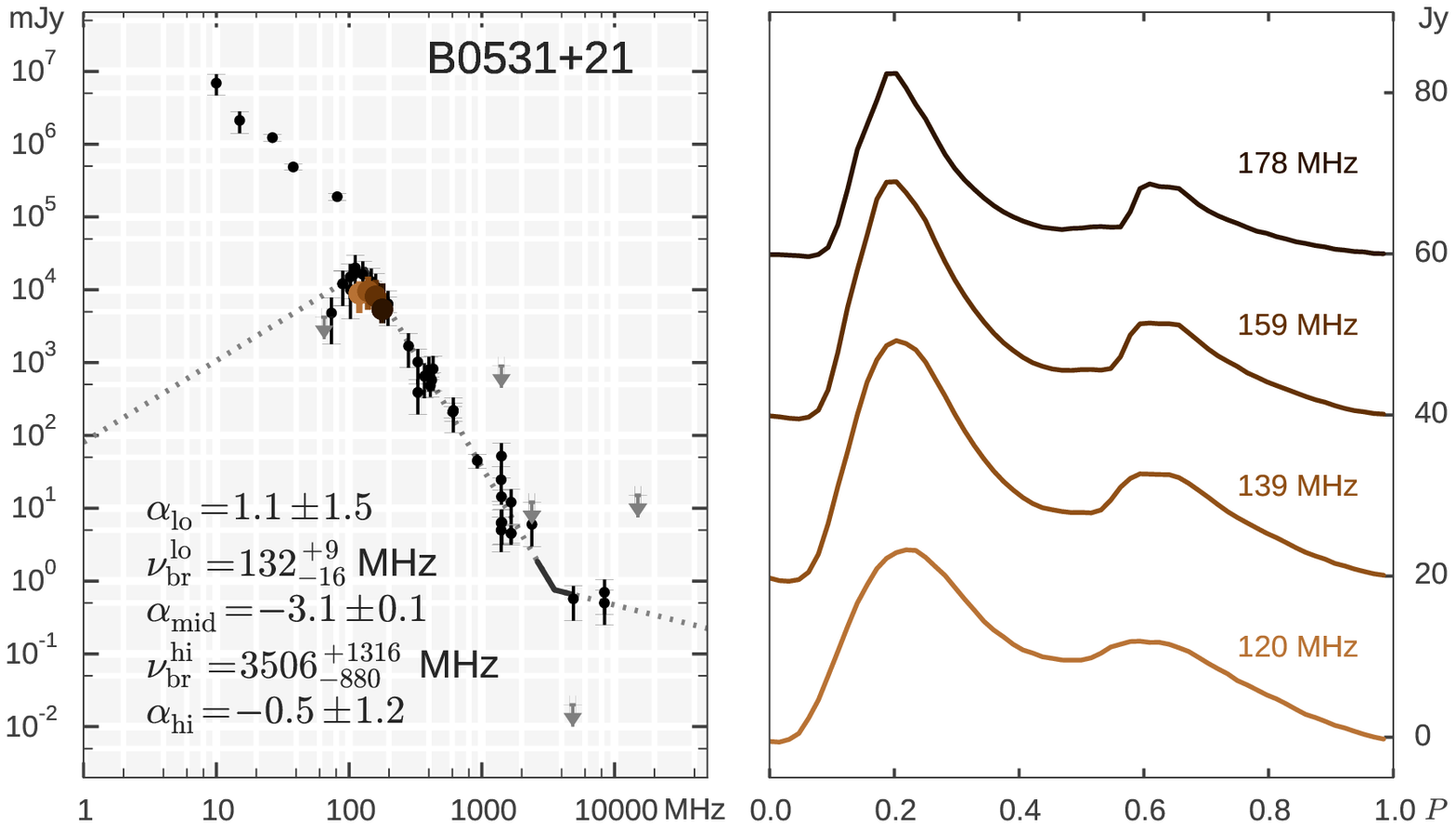}\includegraphics[scale=0.475]{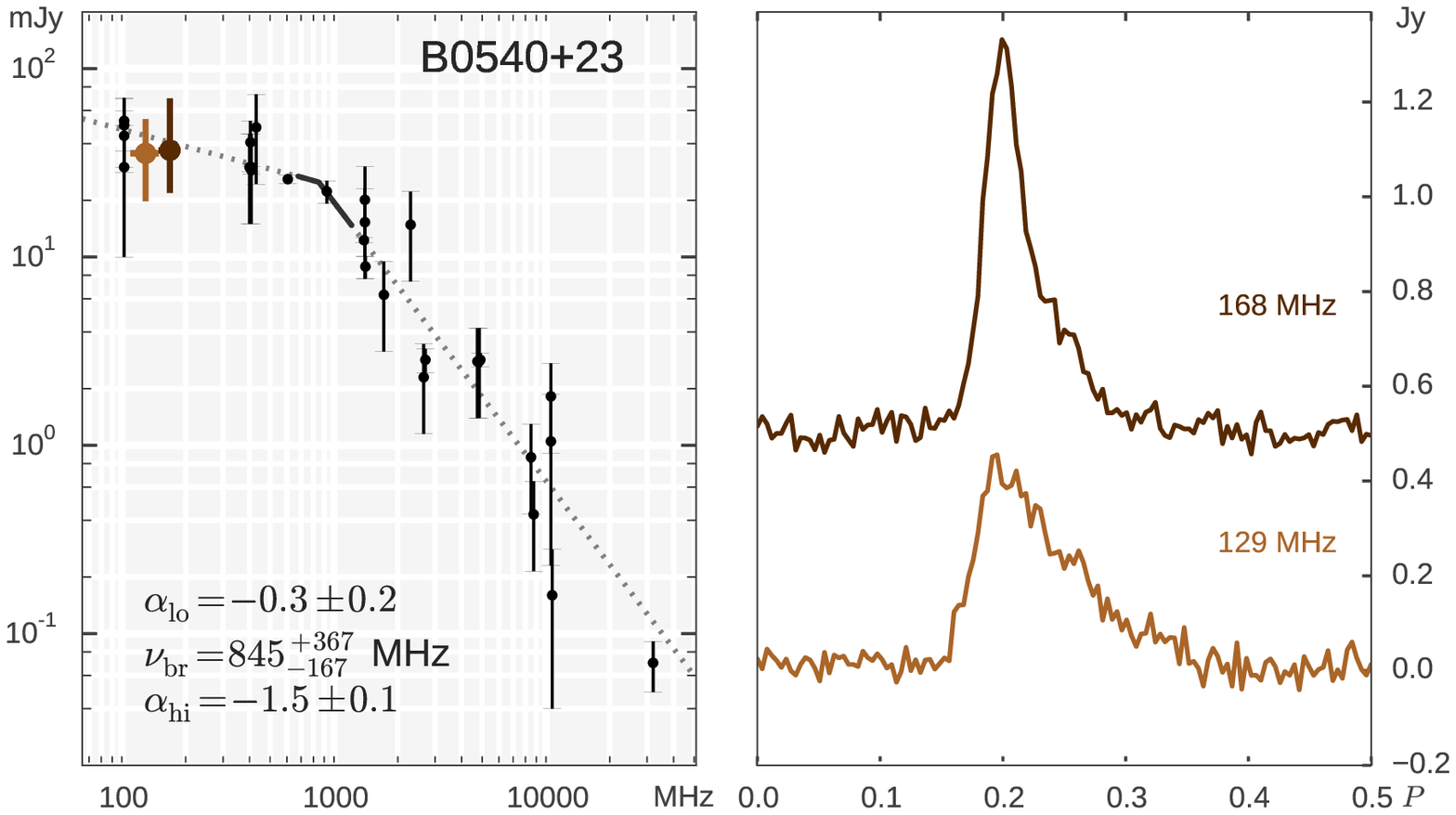}
\caption{See Figure~\ref{fig:prof_sp_1}. For the Crab pulsar (PSR B0531+21, bottom left), the continuum flux density values at 10--80\,MHz from \citet{Bridle1970}
were not included in the fit.}
\label{fig:prof_sp_3}
\end{figure*}

\begin{figure*}
\includegraphics[scale=0.48]{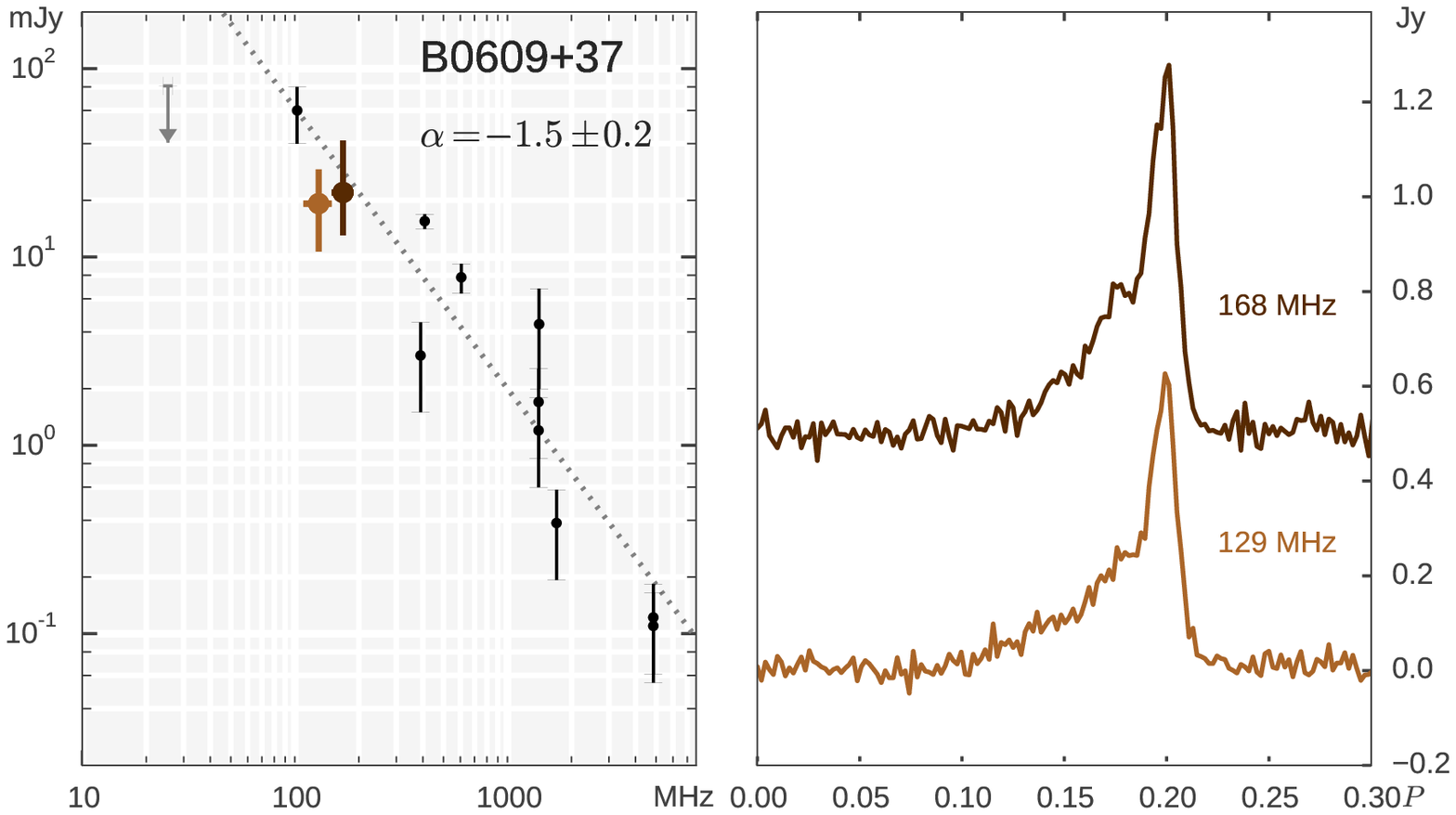}\includegraphics[scale=0.48]{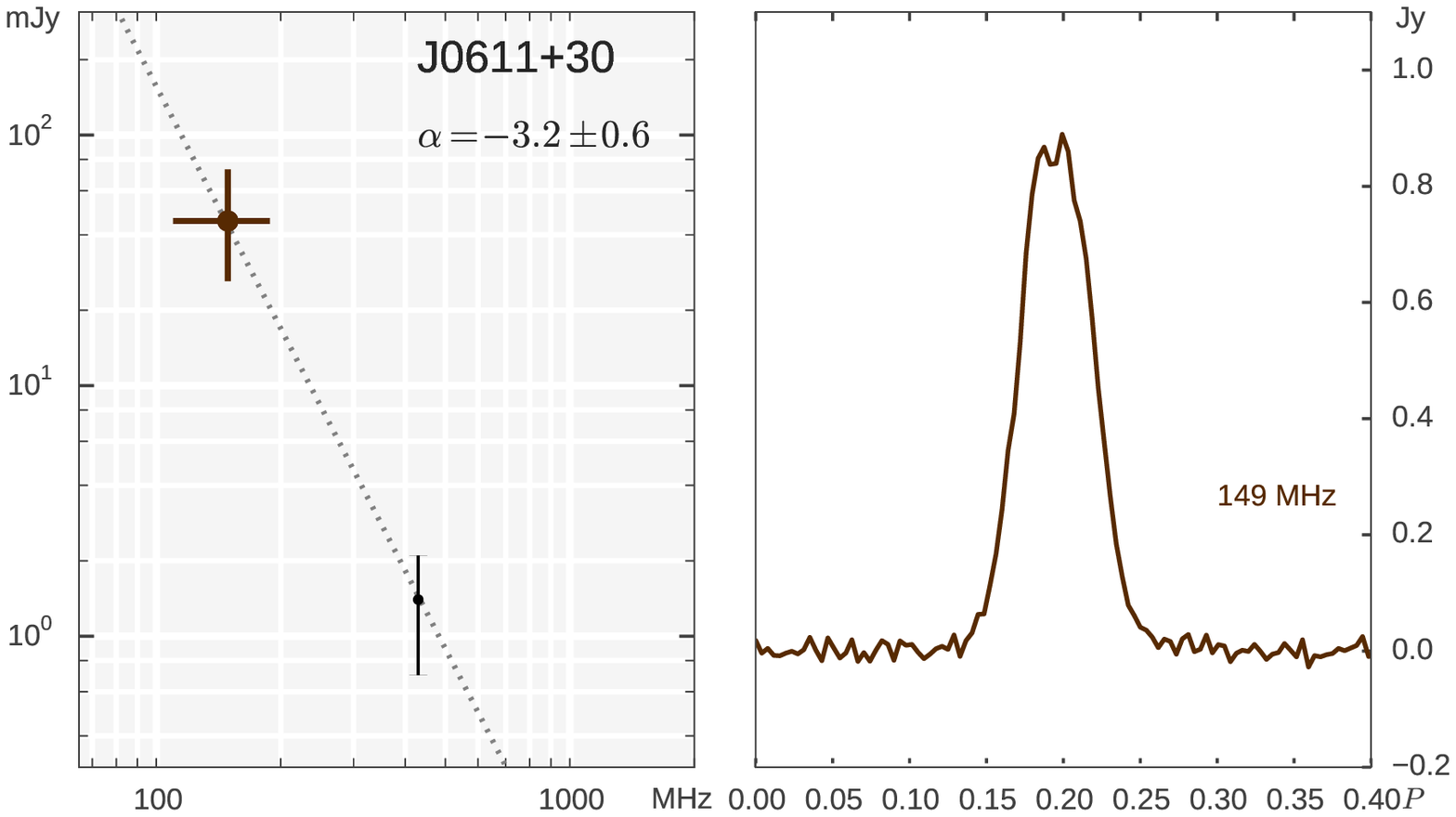}
\includegraphics[scale=0.48]{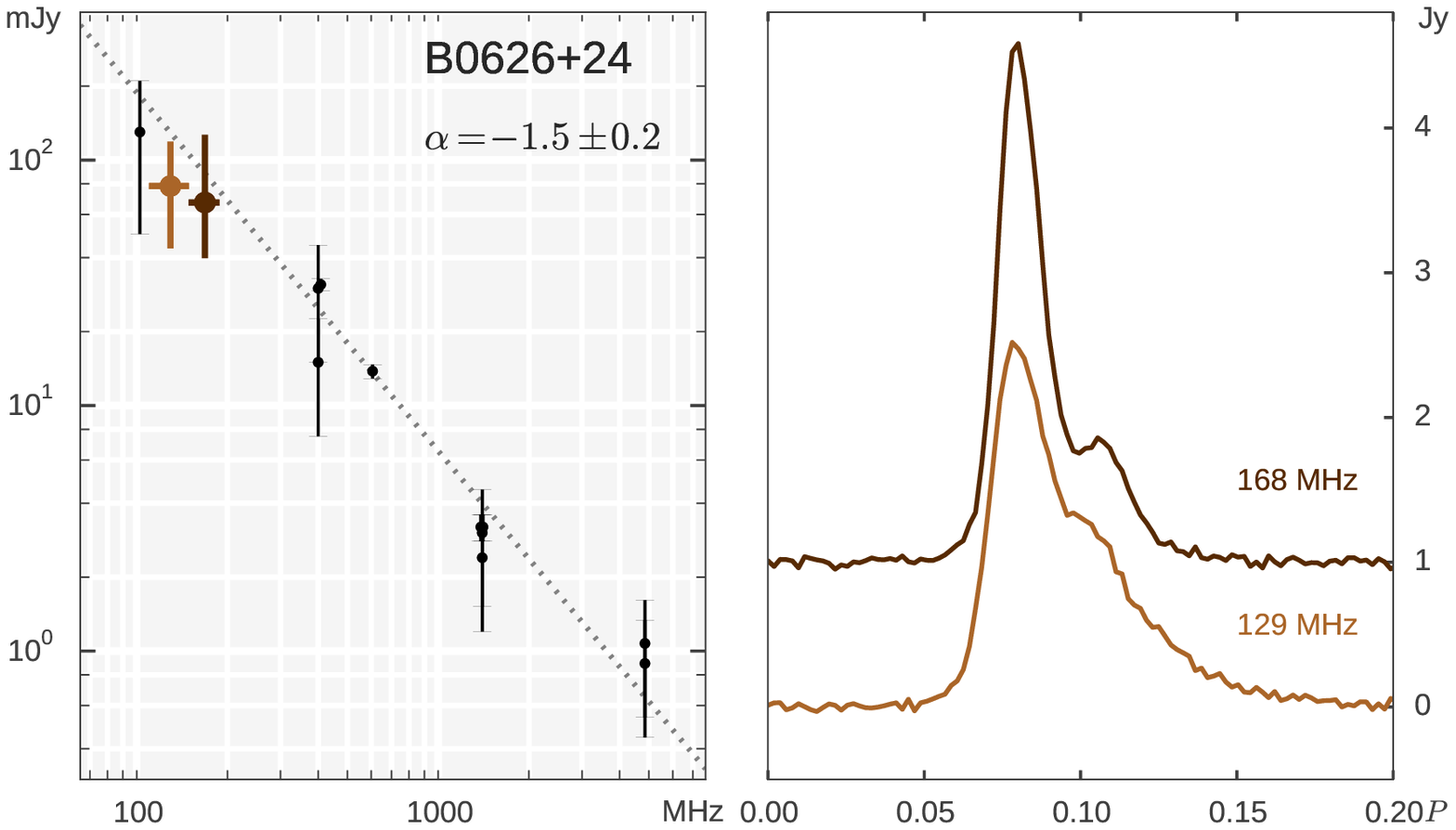}\includegraphics[scale=0.48]{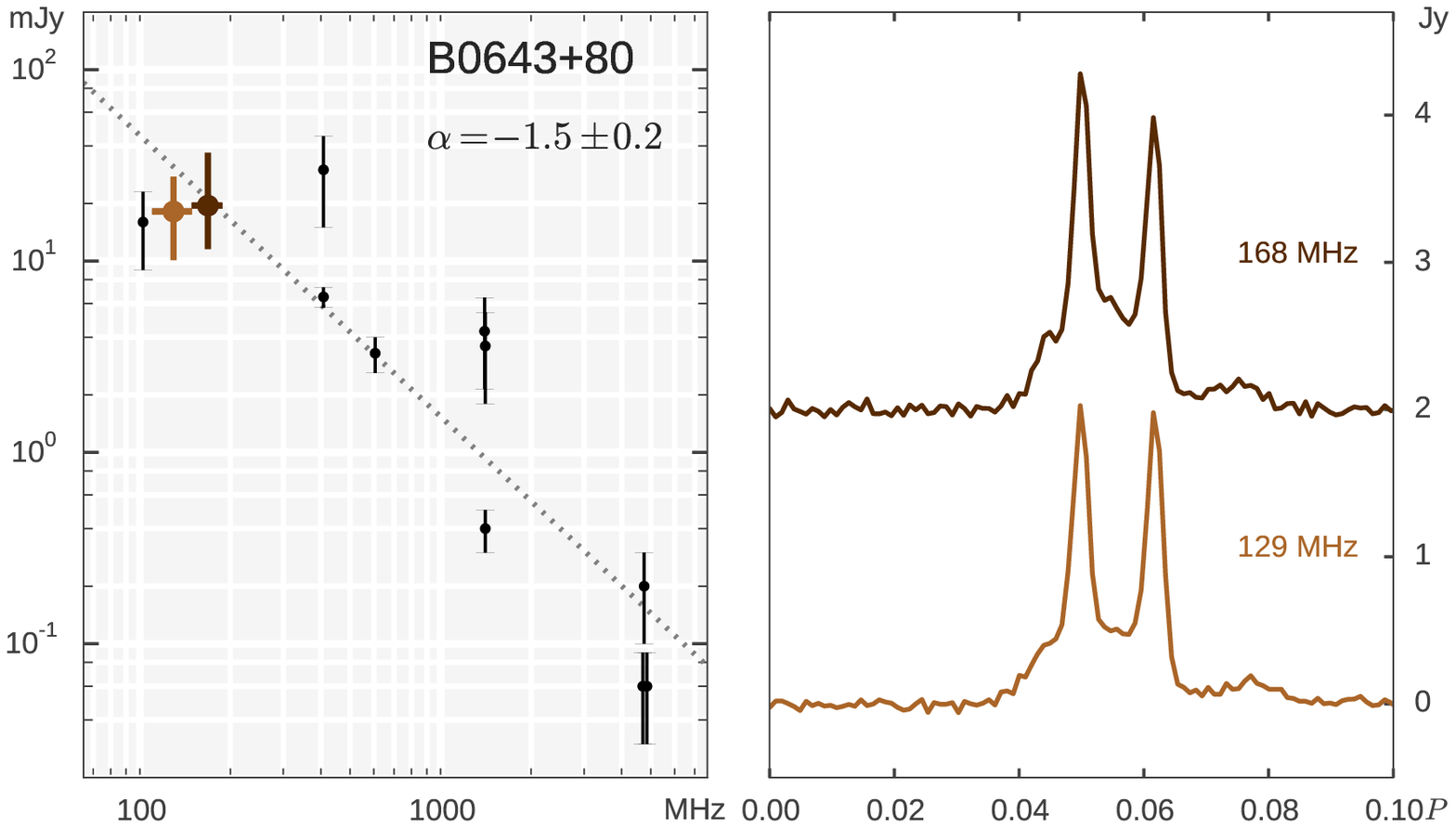}
\includegraphics[scale=0.48]{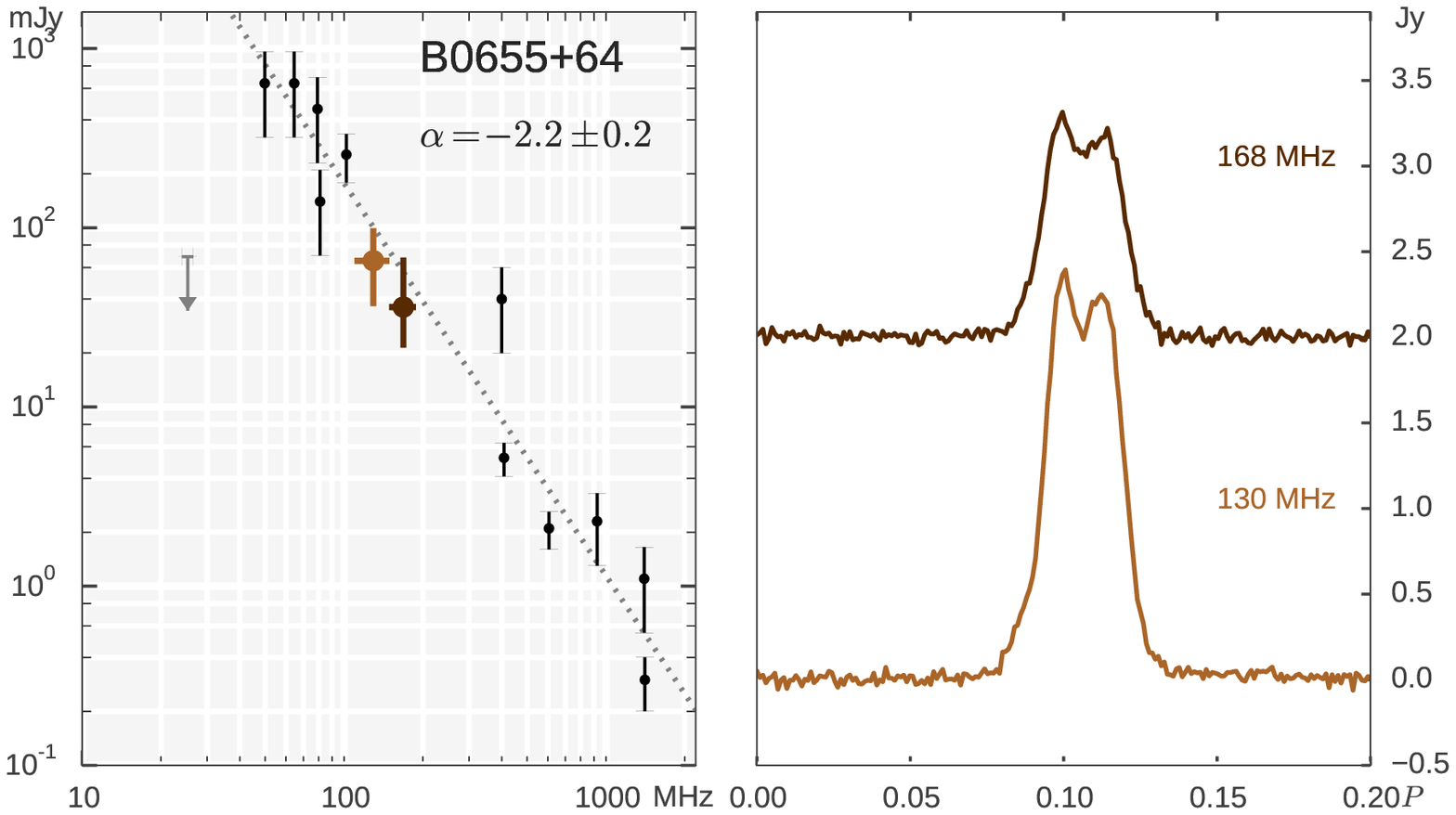}\includegraphics[scale=0.48]{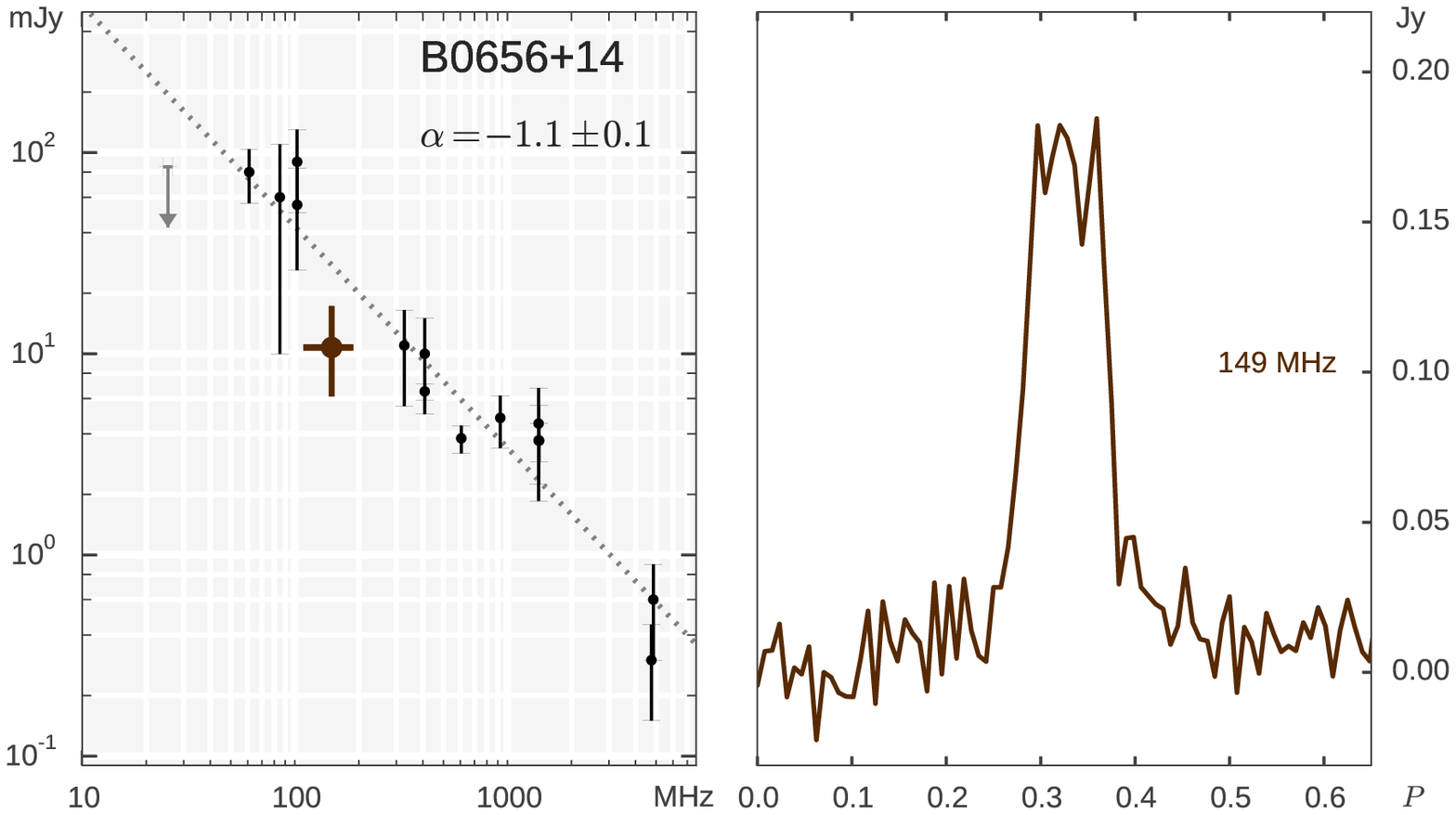}
\includegraphics[scale=0.48]{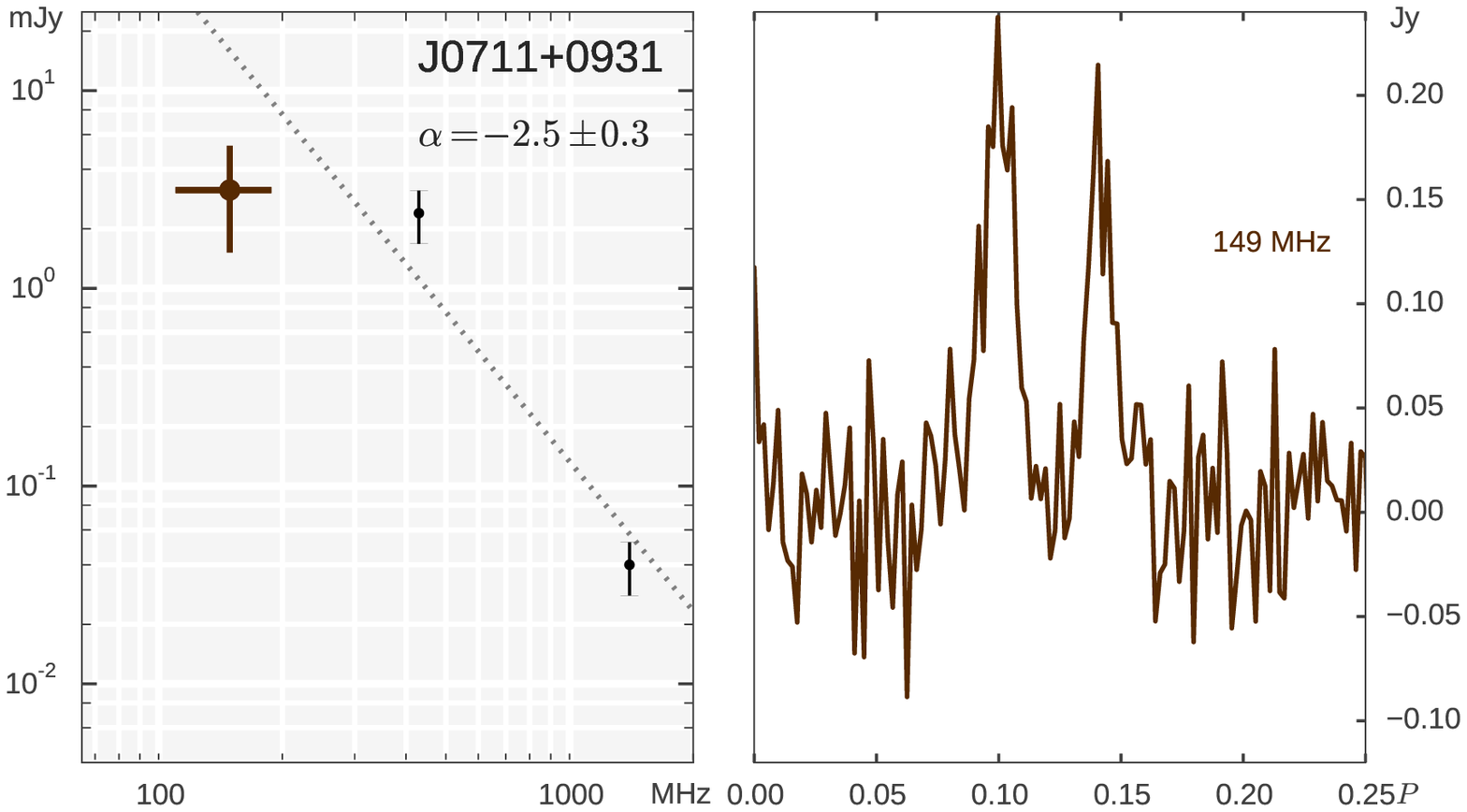}\includegraphics[scale=0.48]{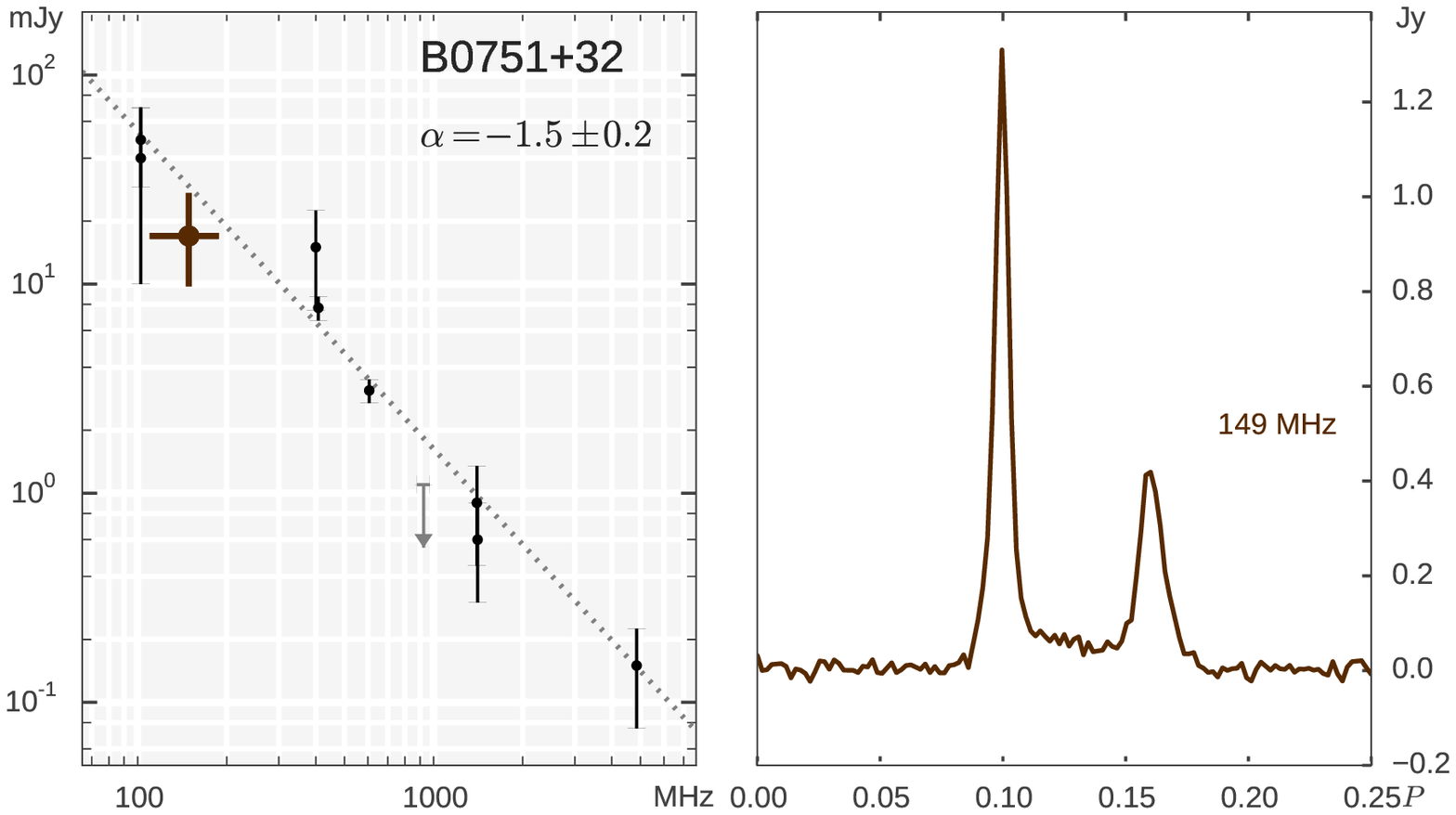}
\includegraphics[scale=0.48]{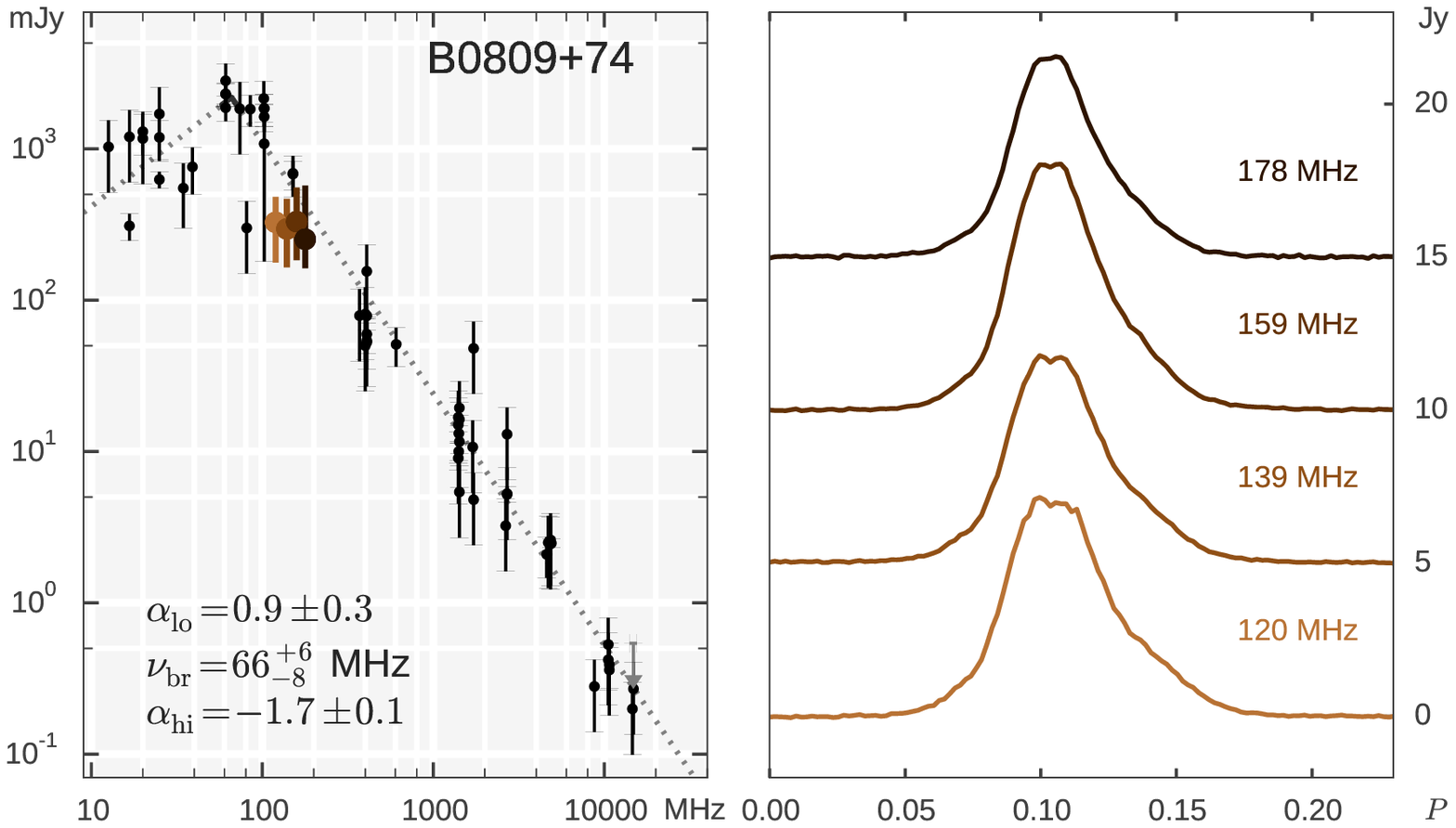}\includegraphics[scale=0.48]{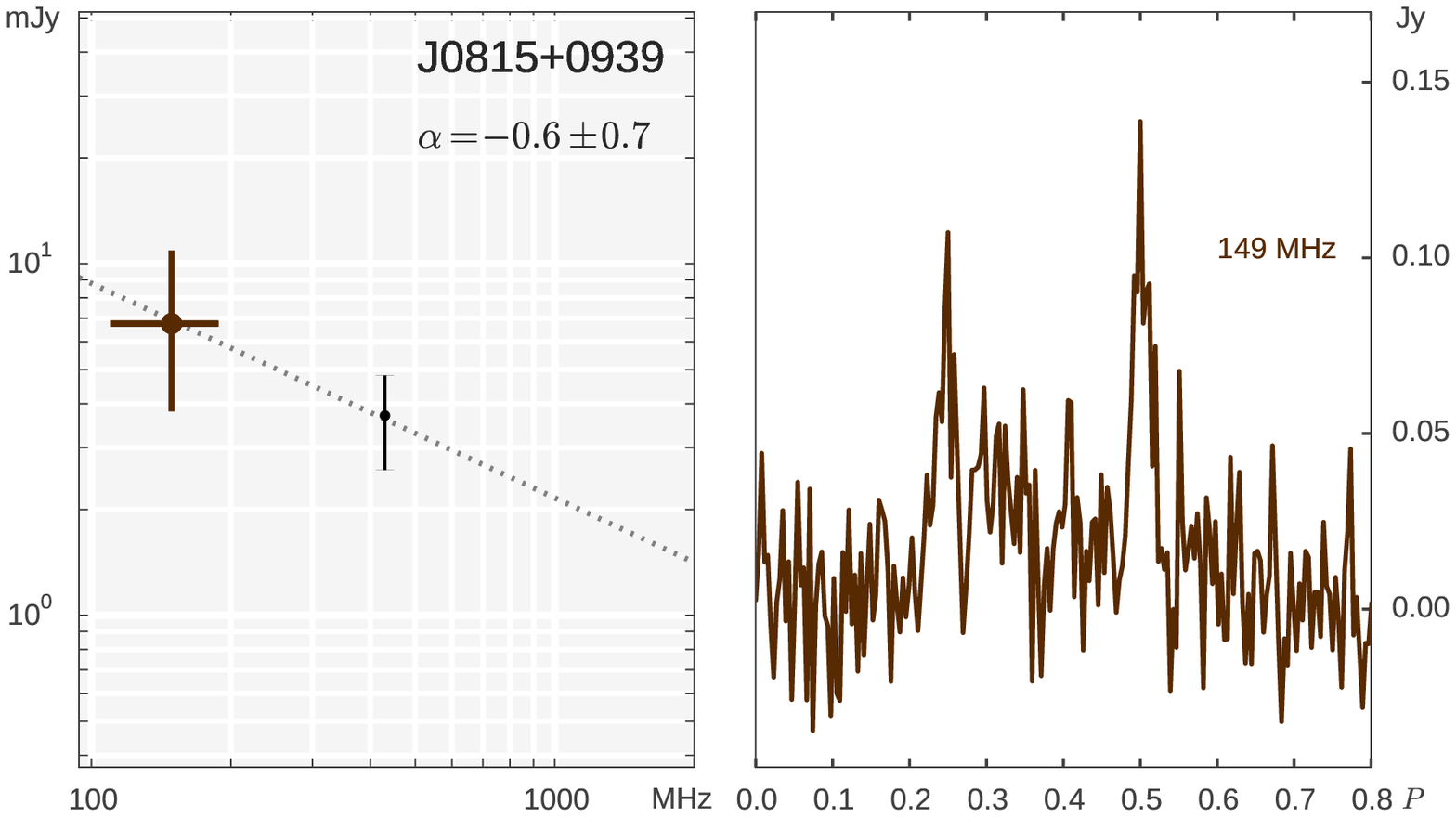}
\caption{See Figure~\ref{fig:prof_sp_1}.}
\label{fig:prof_sp_4}
\end{figure*}

\begin{figure*}
\includegraphics[scale=0.475]{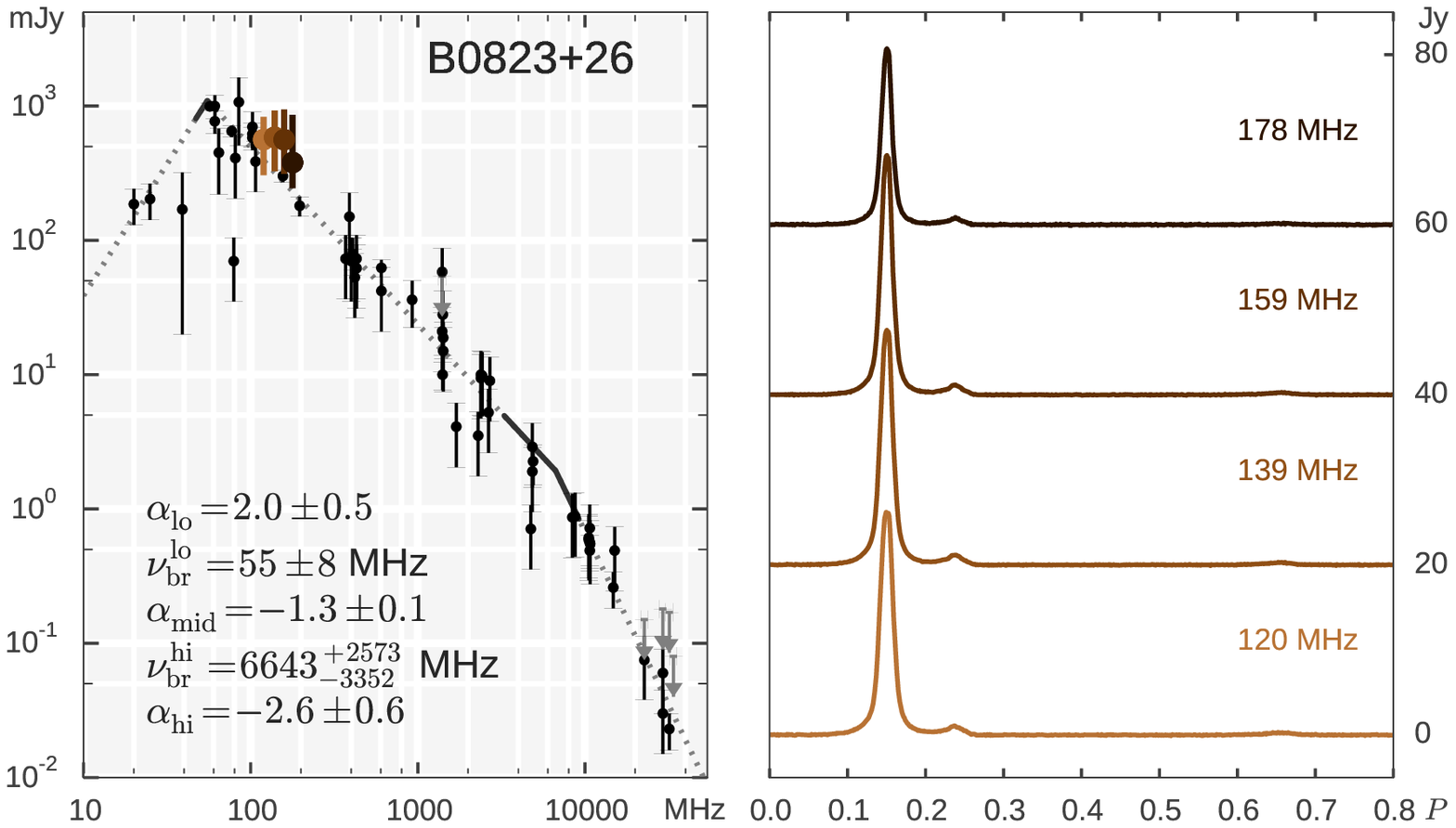}\includegraphics[scale=0.475]{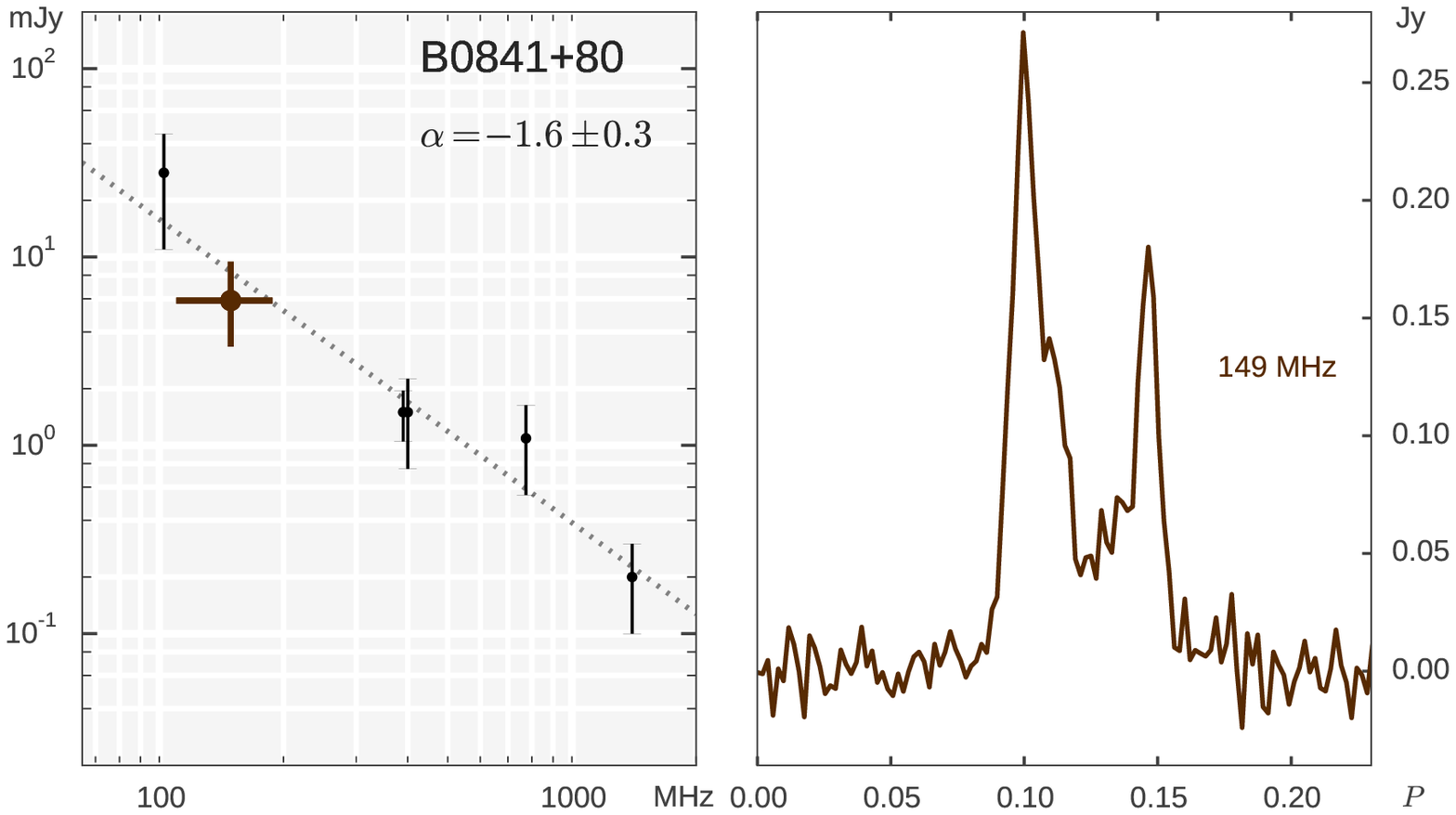}
\includegraphics[scale=0.475]{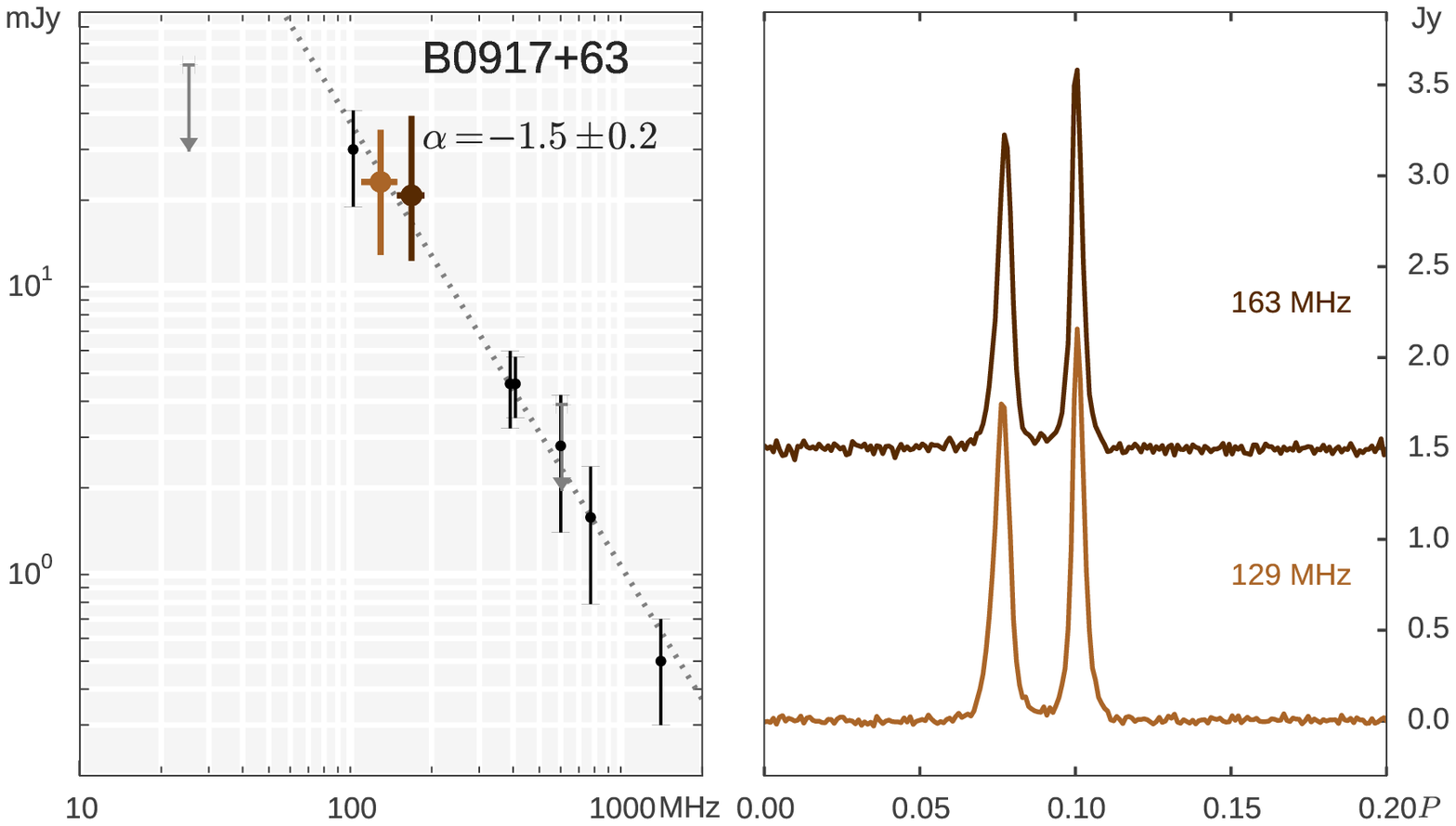}\includegraphics[scale=0.475]{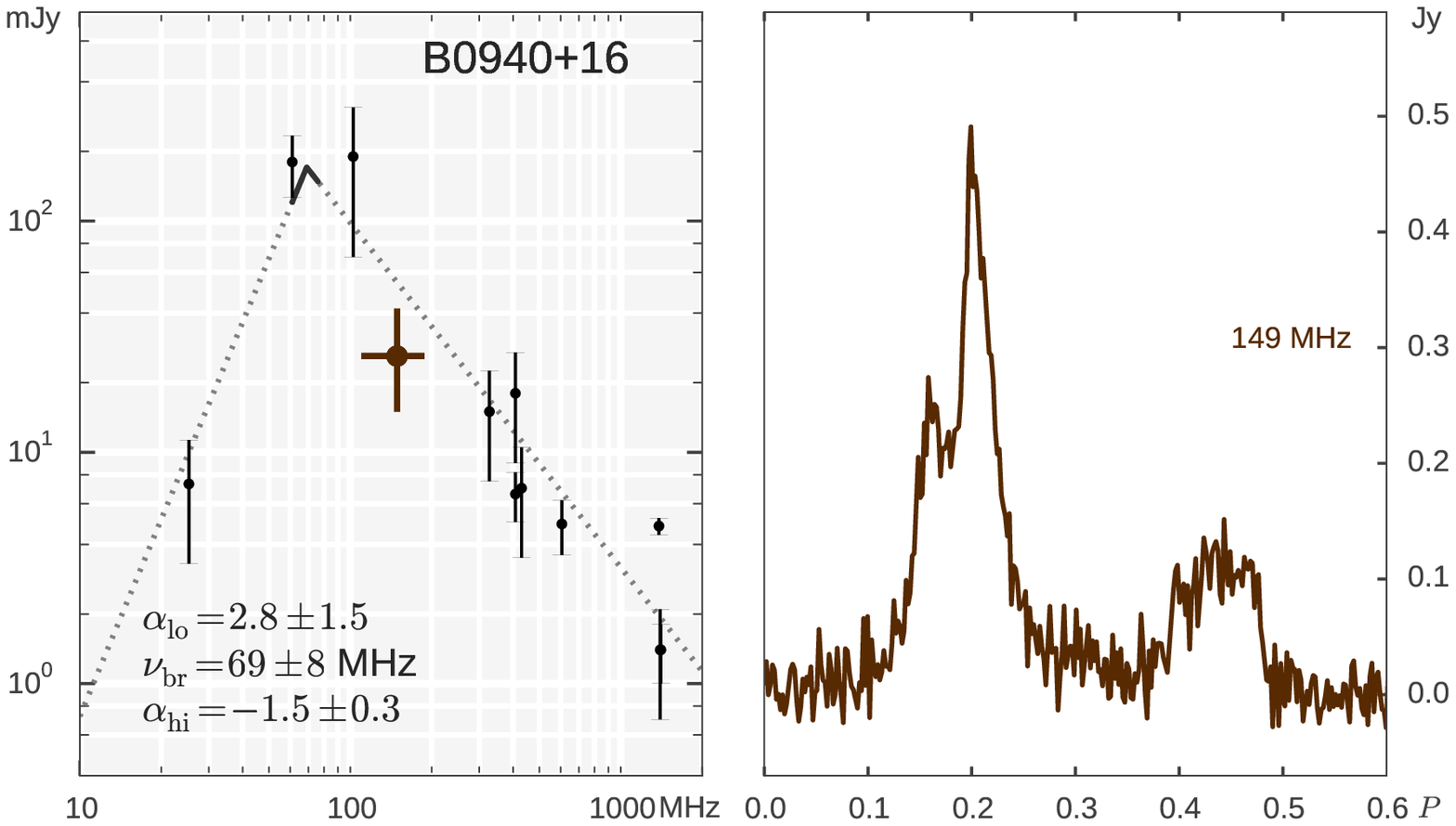}
\includegraphics[scale=0.475]{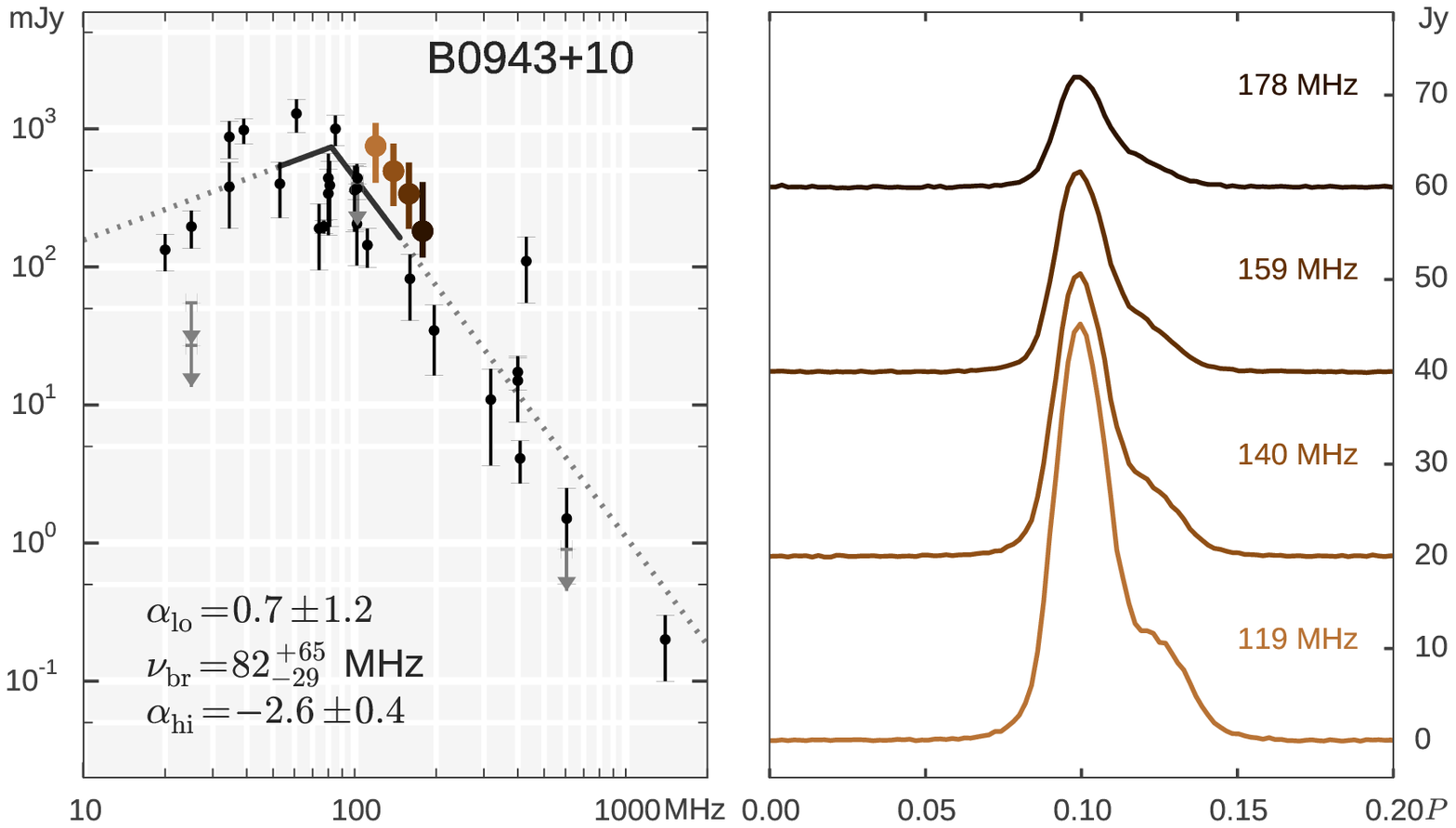}\includegraphics[scale=0.475]{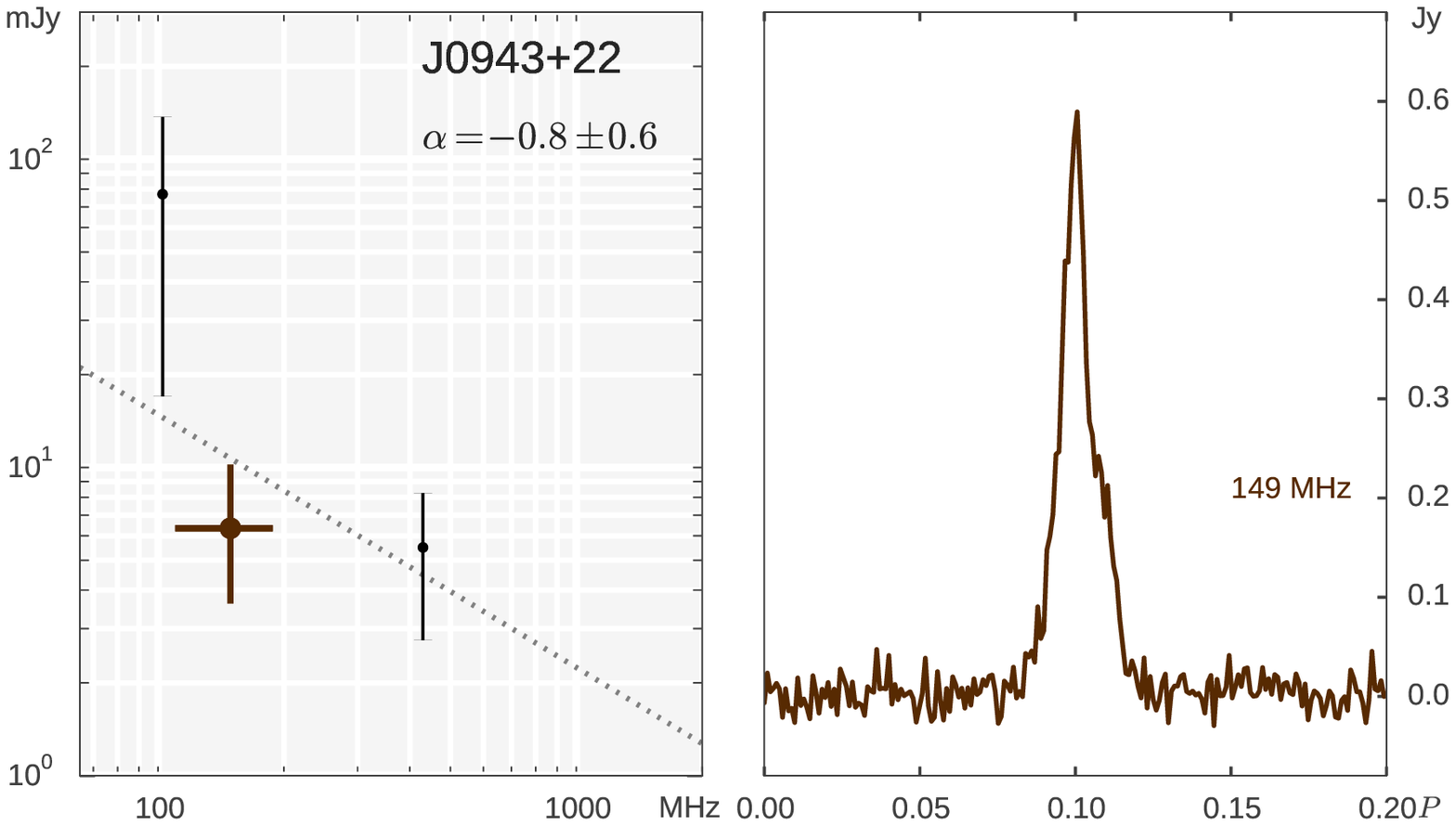}
\includegraphics[scale=0.475]{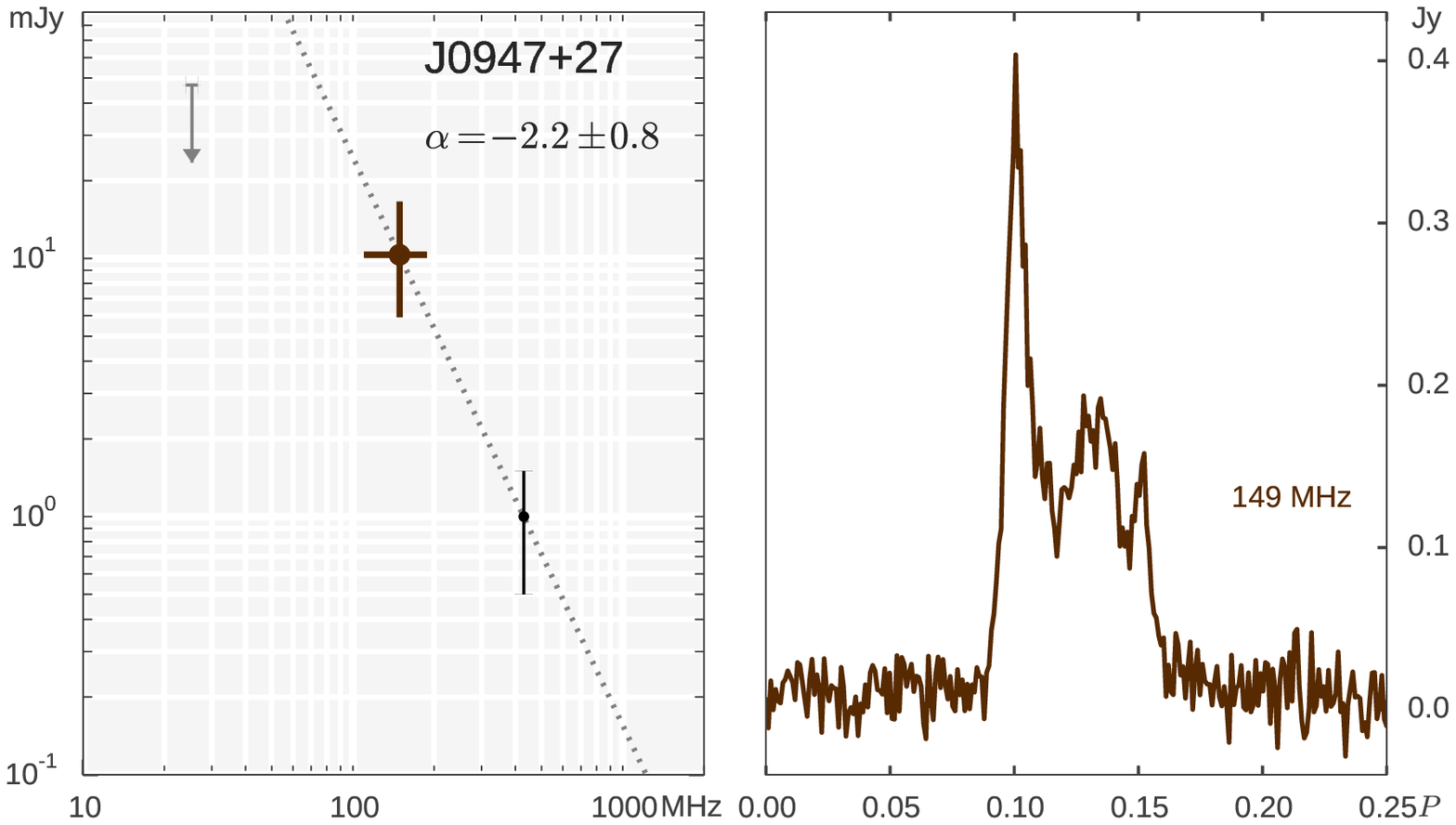}\includegraphics[scale=0.475]{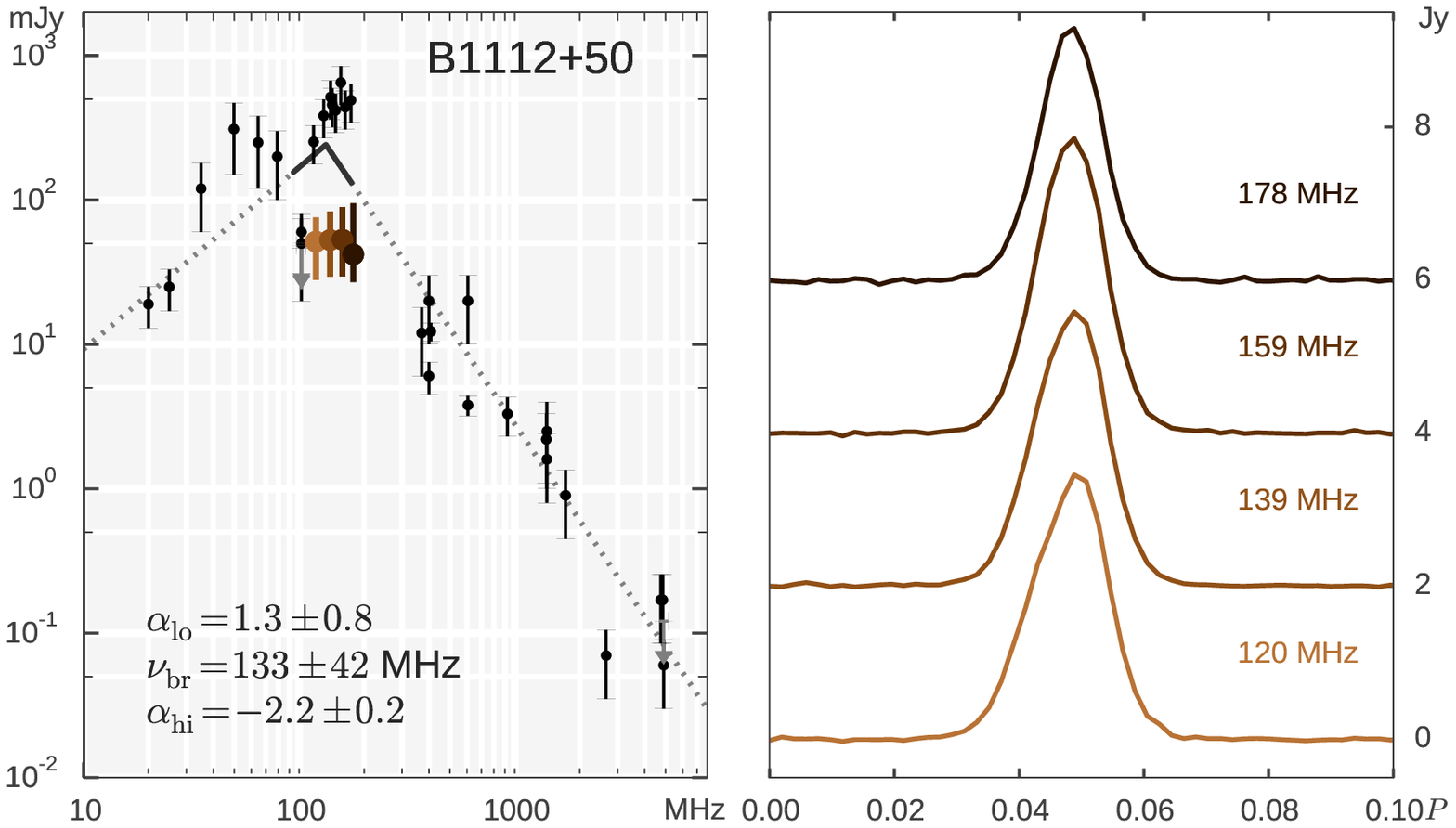}
\includegraphics[scale=0.475]{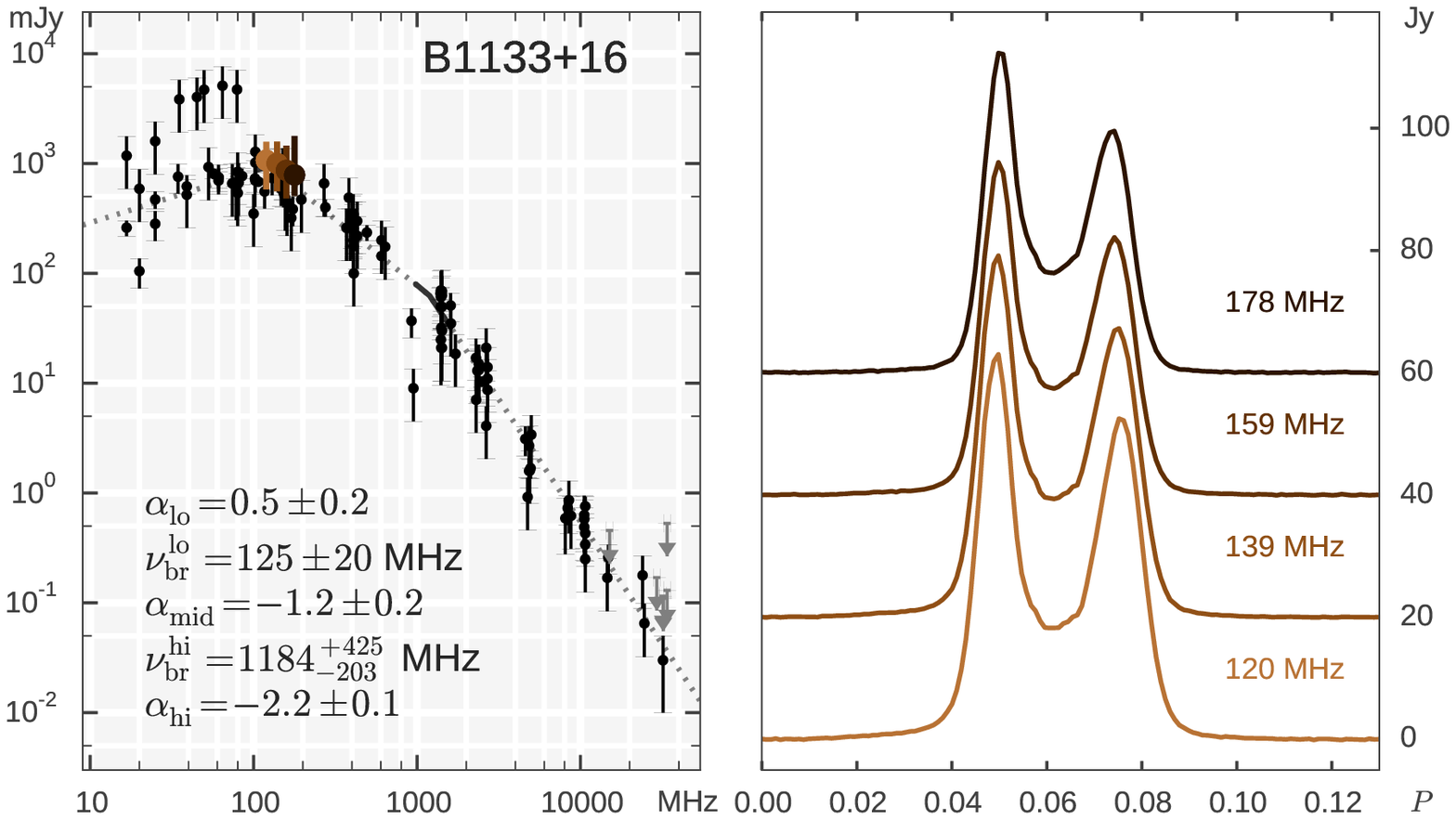}\includegraphics[scale=0.475]{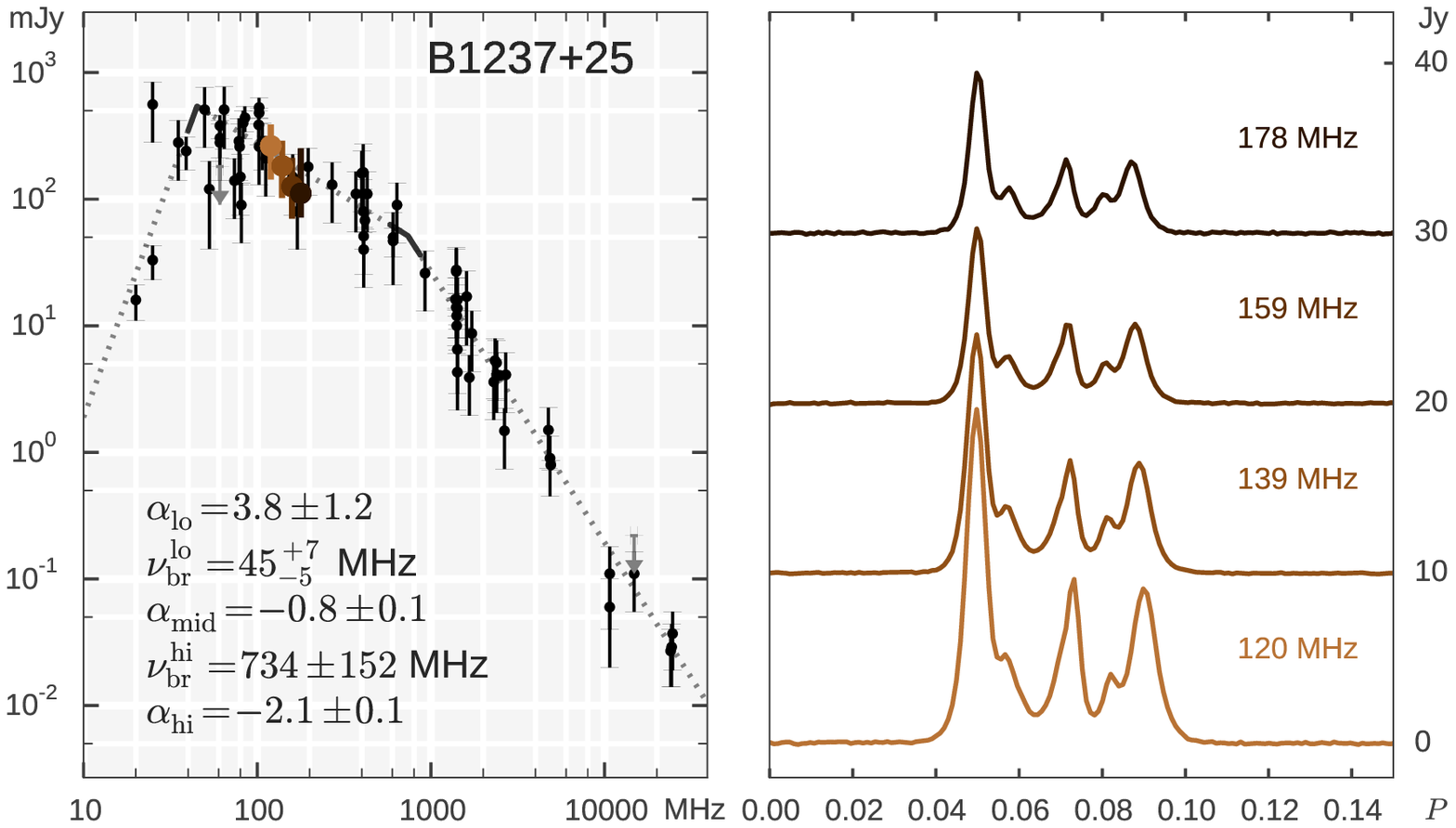}
\caption{See Figure~\ref{fig:prof_sp_1}. For PSR B1133+16 (bottom left), the 35--80\,MHz flux density measurements from \citet{Stovall2015} were
excluded from the fit since they were an order-of-magnitude larger than numerous previous measurements in the same frequency range.}
\label{fig:prof_sp_5}
\end{figure*}

\begin{figure*}
\includegraphics[scale=0.48]{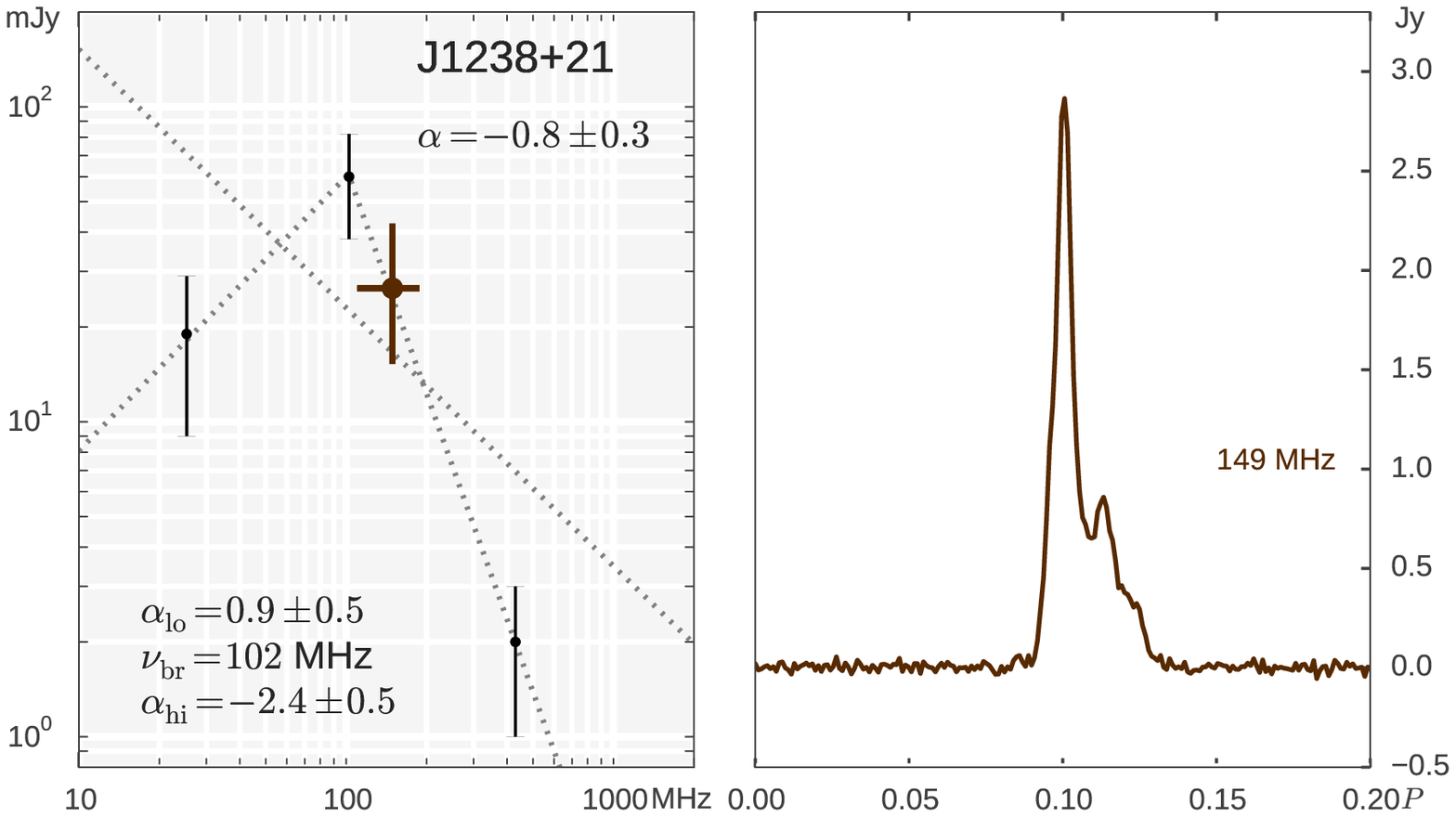}\includegraphics[scale=0.48]{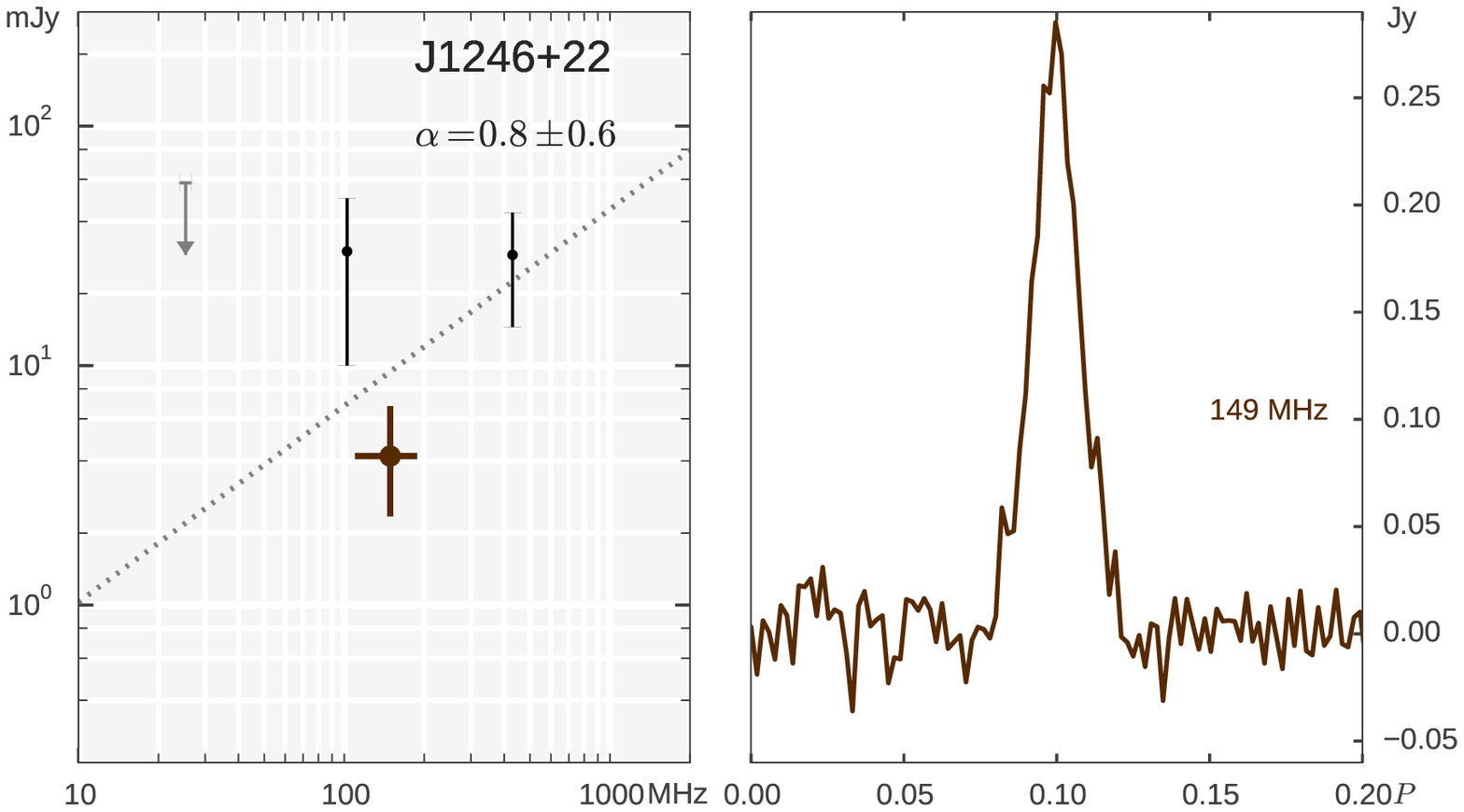}
\includegraphics[scale=0.48]{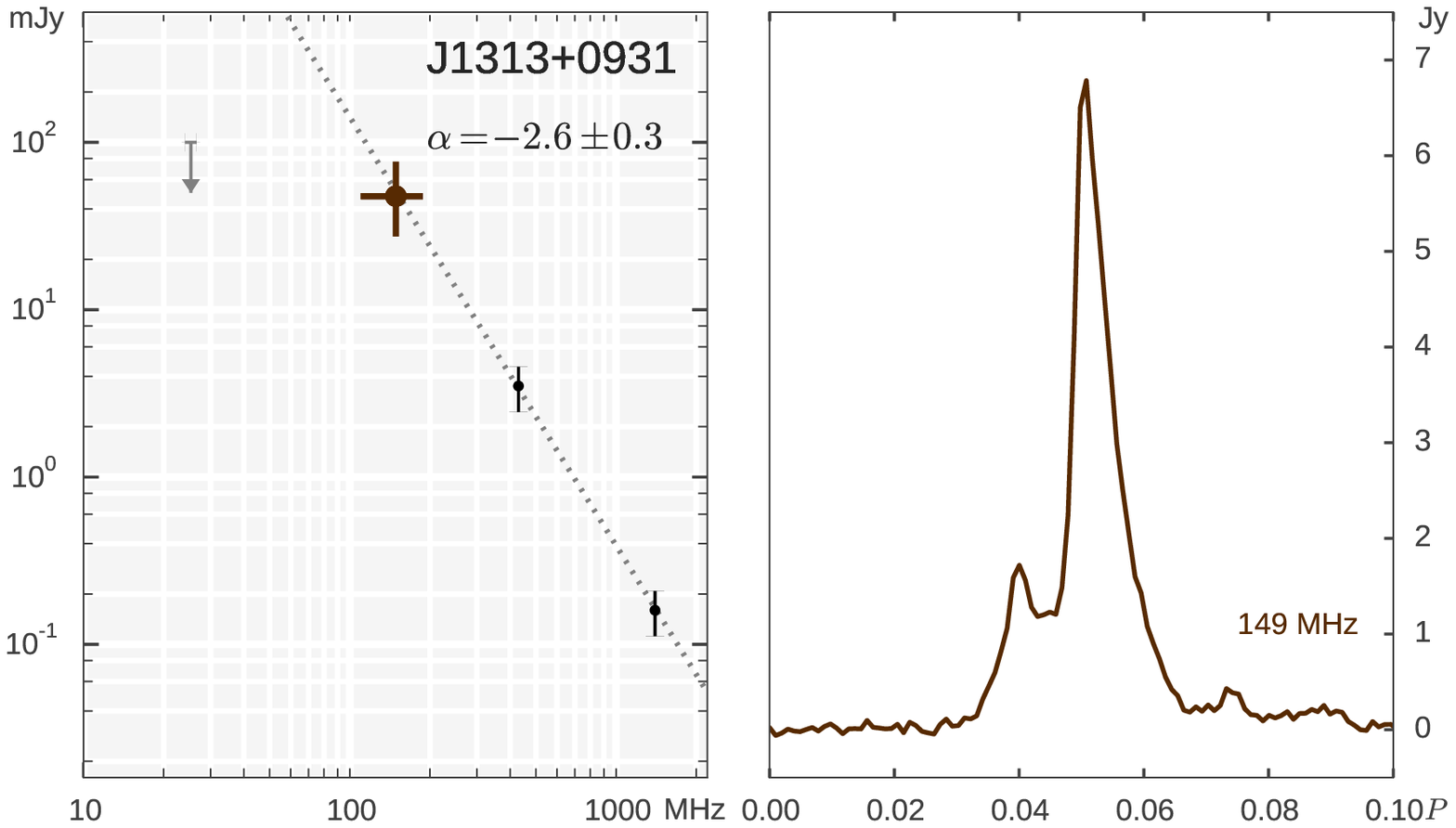}\includegraphics[scale=0.48]{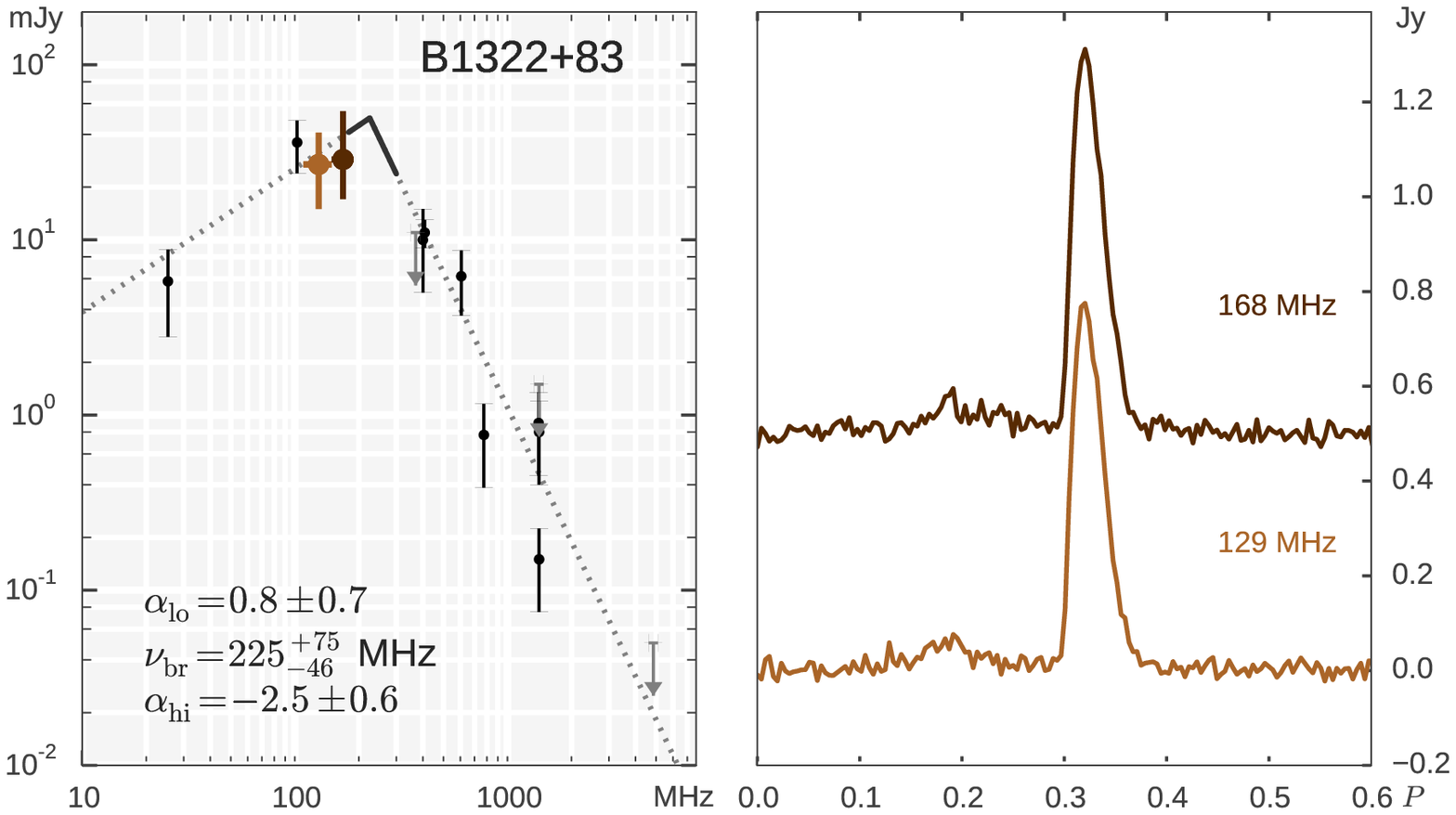}
\includegraphics[scale=0.48]{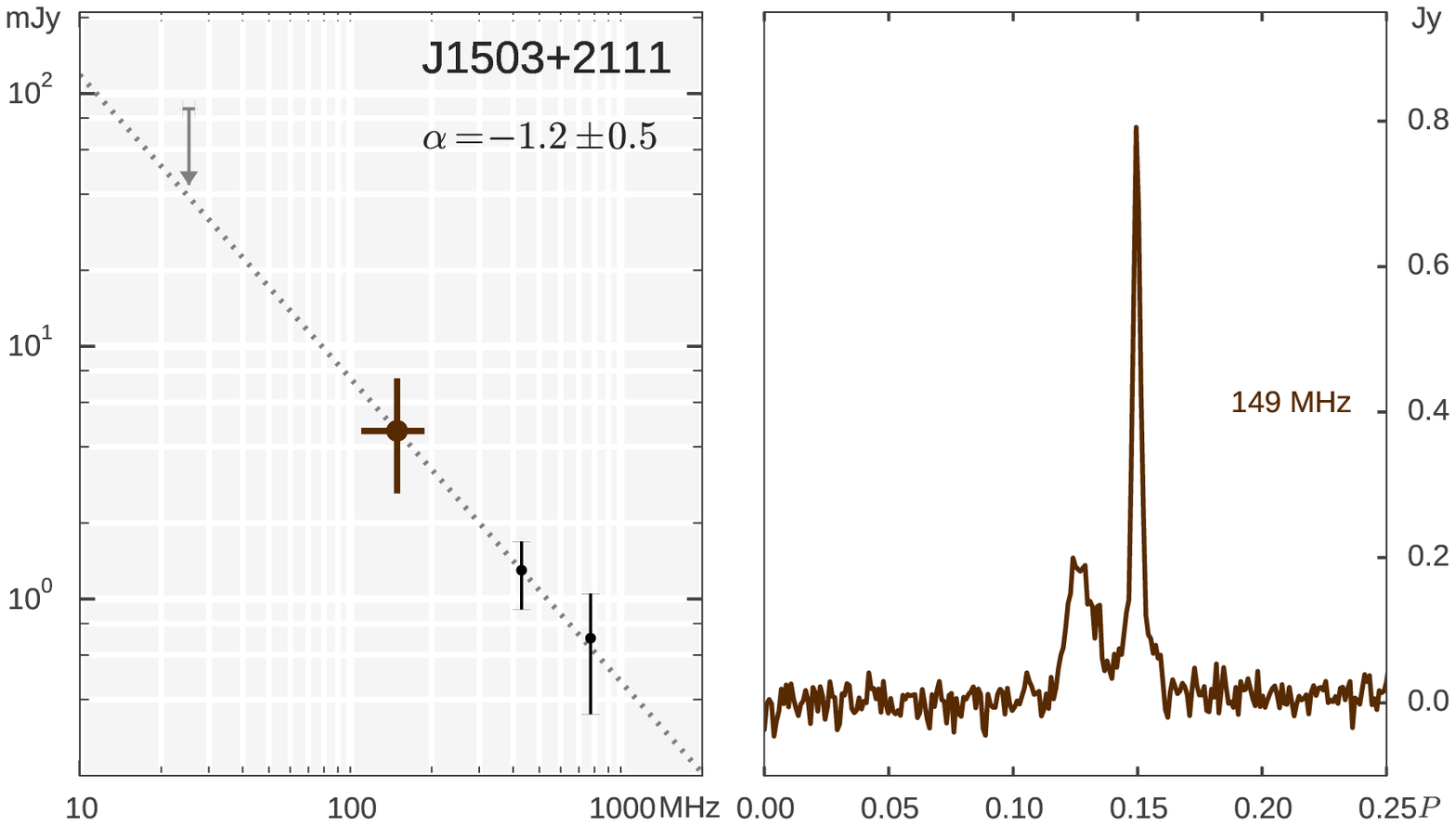}\includegraphics[scale=0.48]{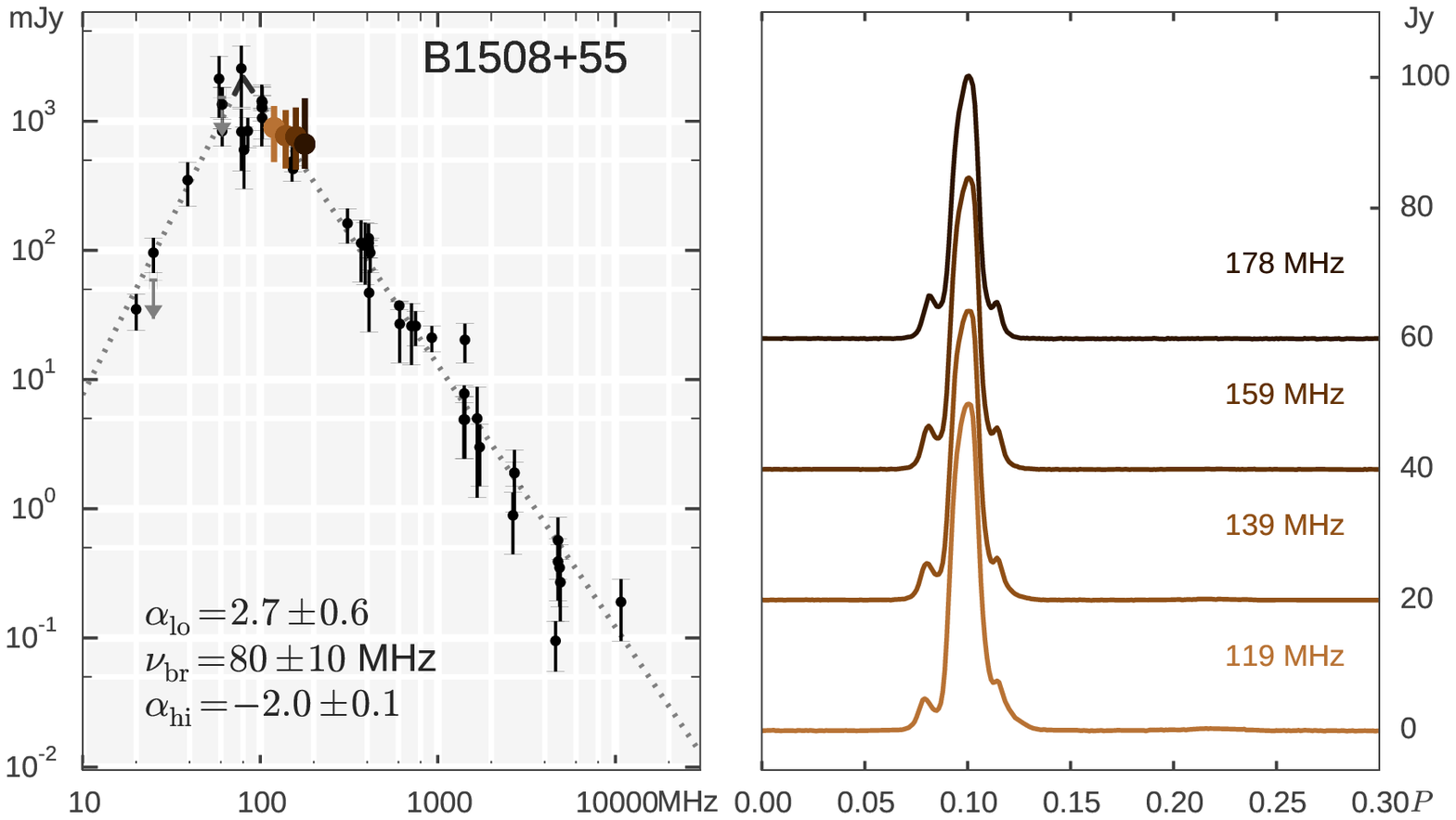}
\includegraphics[scale=0.48]{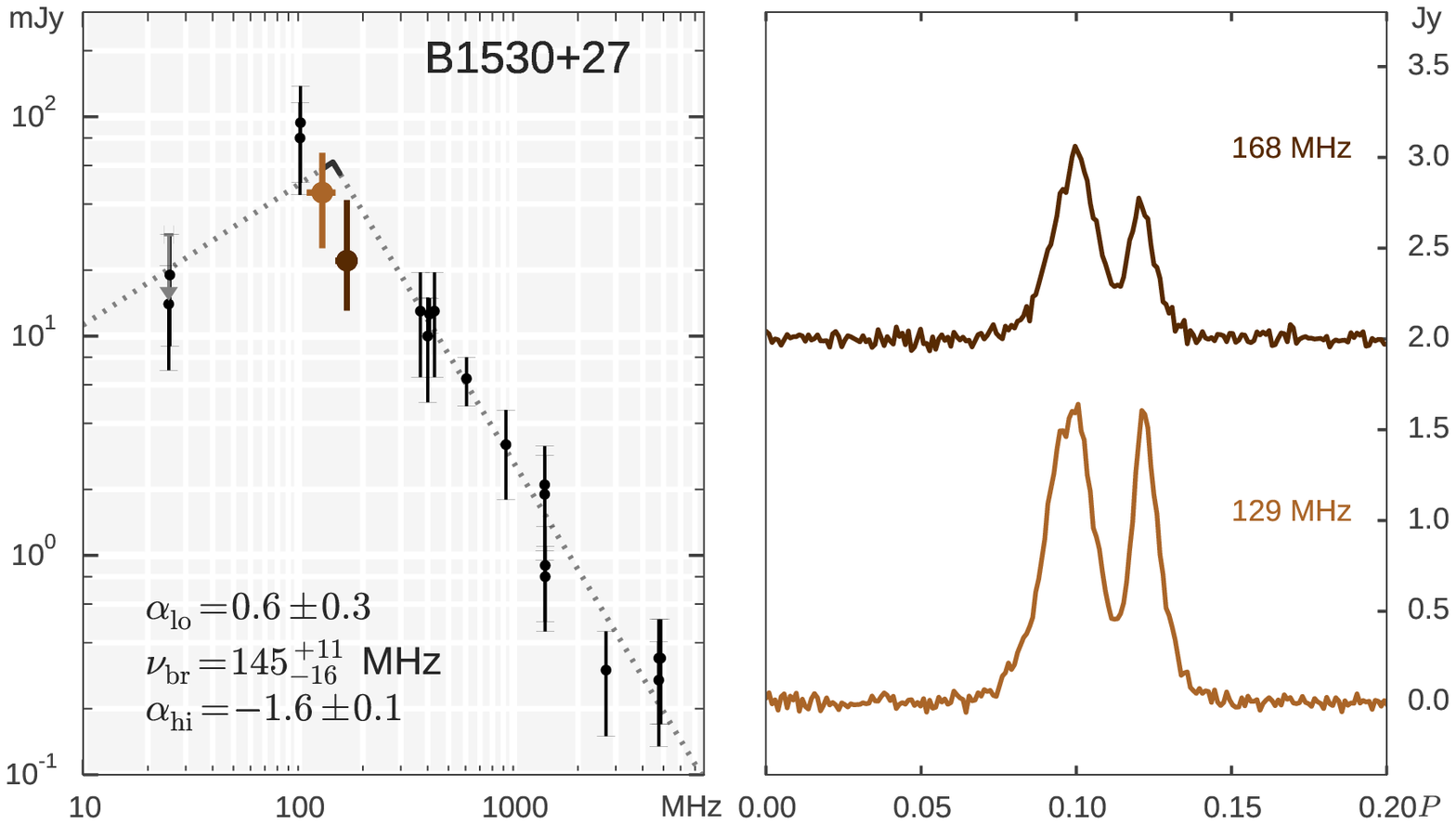}\includegraphics[scale=0.48]{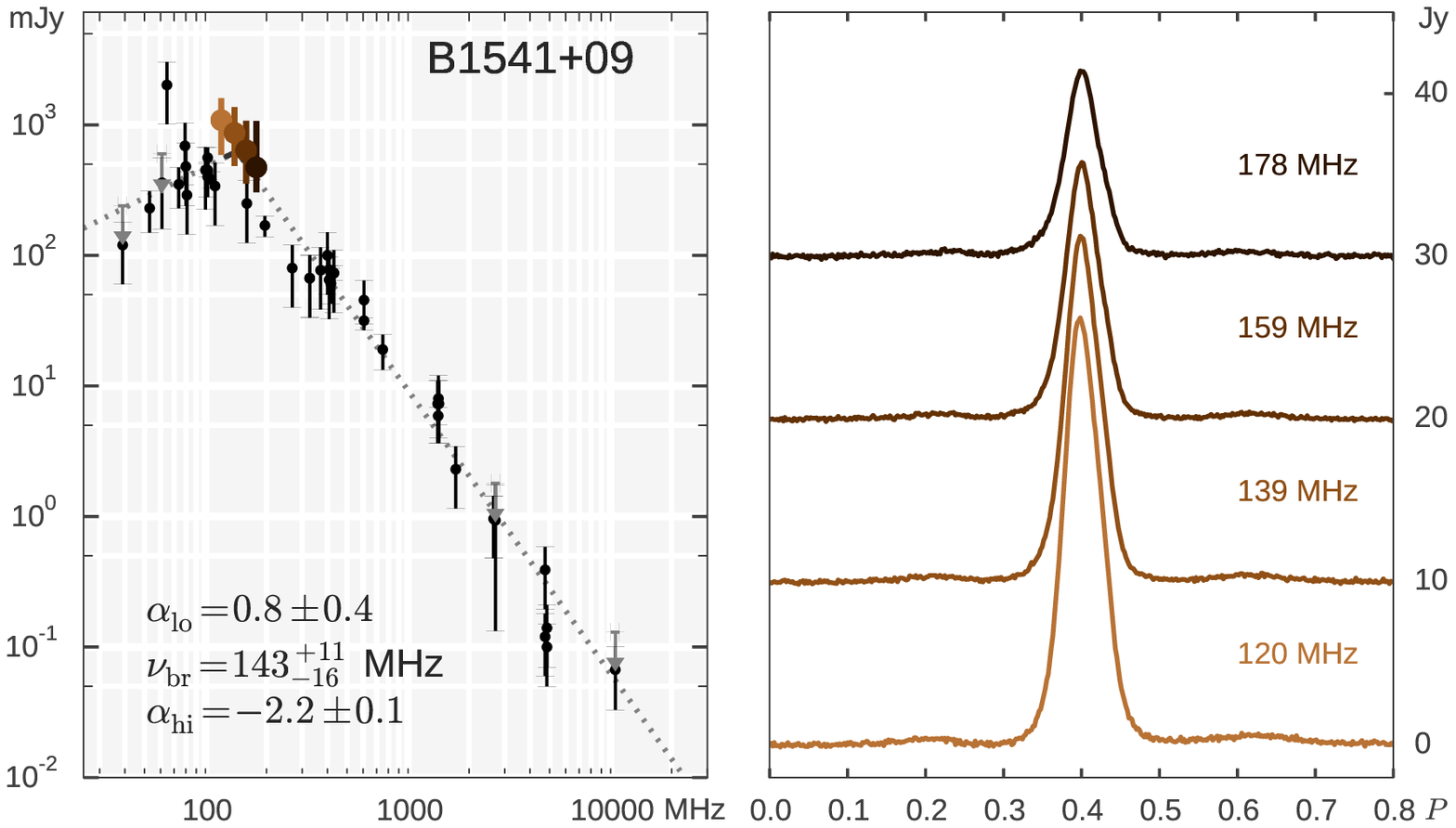}
\includegraphics[scale=0.48]{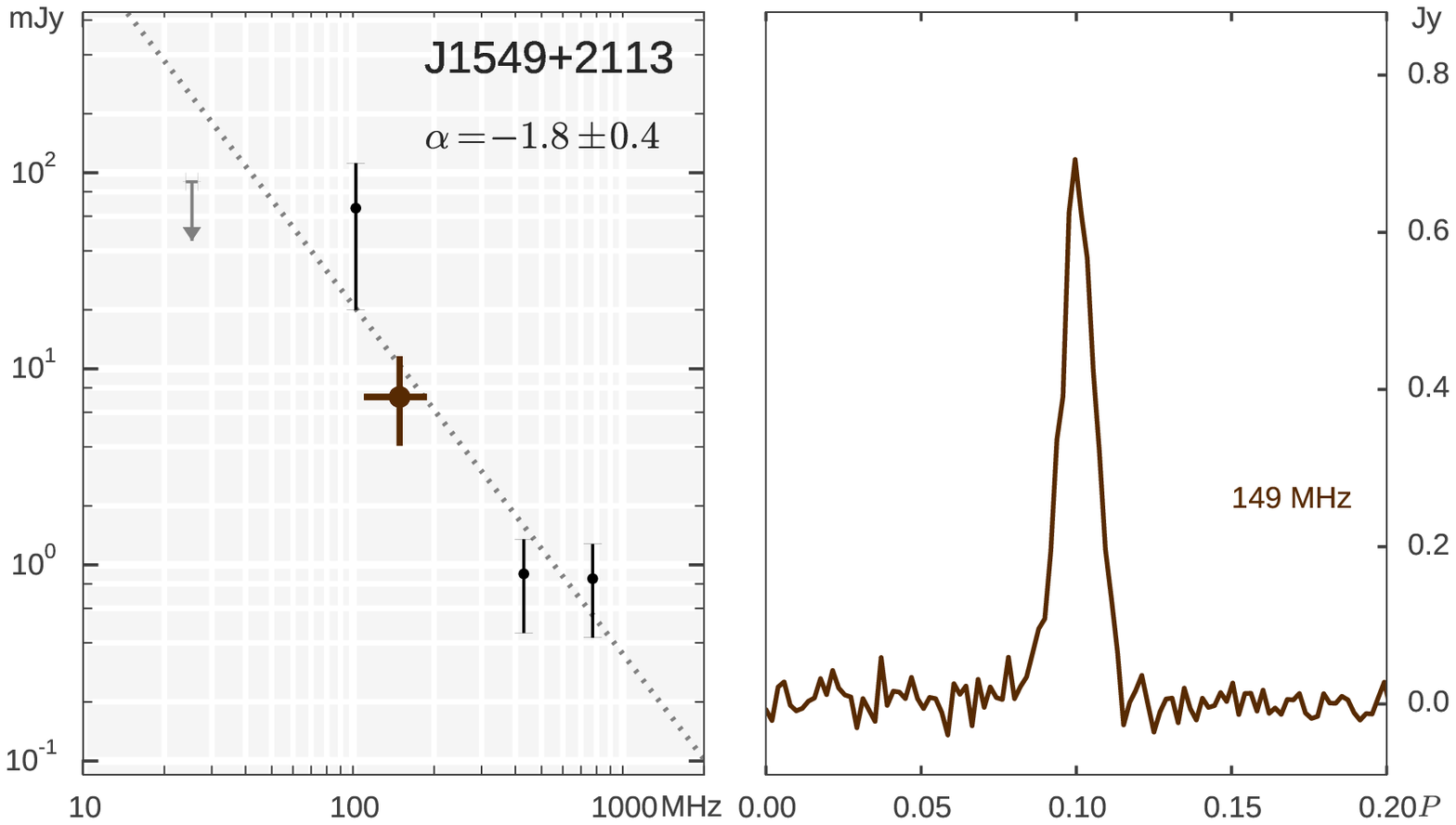}\includegraphics[scale=0.48]{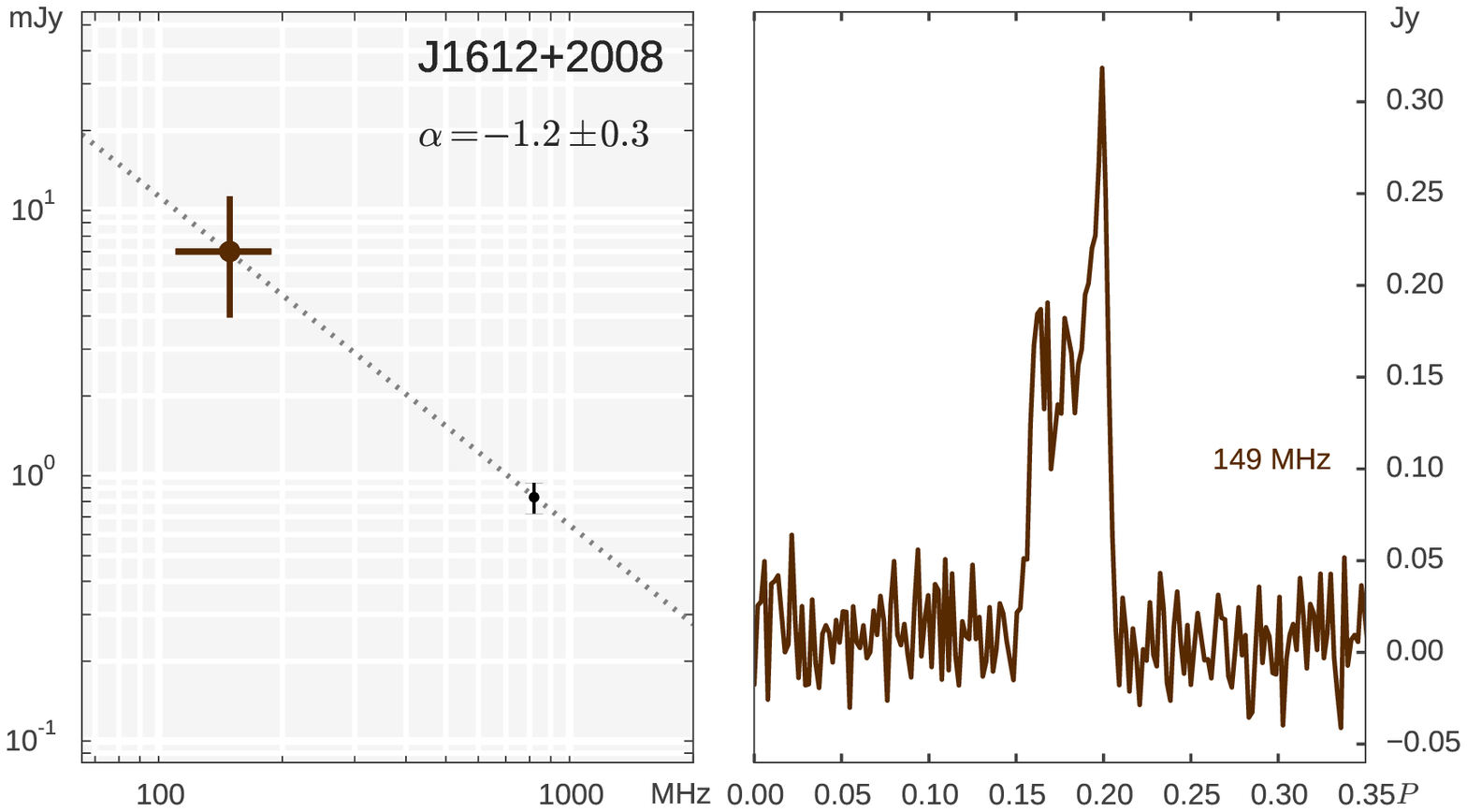}
\caption{See Figure~\ref{fig:prof_sp_1}.}
\label{fig:prof_sp_6}
\end{figure*}

\begin{figure*}
\includegraphics[scale=0.48]{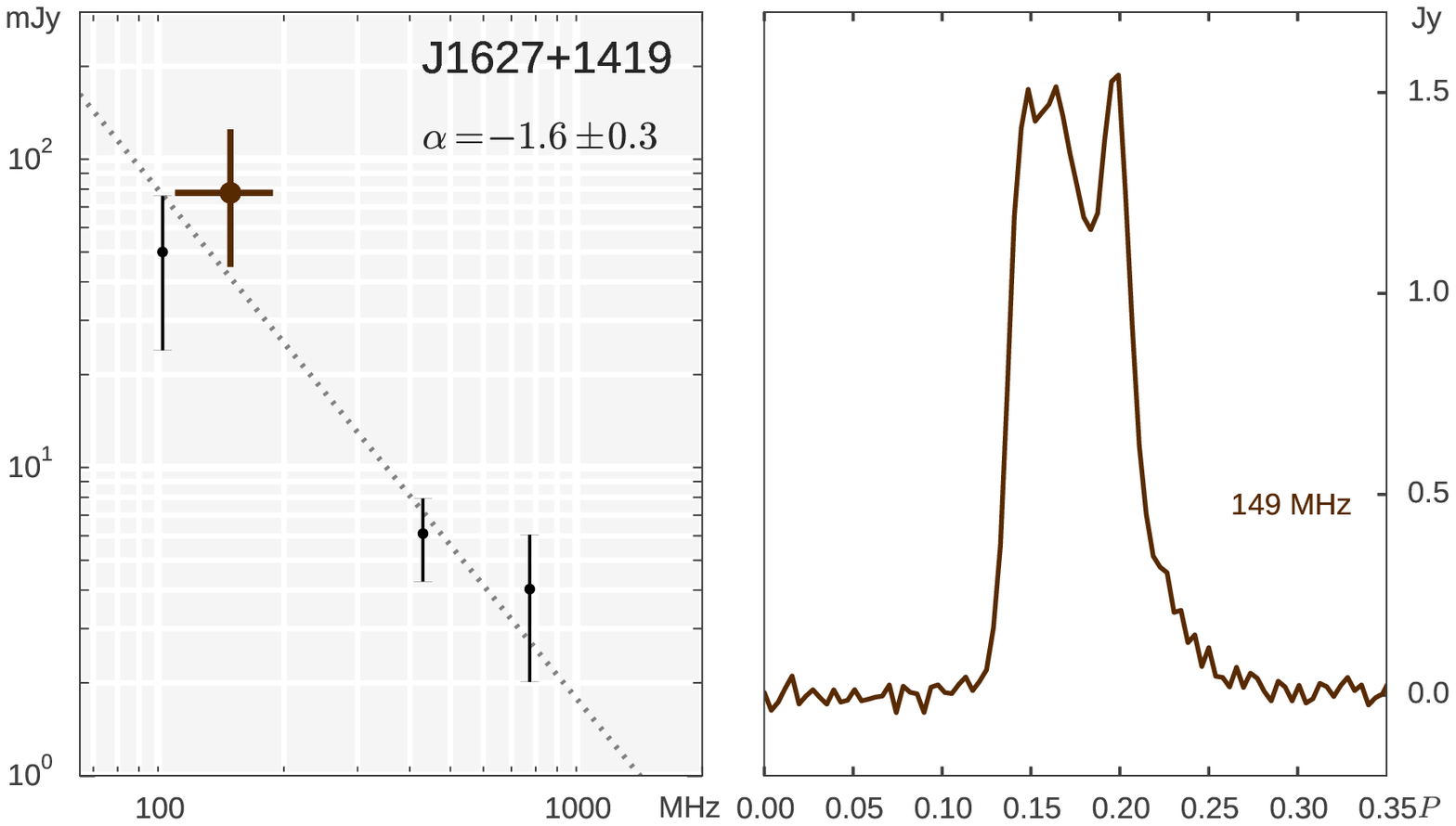}\includegraphics[scale=0.48]{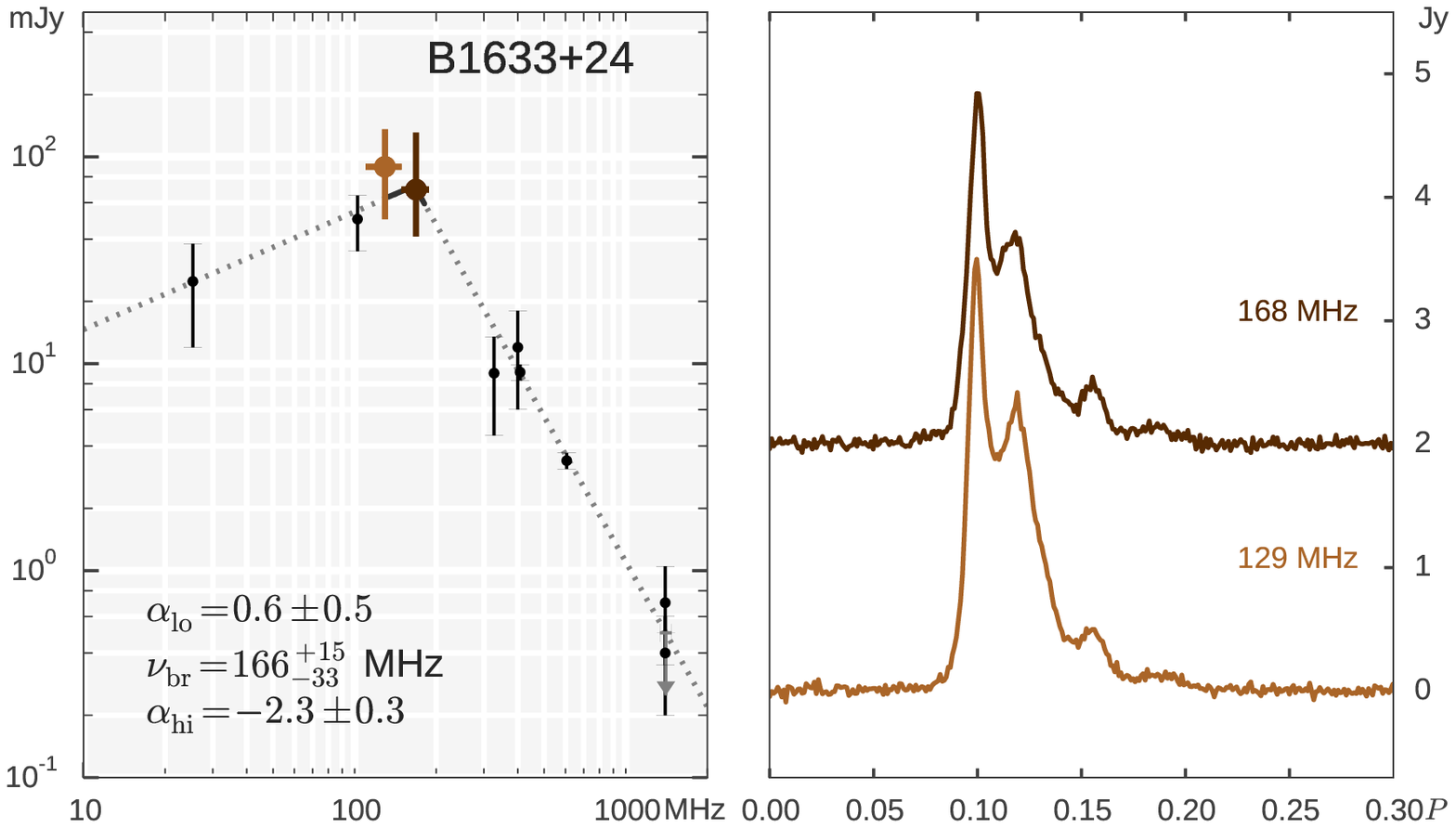}
\includegraphics[scale=0.48]{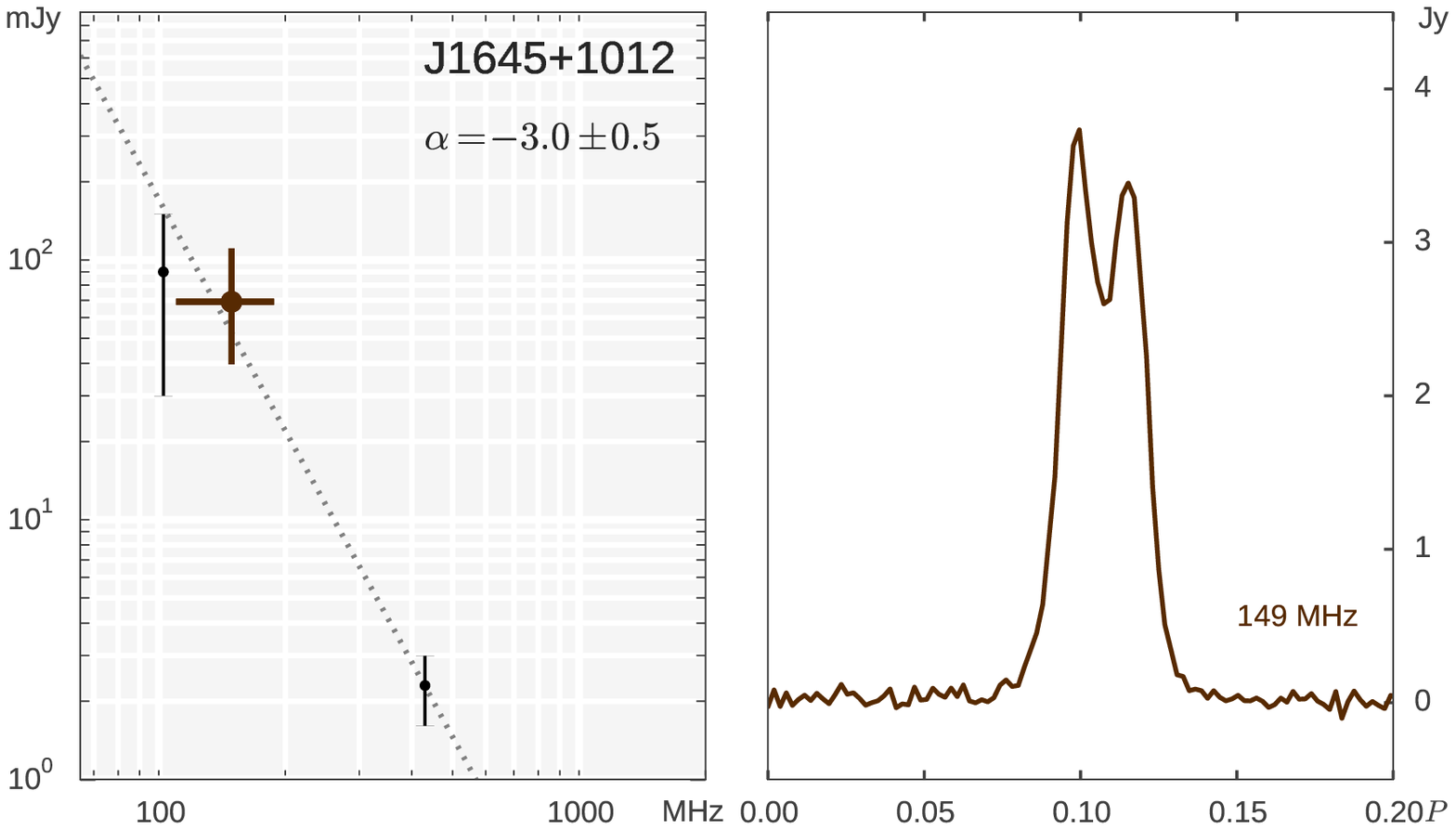}\includegraphics[scale=0.48]{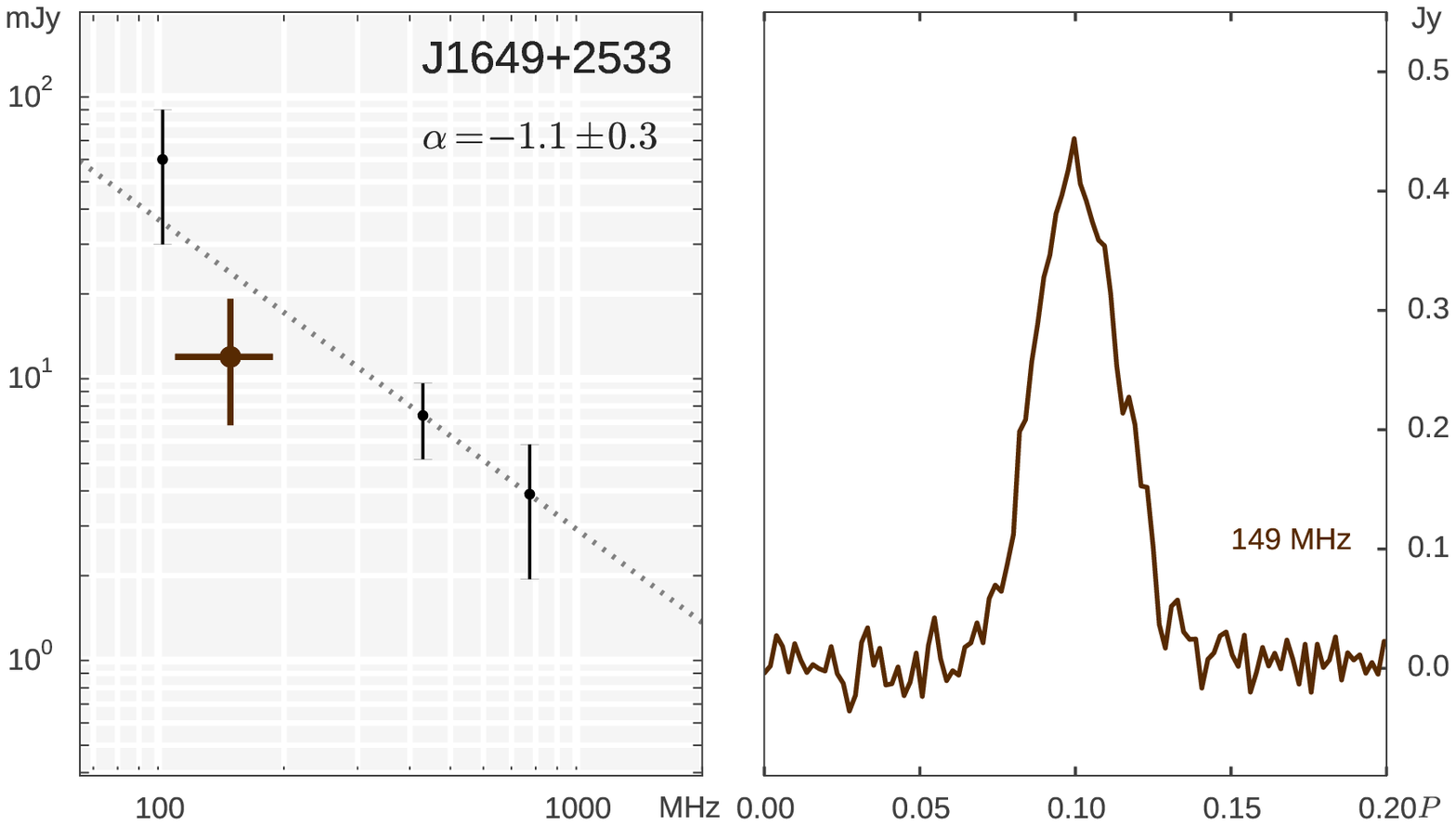}
\includegraphics[scale=0.48]{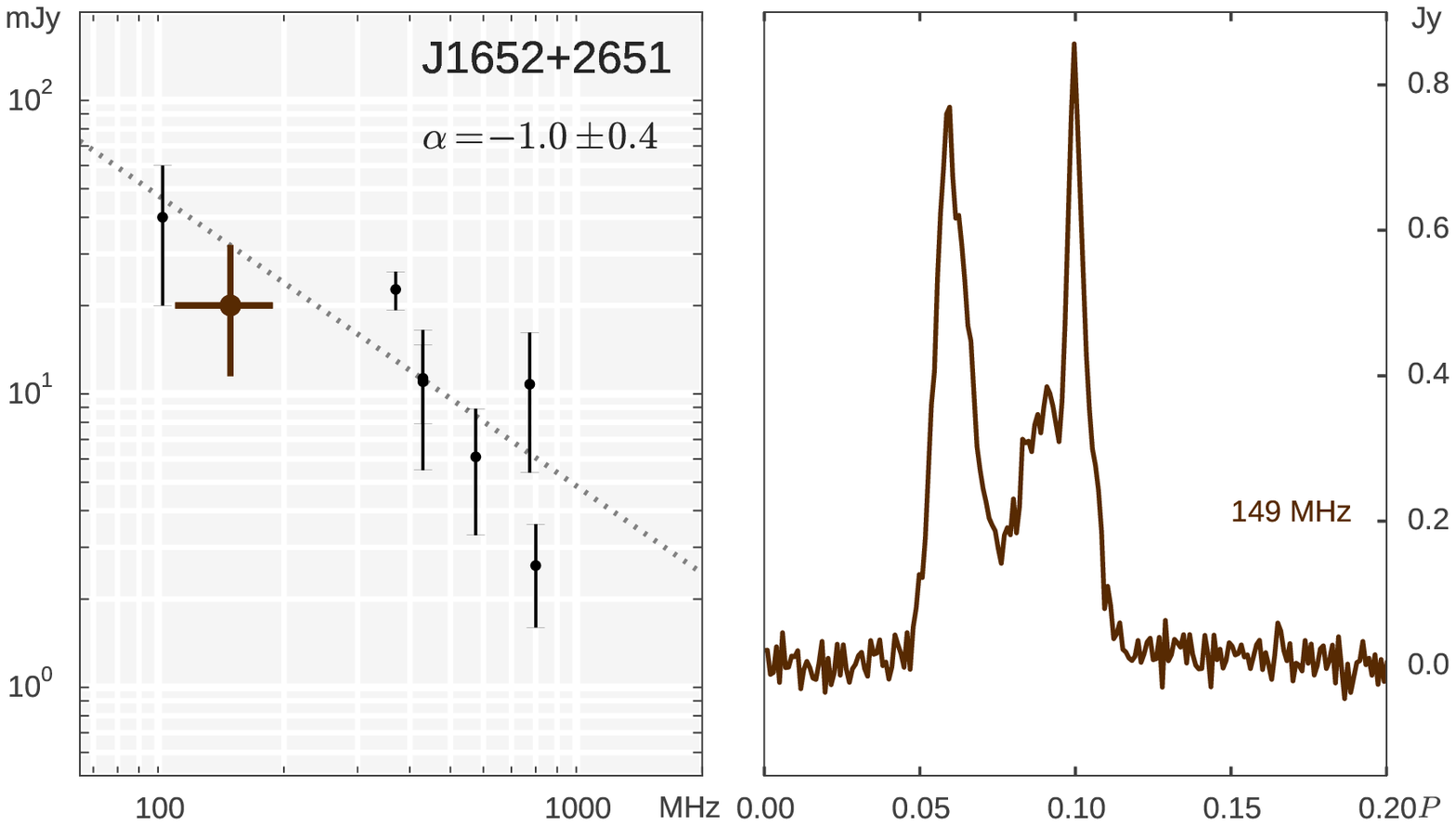}\includegraphics[scale=0.48]{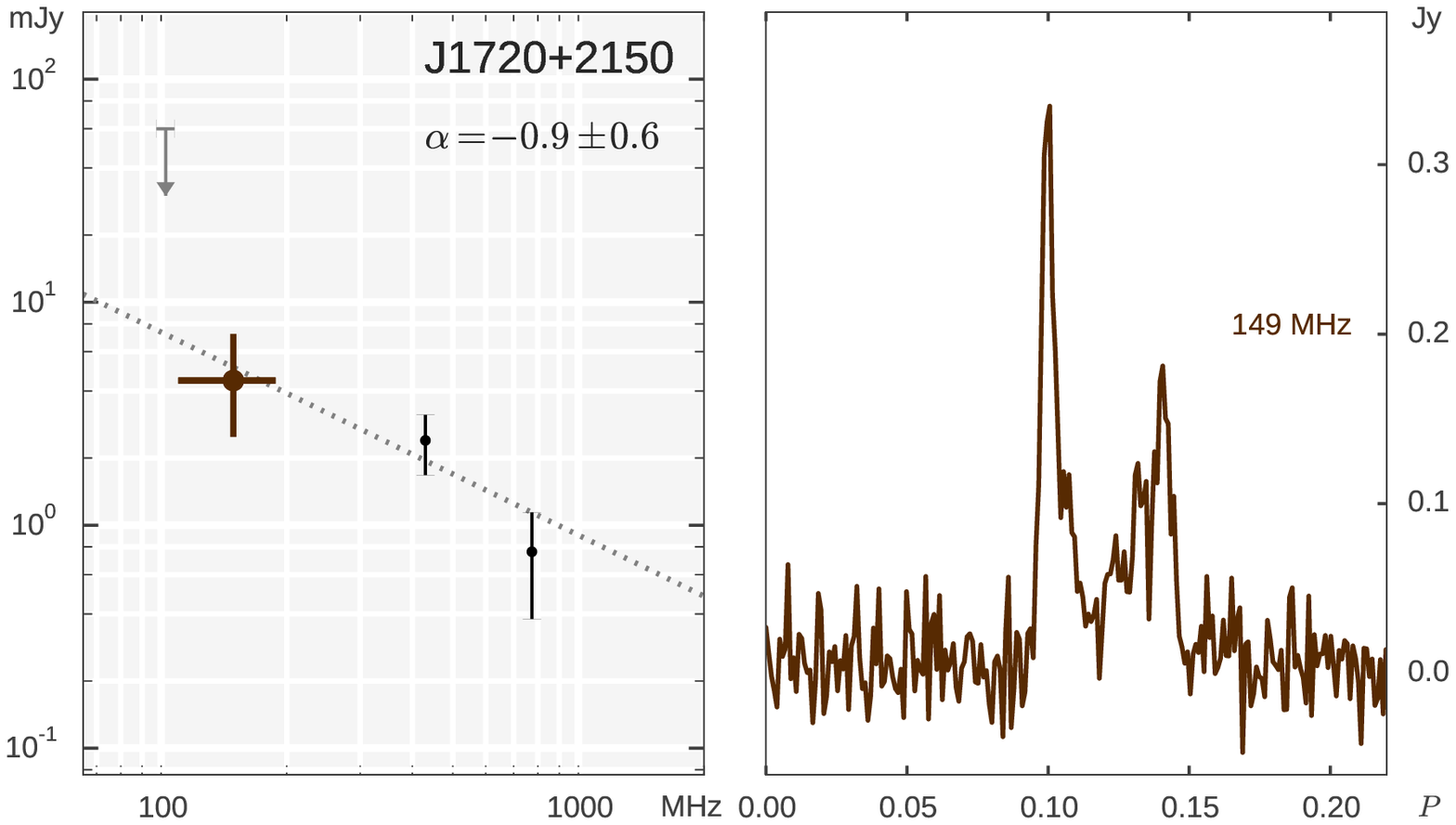}
\includegraphics[scale=0.48]{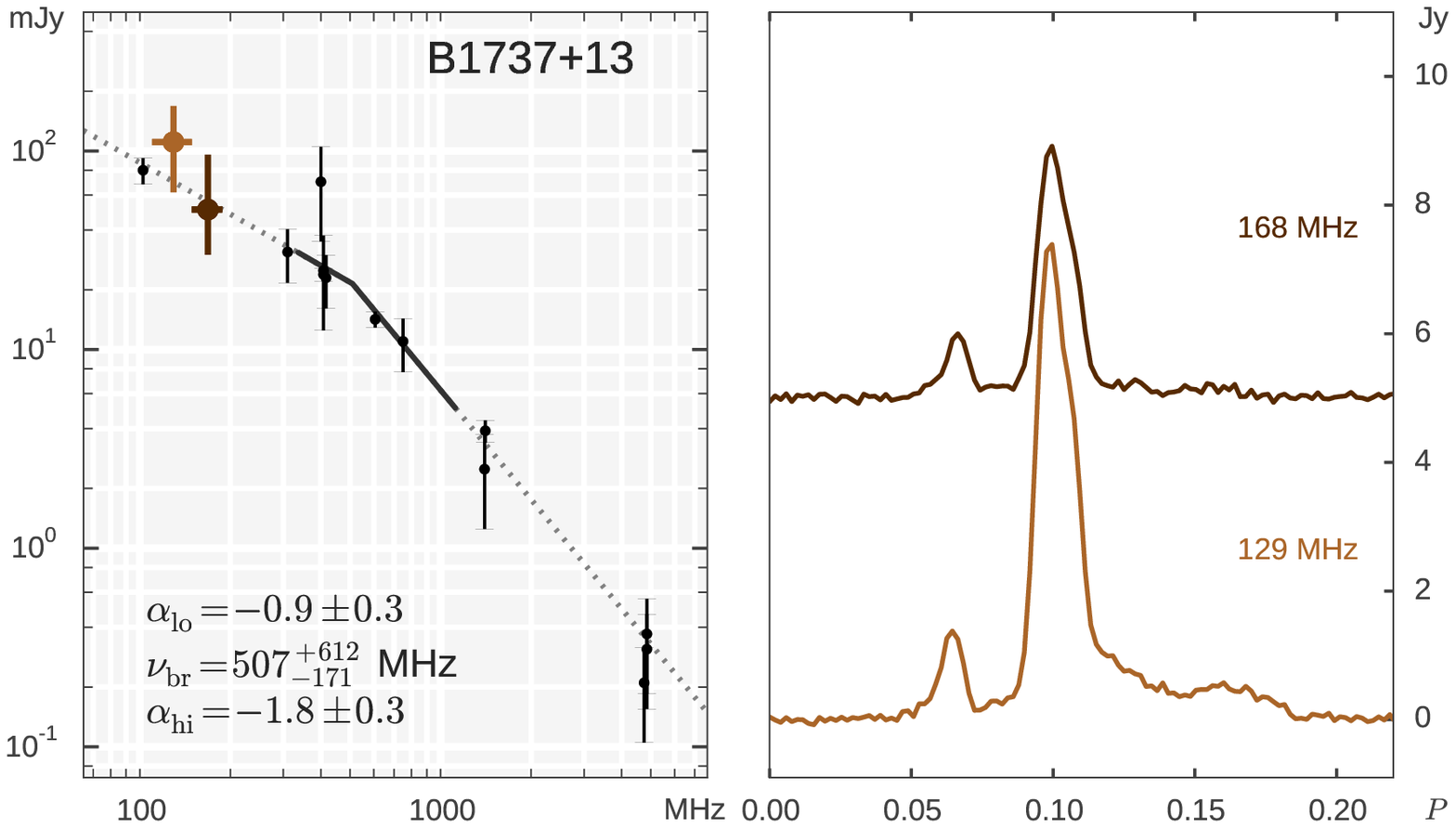}\includegraphics[scale=0.48]{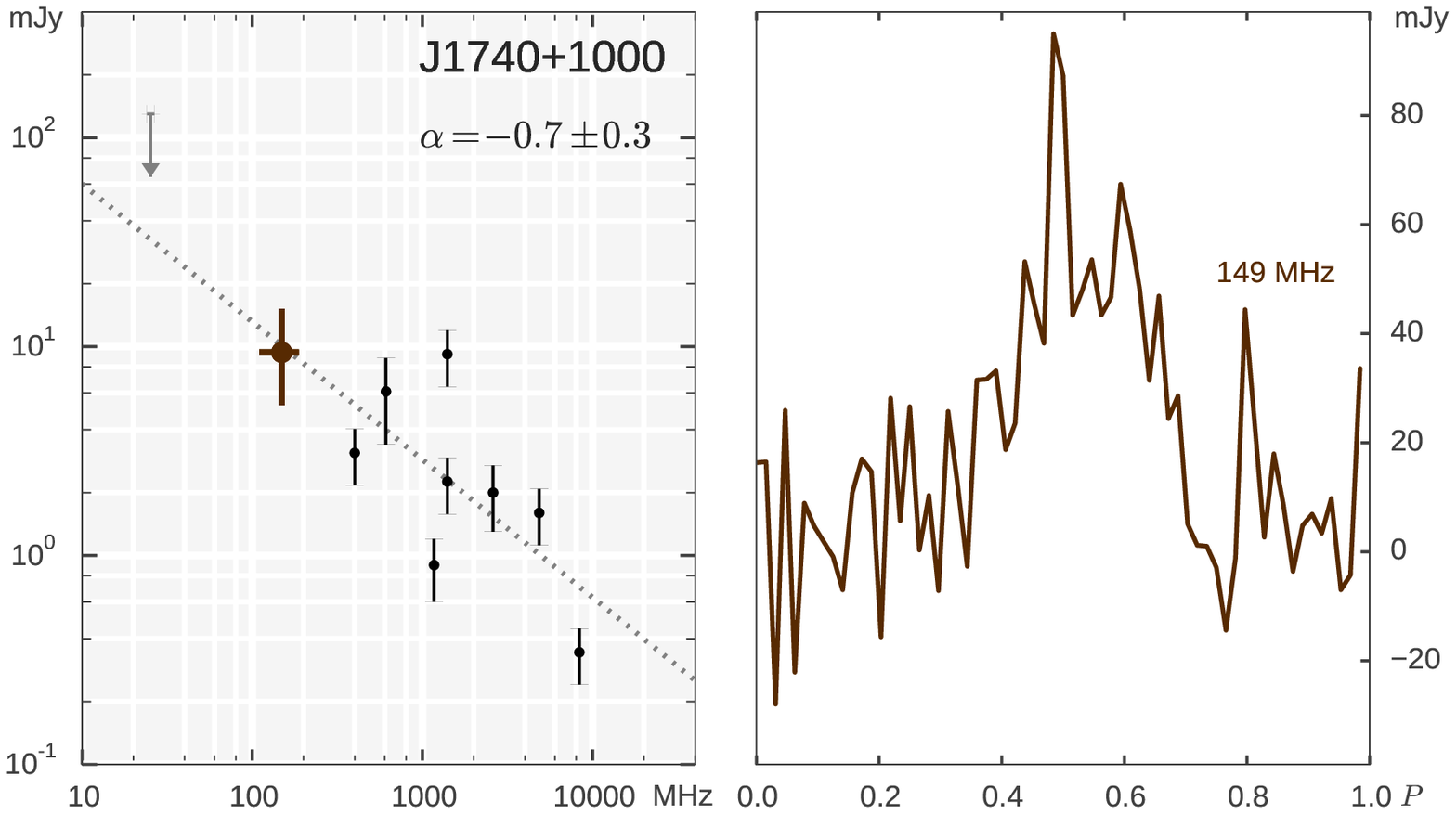}
\includegraphics[scale=0.48]{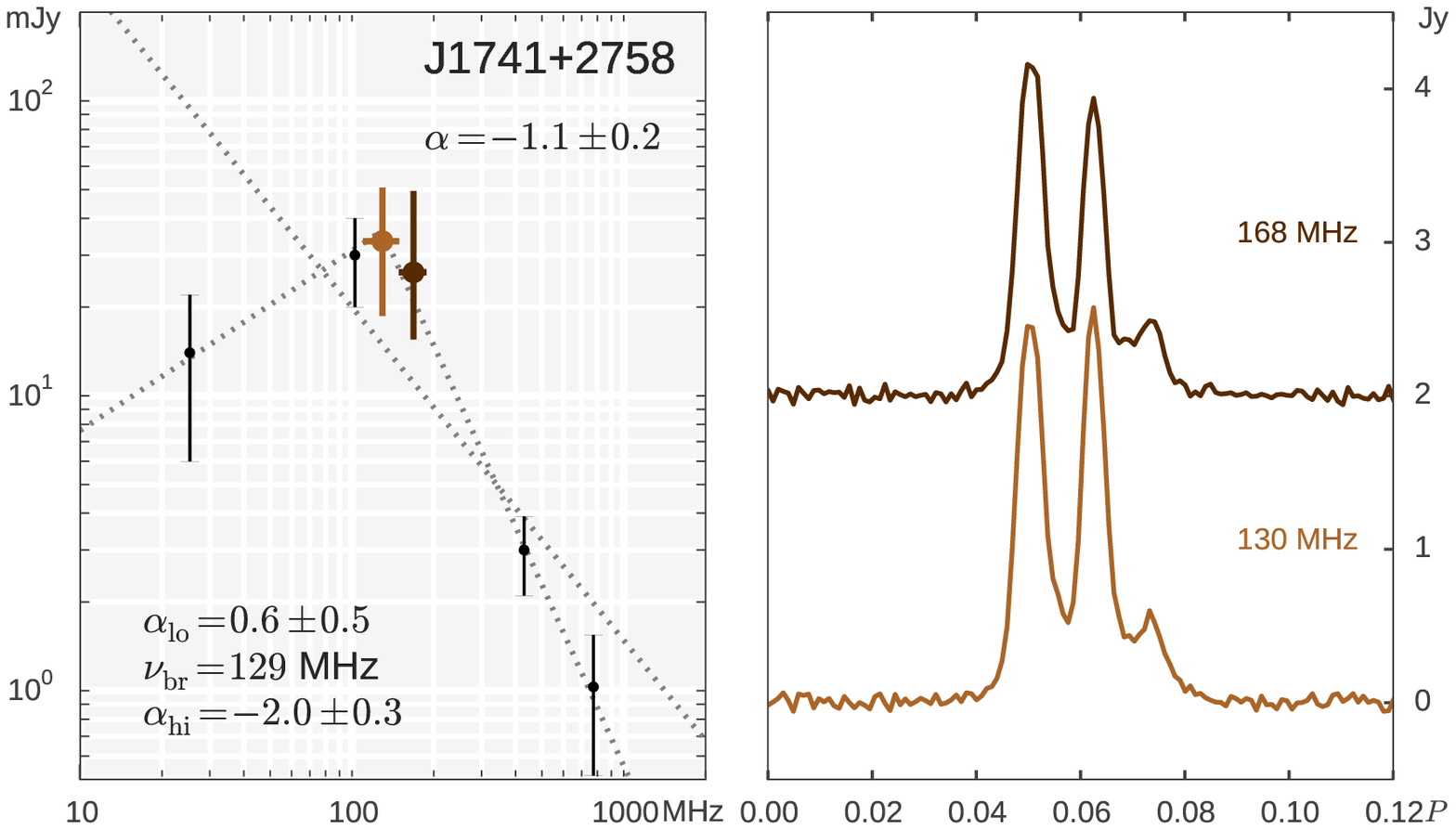}\includegraphics[scale=0.48]{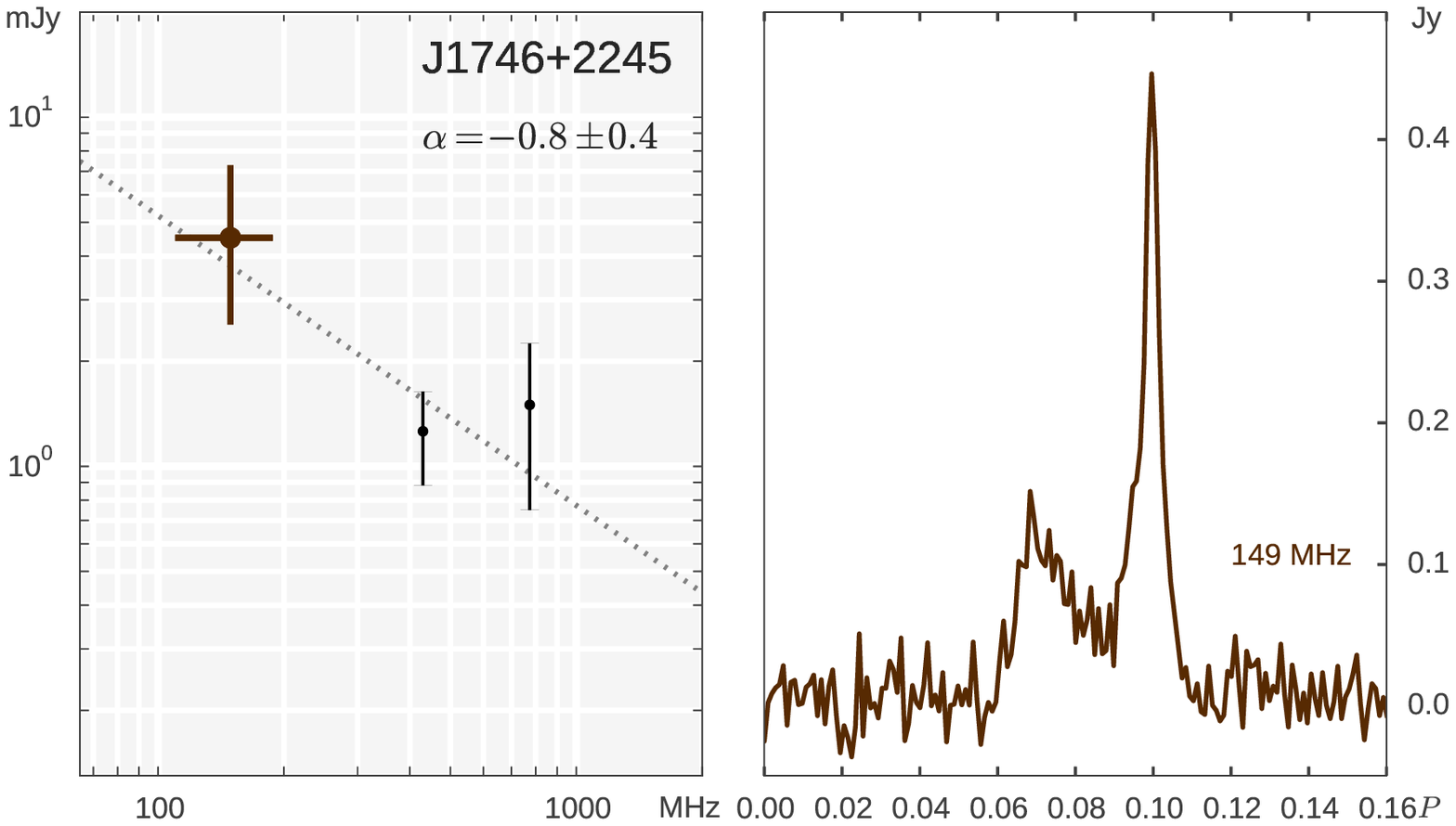}
\caption{See Figure~\ref{fig:prof_sp_1}.}
\label{fig:prof_sp_7}
\end{figure*}

\begin{figure*}
\includegraphics[scale=0.48]{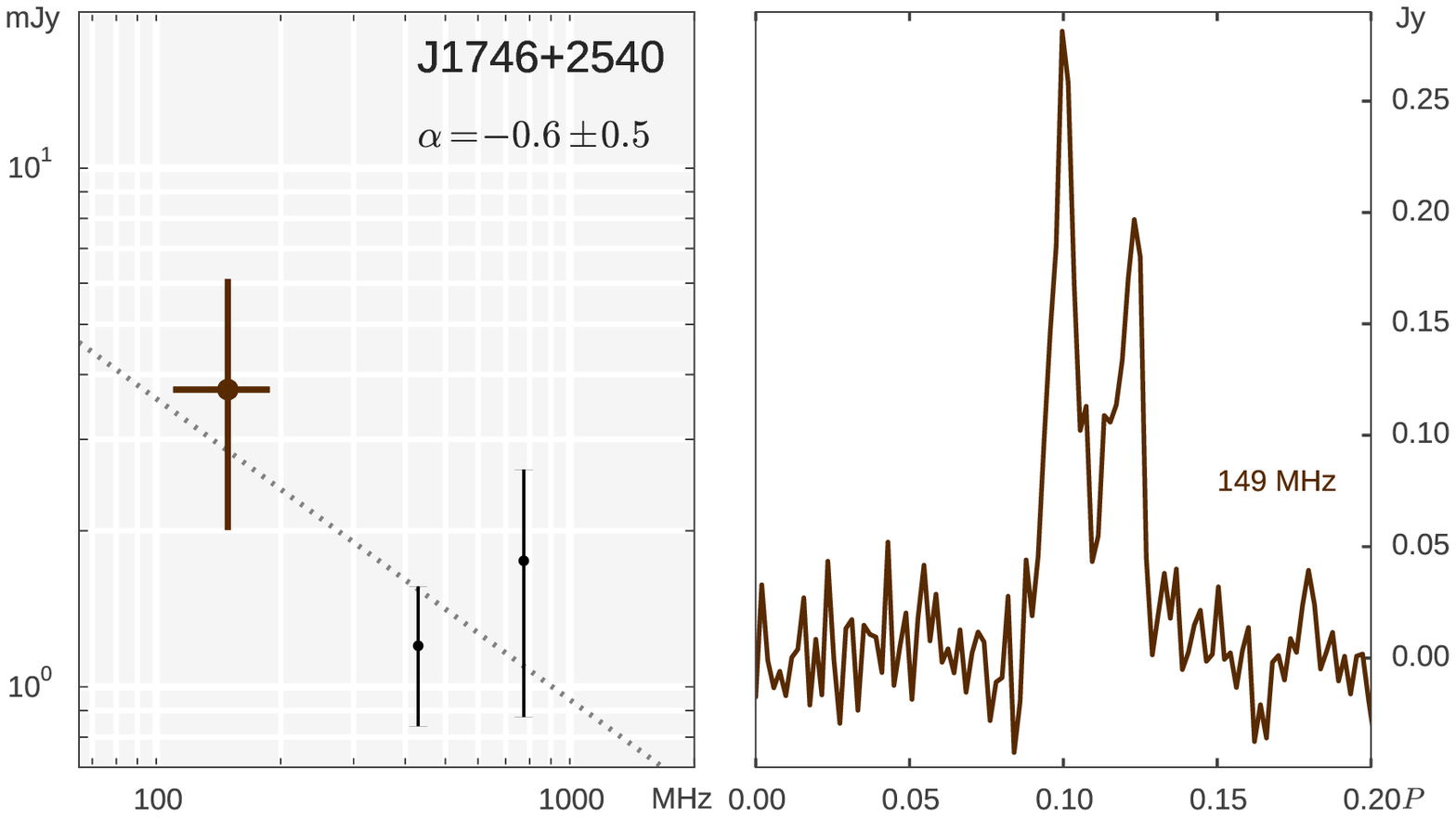}\includegraphics[scale=0.48]{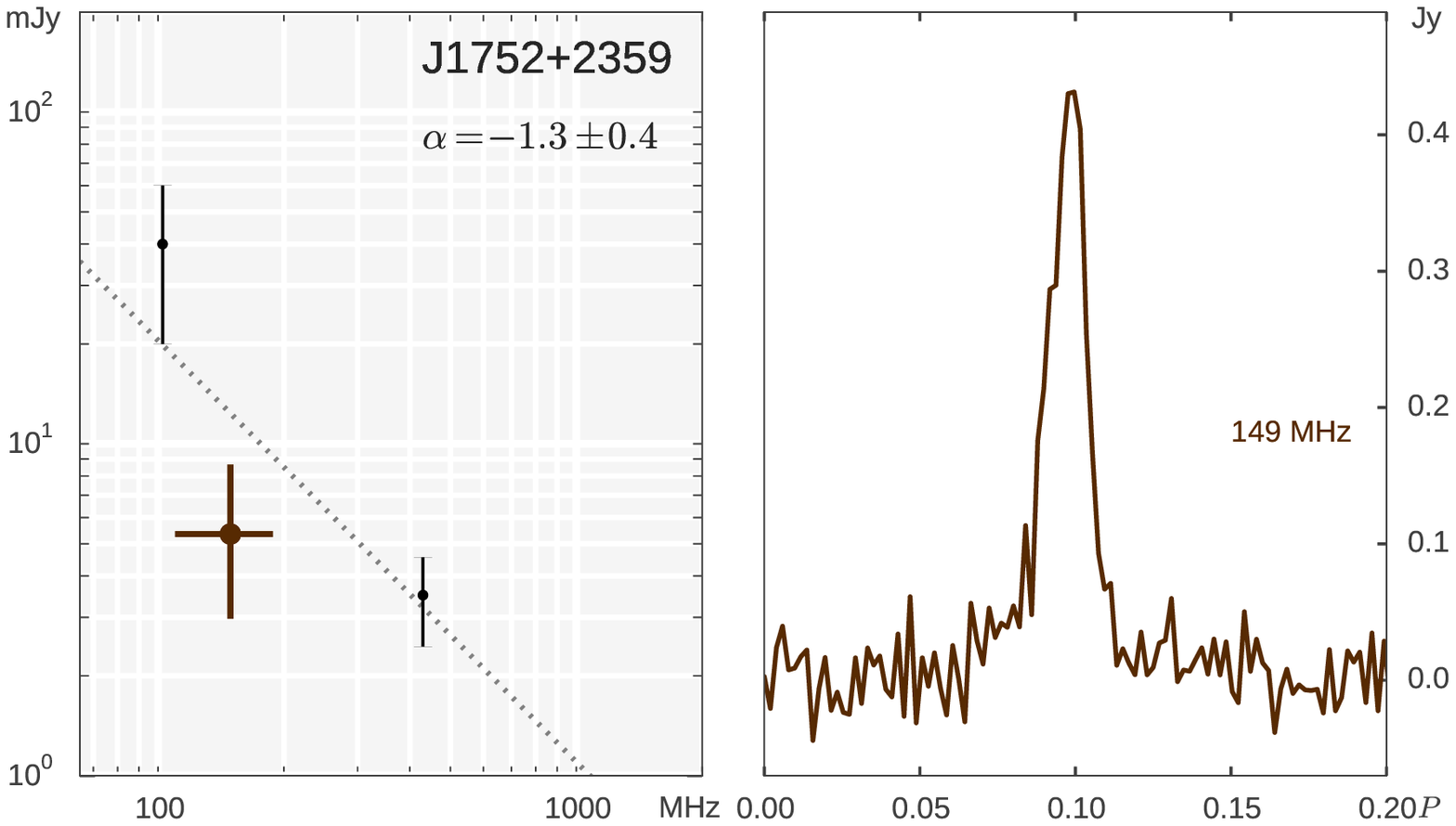}
\includegraphics[scale=0.48]{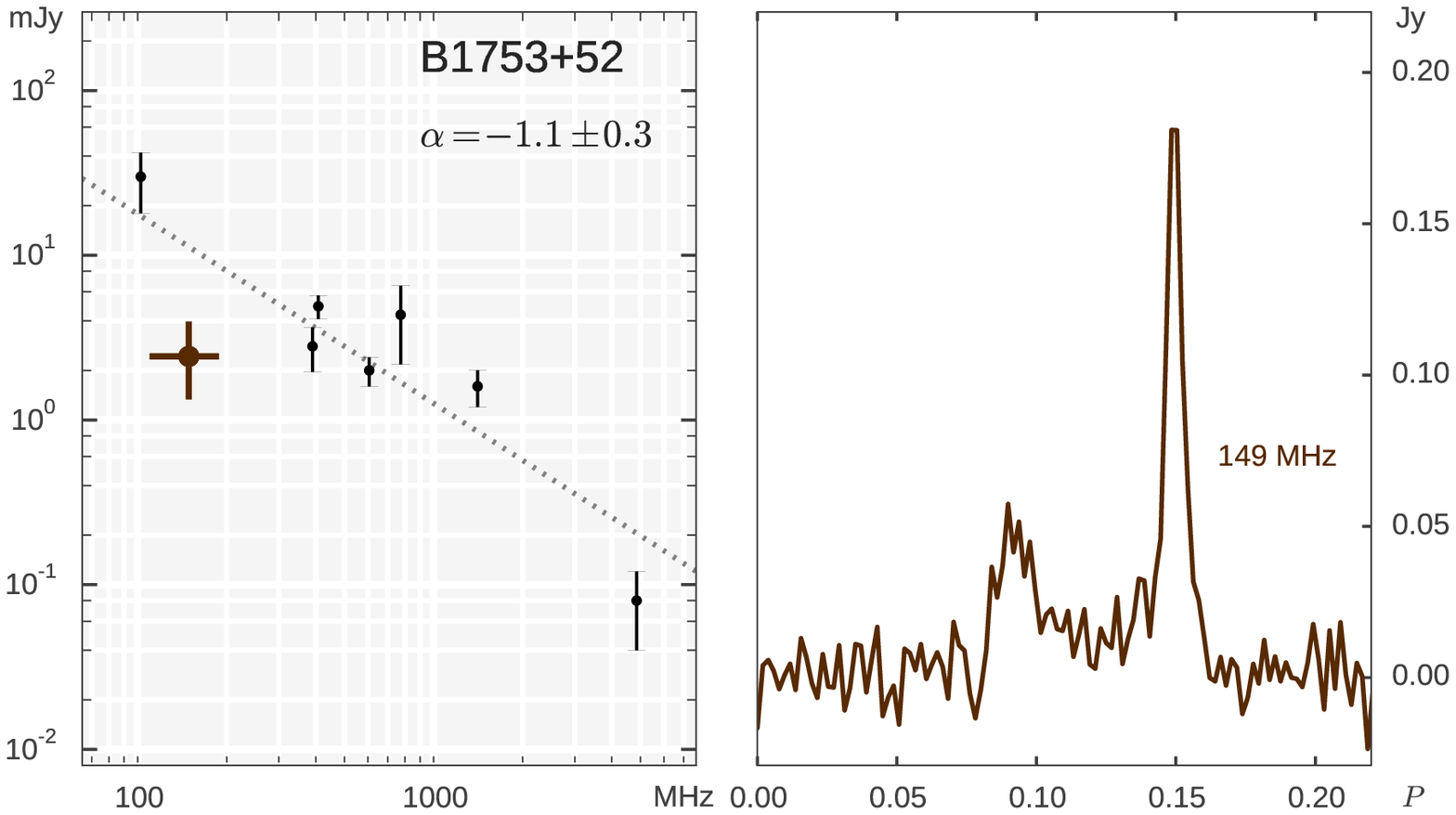}\includegraphics[scale=0.48]{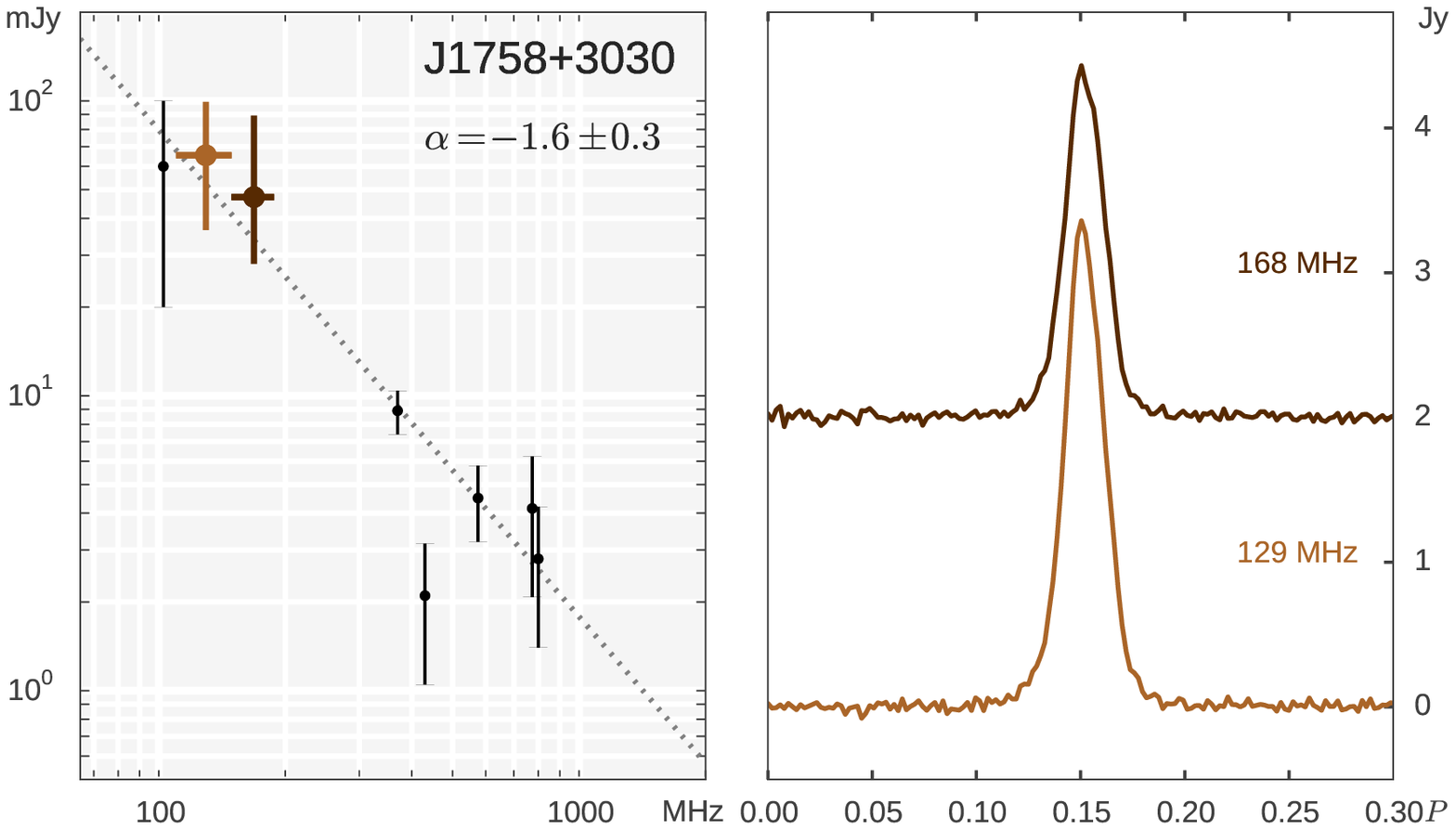}
\includegraphics[scale=0.48]{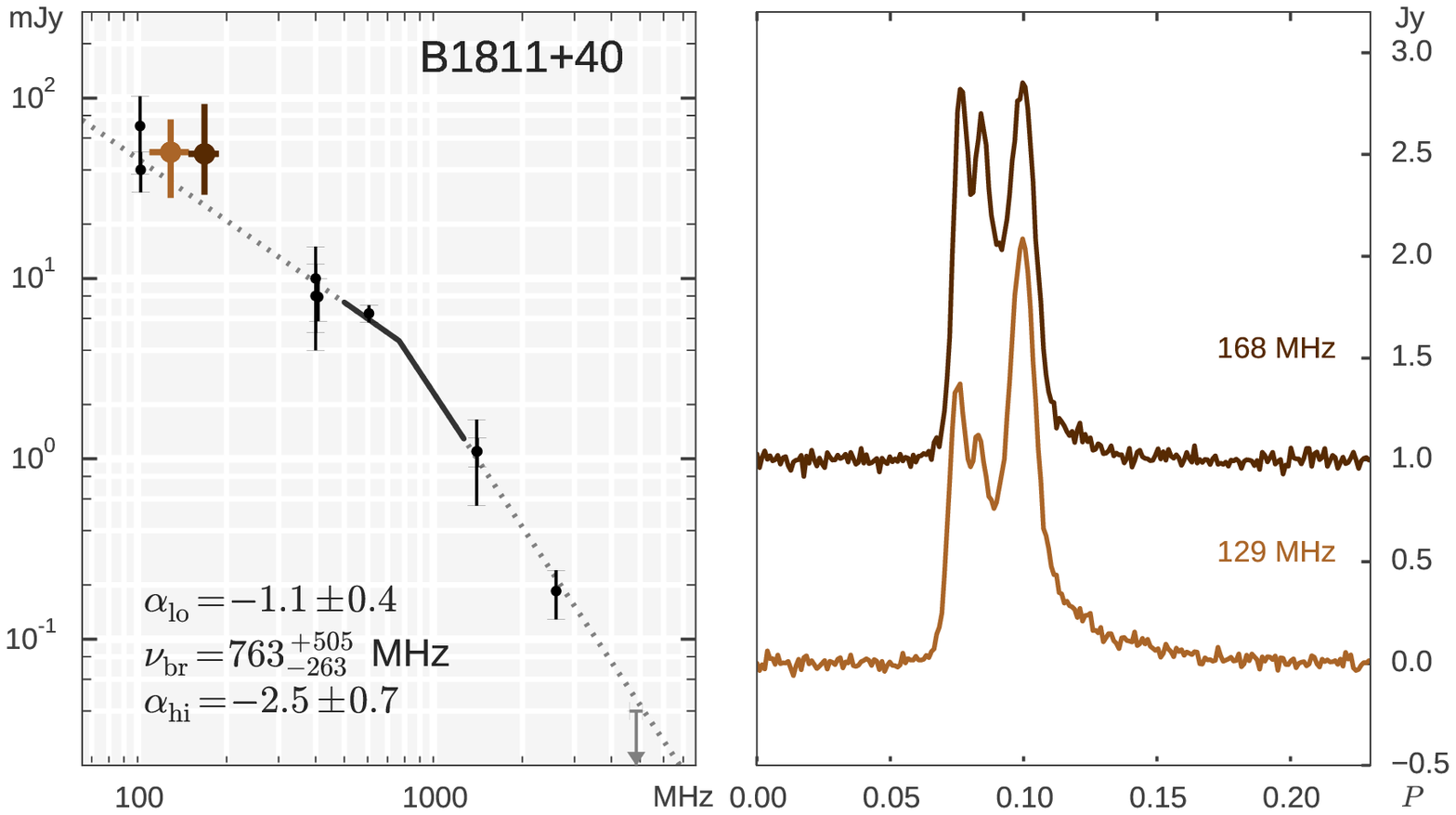}\includegraphics[scale=0.48]{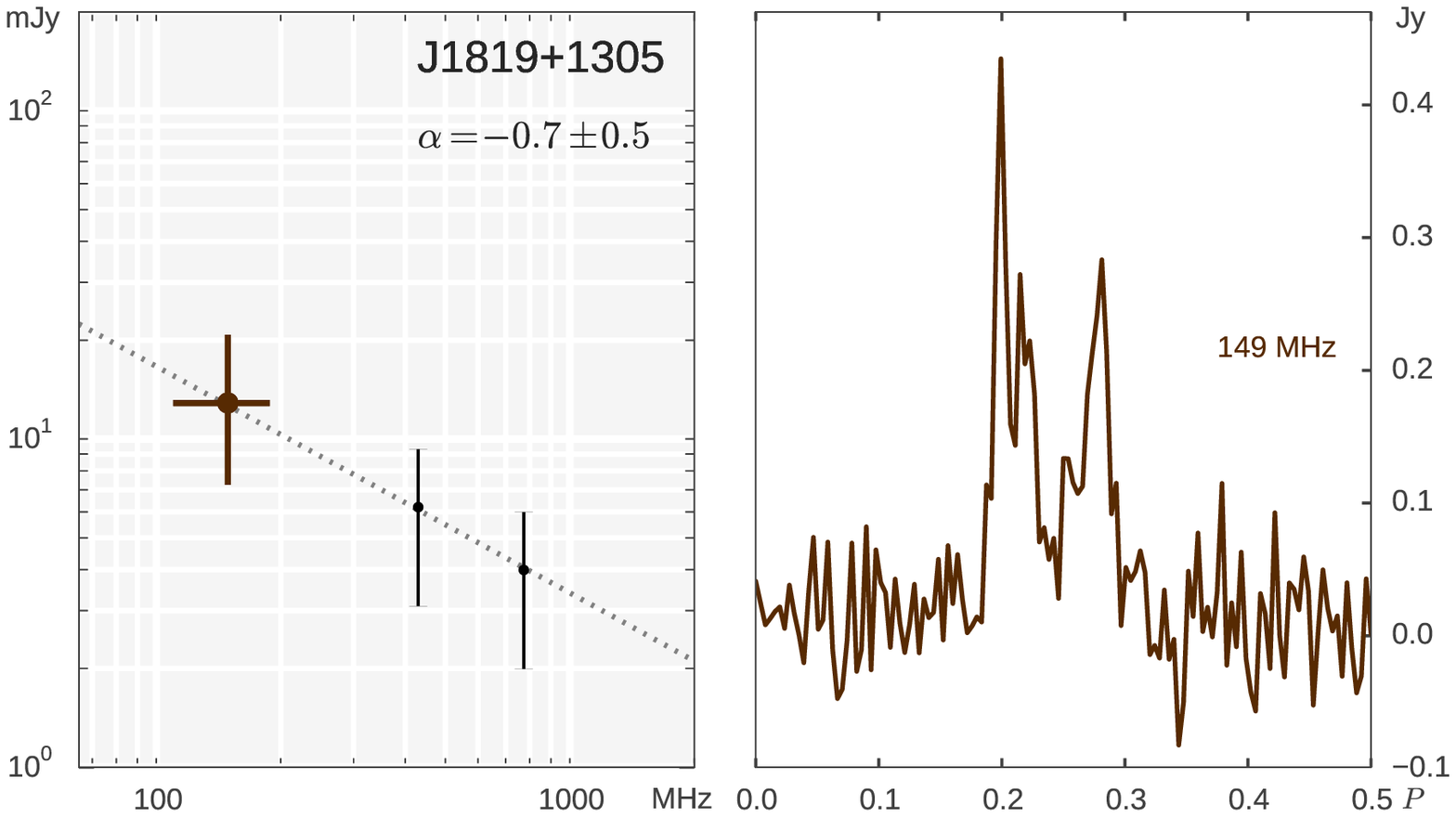}
\includegraphics[scale=0.48]{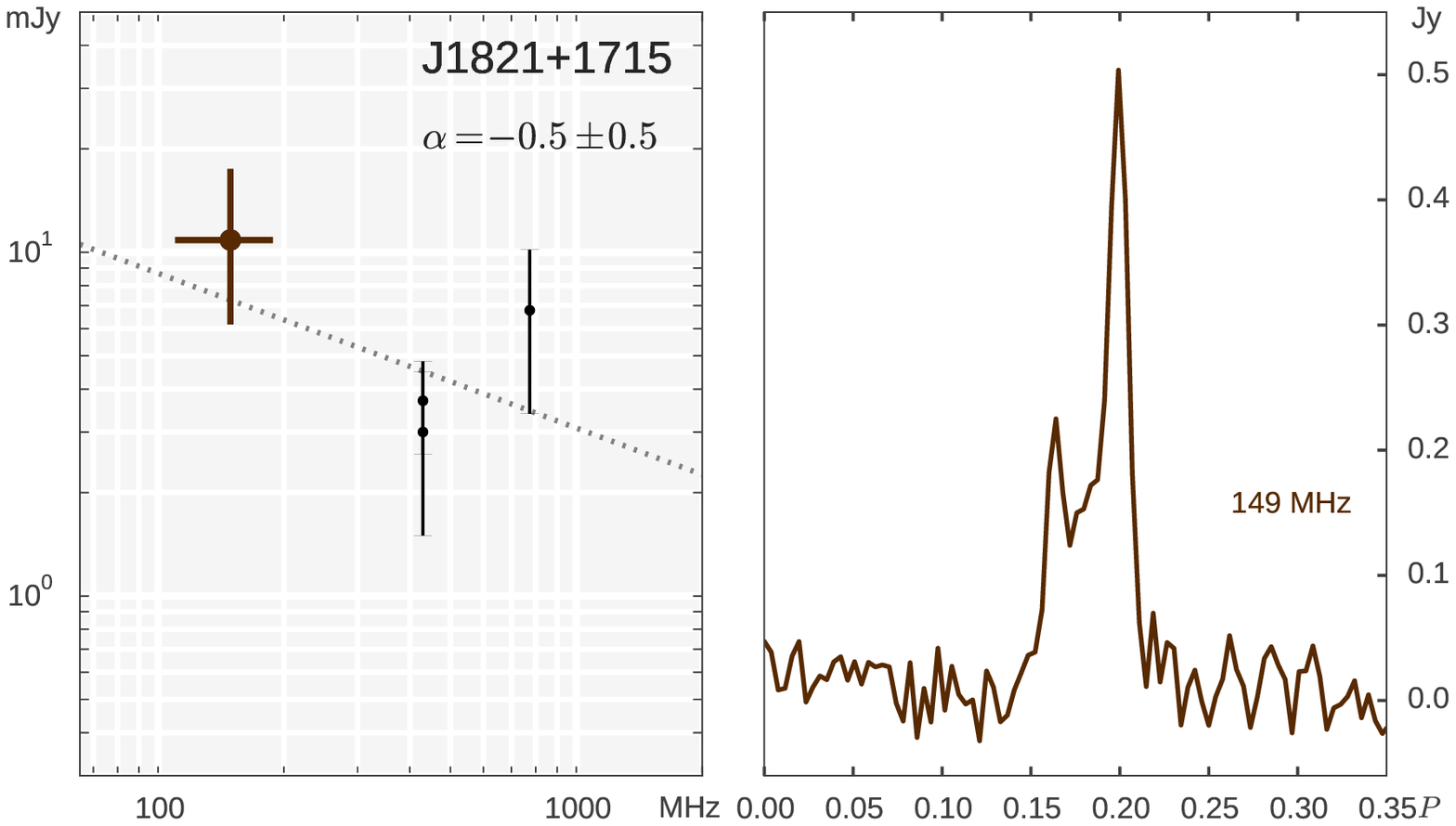}\includegraphics[scale=0.48]{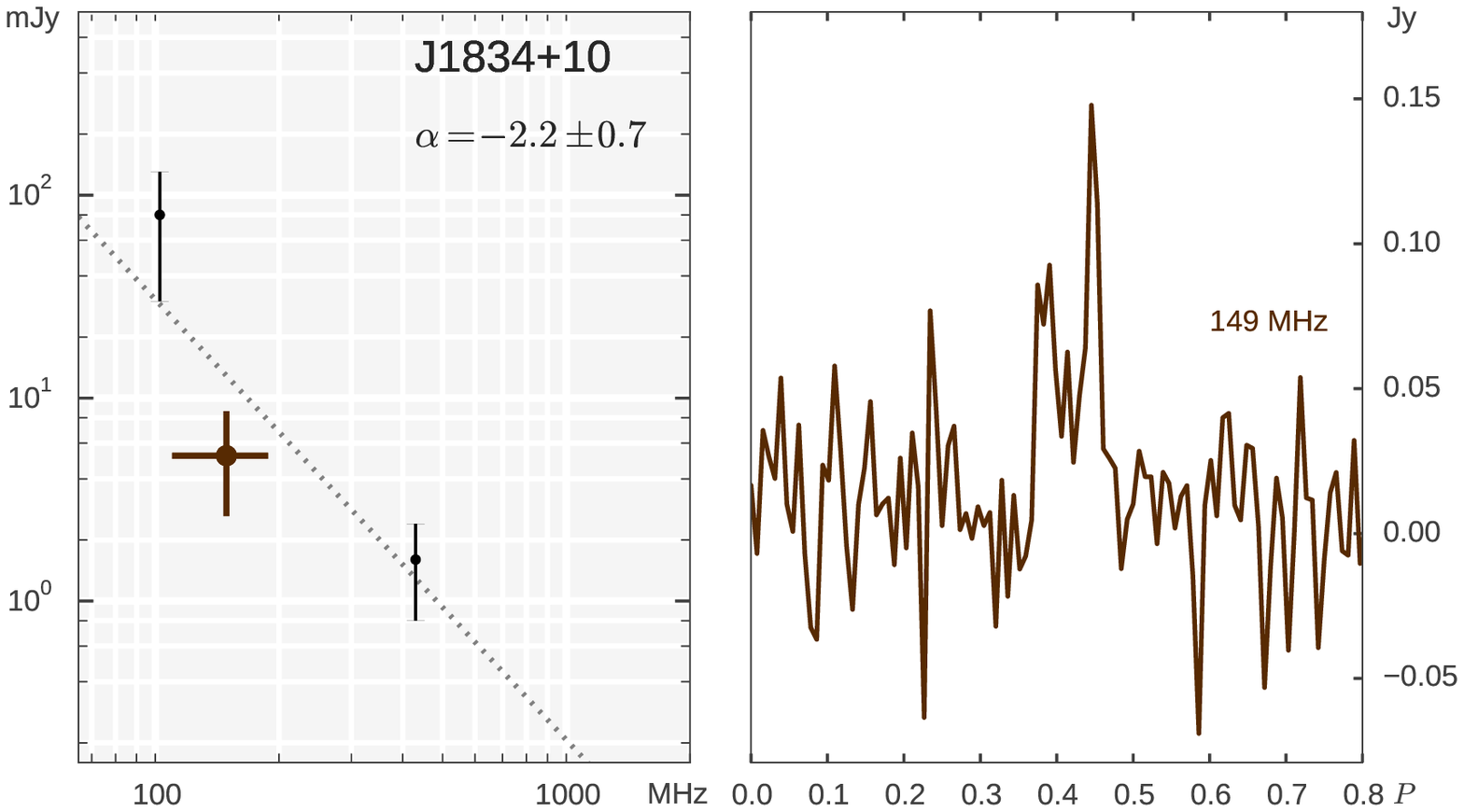}
\includegraphics[scale=0.48]{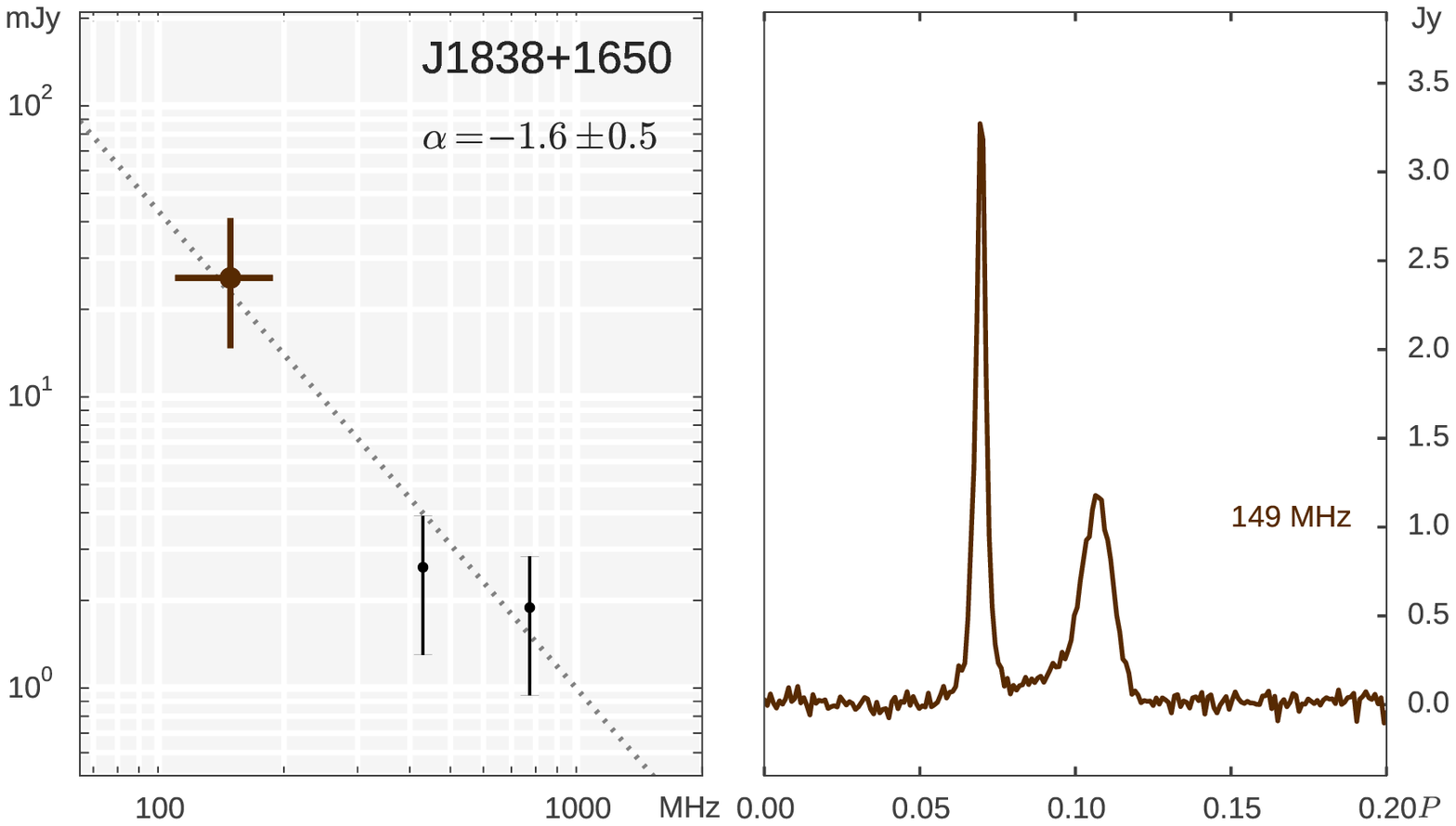}\includegraphics[scale=0.48]{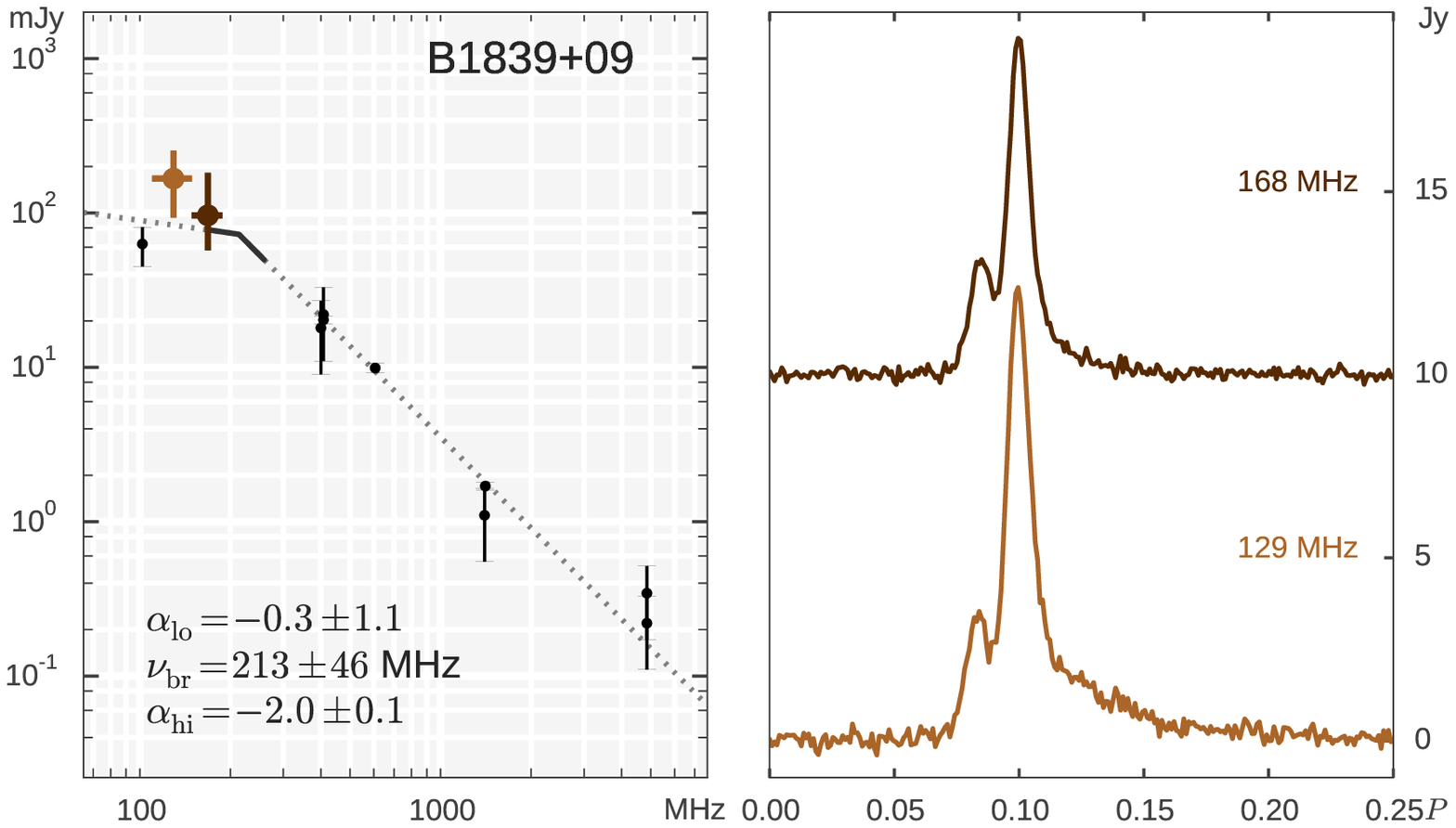}
\caption{See Figure~\ref{fig:prof_sp_1}.}
\label{fig:prof_sp_8}
\end{figure*}

\begin{figure*}
\includegraphics[scale=0.48]{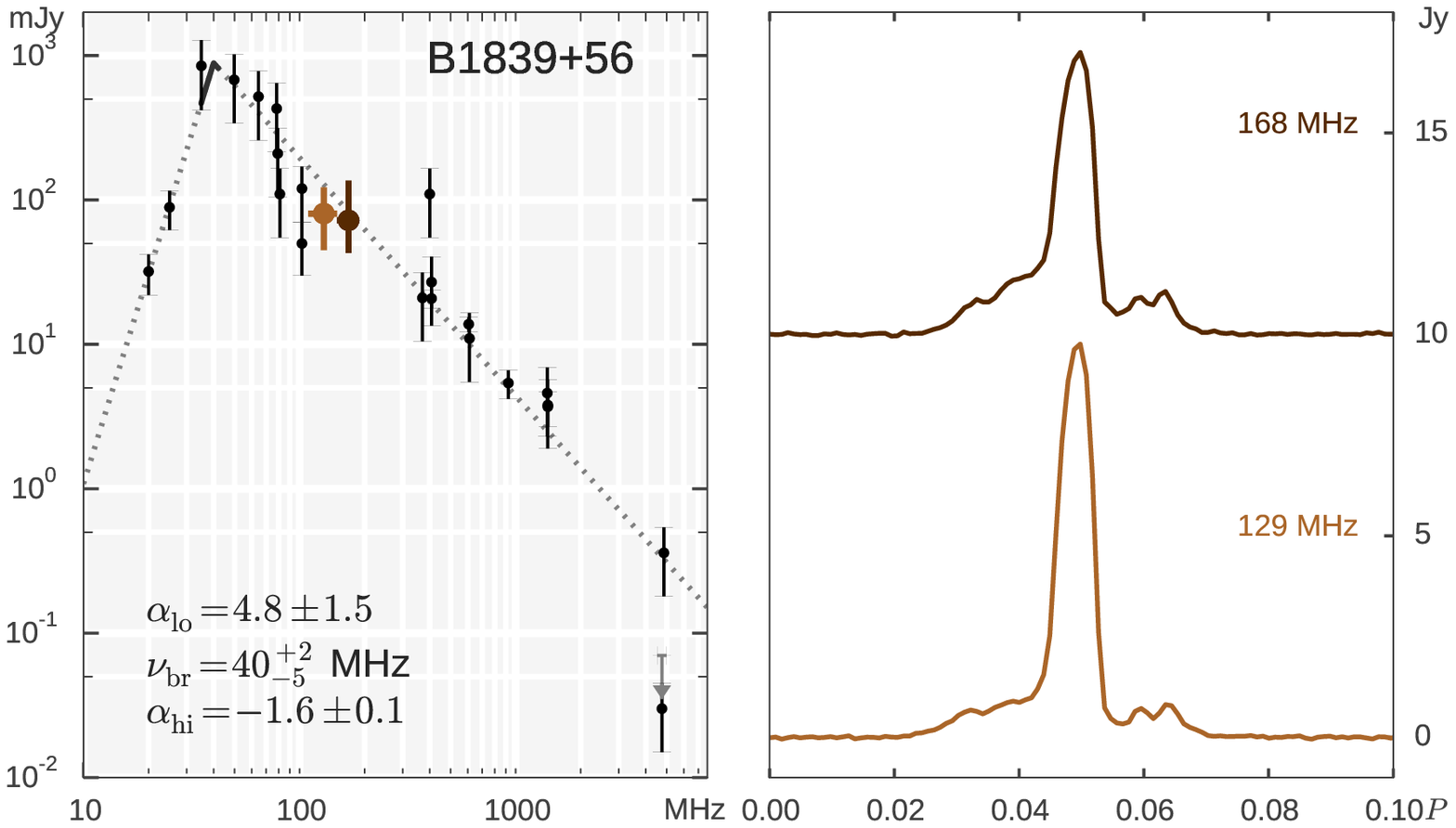}\includegraphics[scale=0.48]{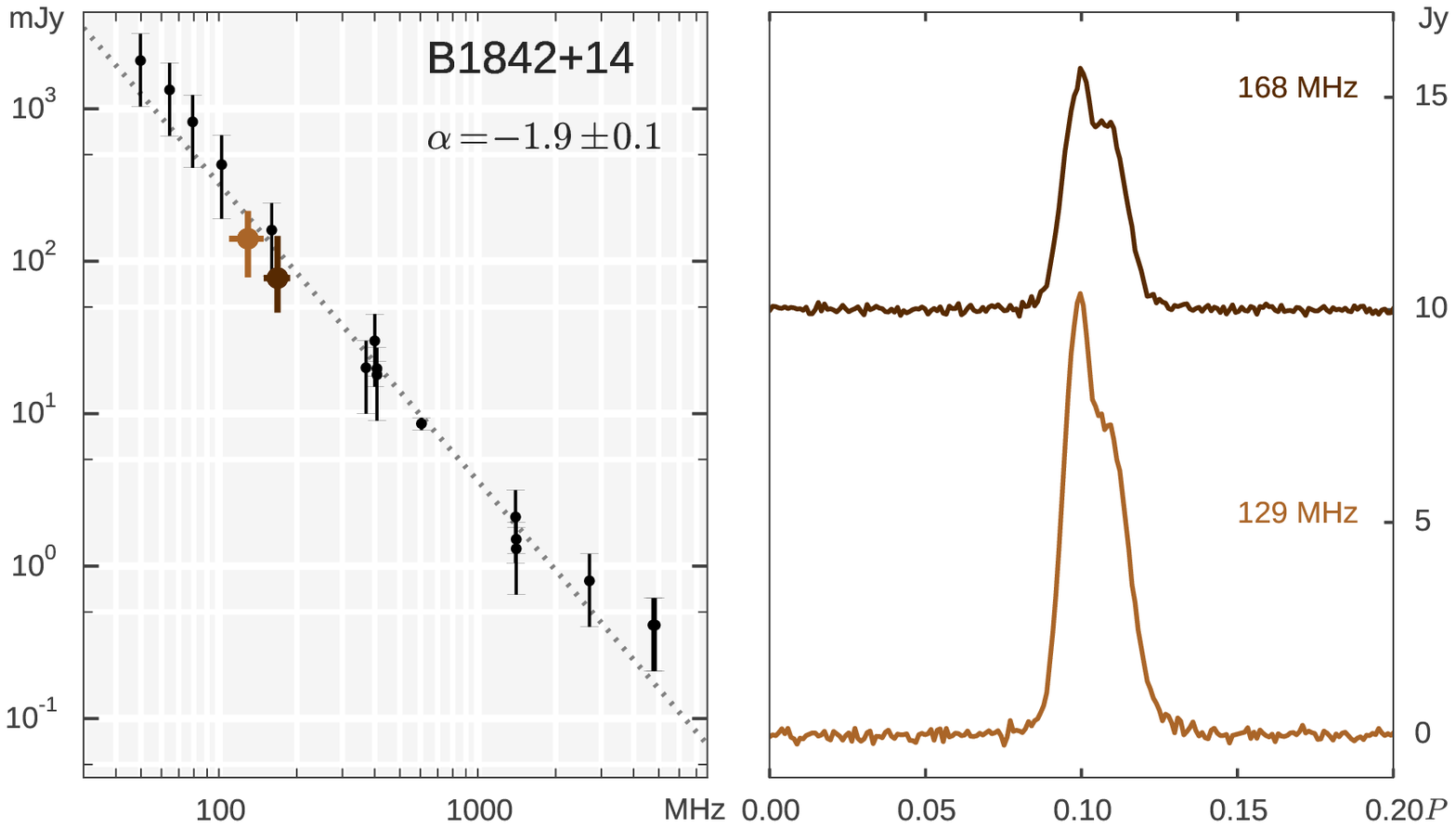}
\includegraphics[scale=0.48]{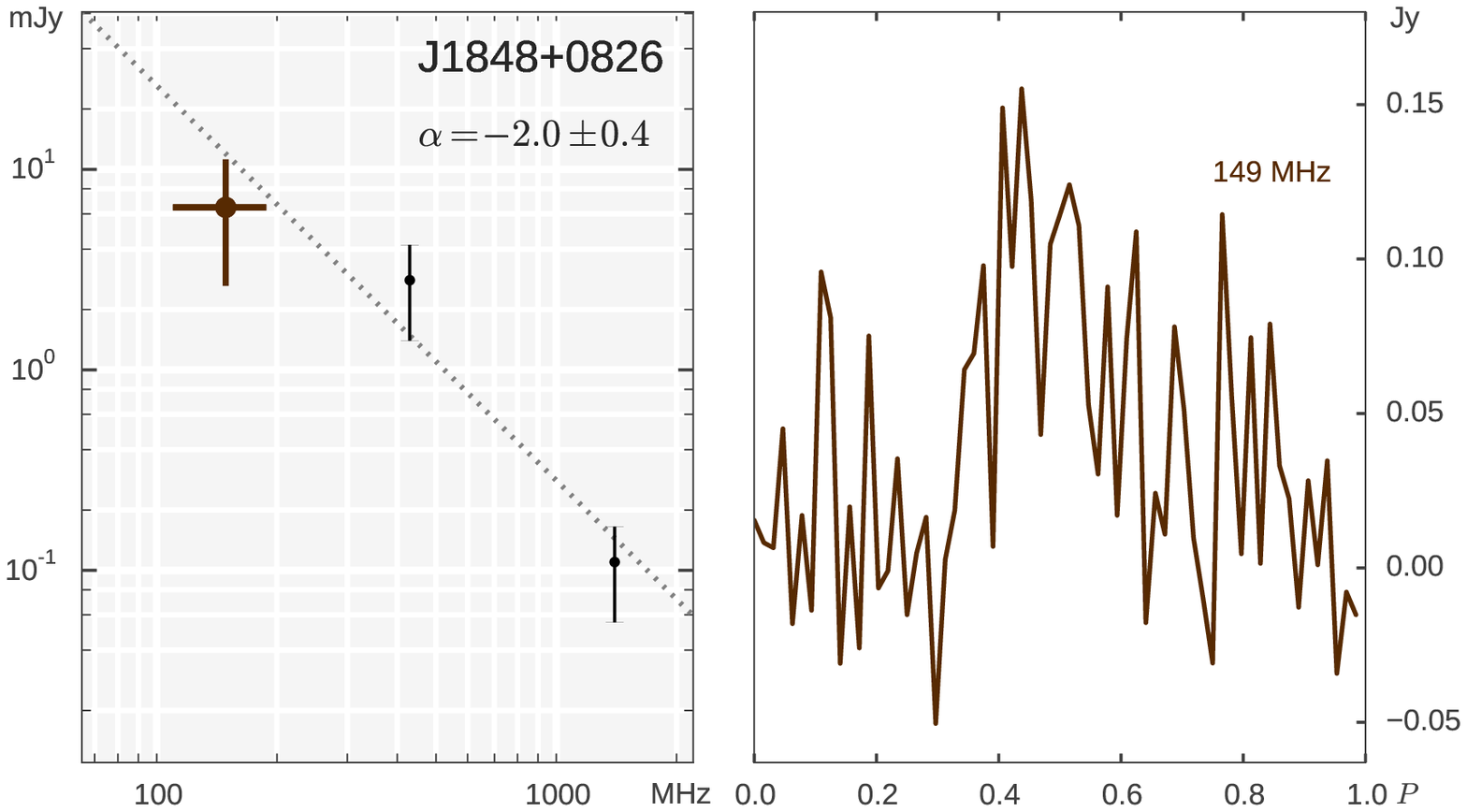}\includegraphics[scale=0.48]{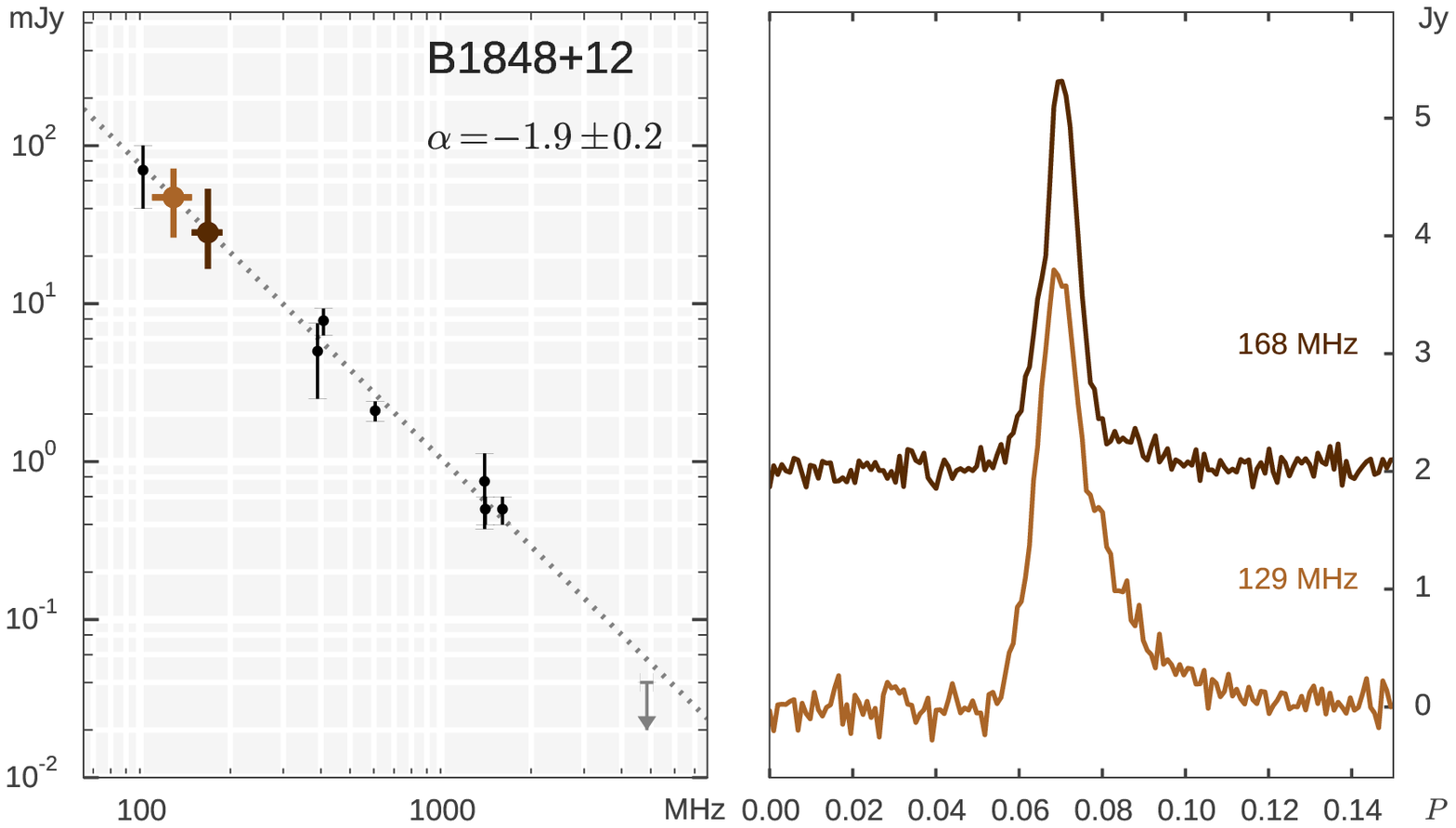}
\includegraphics[scale=0.48]{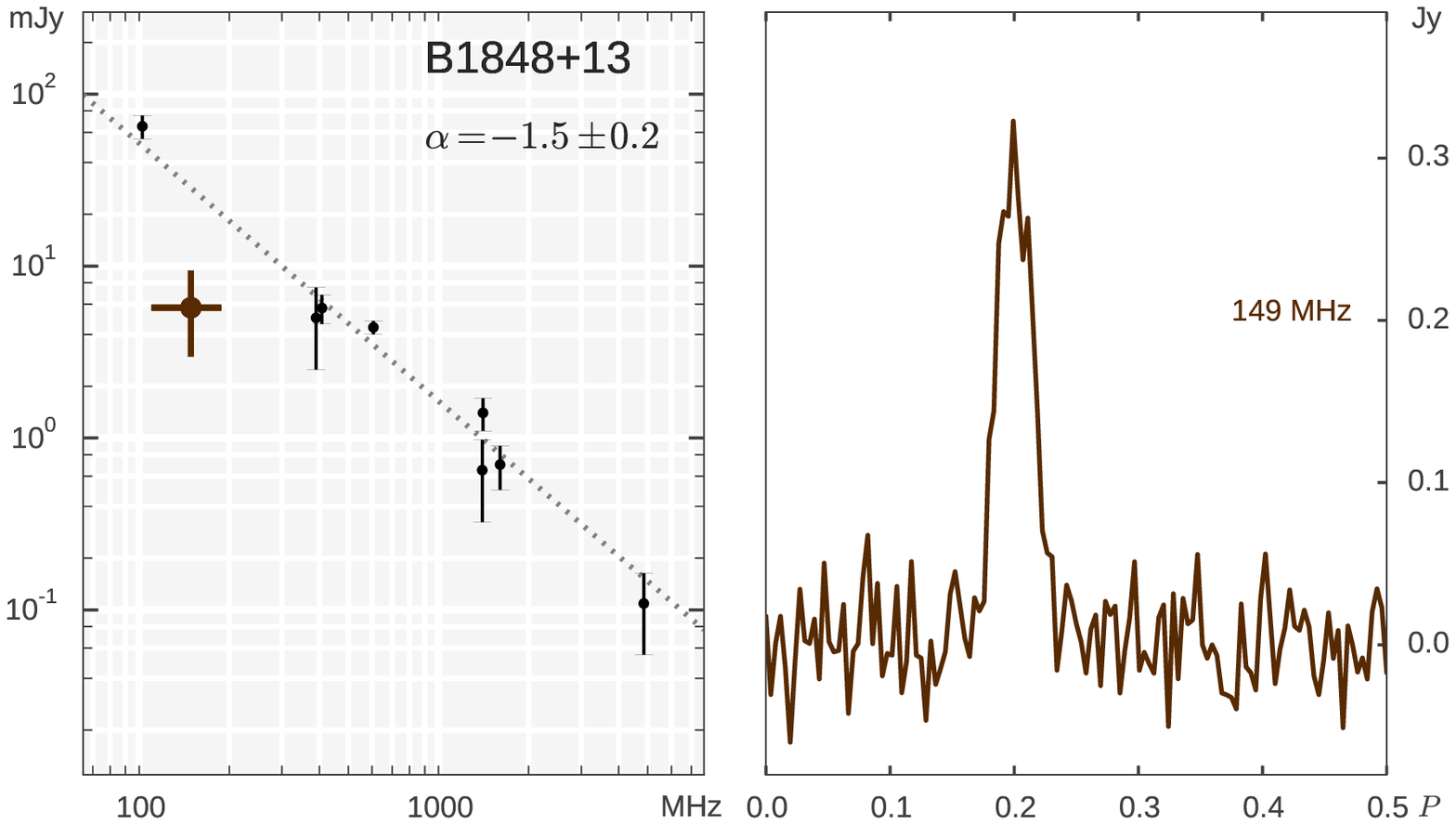}\includegraphics[scale=0.48]{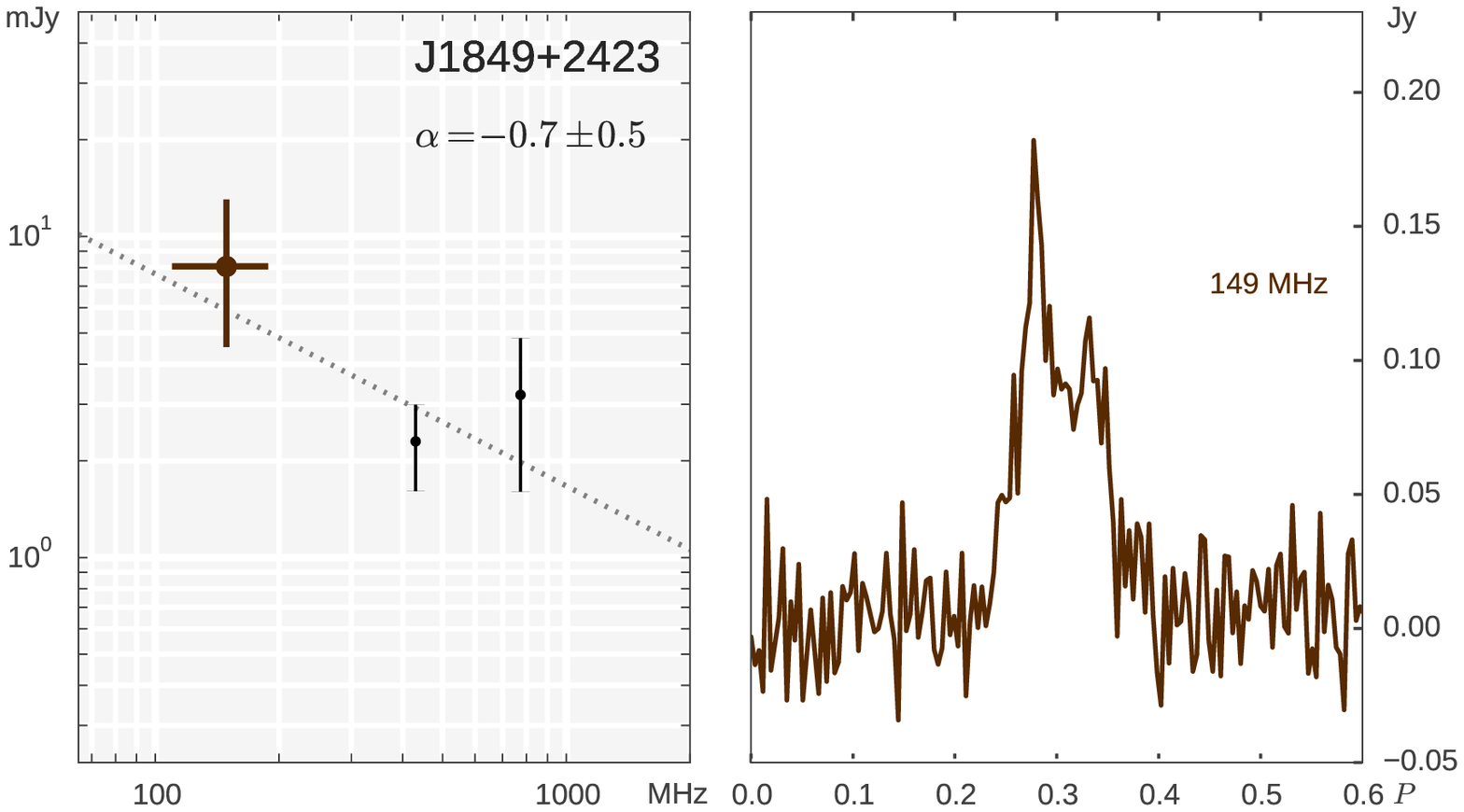}
\includegraphics[scale=0.48]{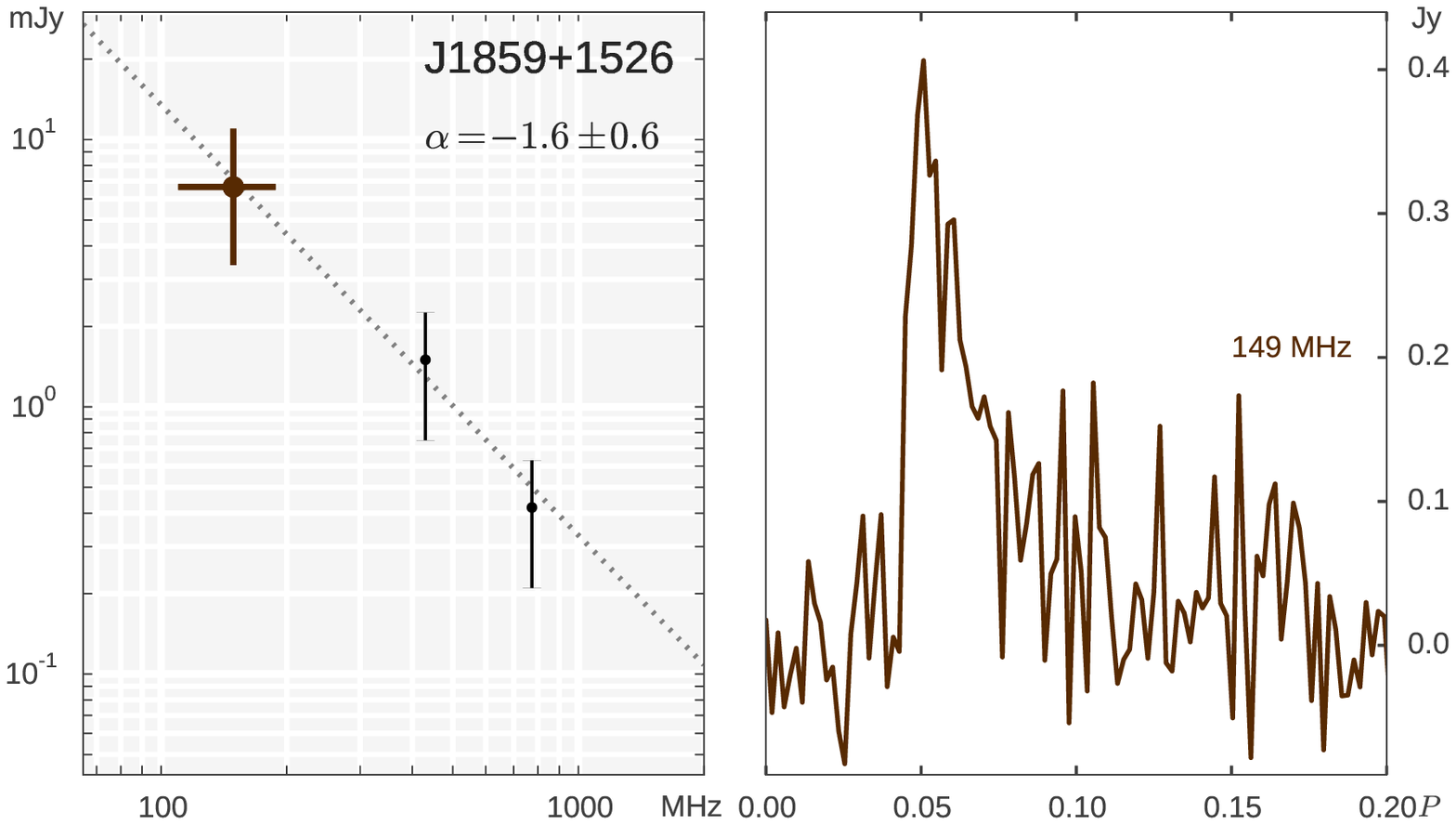}\includegraphics[scale=0.48]{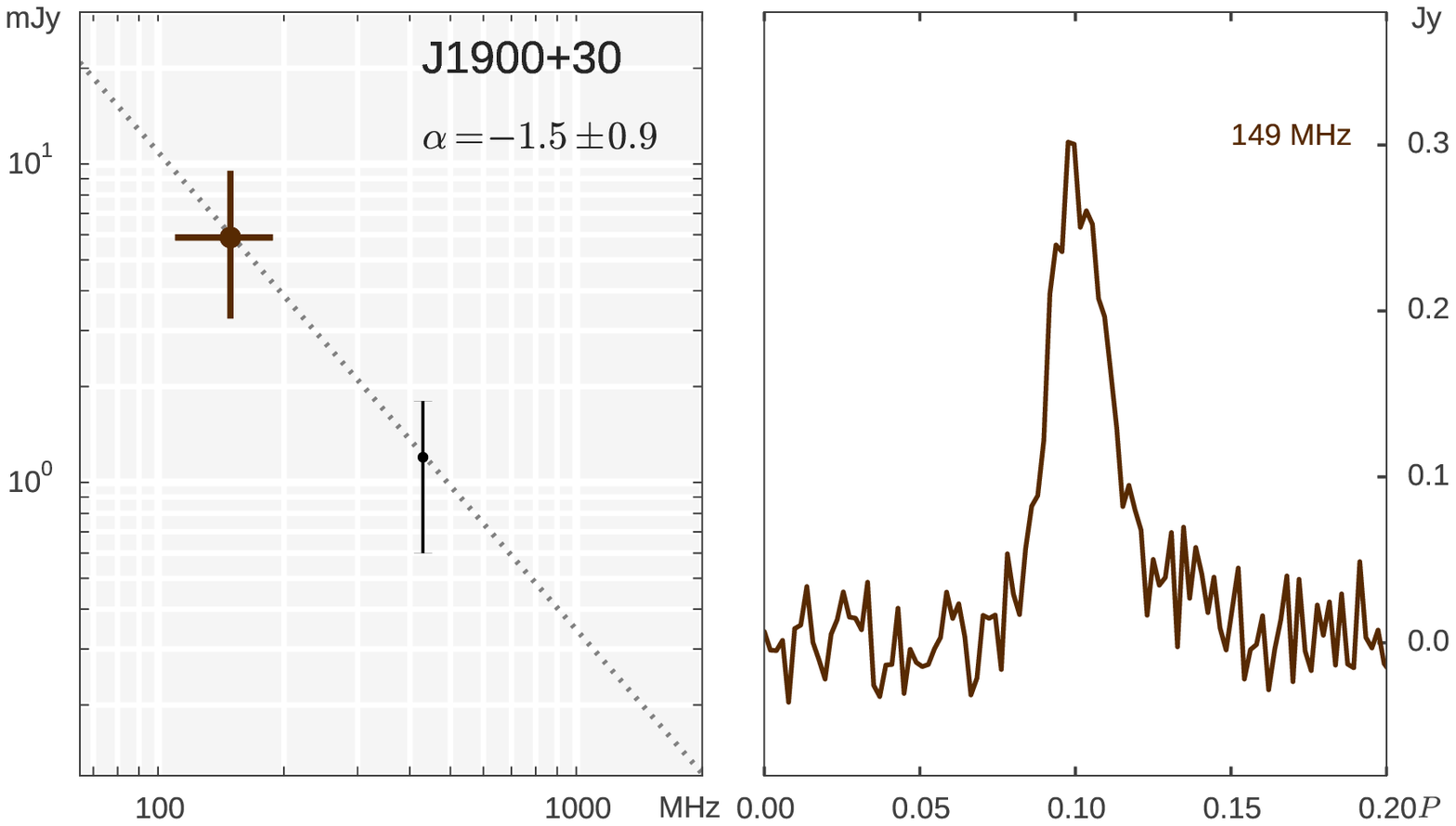}
\includegraphics[scale=0.48]{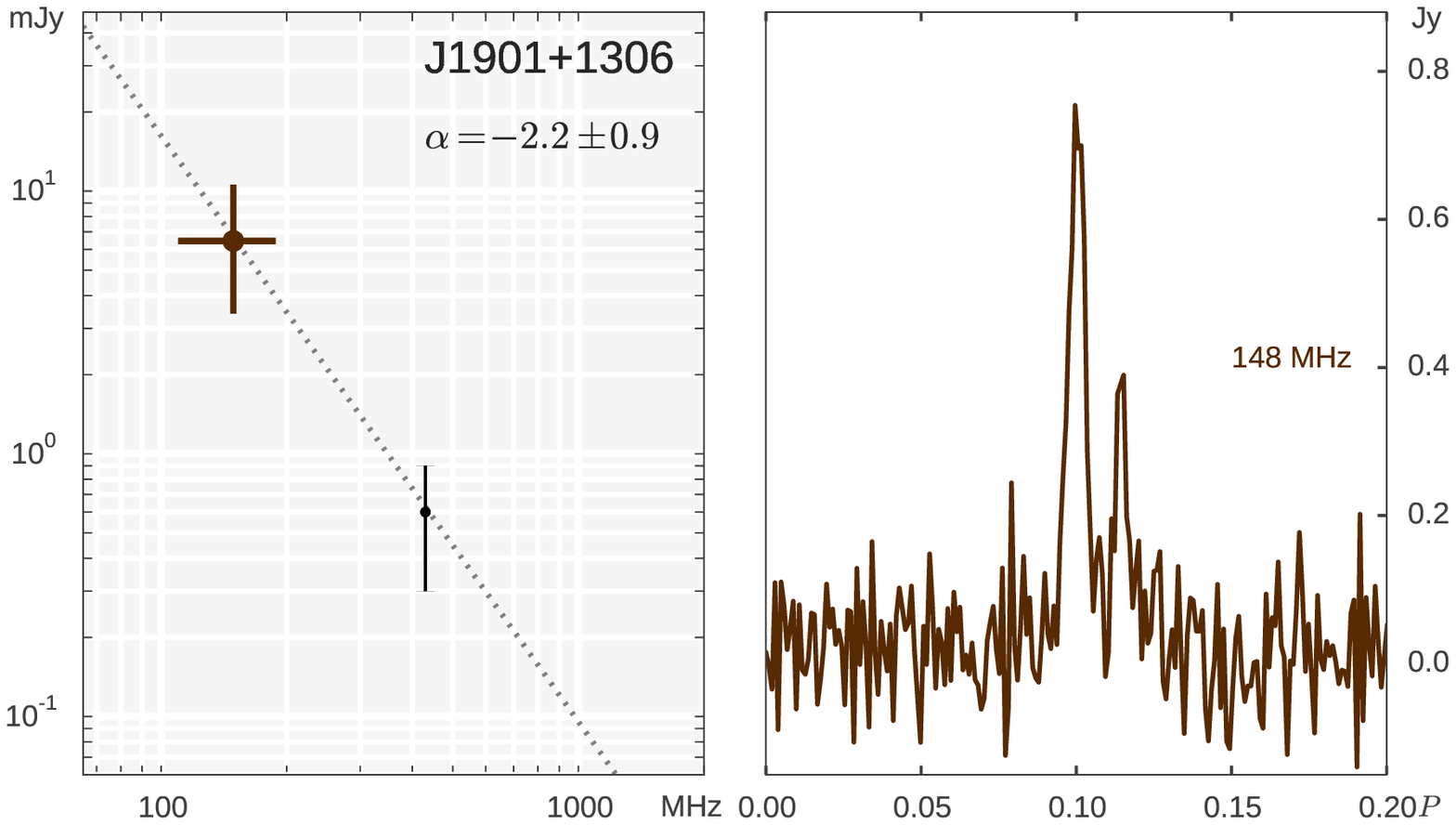}\includegraphics[scale=0.48]{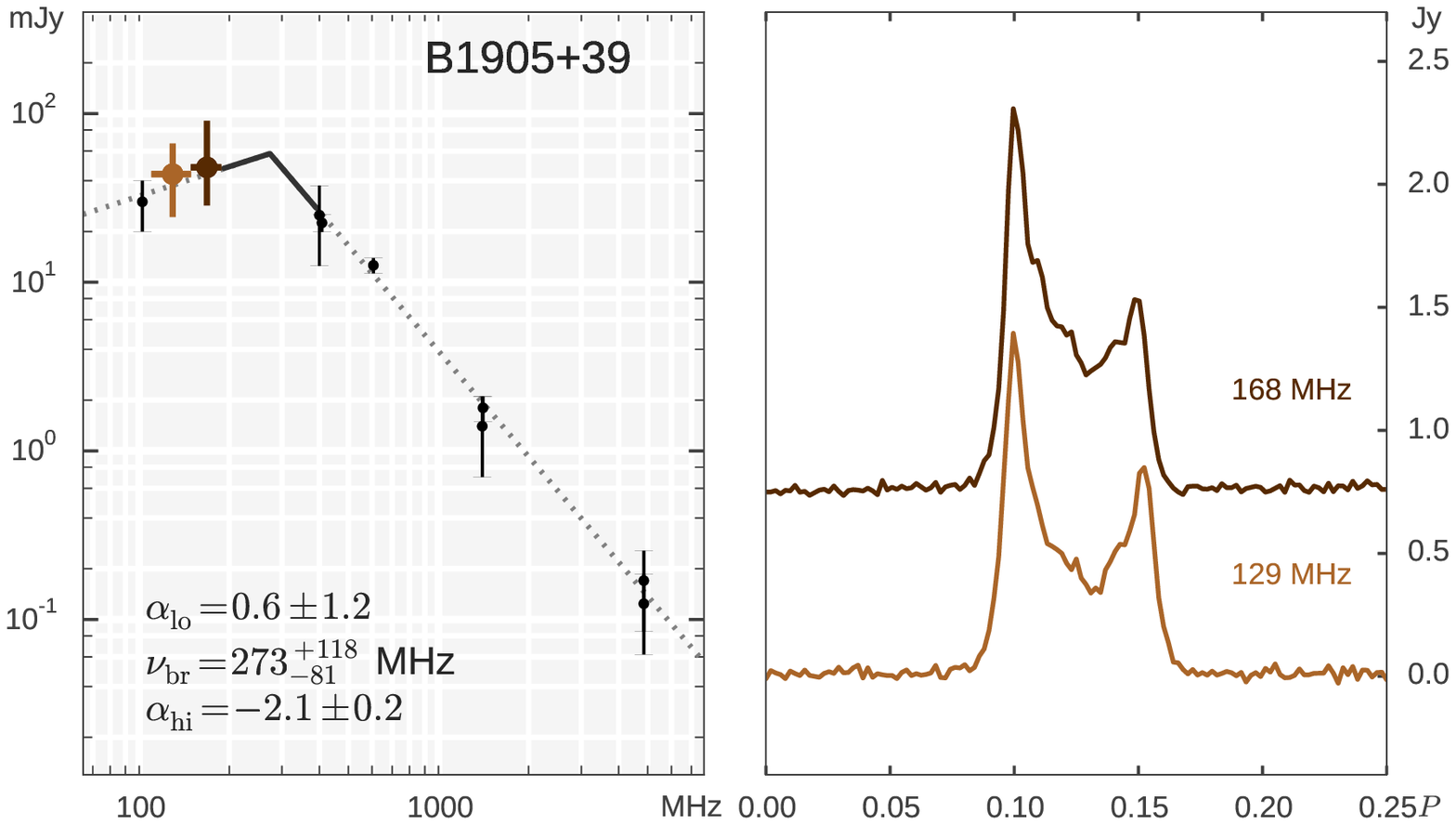}
\caption{See Figure~\ref{fig:prof_sp_1}.}
\label{fig:prof_sp_9}
\end{figure*}

\begin{figure*}
\includegraphics[scale=0.48]{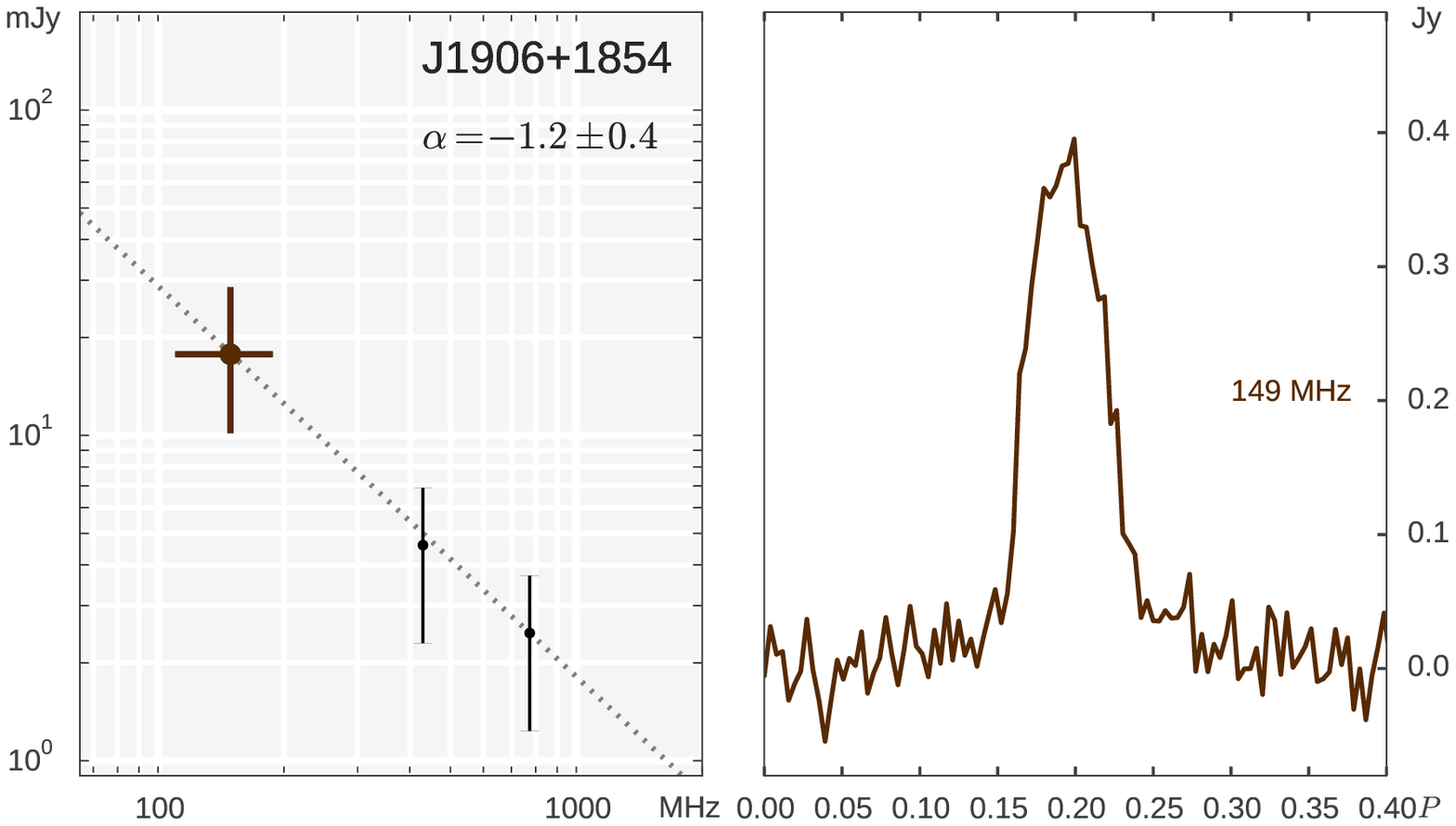}\includegraphics[scale=0.48]{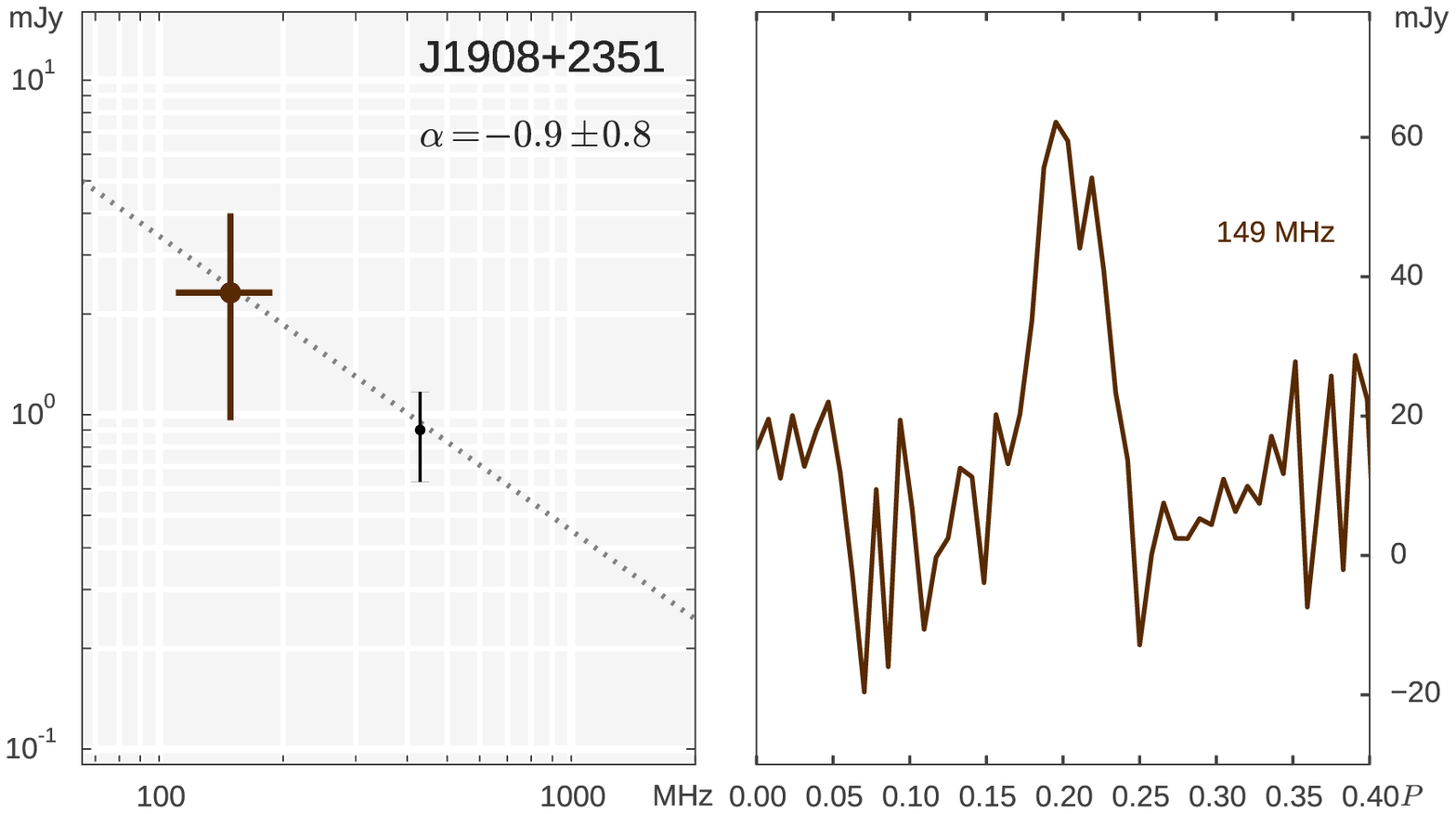}
\includegraphics[scale=0.48]{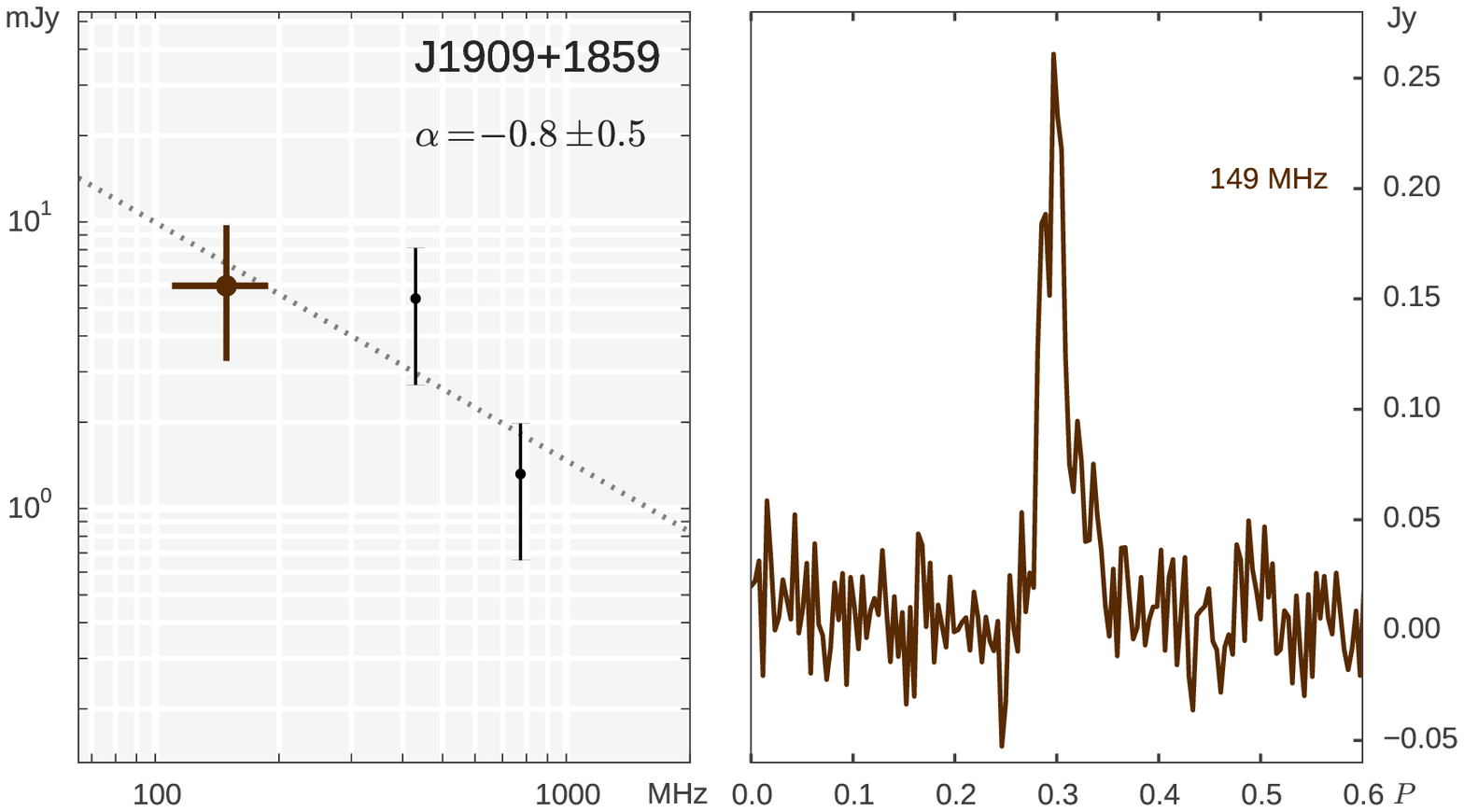}\includegraphics[scale=0.48]{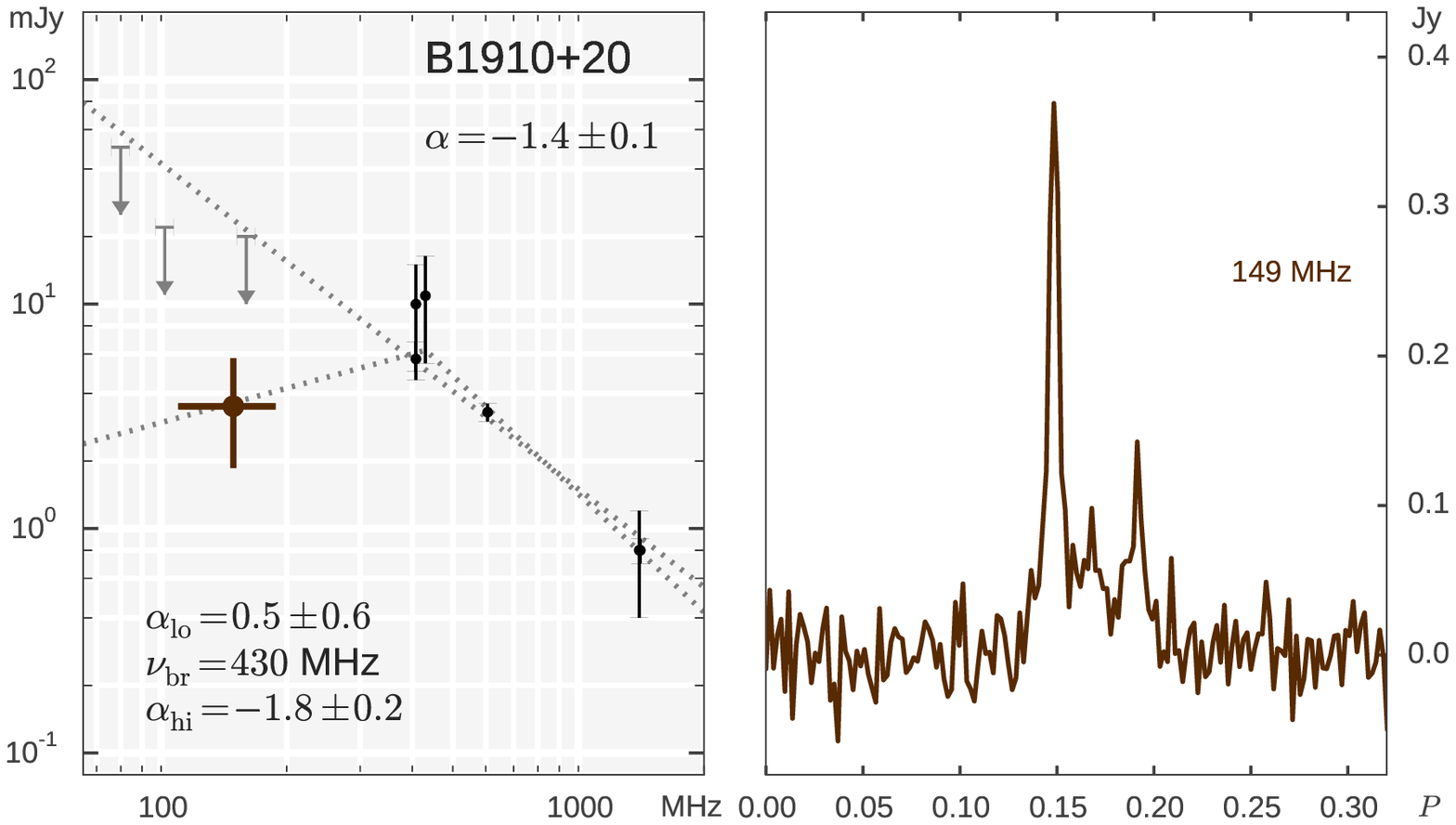}
\includegraphics[scale=0.48]{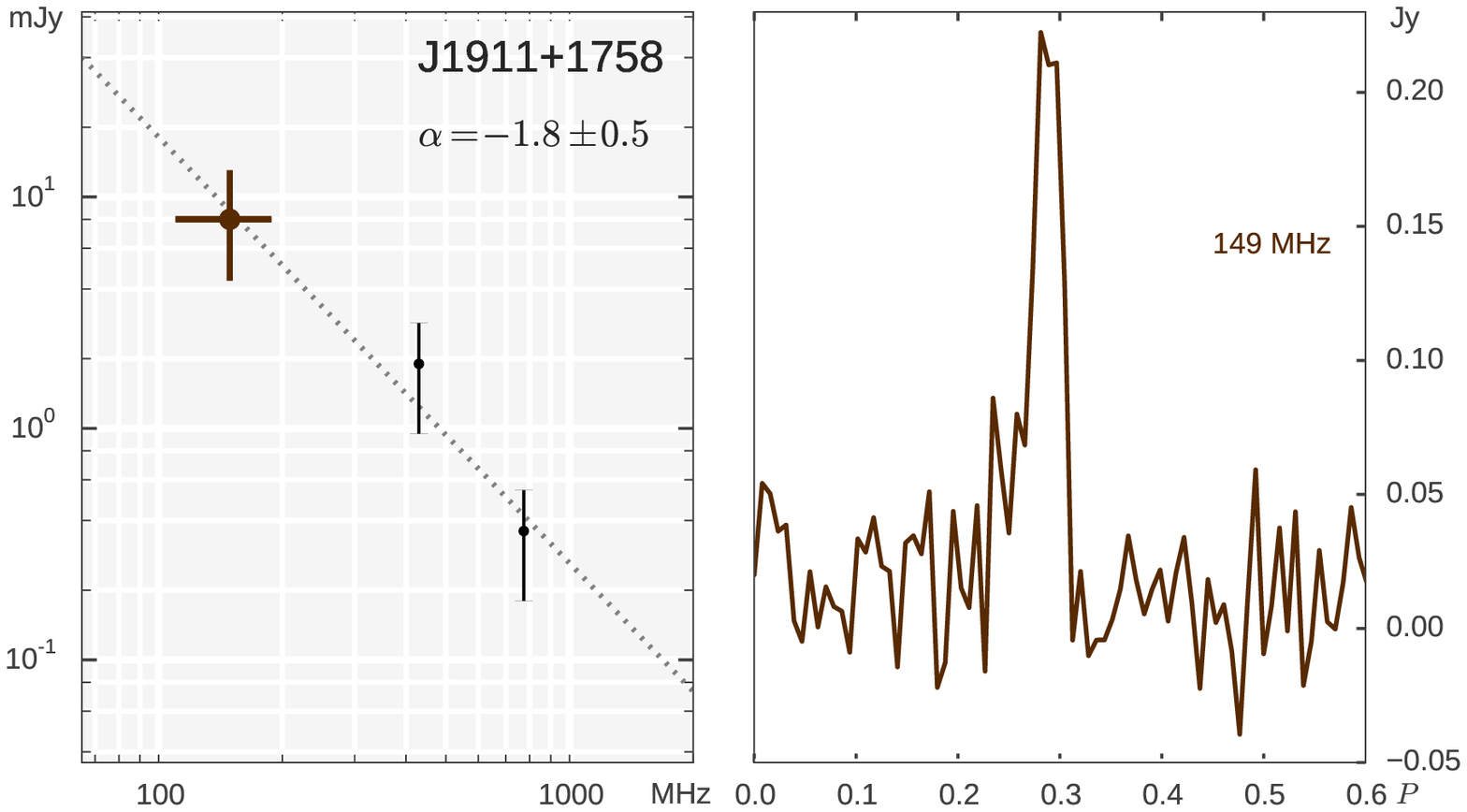}\includegraphics[scale=0.48]{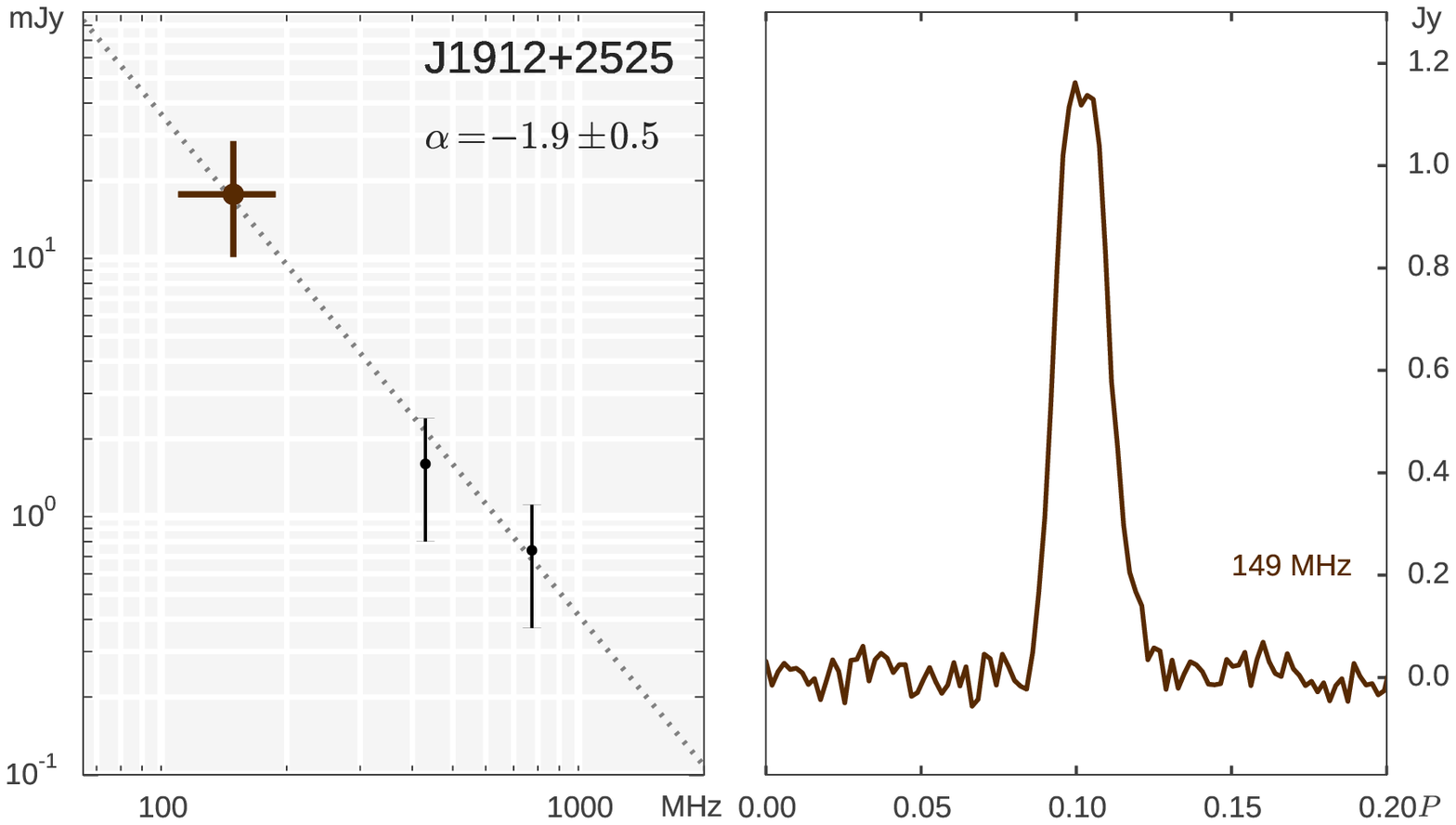}
\includegraphics[scale=0.48]{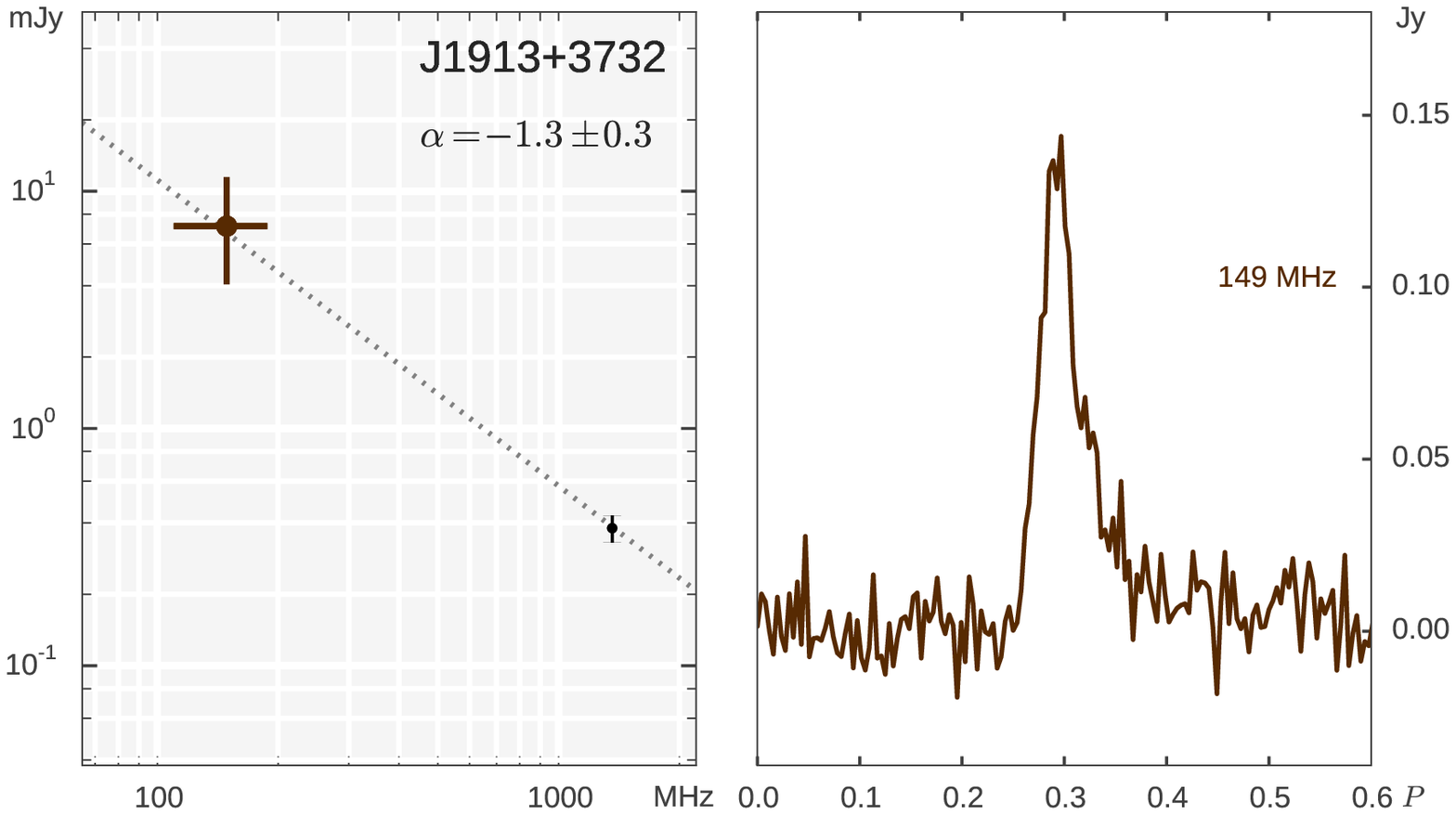}\includegraphics[scale=0.48]{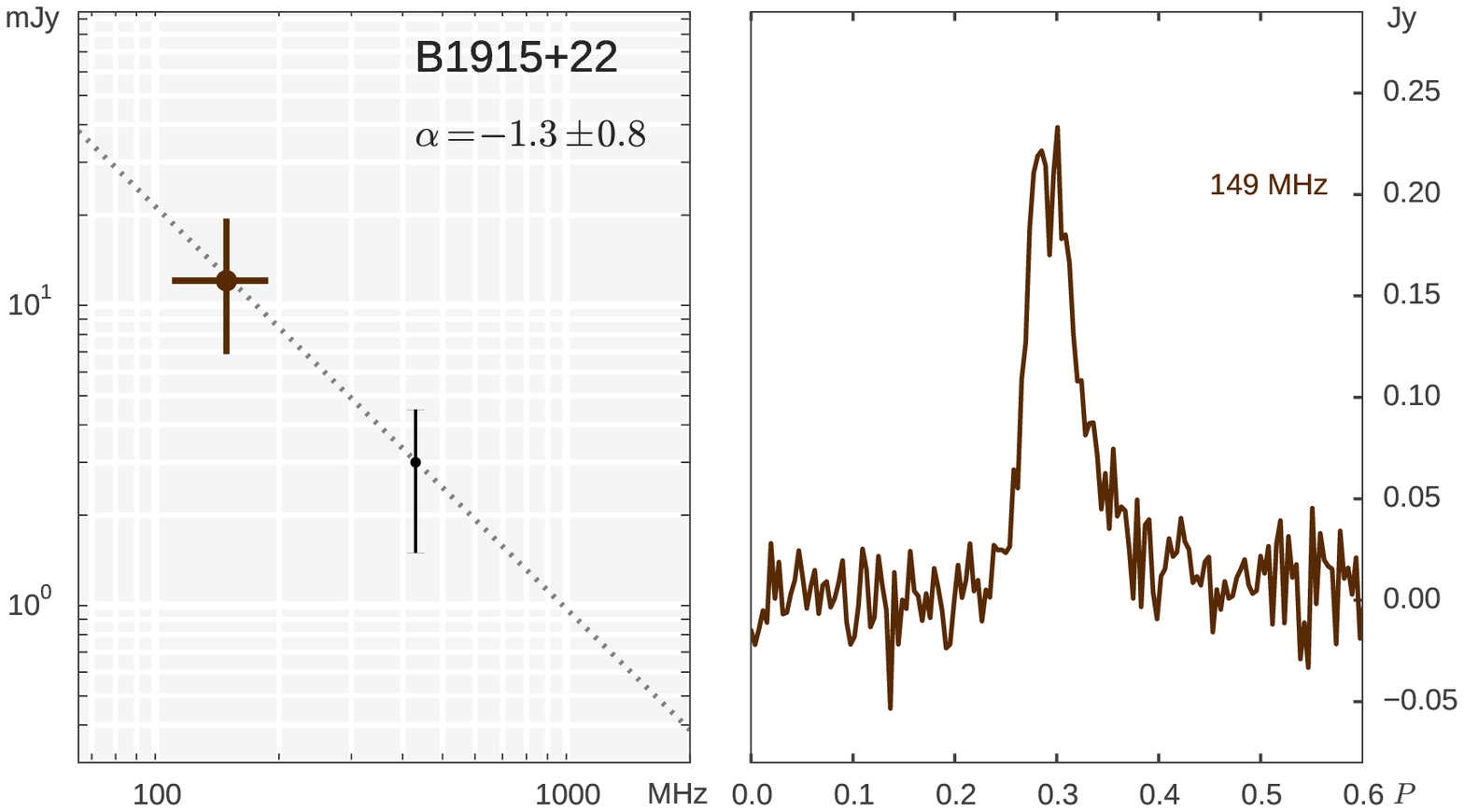}
\includegraphics[scale=0.48]{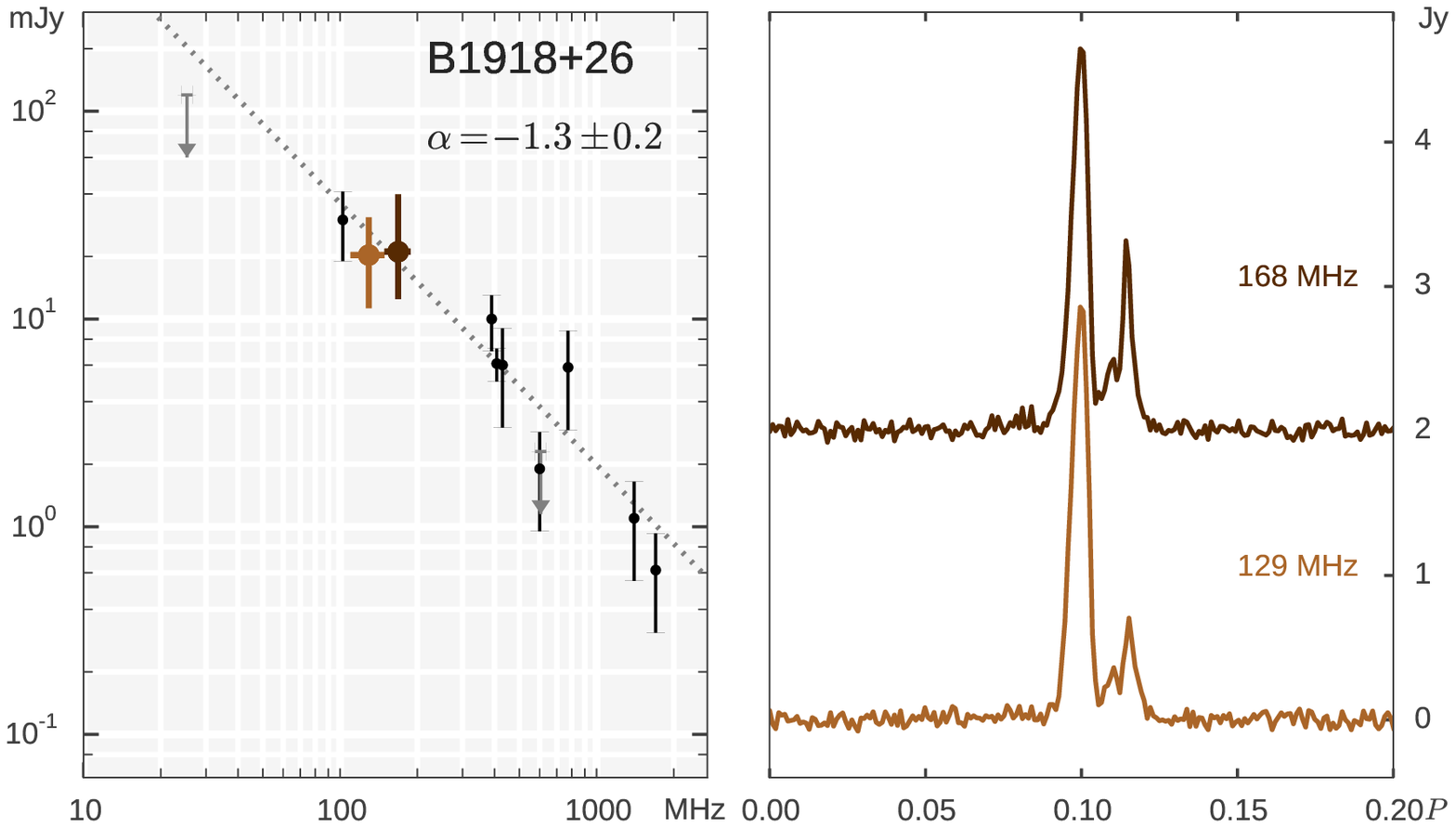}\includegraphics[scale=0.48]{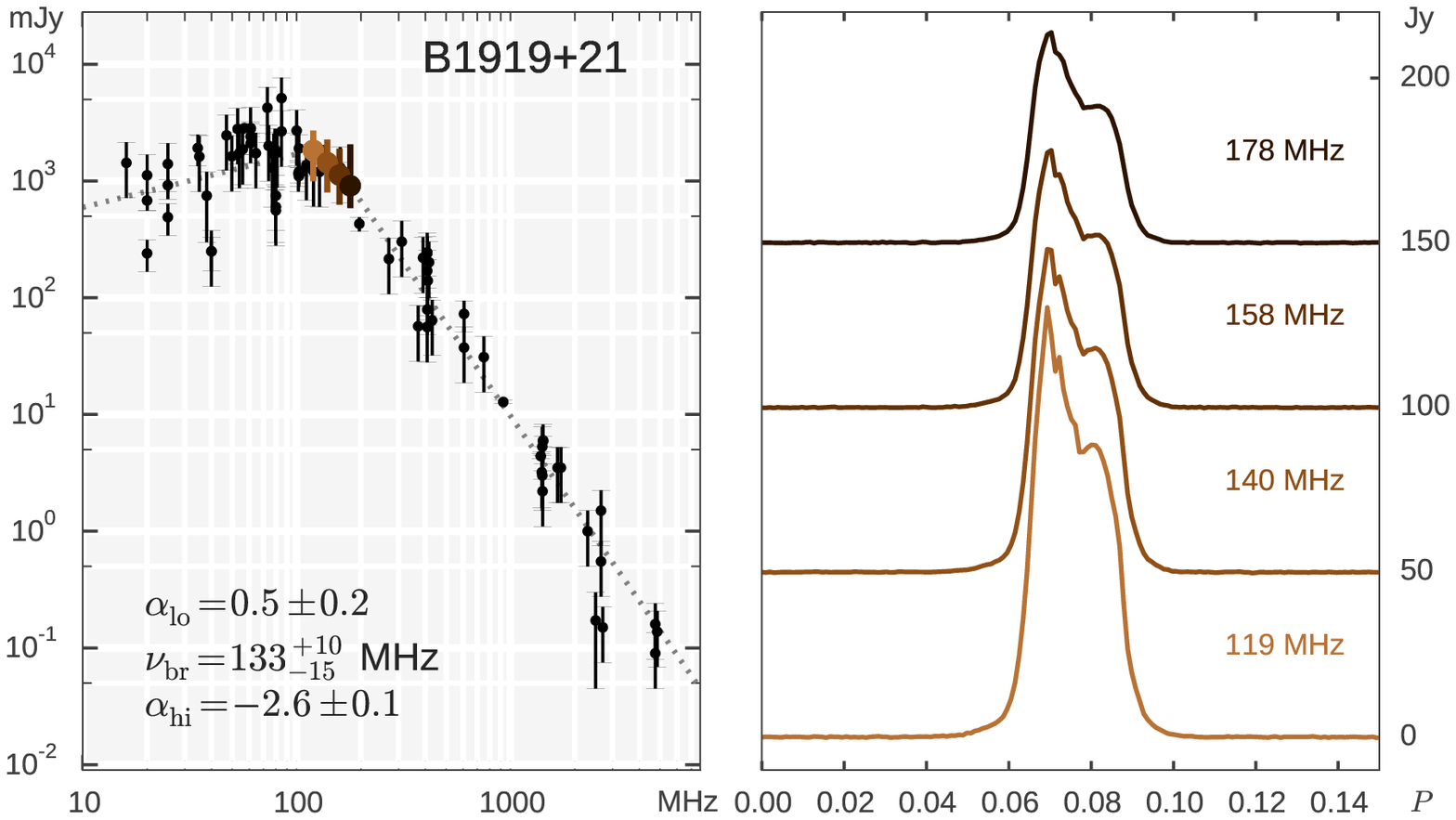}
\caption{See Figure~\ref{fig:prof_sp_1}.}
\label{fig:prof_sp_10}
\end{figure*}

\begin{figure*}
\includegraphics[scale=0.48]{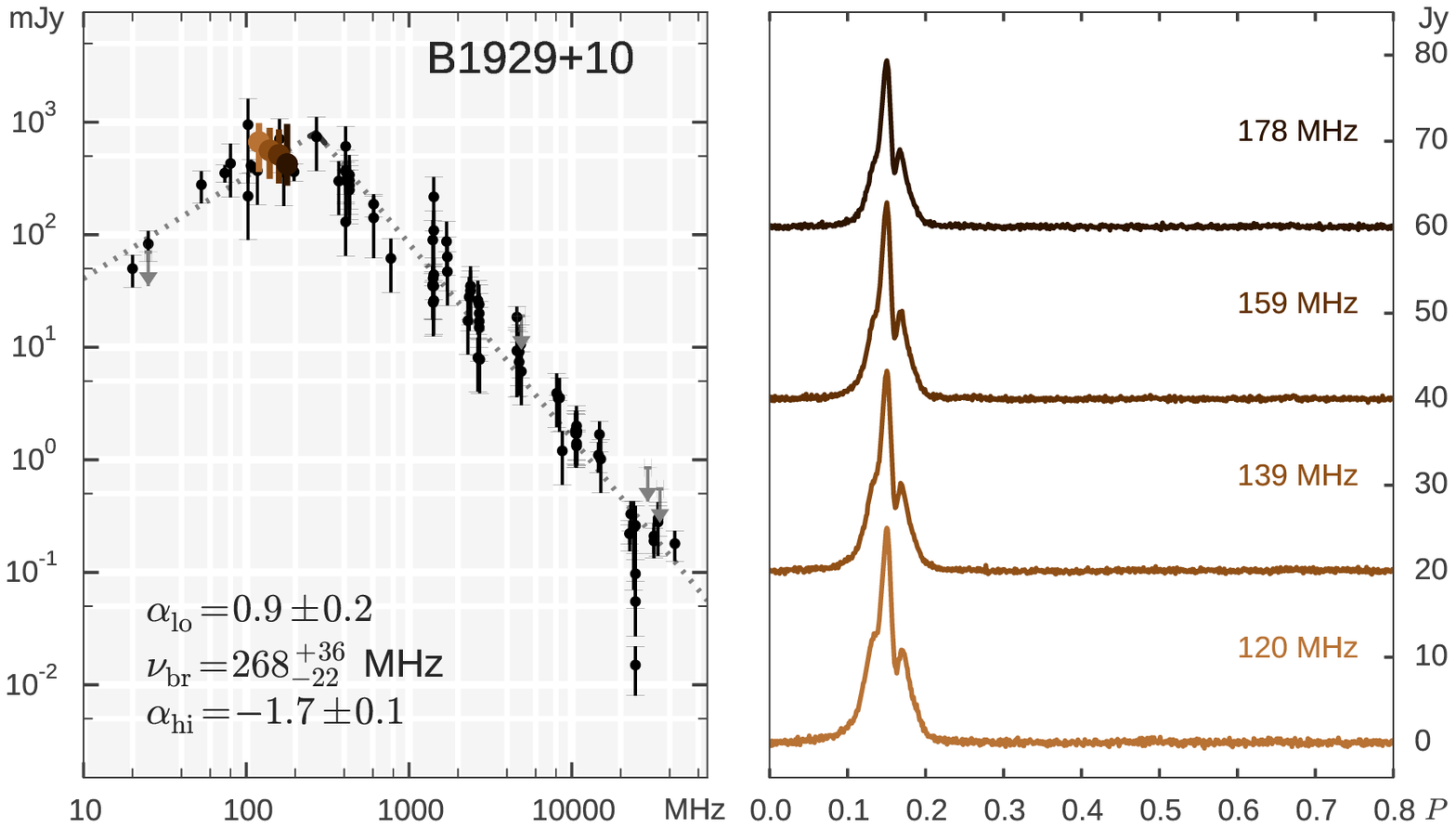}\includegraphics[scale=0.48]{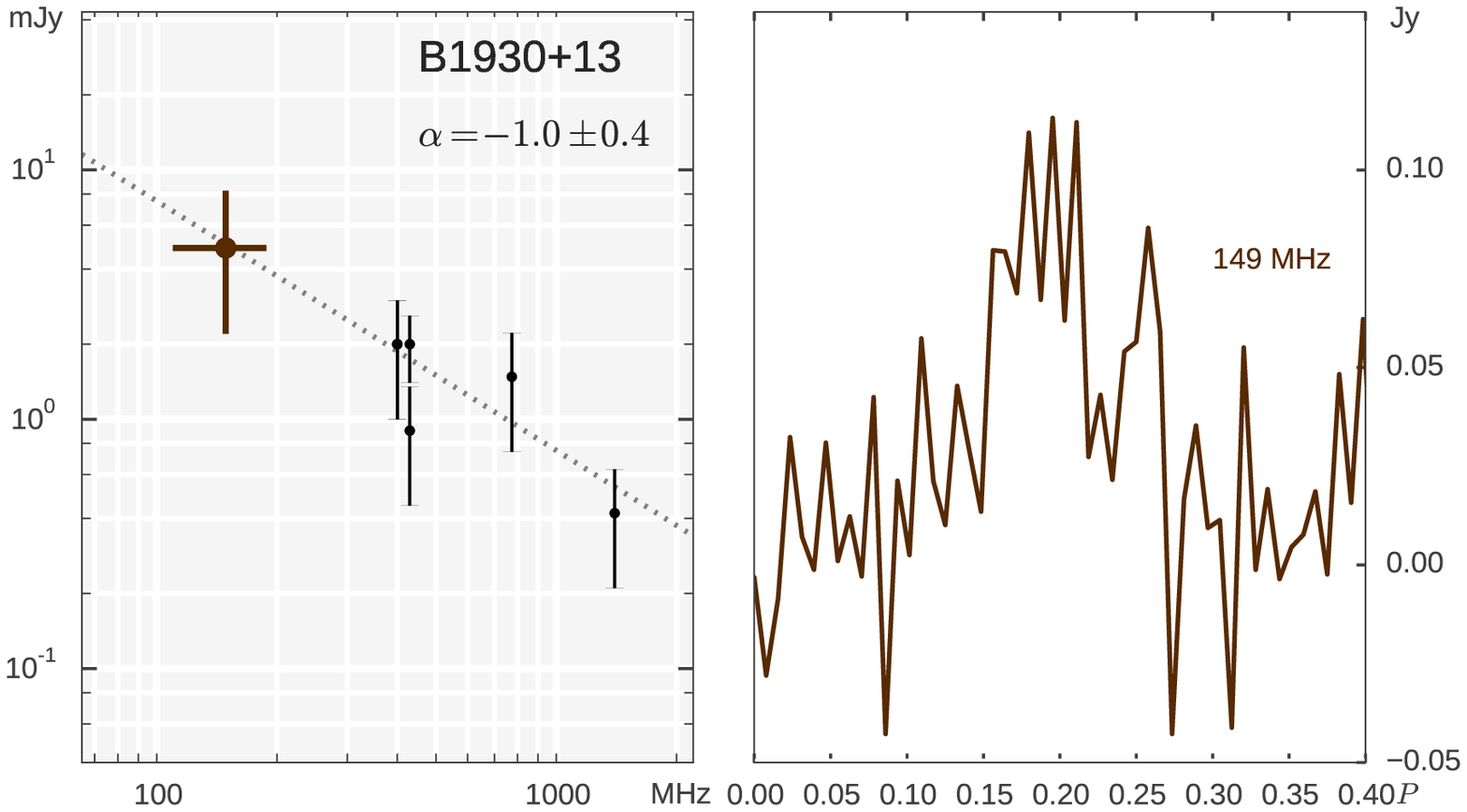}
\includegraphics[scale=0.48]{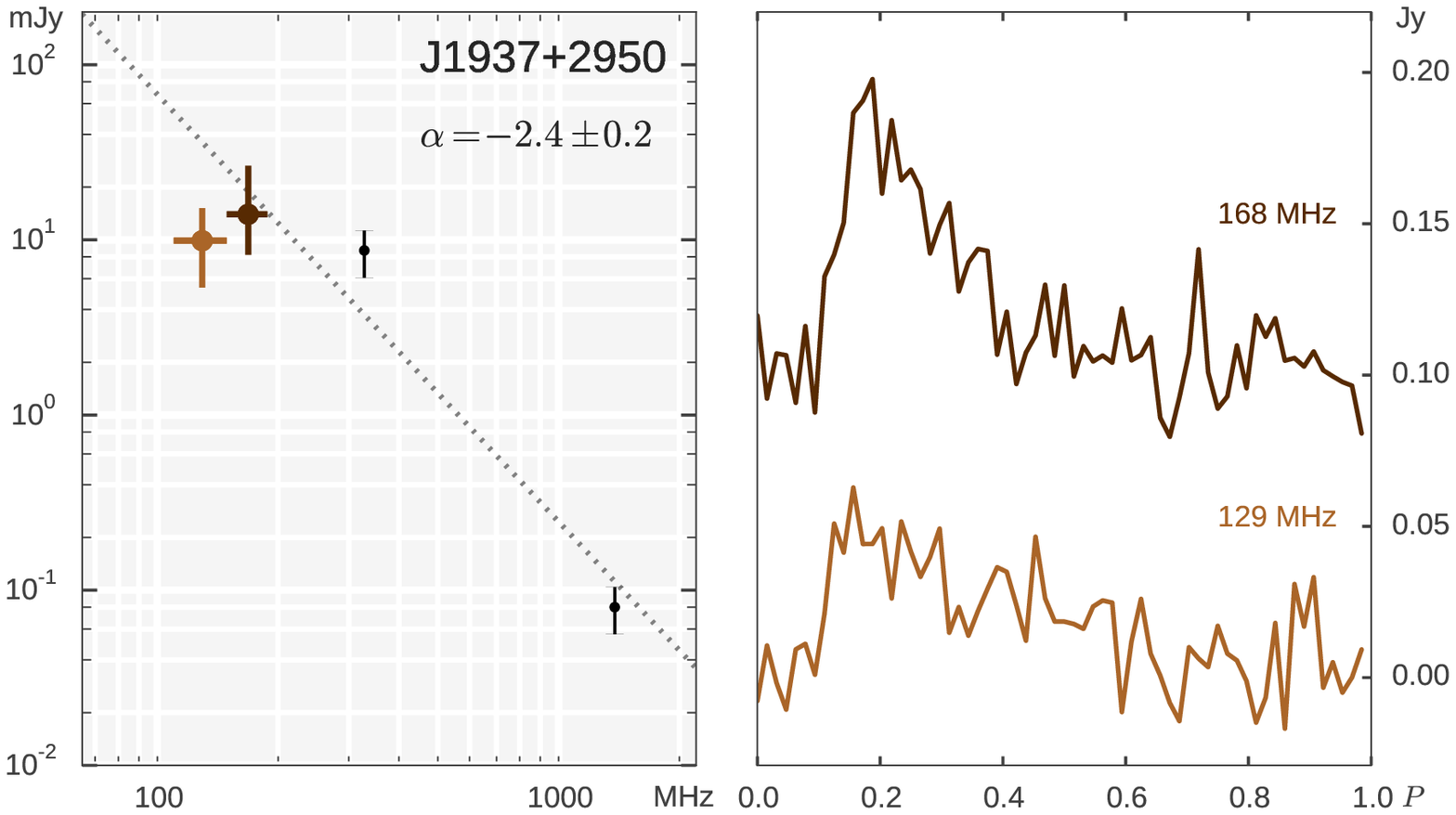}\includegraphics[scale=0.48]{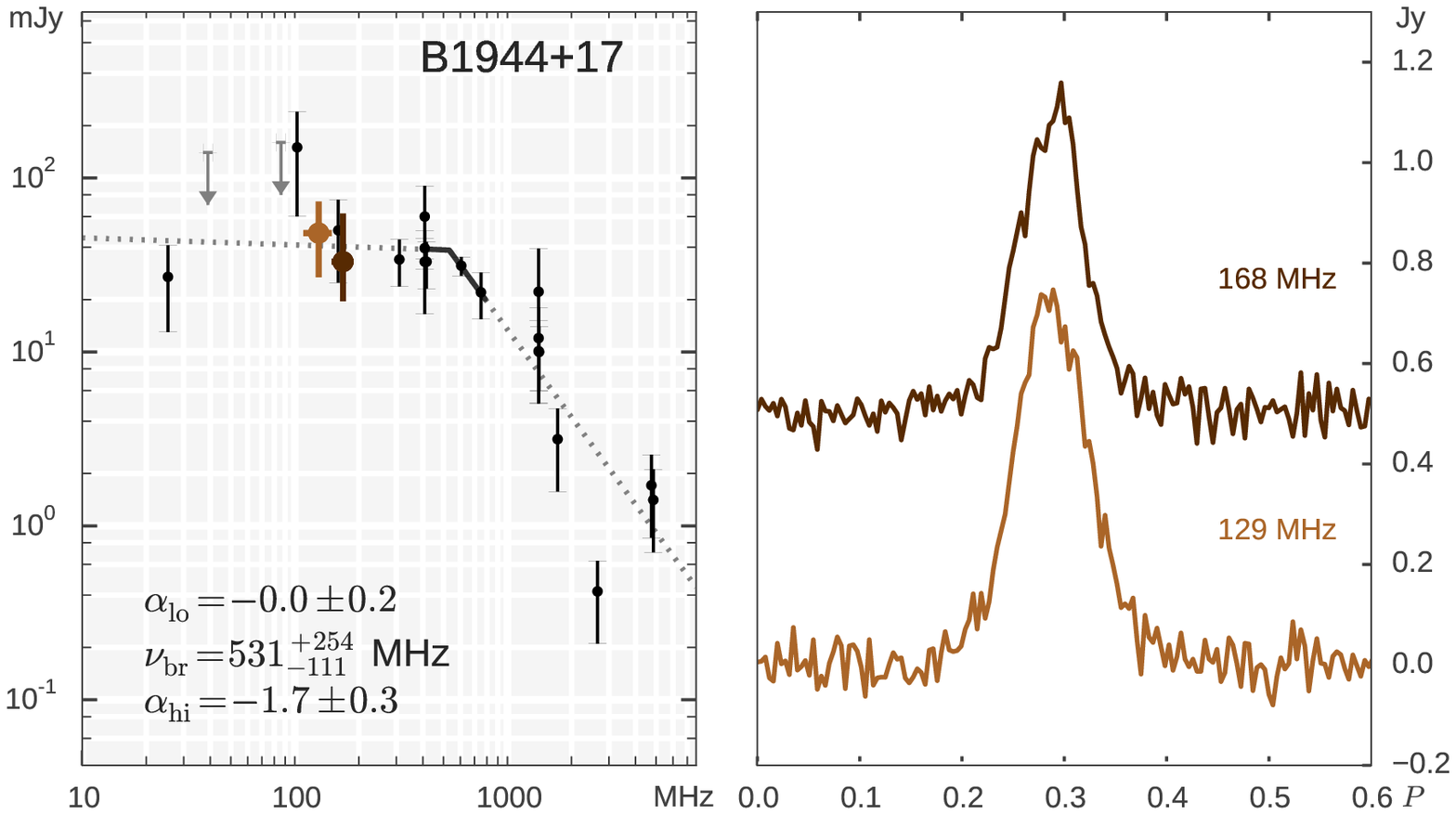}
\includegraphics[scale=0.48]{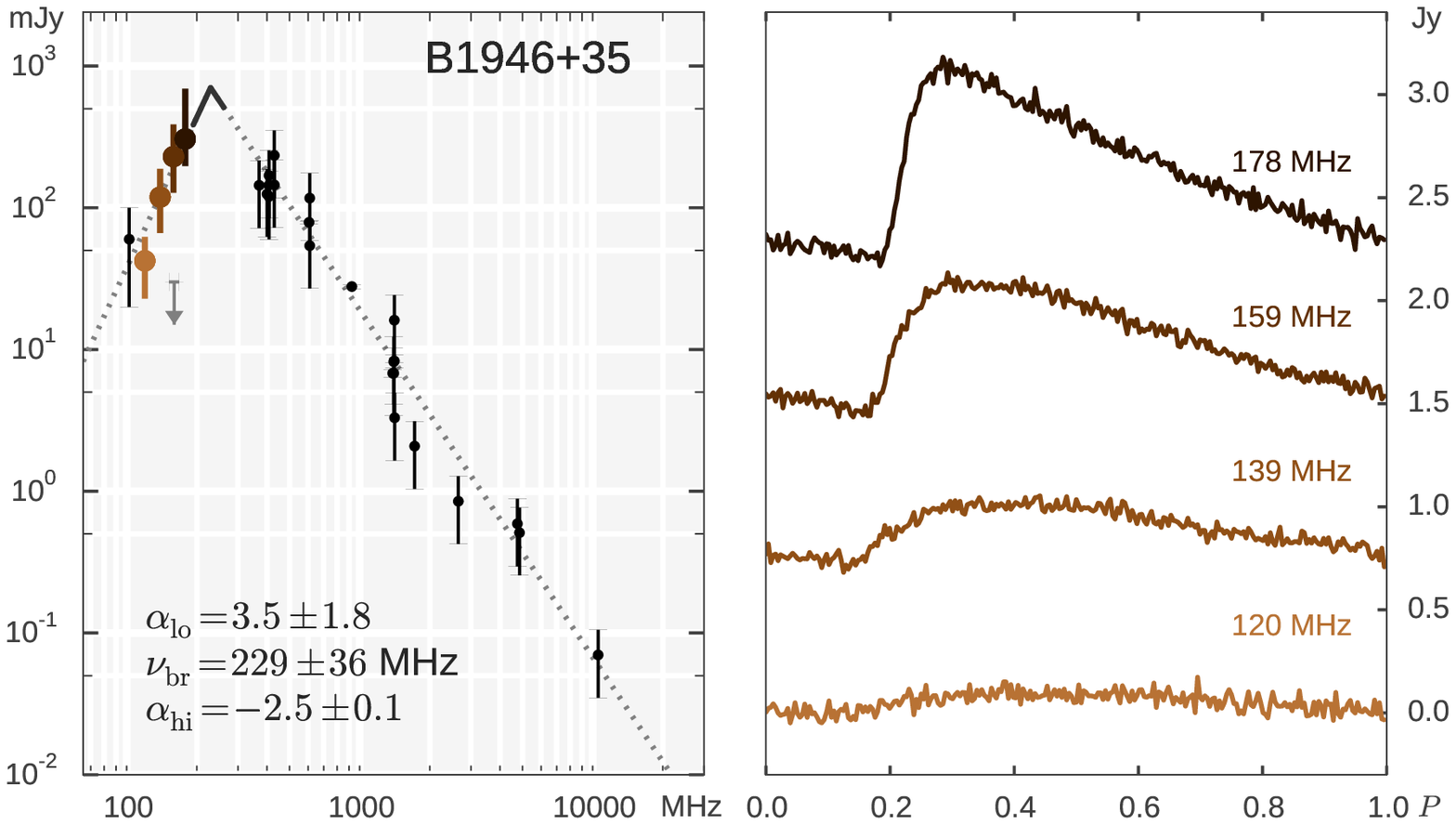}\includegraphics[scale=0.48]{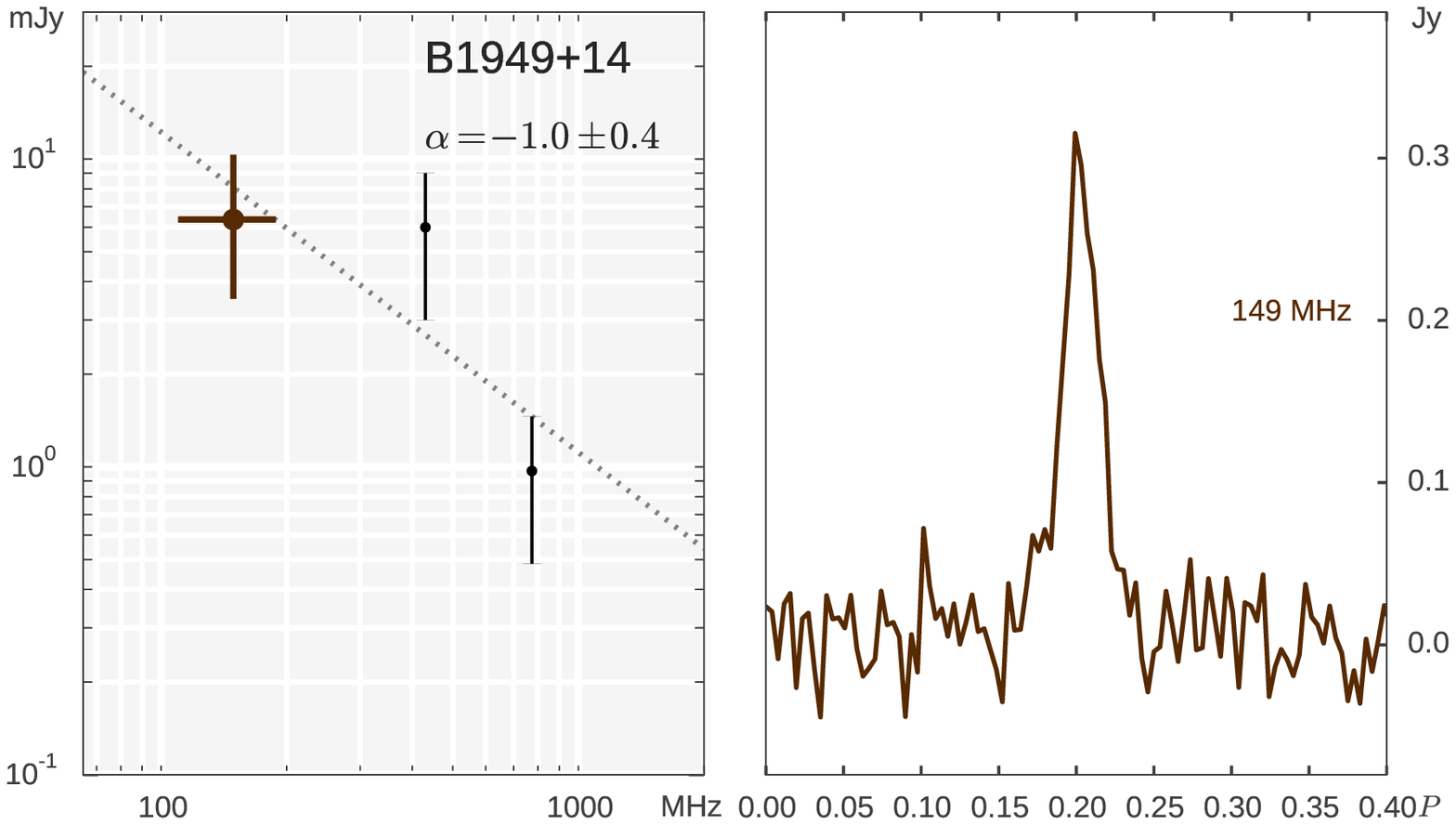}
\includegraphics[scale=0.48]{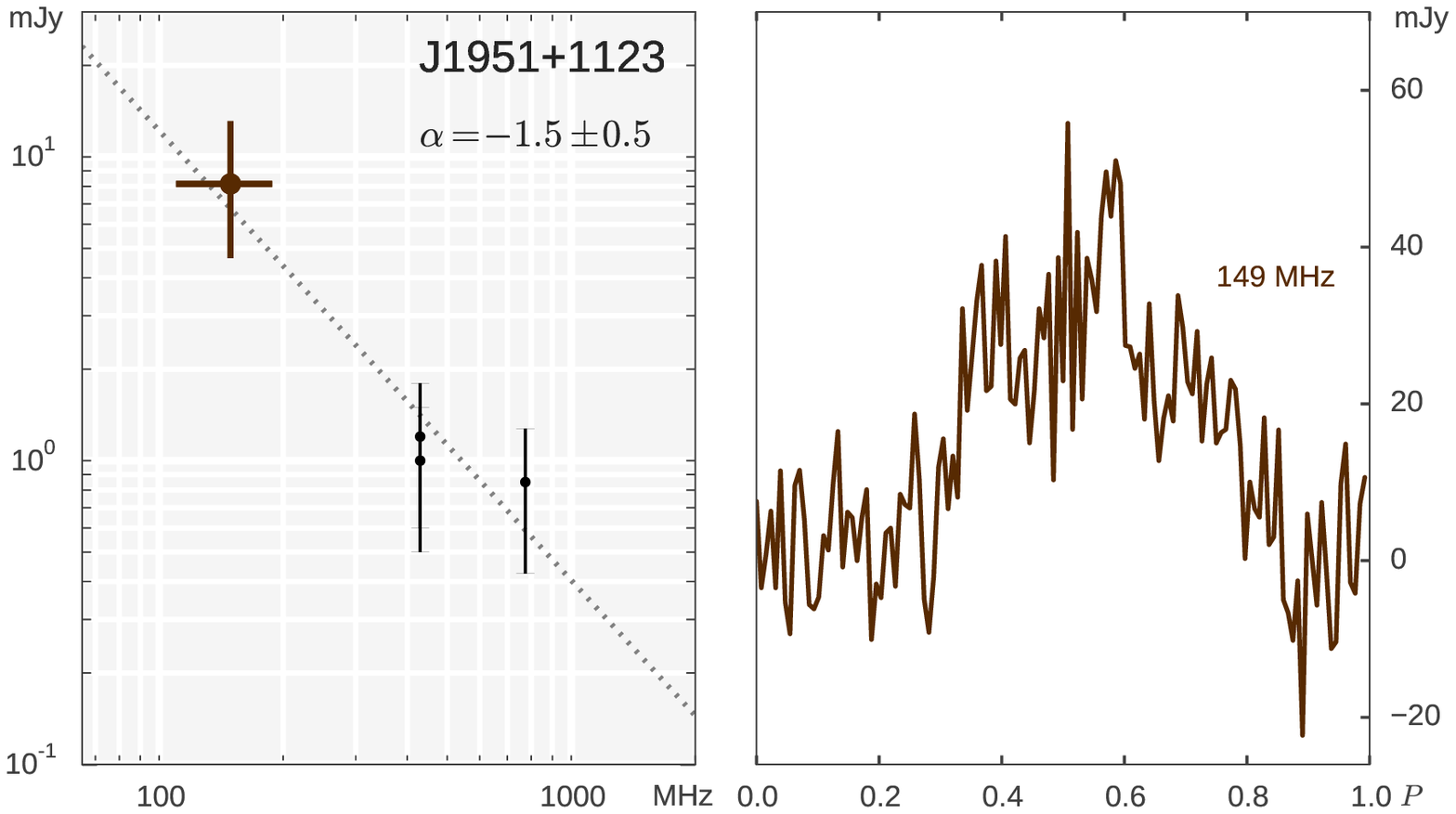}\includegraphics[scale=0.48]{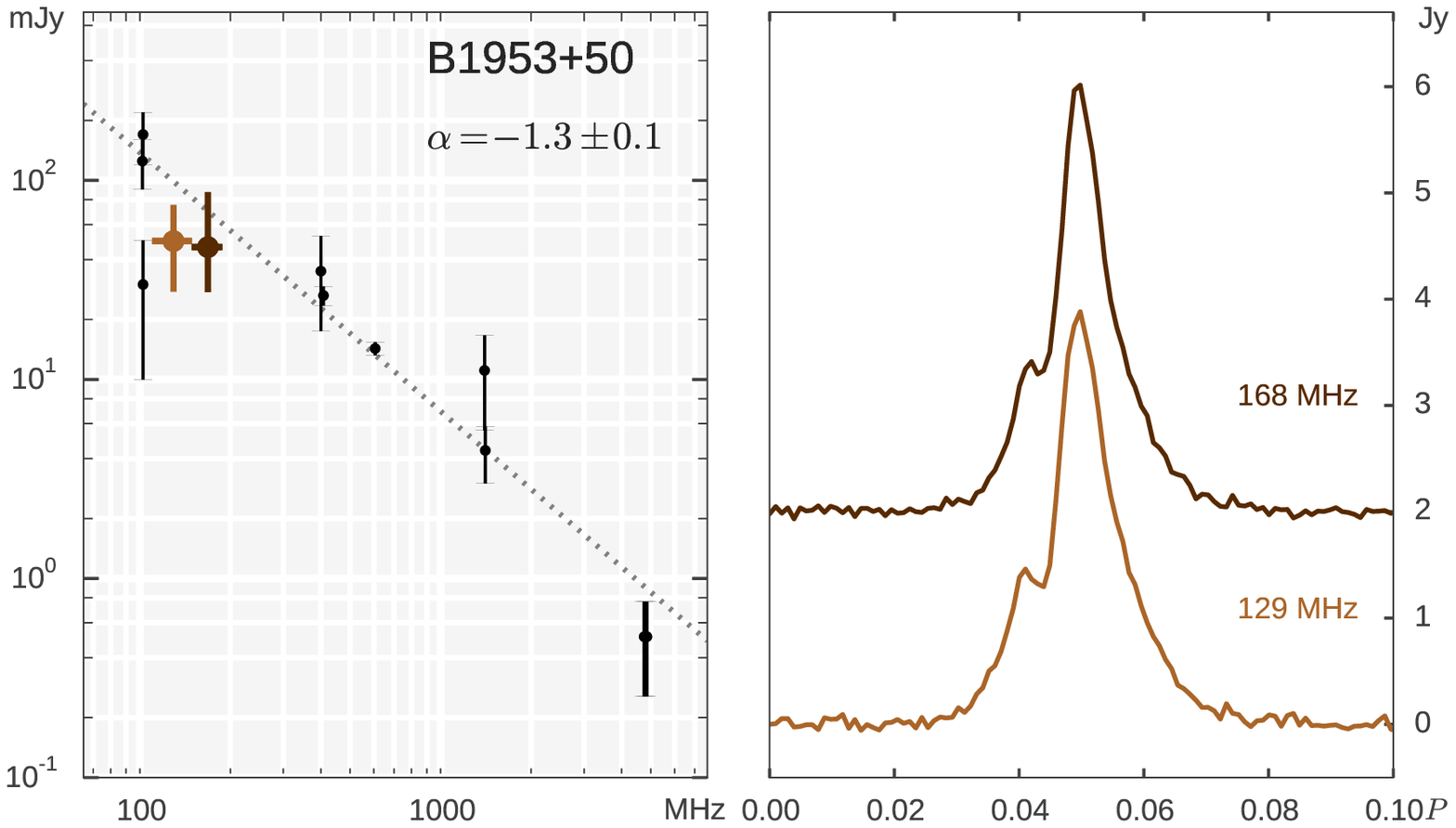}
\includegraphics[scale=0.48]{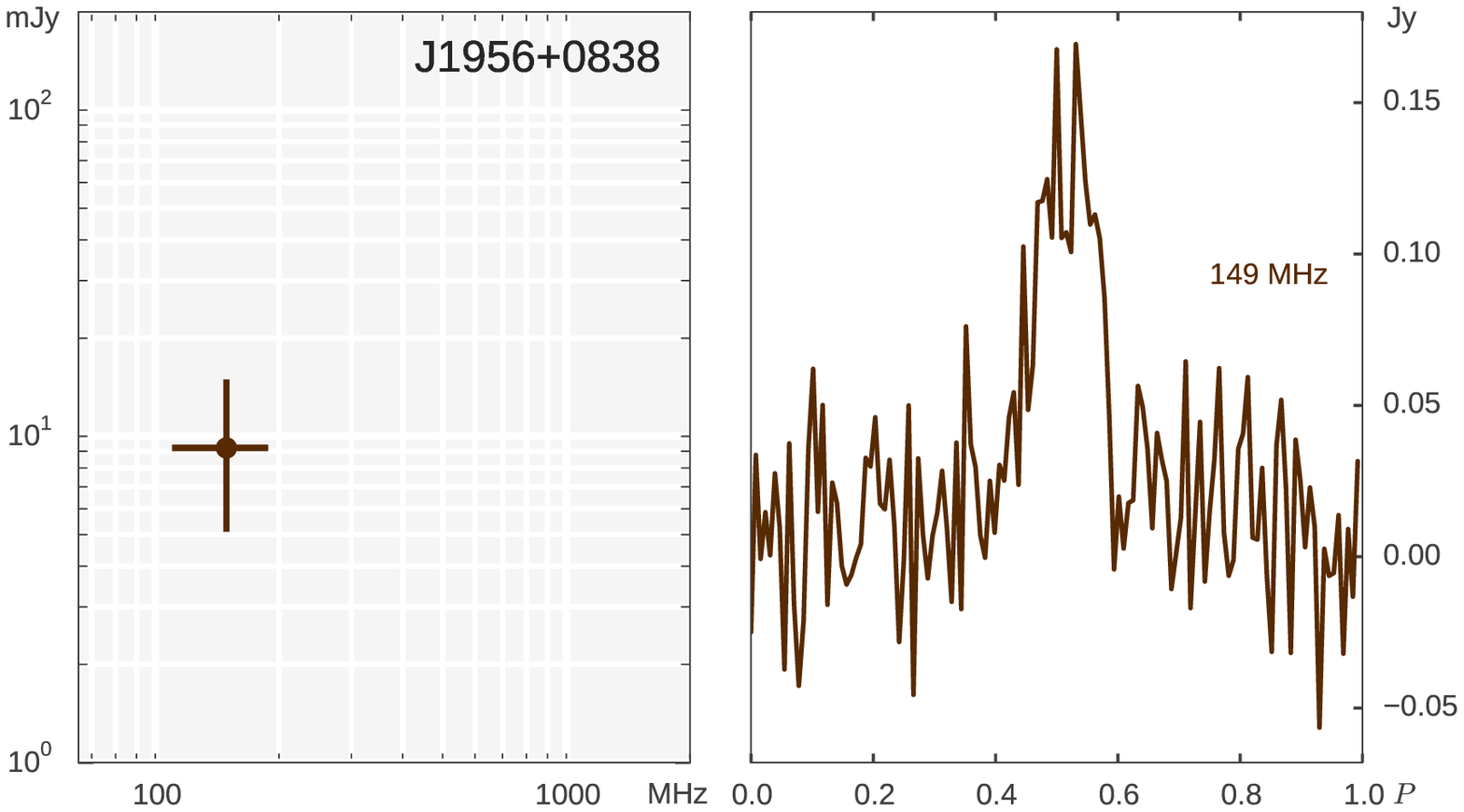}\includegraphics[scale=0.48]{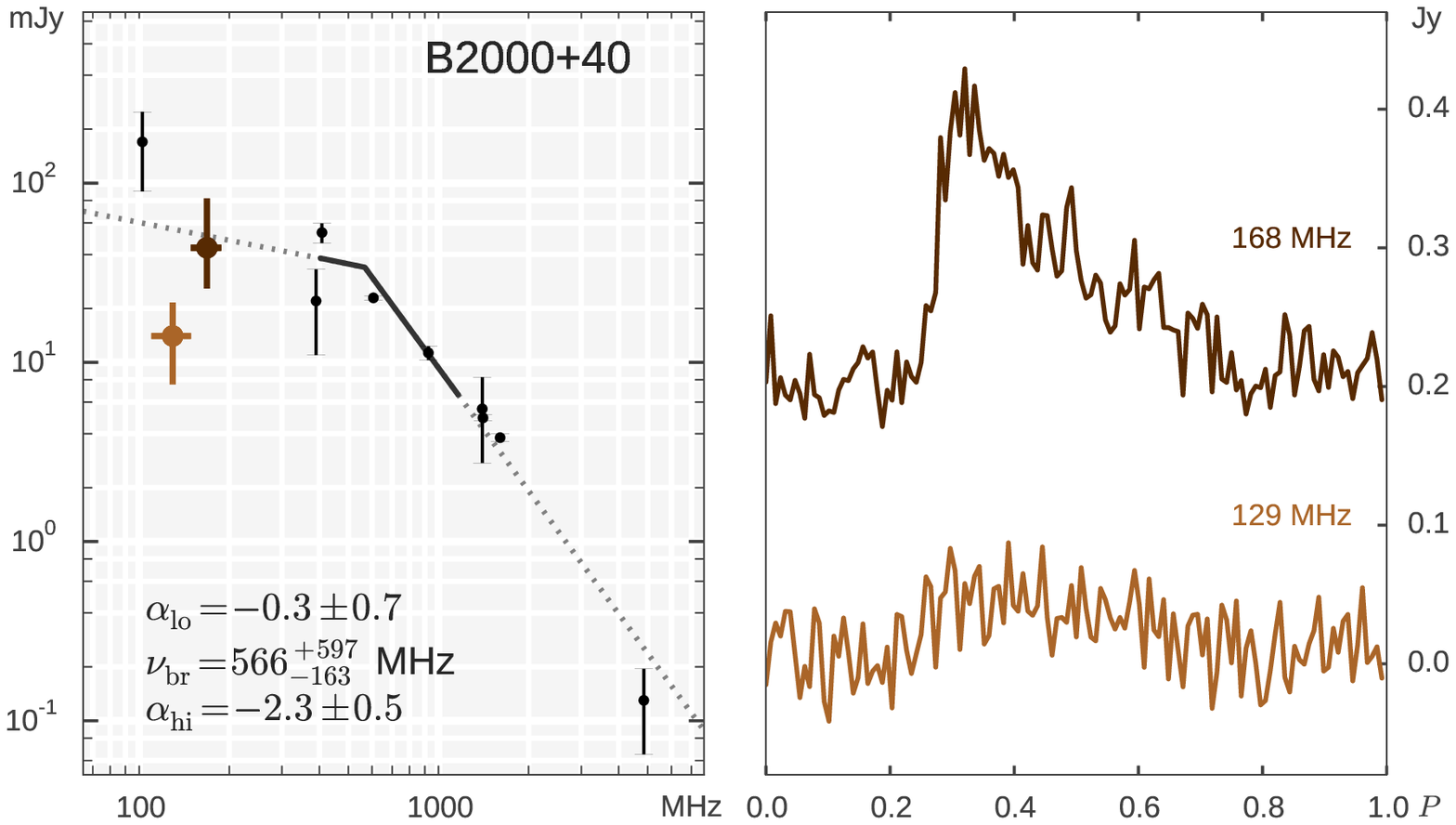}
\caption{See Figure~\ref{fig:prof_sp_1}.}
\label{fig:prof_sp_11}
\end{figure*}

\begin{figure*}
\includegraphics[scale=0.48]{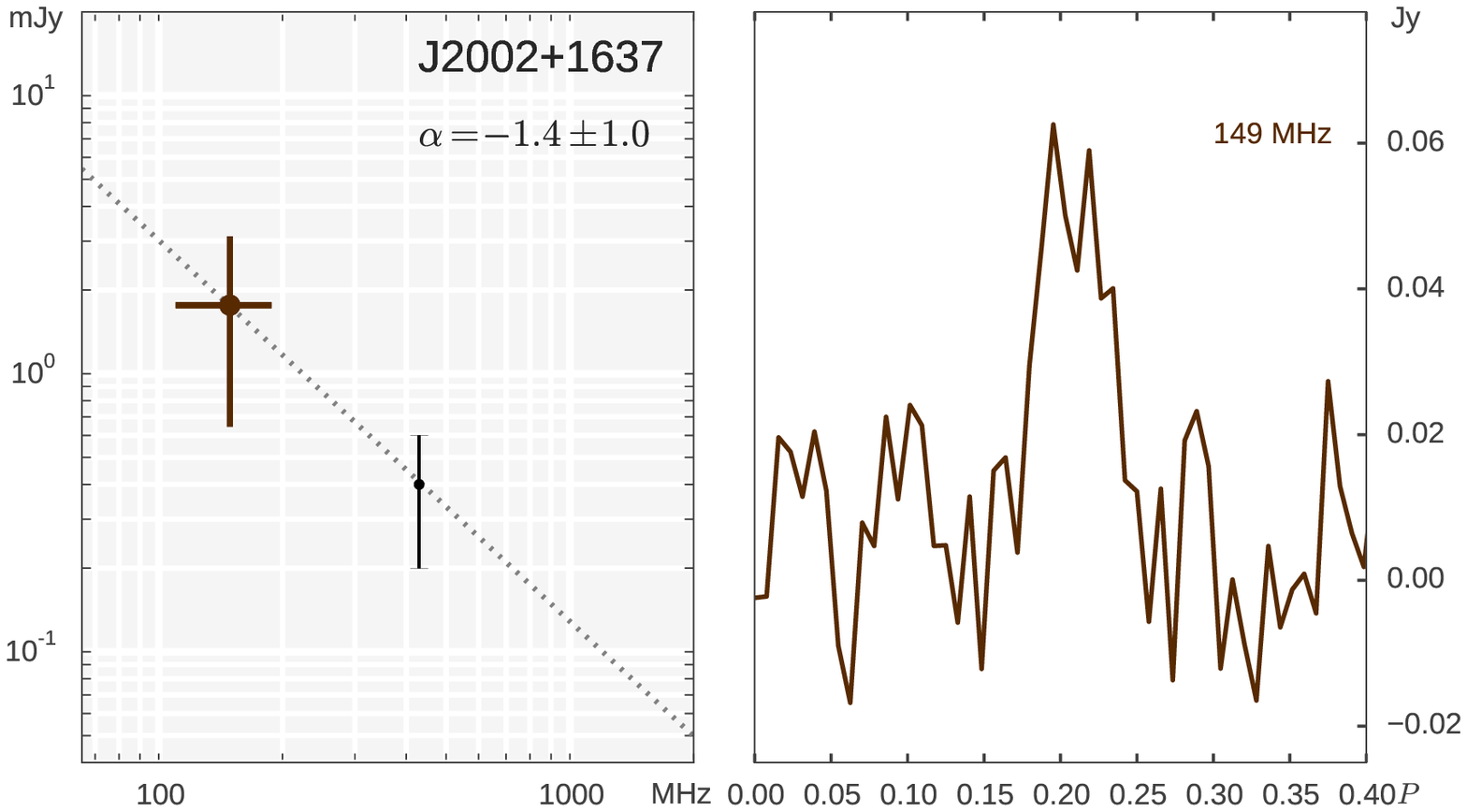}\includegraphics[scale=0.48]{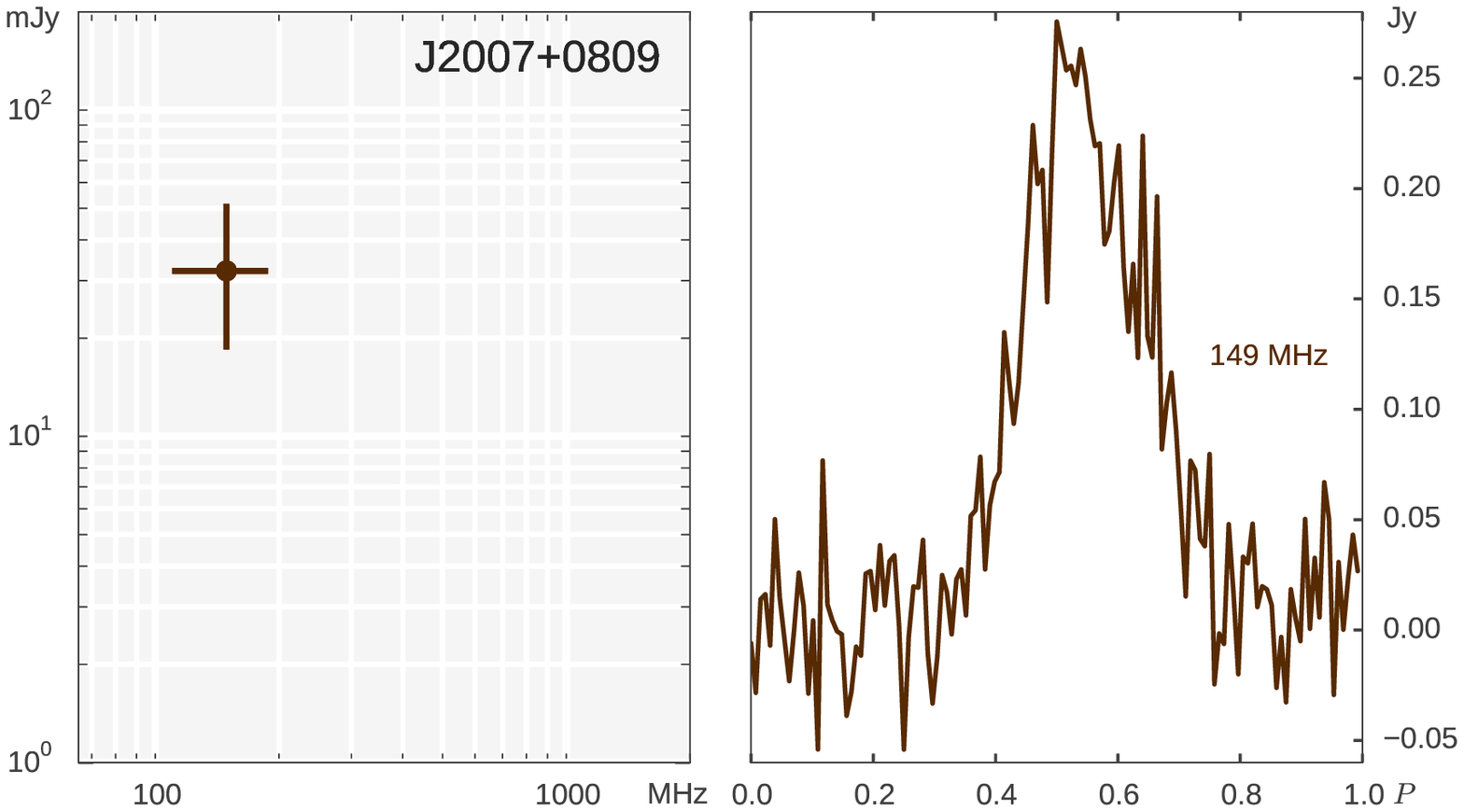}
\includegraphics[scale=0.48]{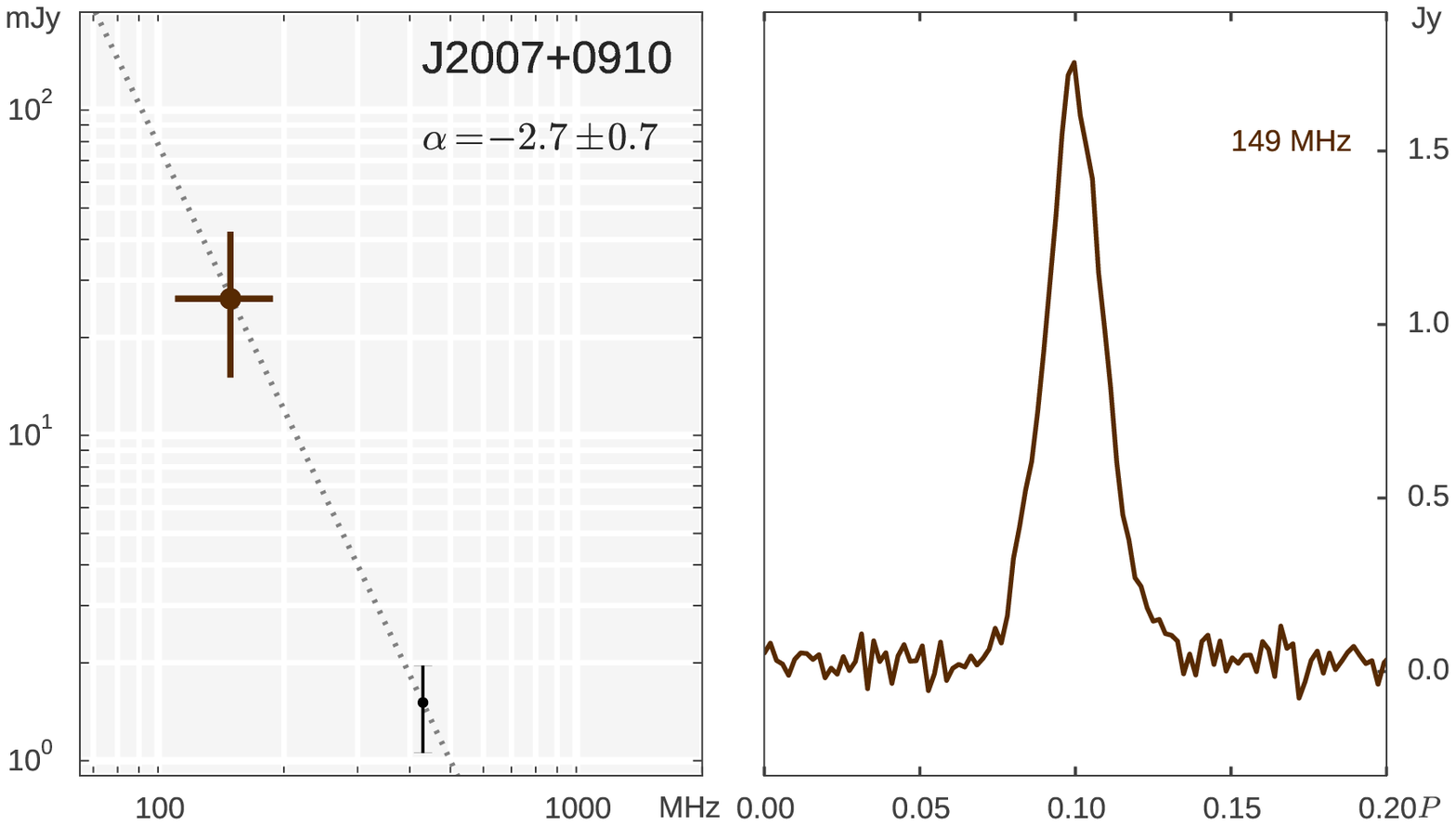}\includegraphics[scale=0.48]{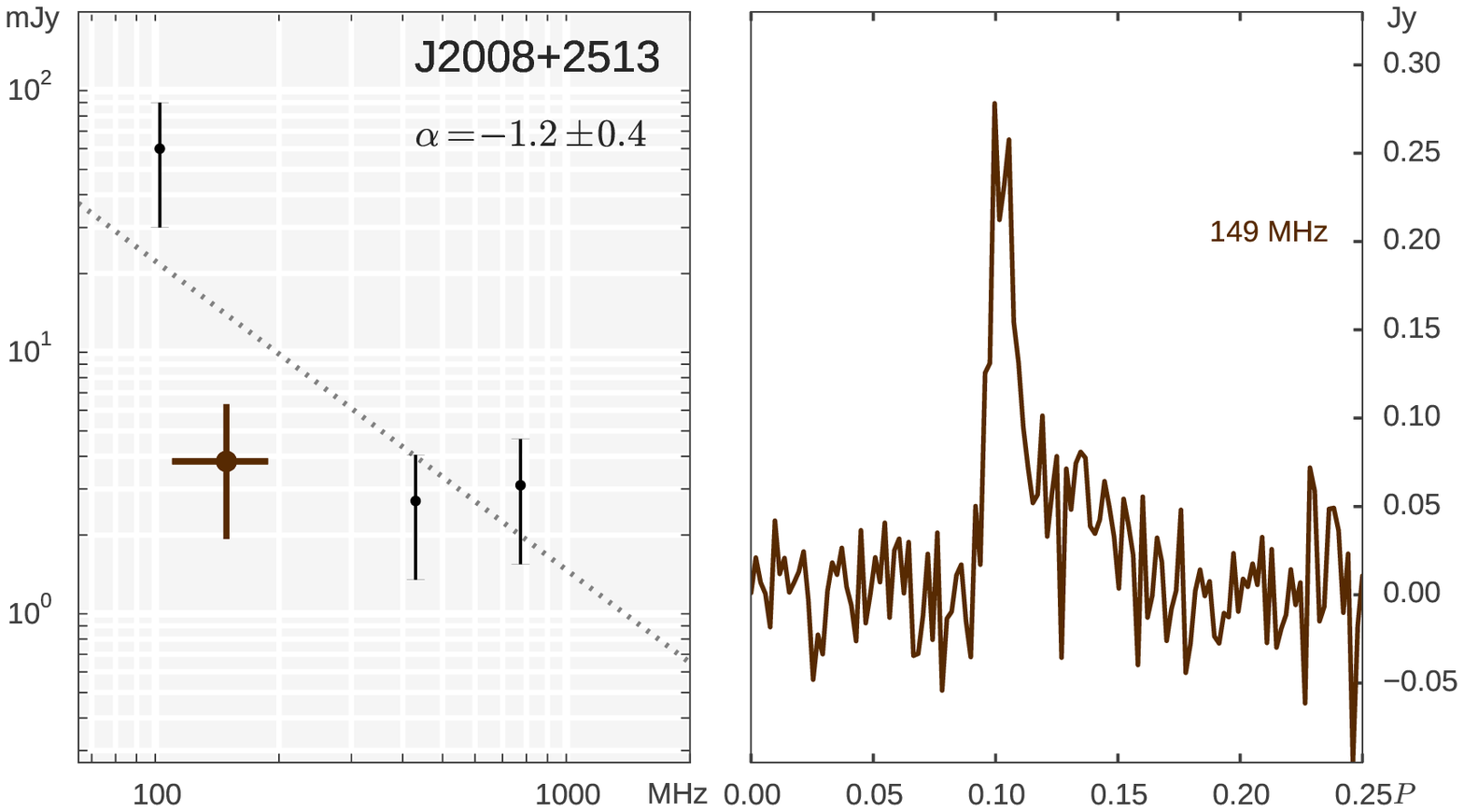}
\includegraphics[scale=0.48]{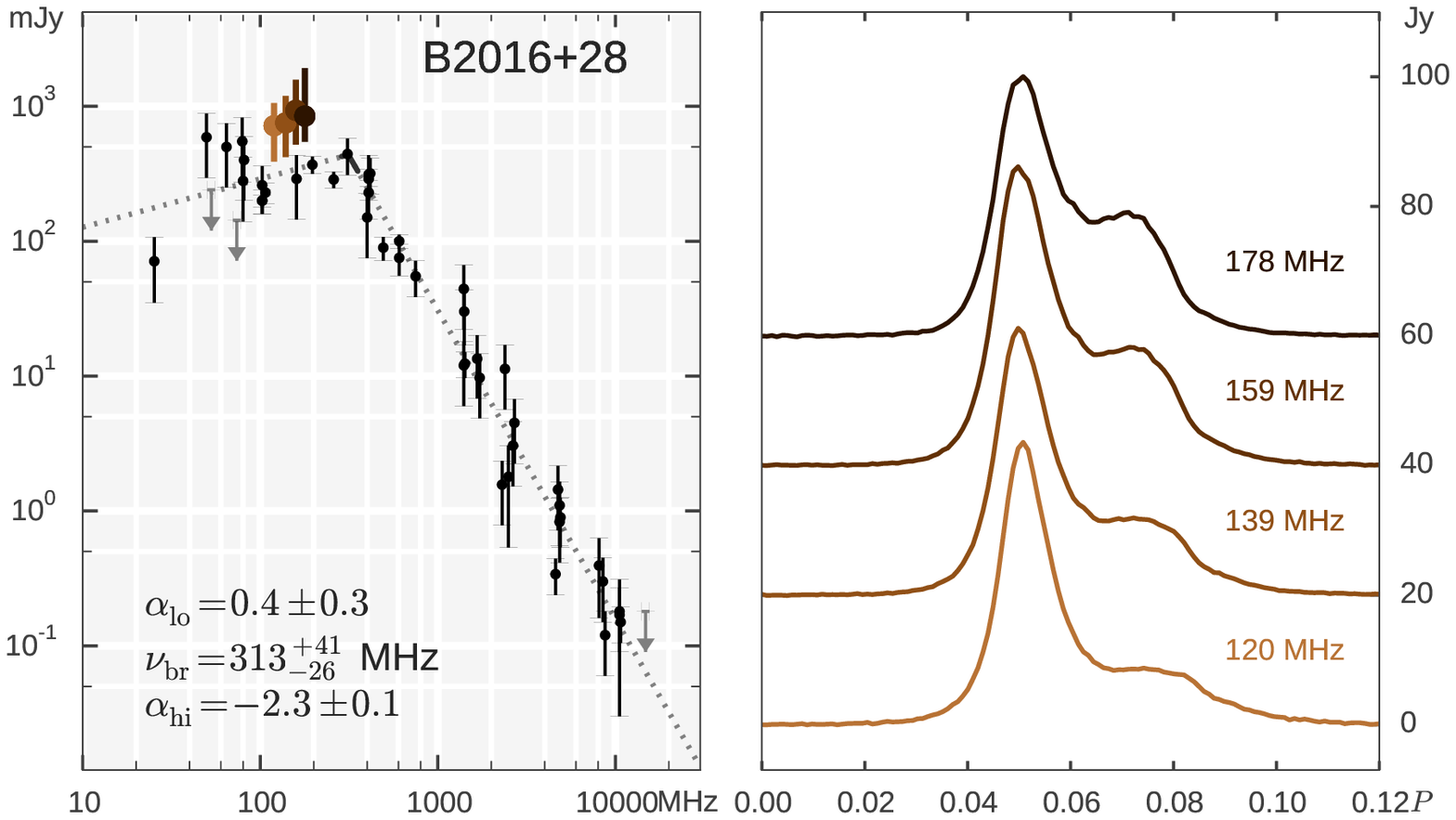}\includegraphics[scale=0.48]{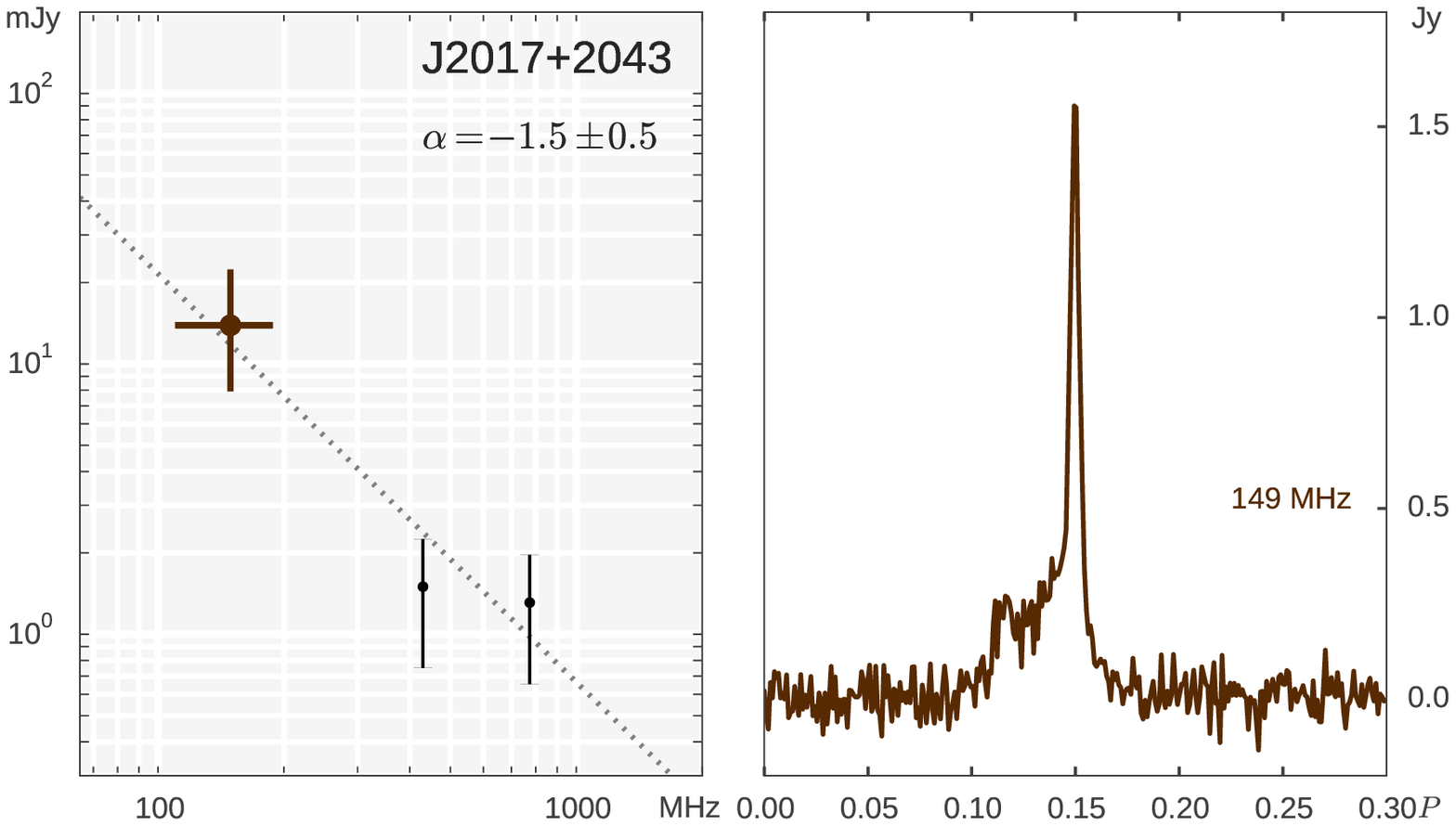}
\includegraphics[scale=0.48]{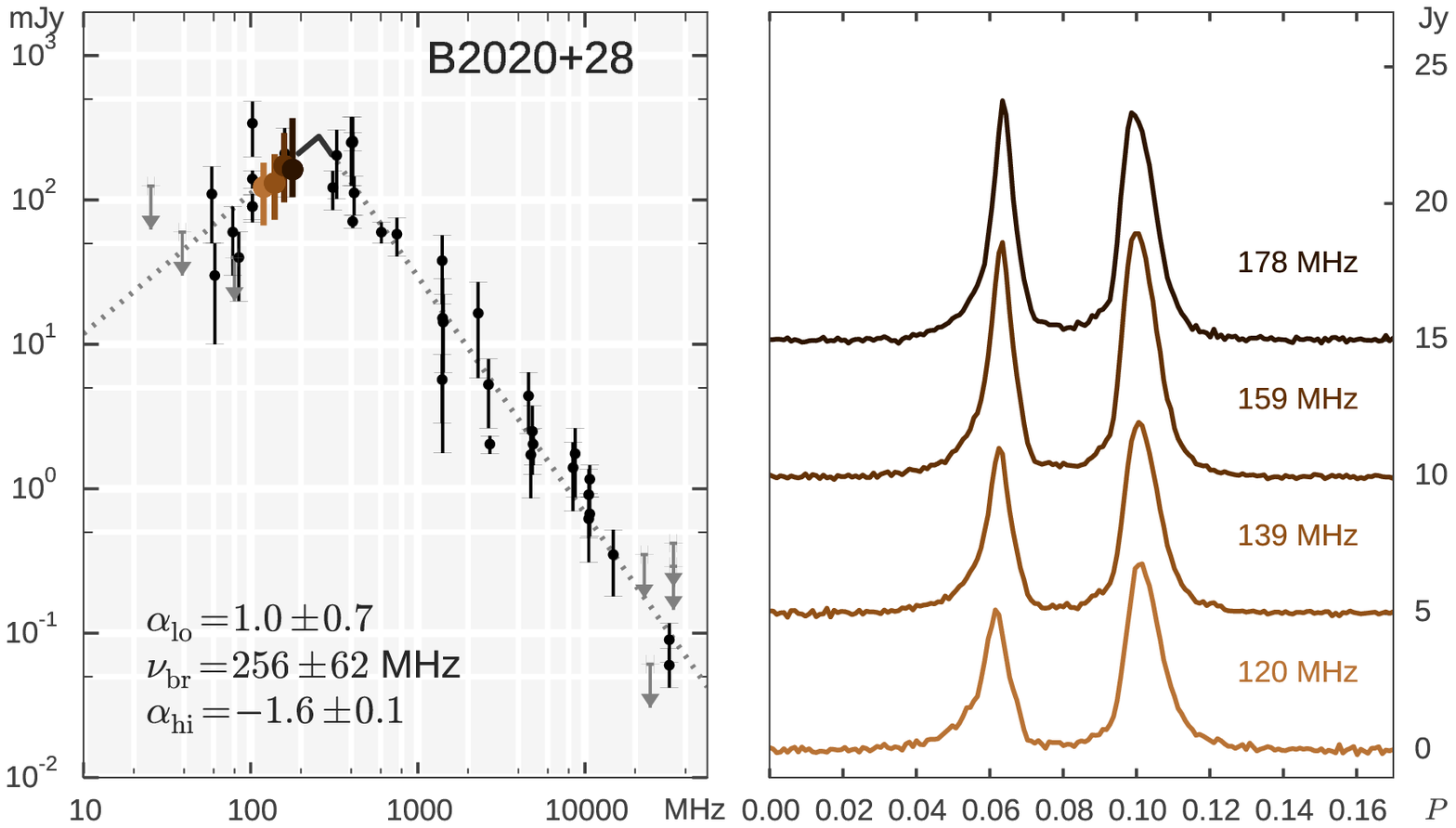}\includegraphics[scale=0.48]{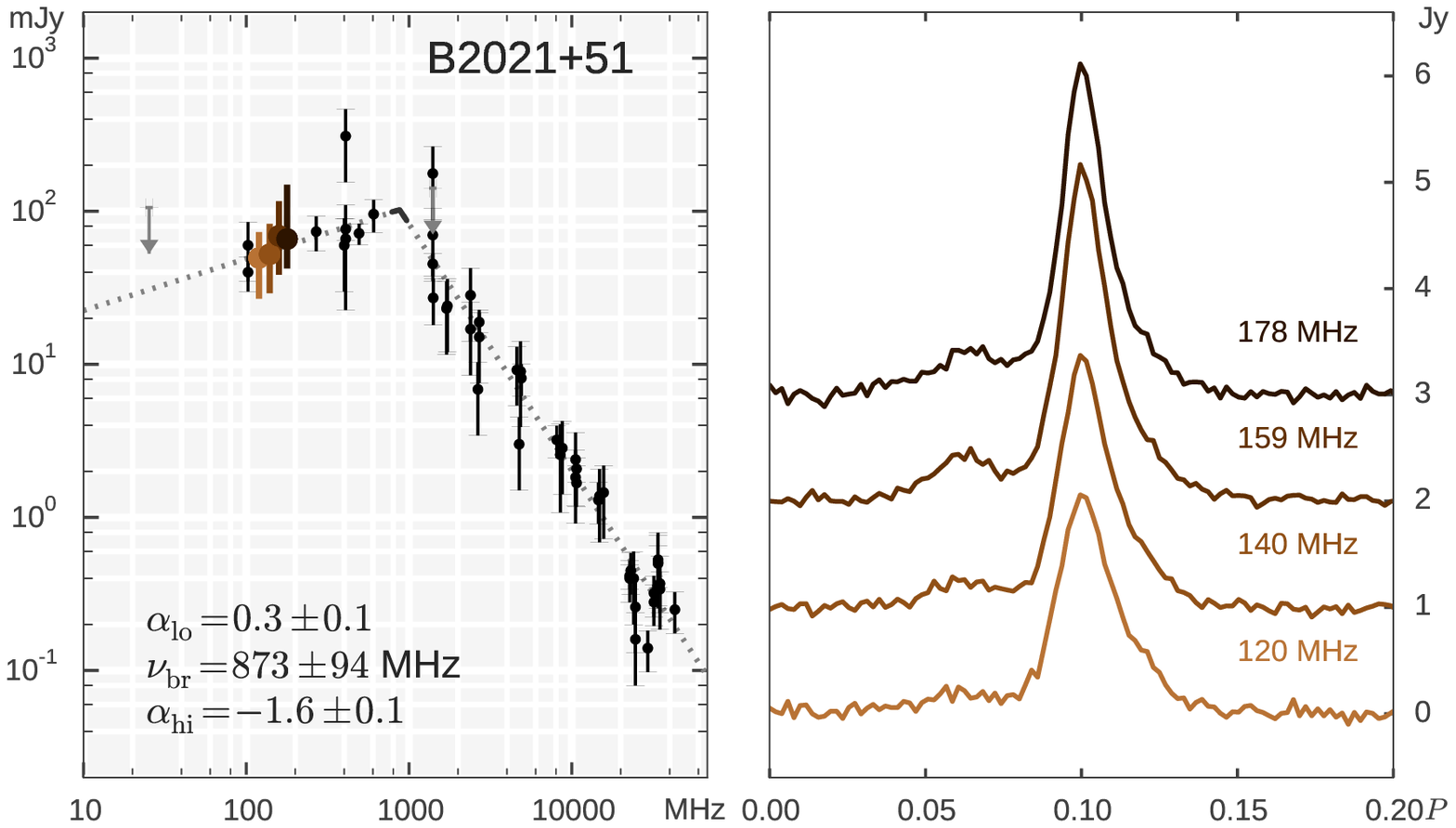}
\includegraphics[scale=0.48]{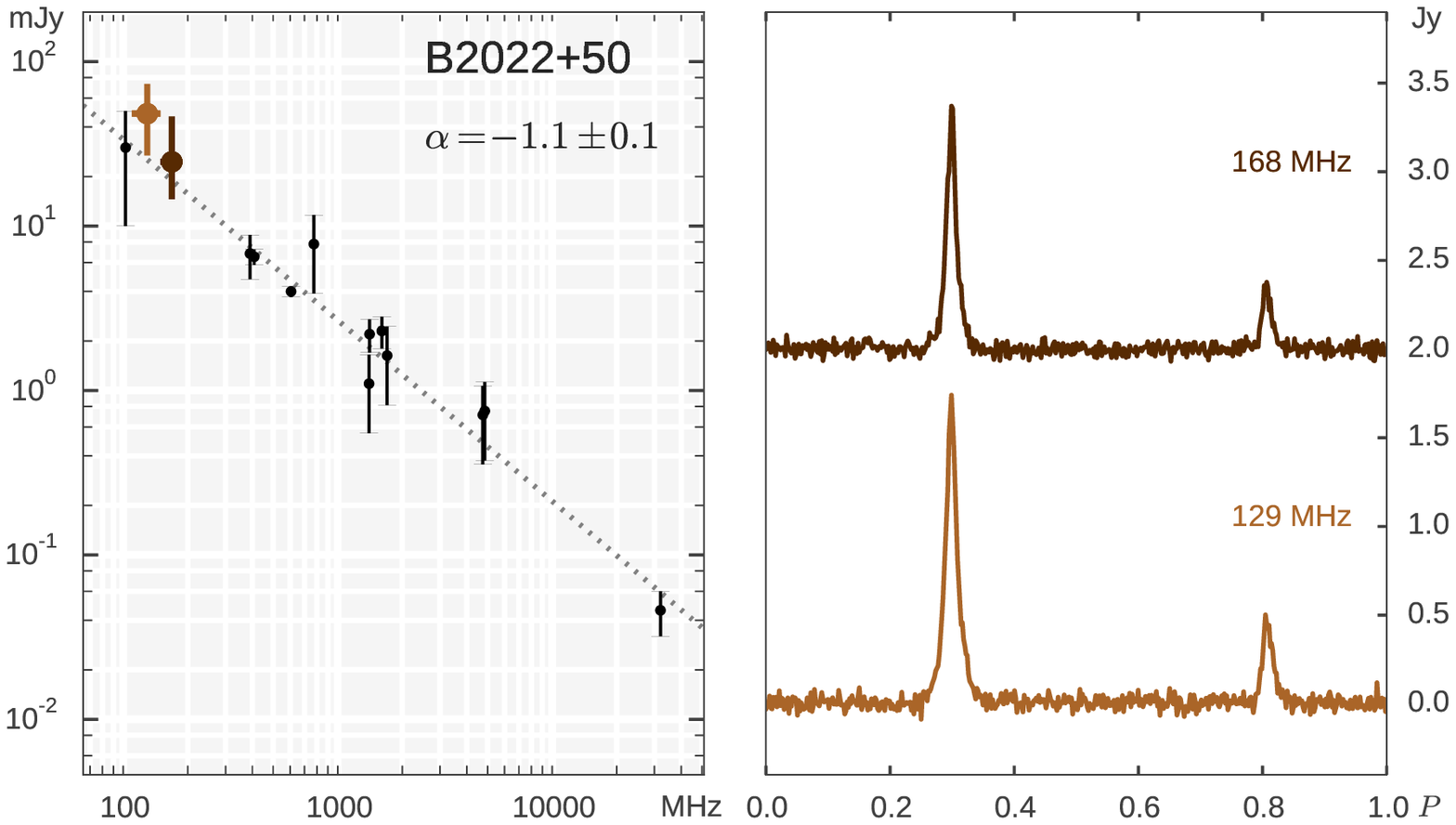}\includegraphics[scale=0.48]{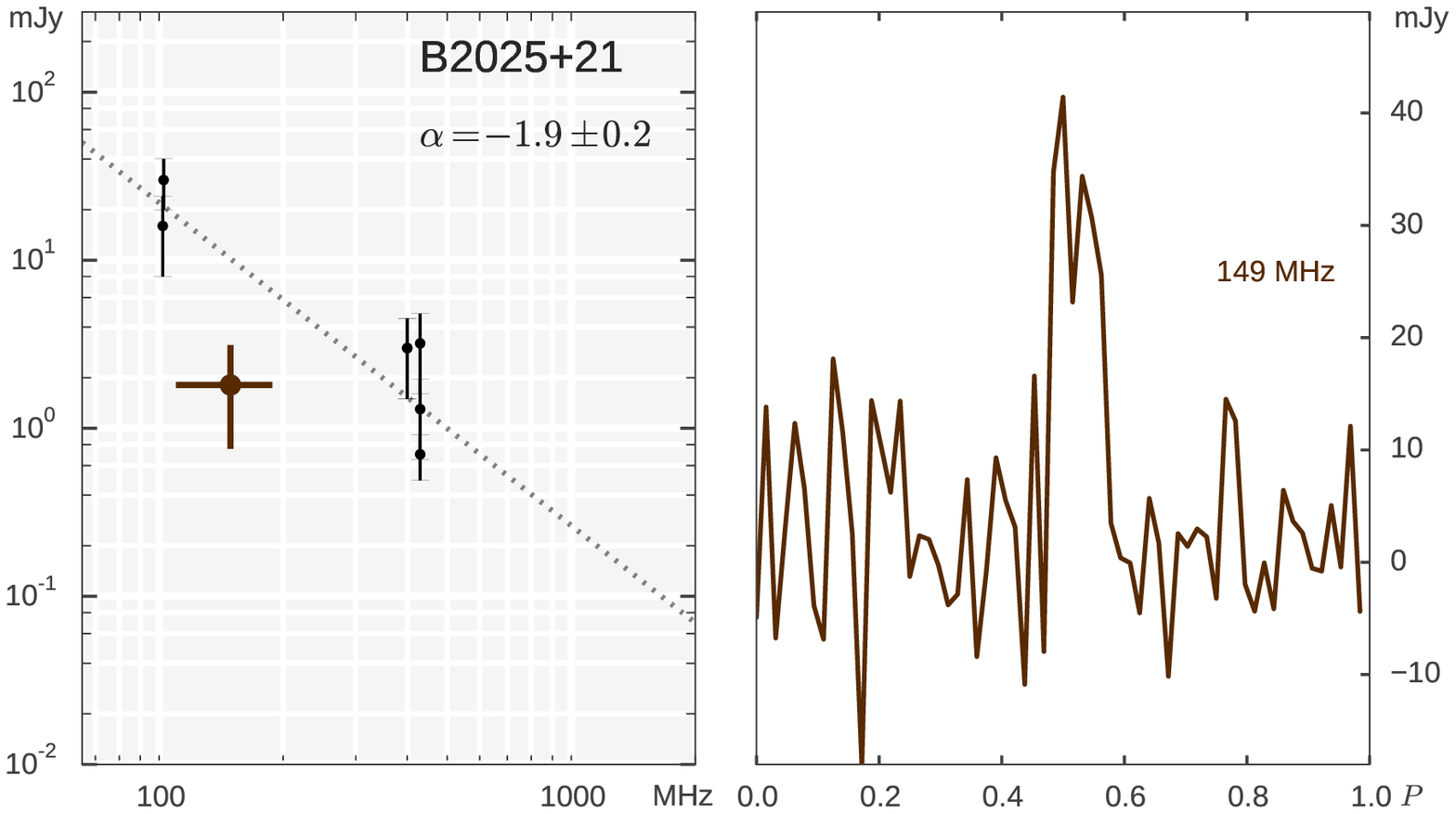}
\caption{See Figure~\ref{fig:prof_sp_1}.}
\label{fig:prof_sp_12}
\end{figure*}

\begin{figure*}
\includegraphics[scale=0.48]{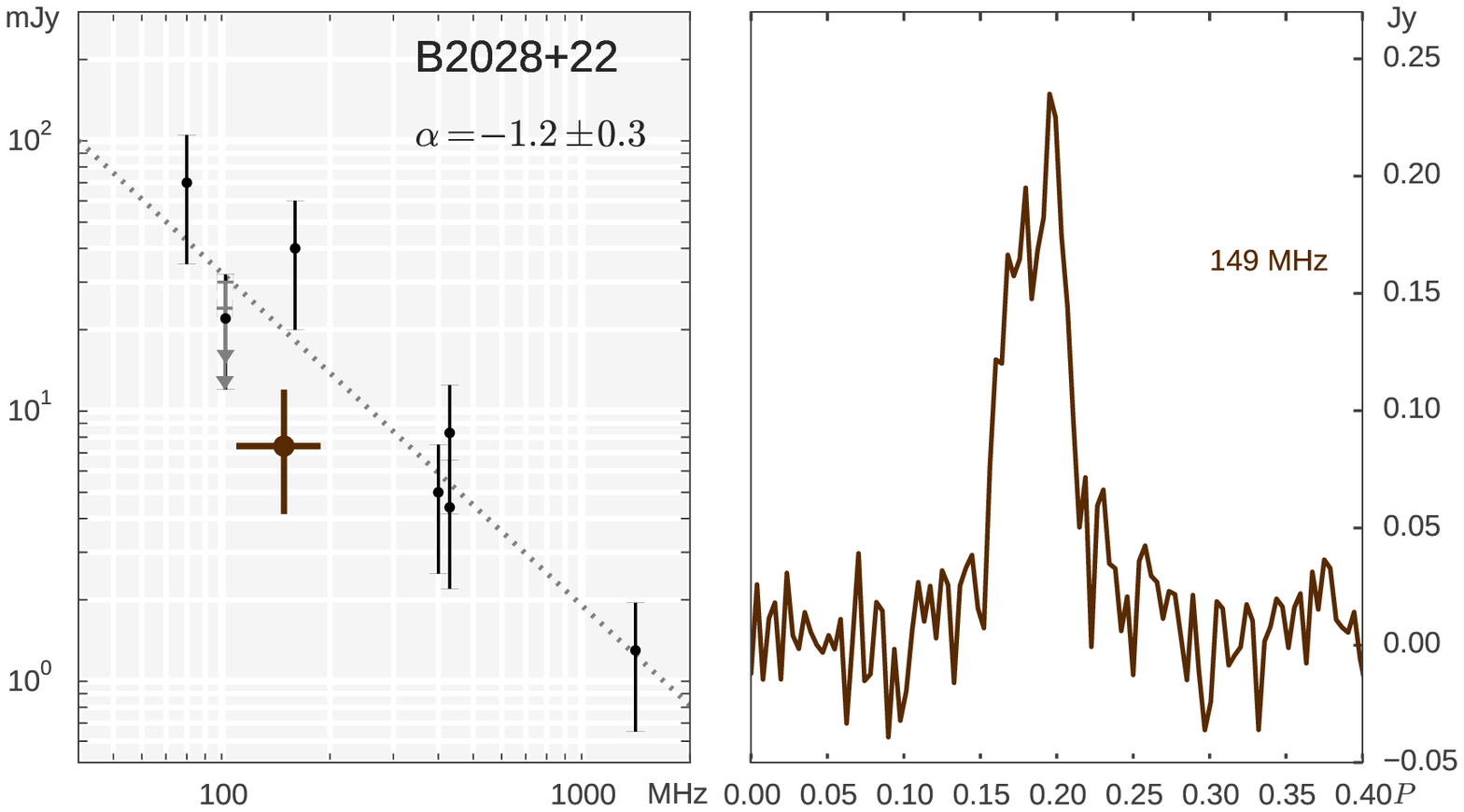}\includegraphics[scale=0.48]{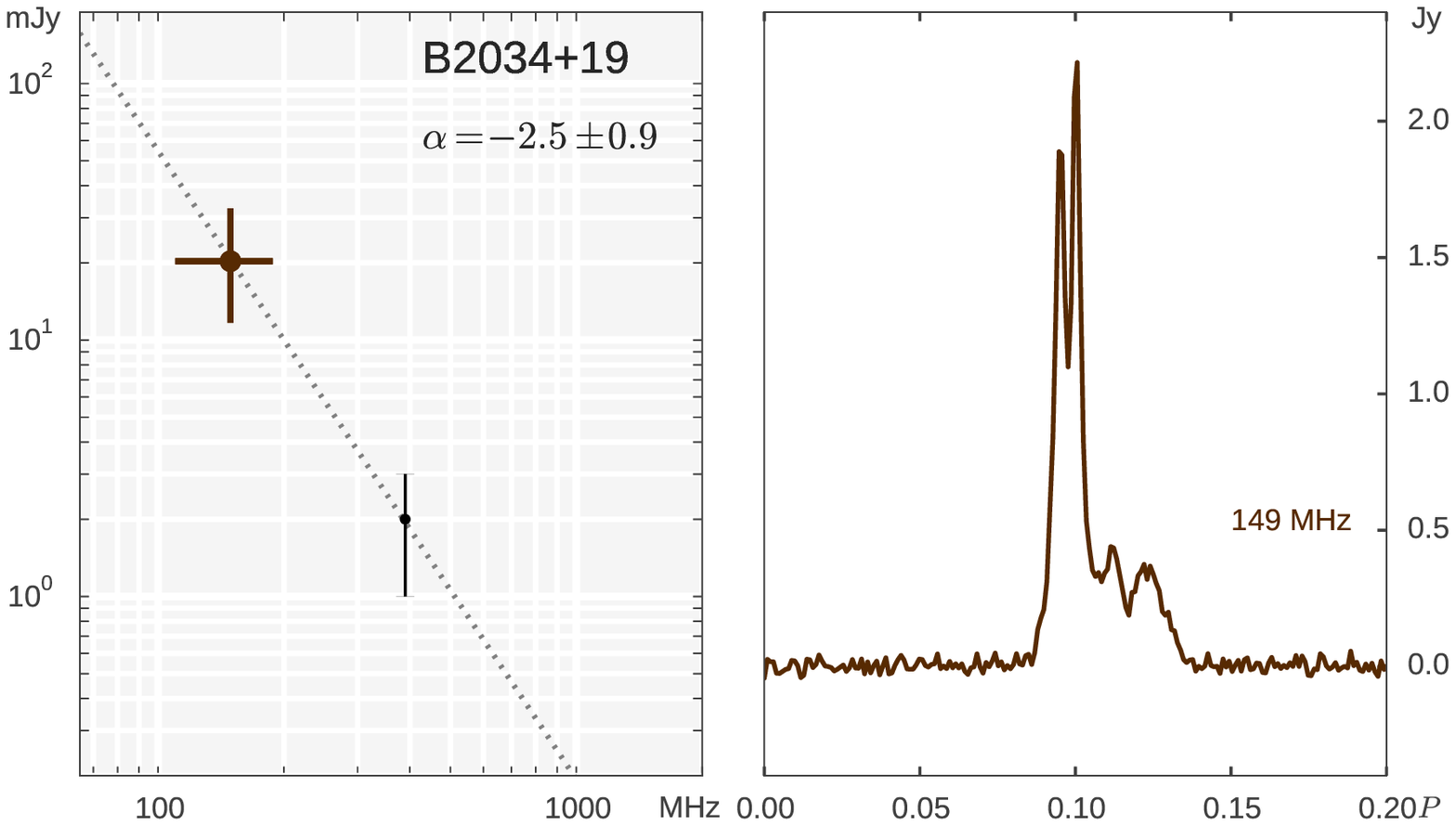}
\includegraphics[scale=0.48]{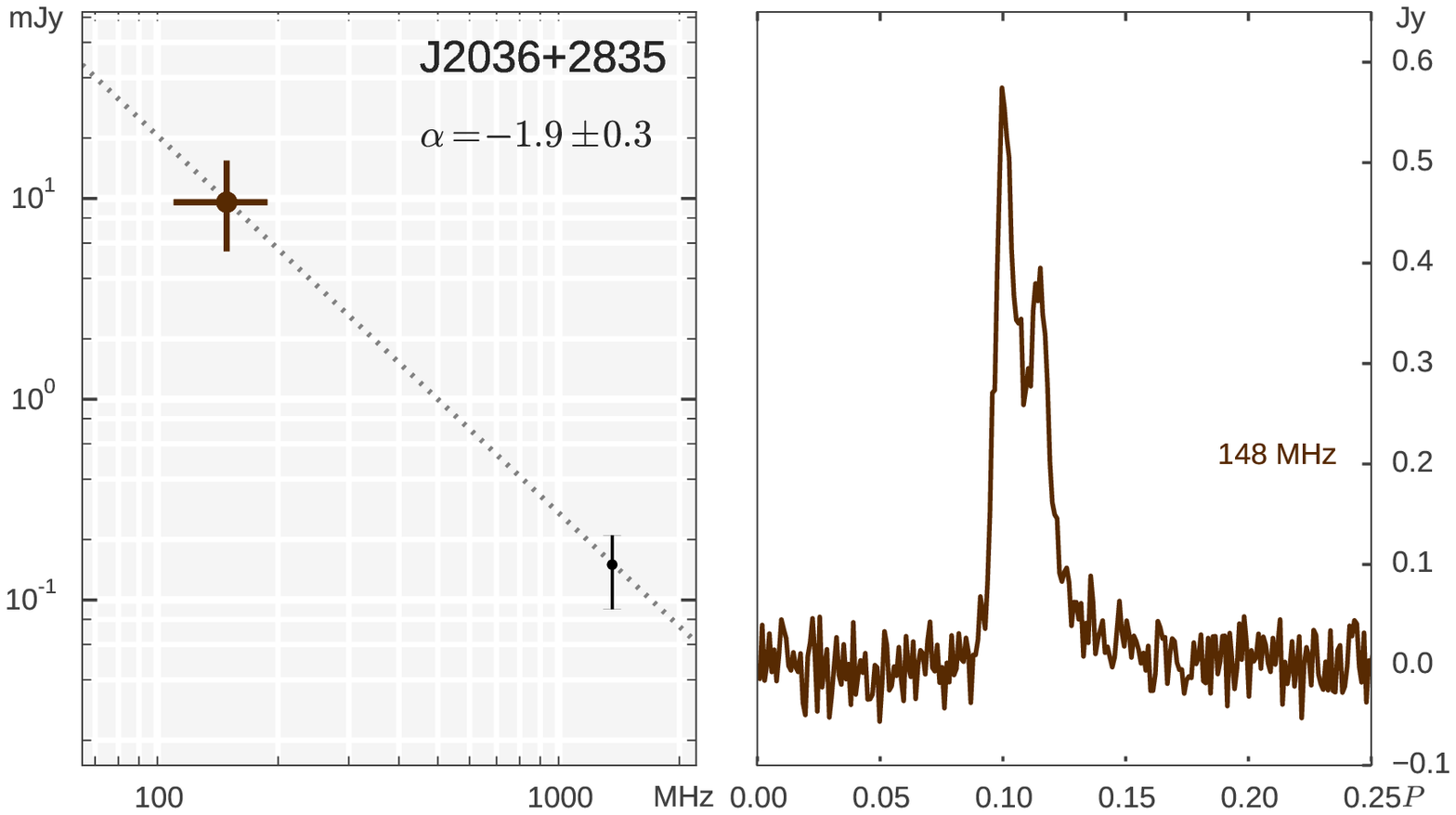}\includegraphics[scale=0.48]{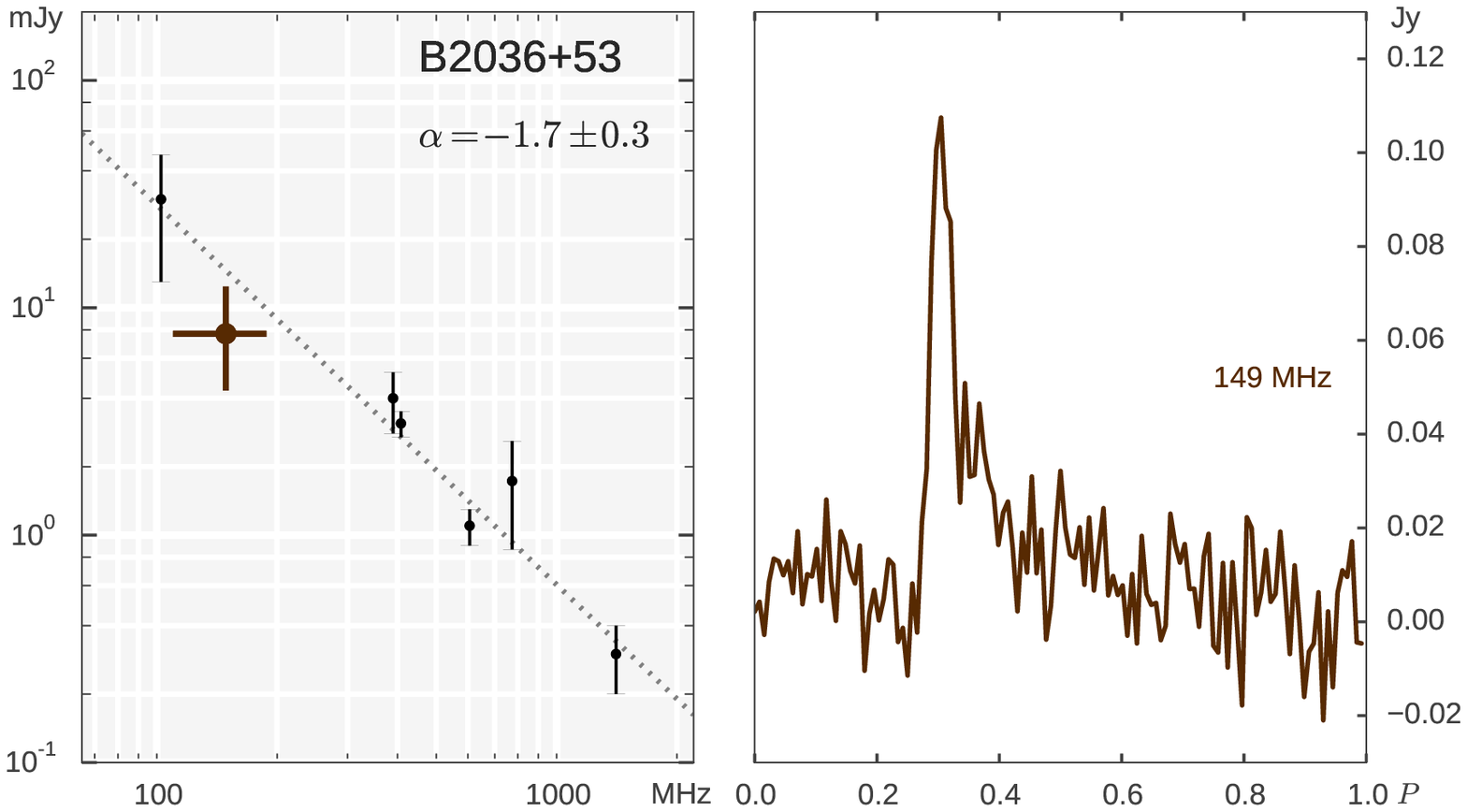}
\includegraphics[scale=0.48]{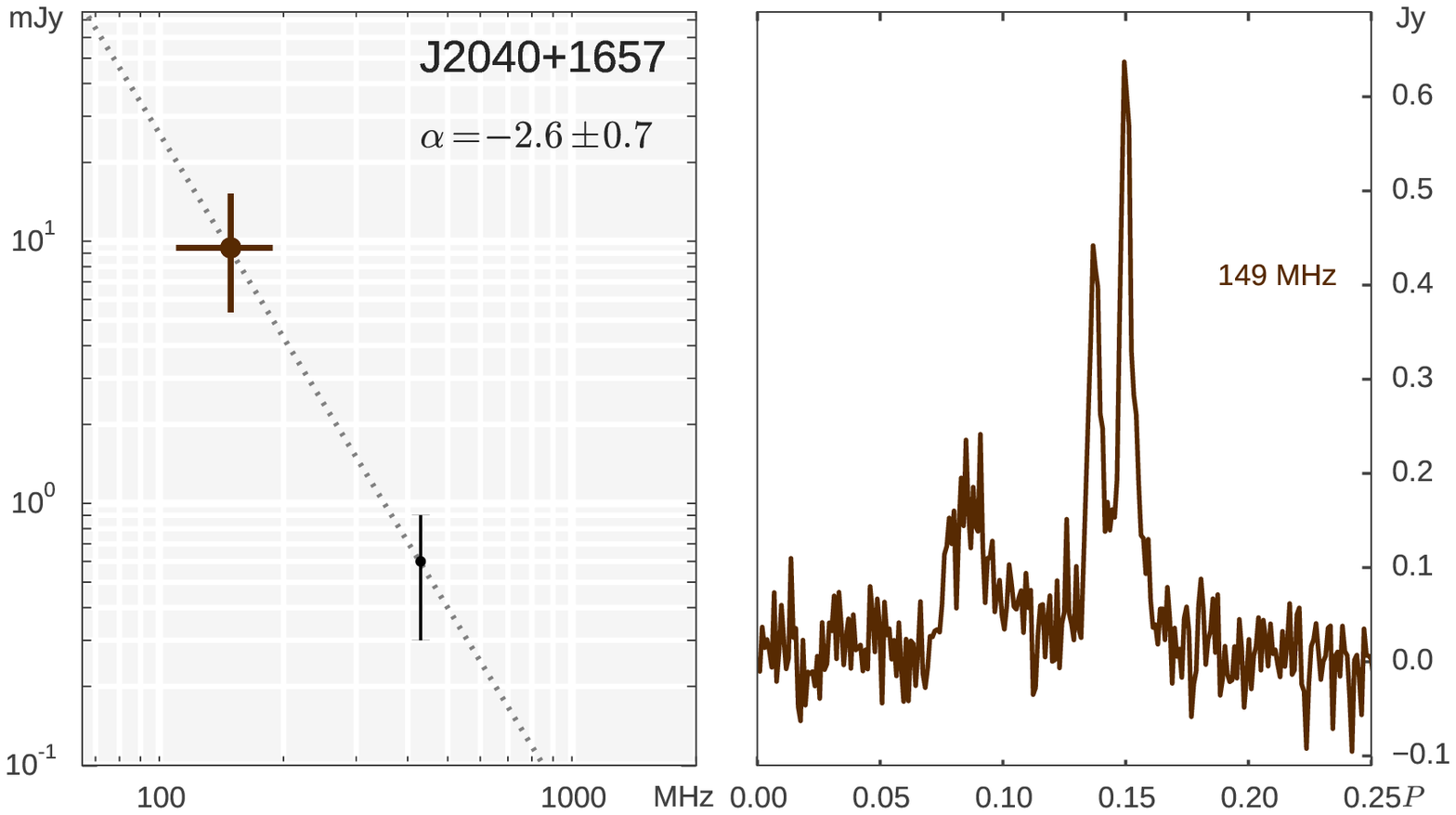}\includegraphics[scale=0.48]{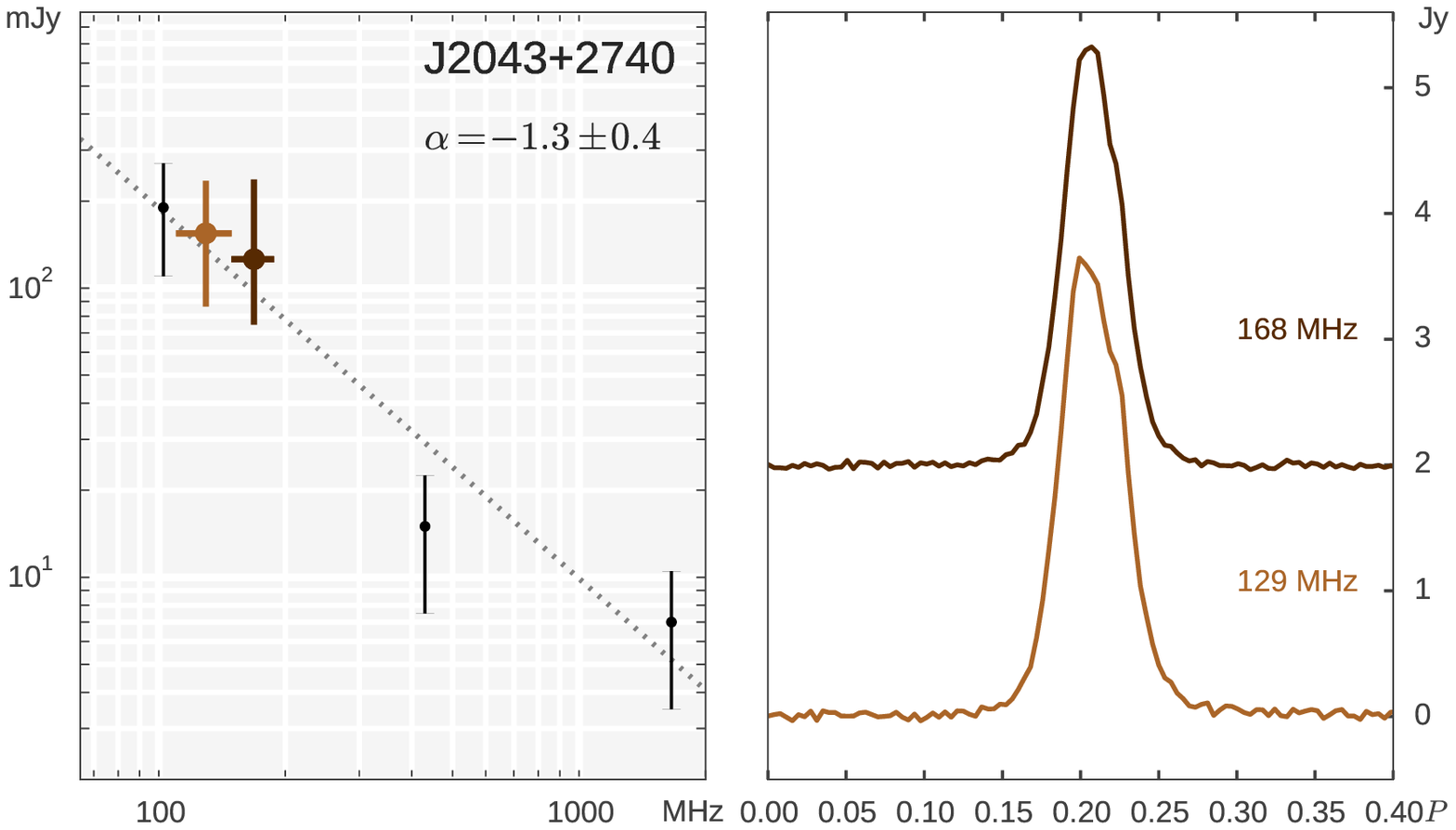}
\includegraphics[scale=0.48]{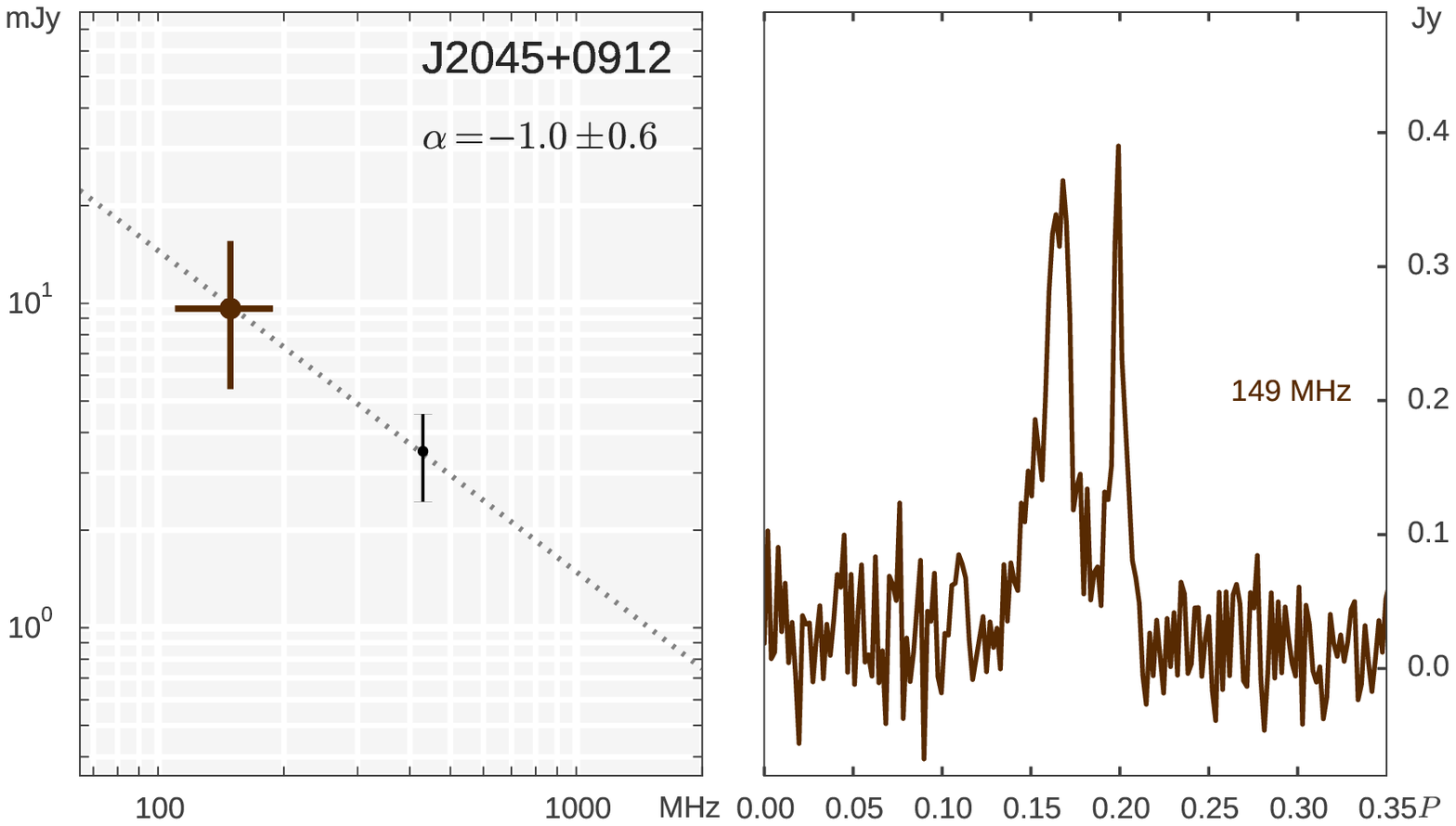}\includegraphics[scale=0.48]{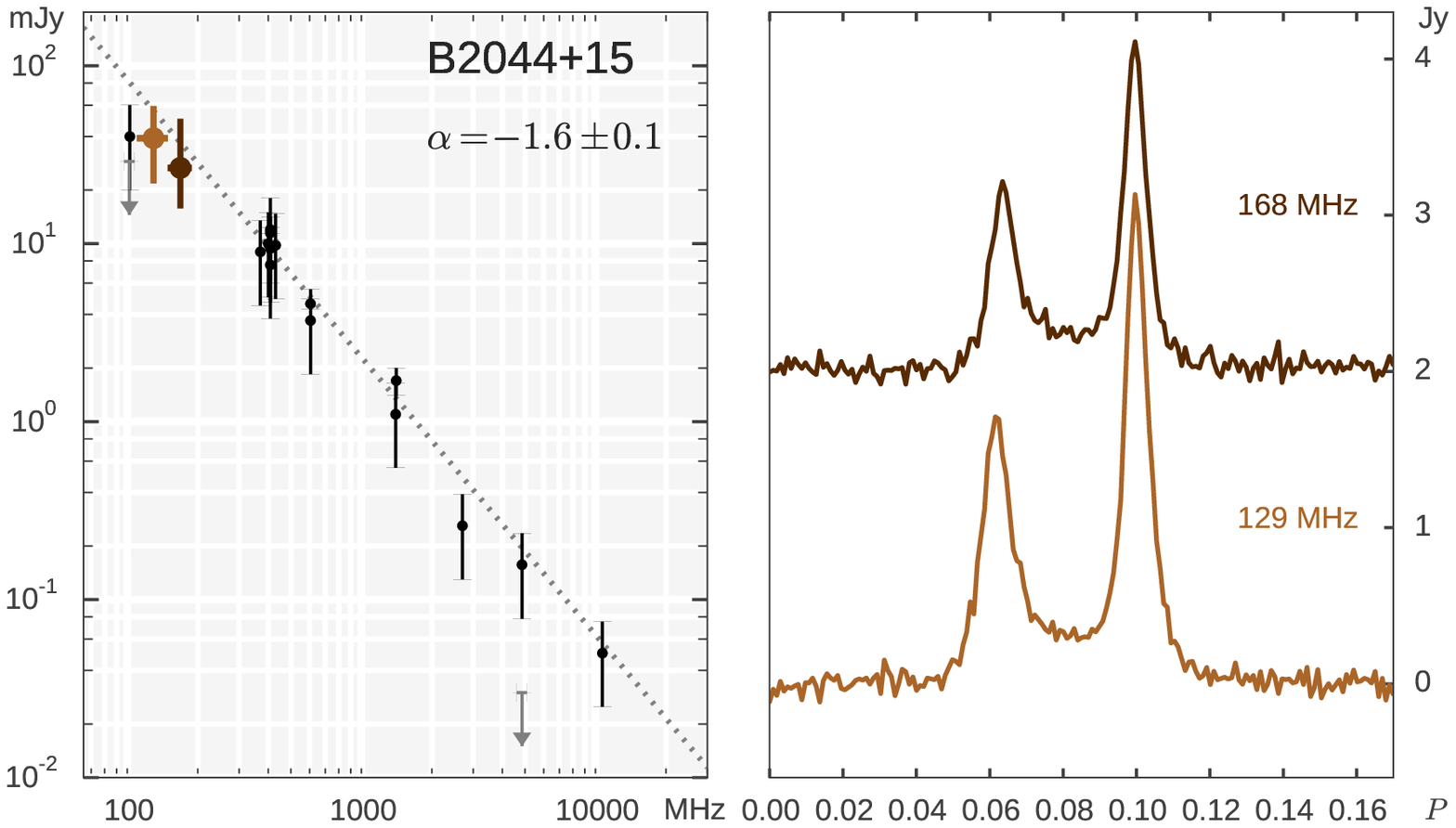}
\includegraphics[scale=0.48]{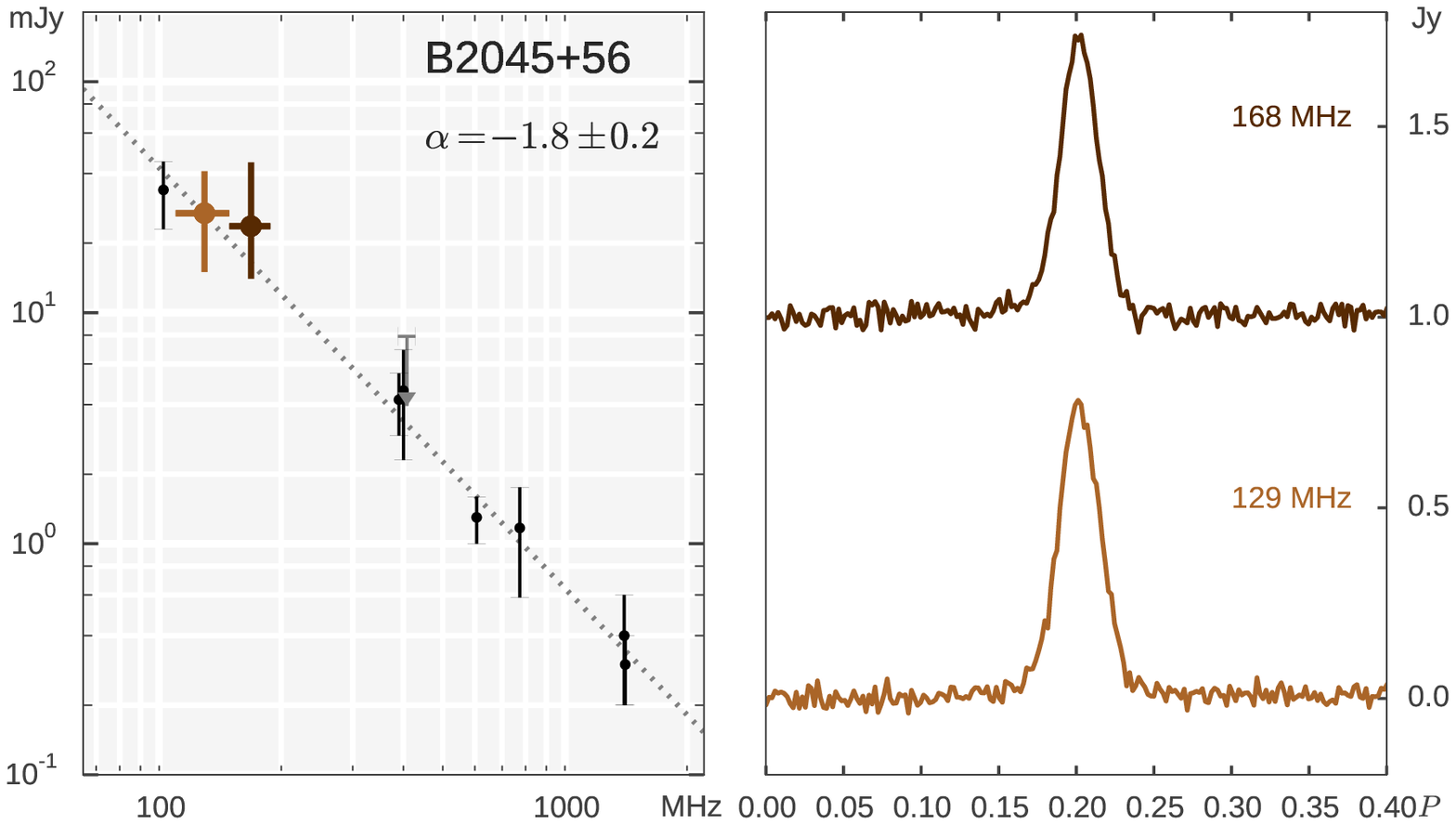}\includegraphics[scale=0.48]{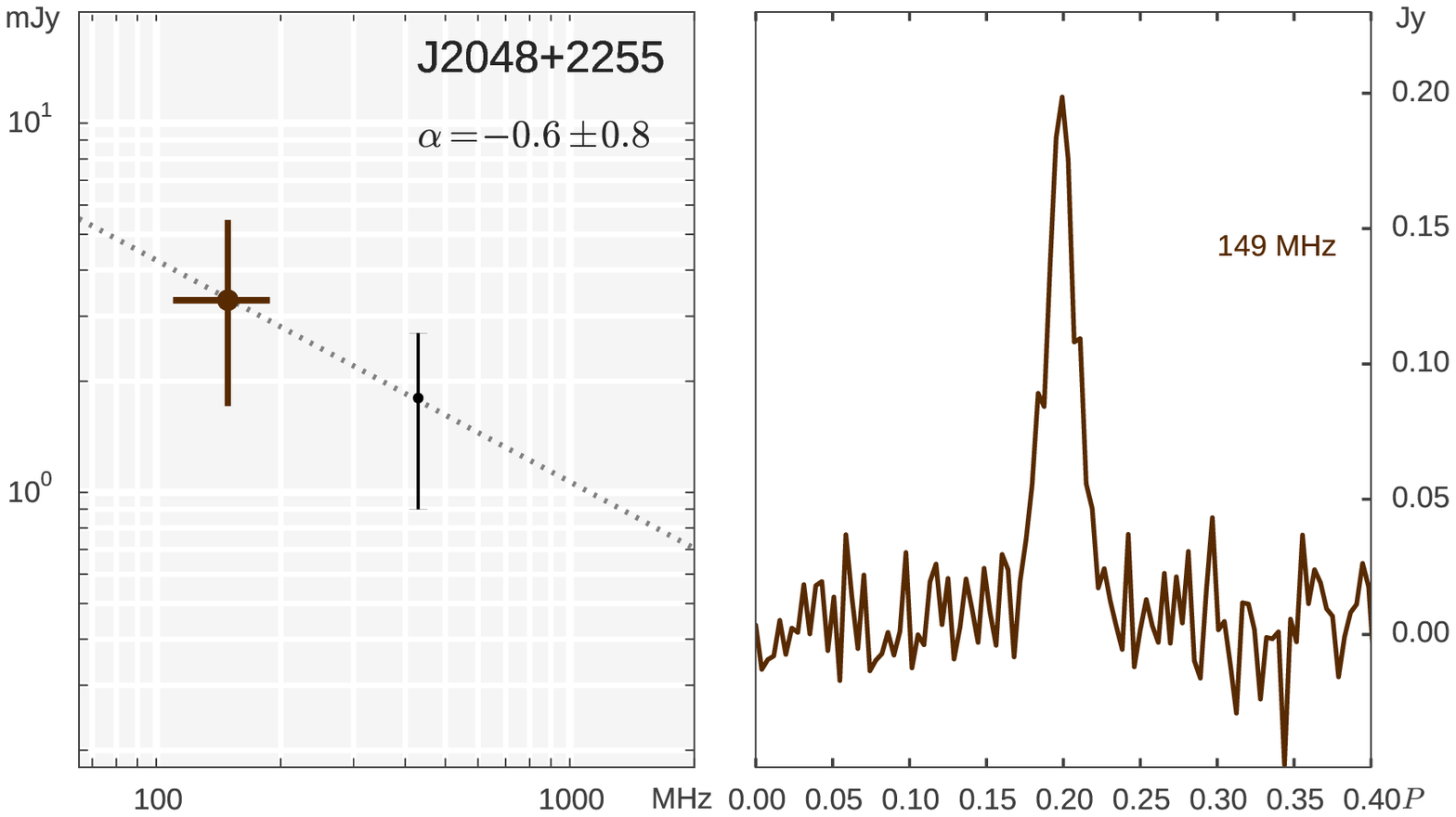}
\caption{See Figure~\ref{fig:prof_sp_1}.}
\label{fig:prof_sp_13}
\end{figure*}

\begin{figure*}
\includegraphics[scale=0.48]{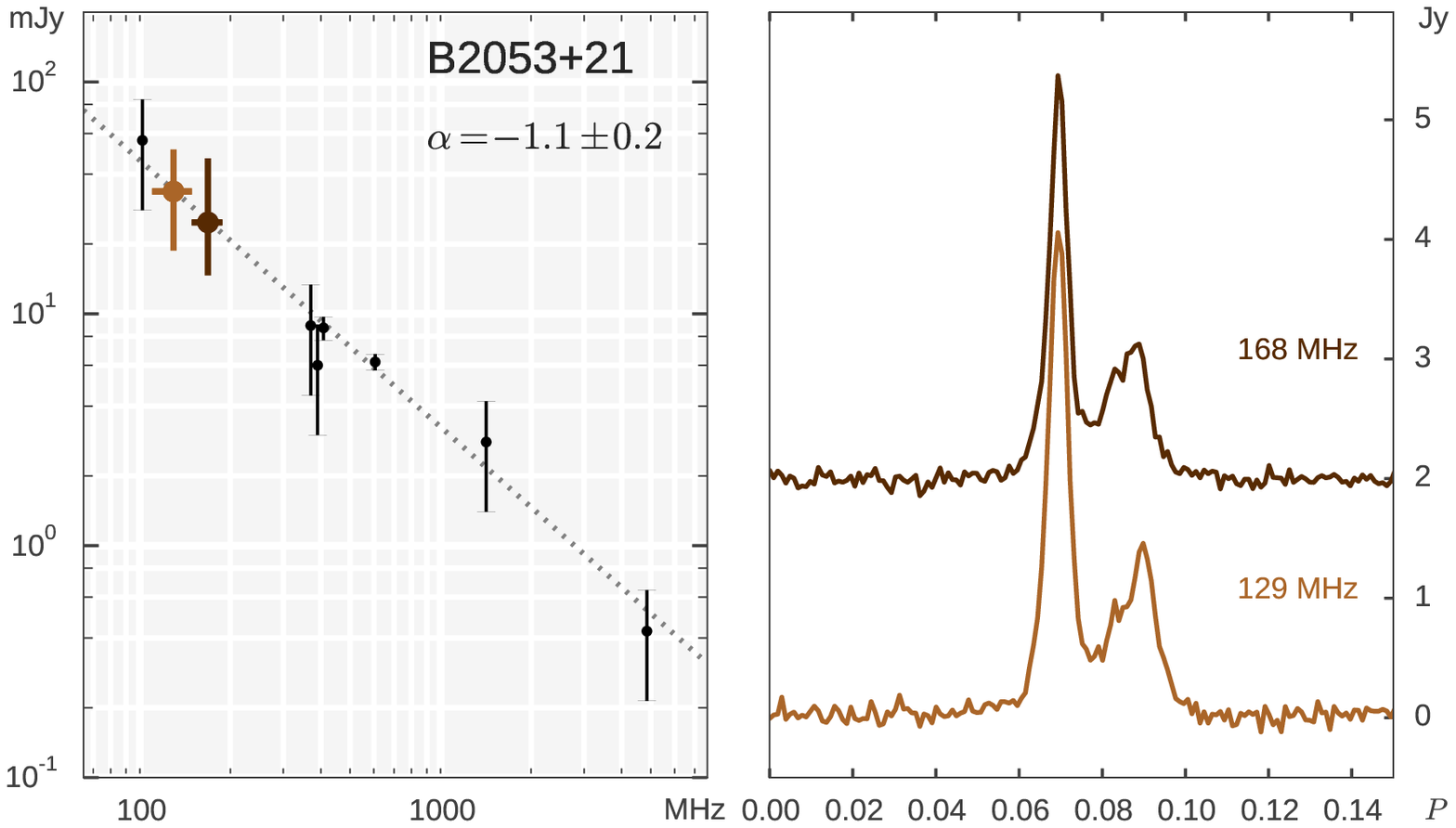}\includegraphics[scale=0.48]{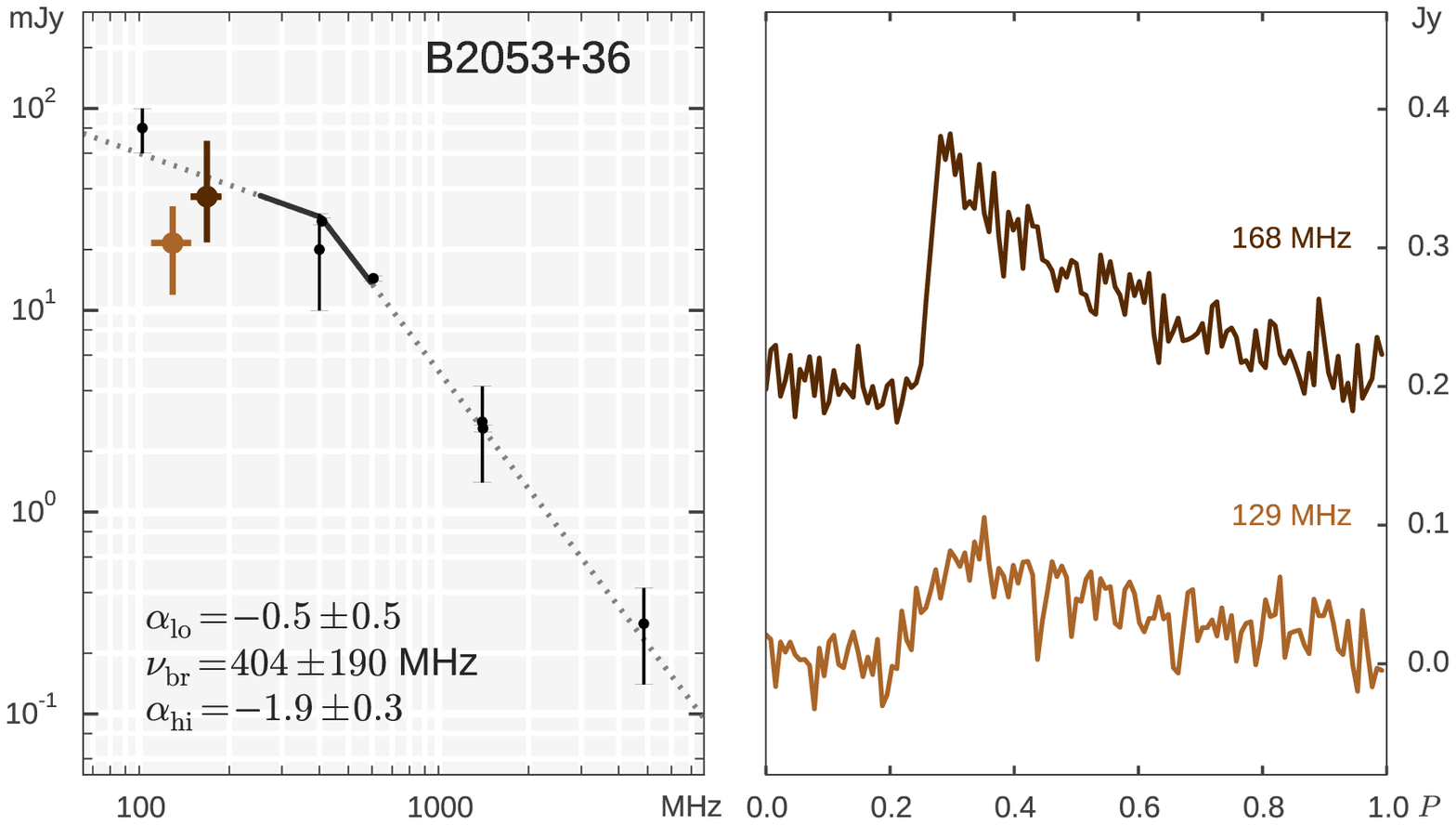}
\includegraphics[scale=0.48]{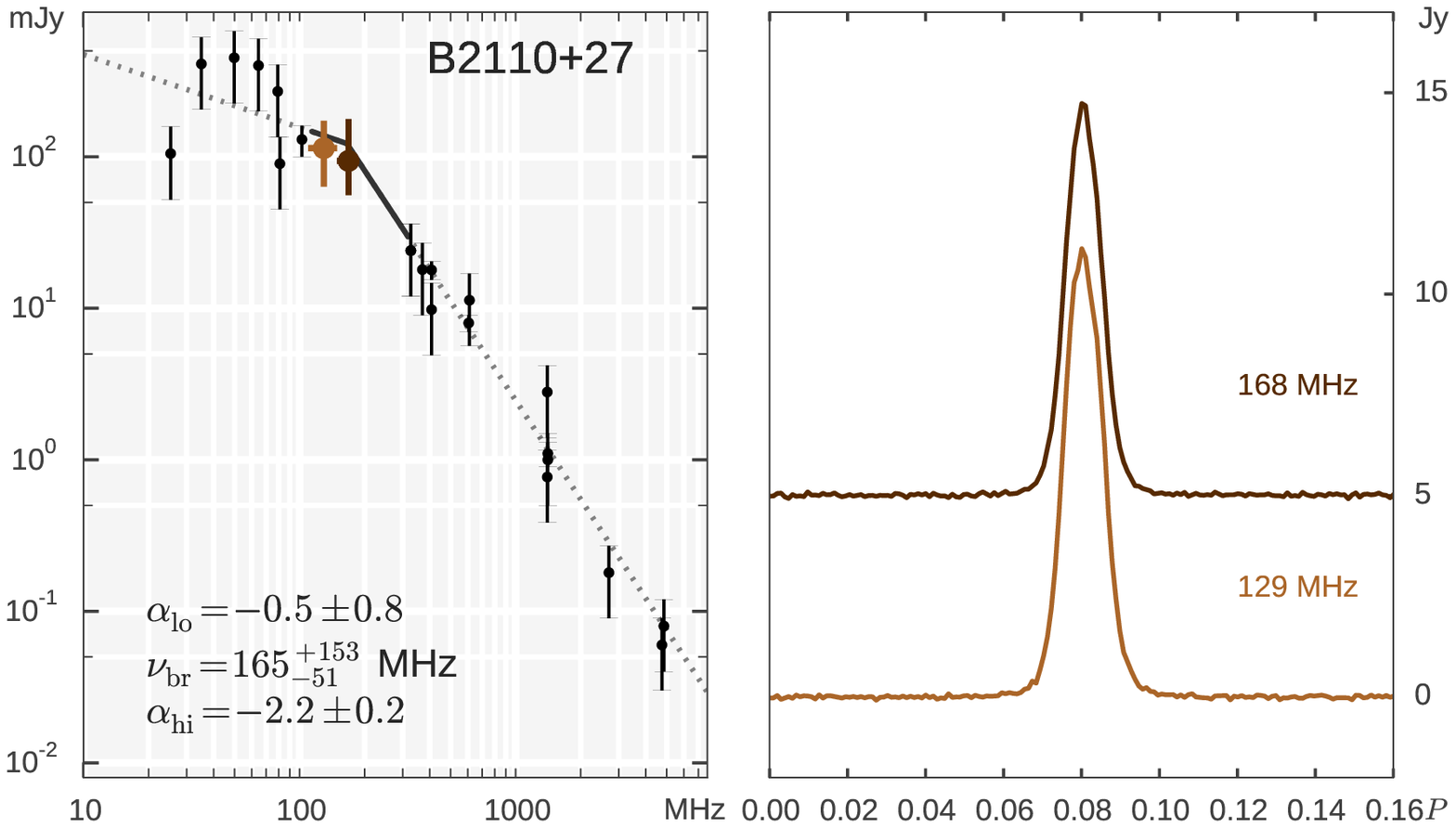}\includegraphics[scale=0.48]{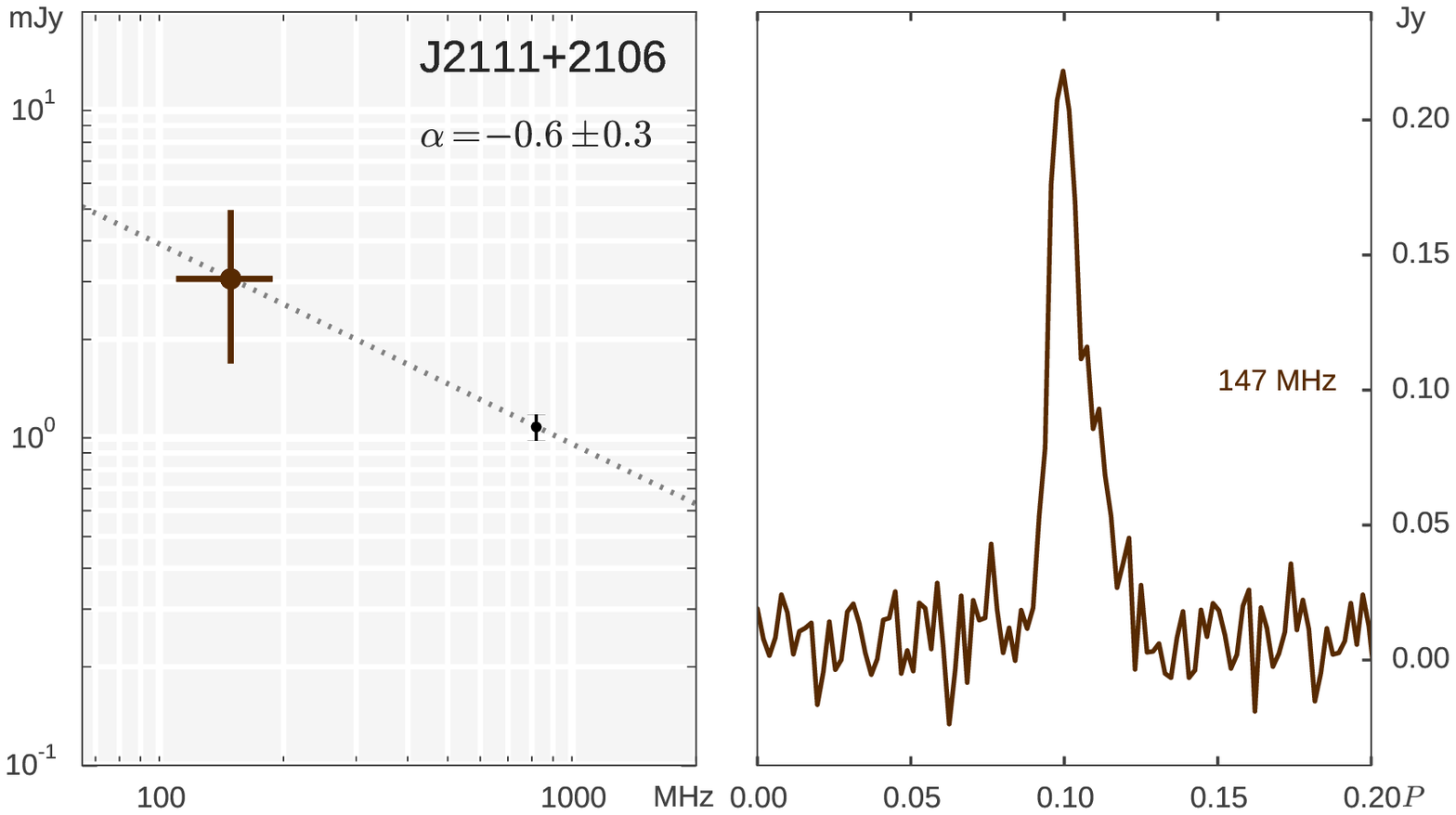}
\includegraphics[scale=0.48]{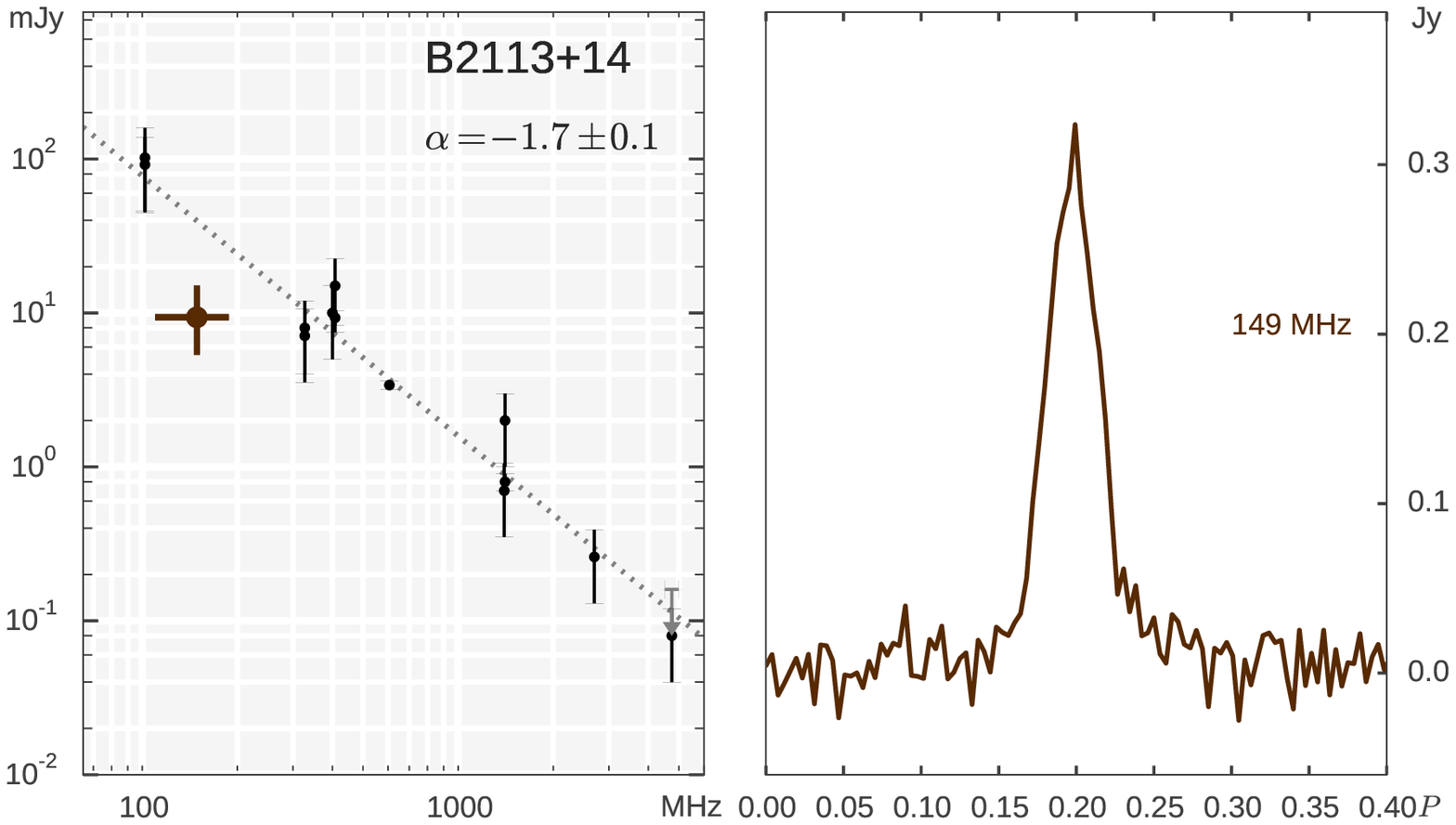}\includegraphics[scale=0.48]{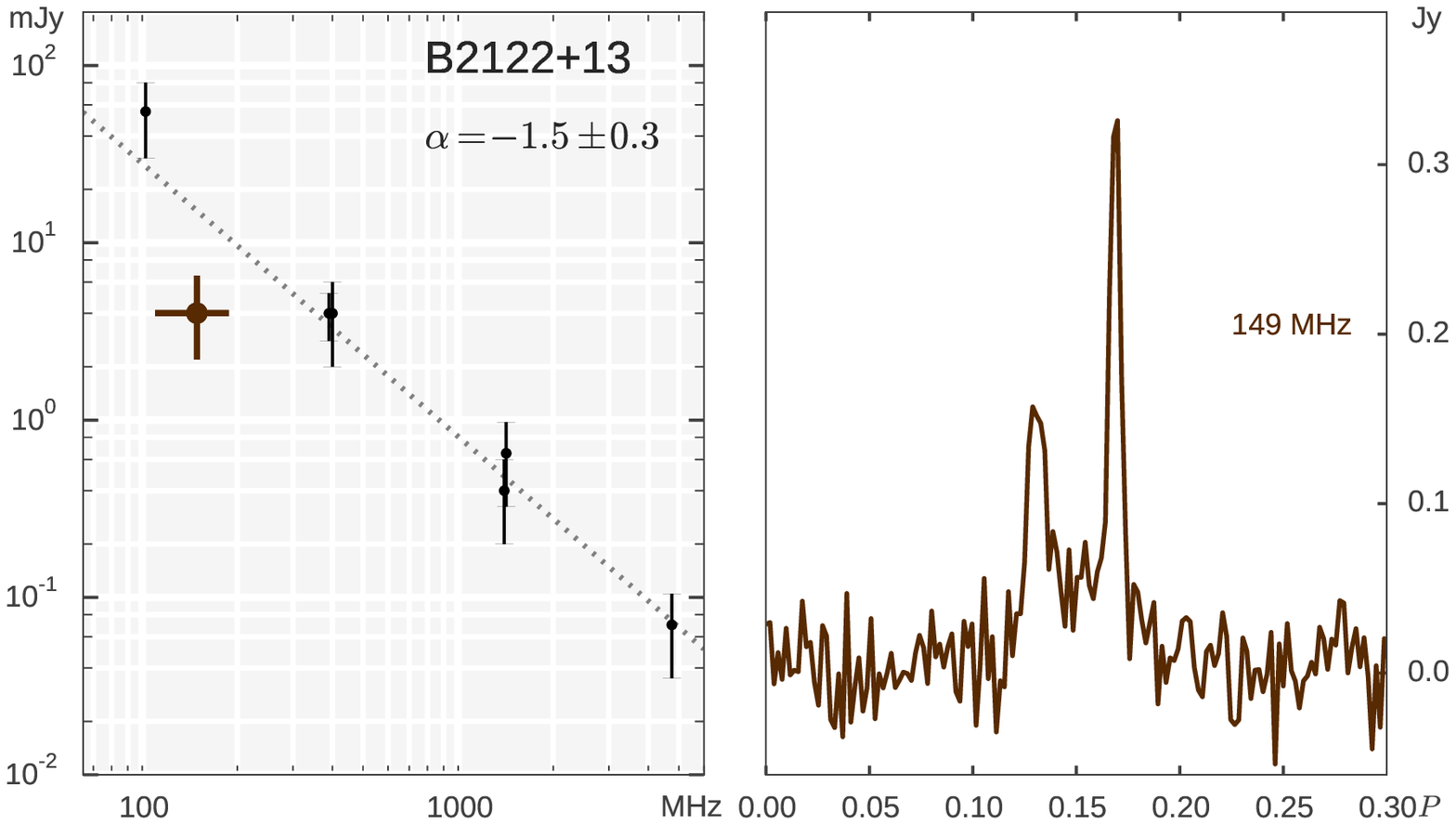}
\includegraphics[scale=0.48]{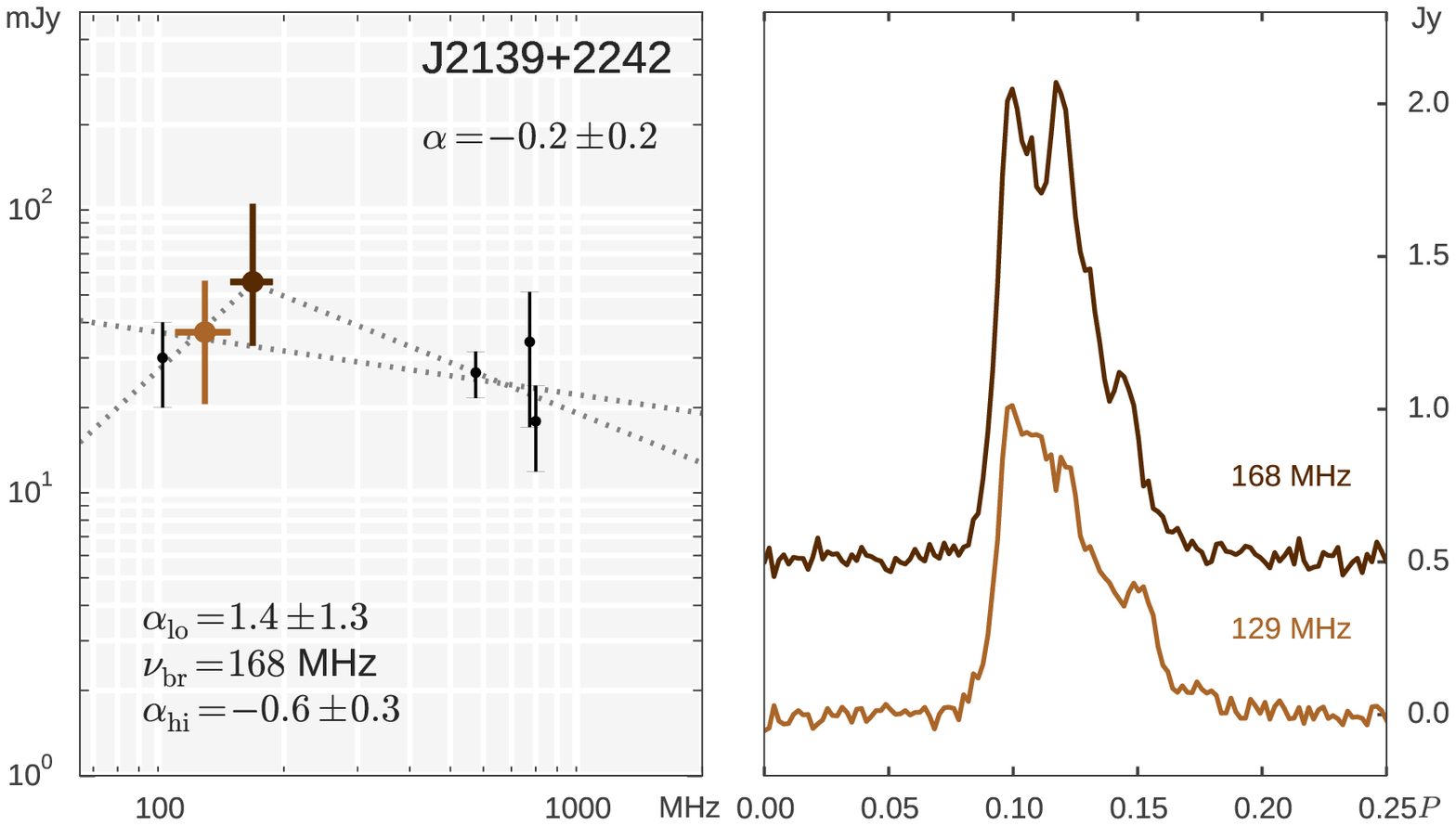}\includegraphics[scale=0.48]{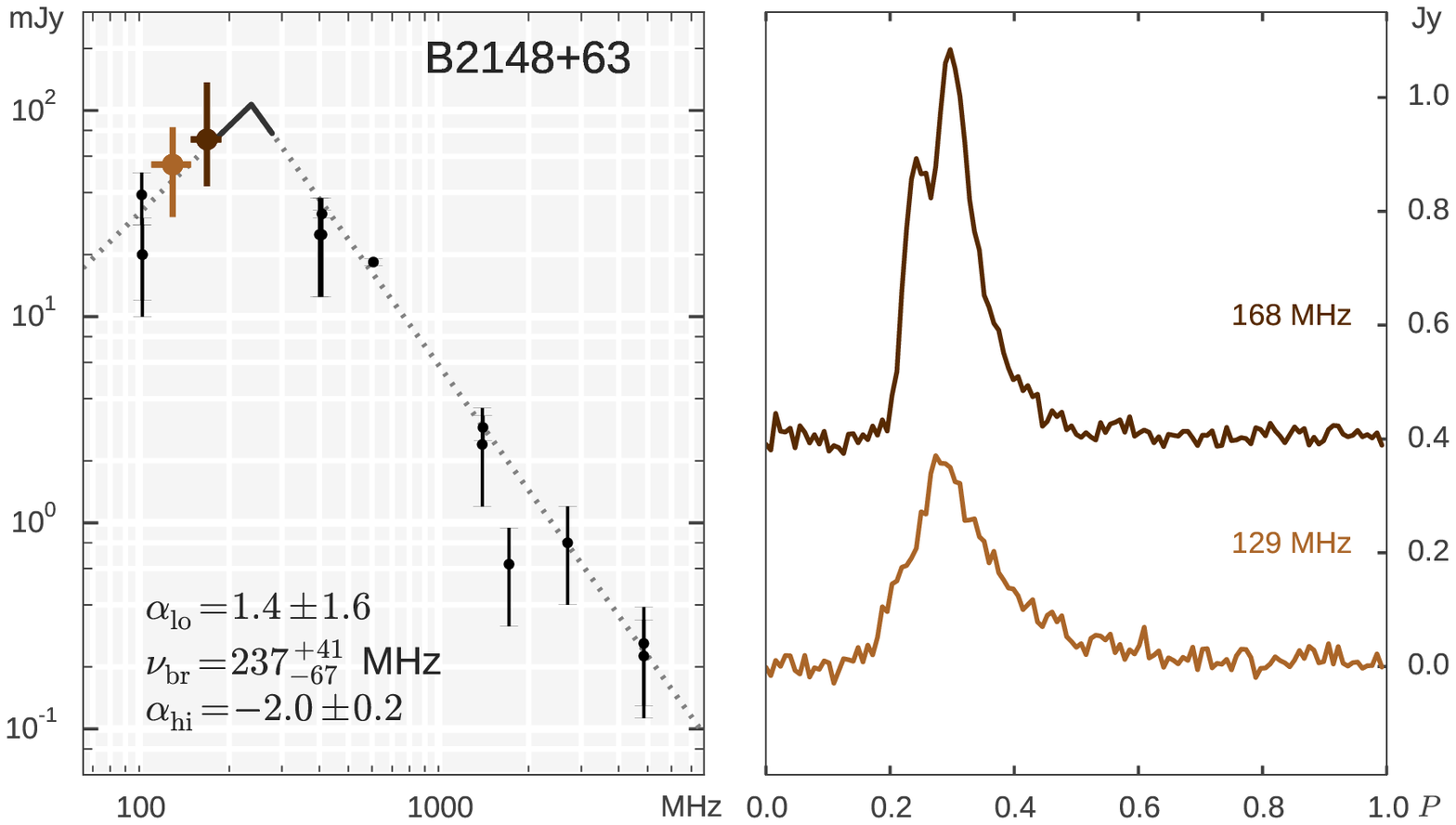}
\includegraphics[scale=0.48]{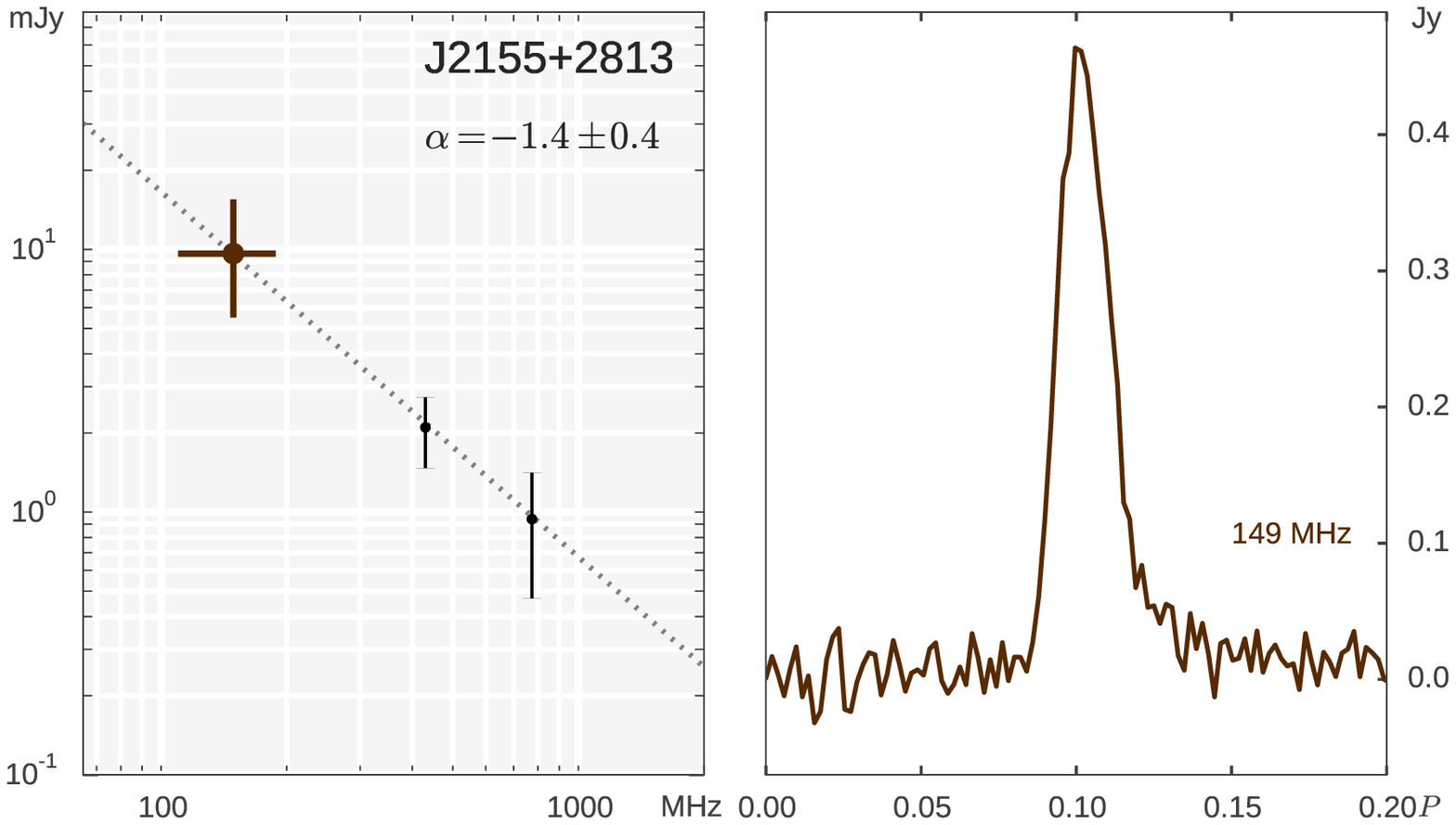}\includegraphics[scale=0.48]{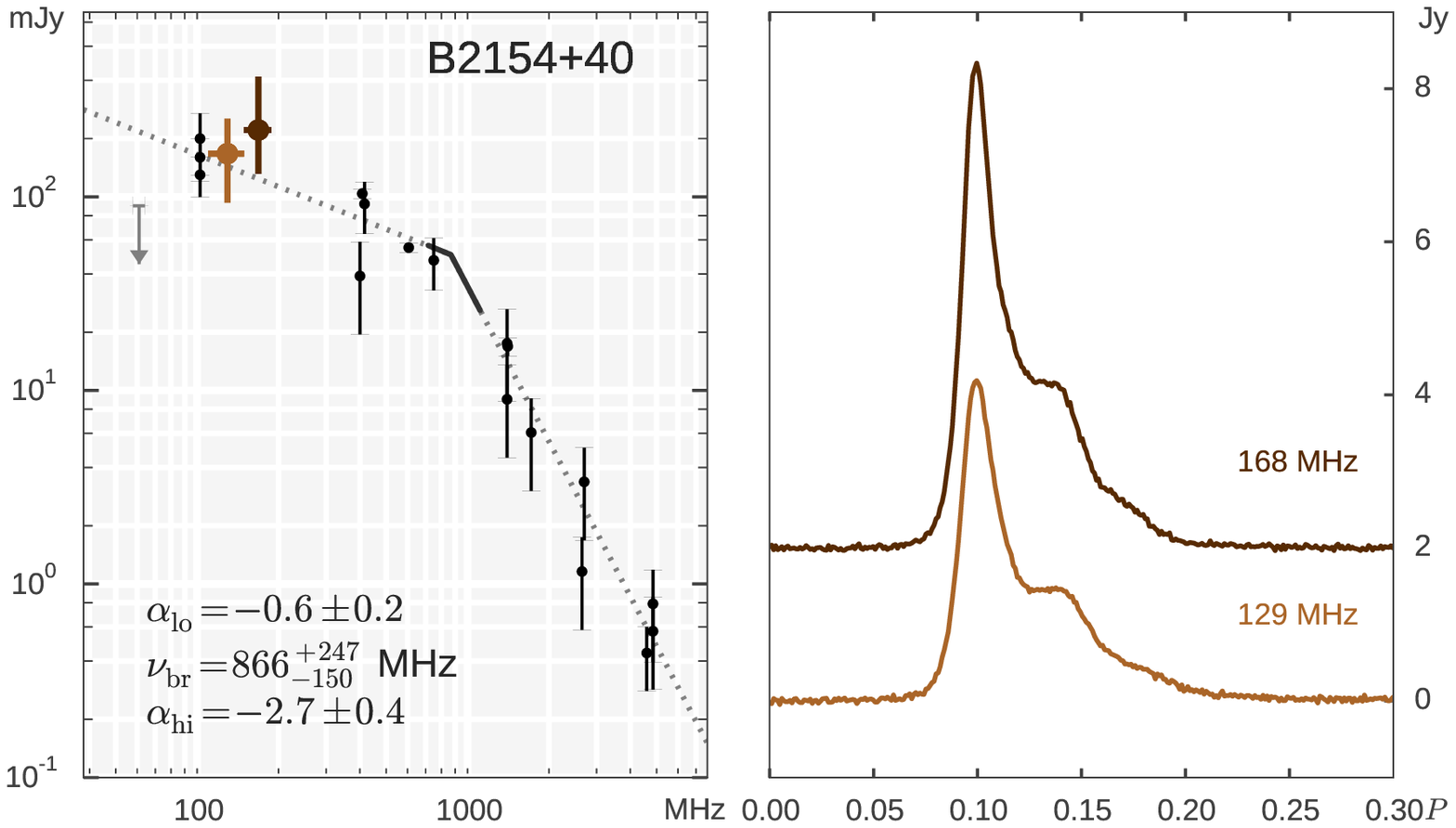}
\caption{See Figure~\ref{fig:prof_sp_1}.}
\label{fig:prof_sp_14}
\end{figure*}

\begin{figure*}
\includegraphics[scale=0.48]{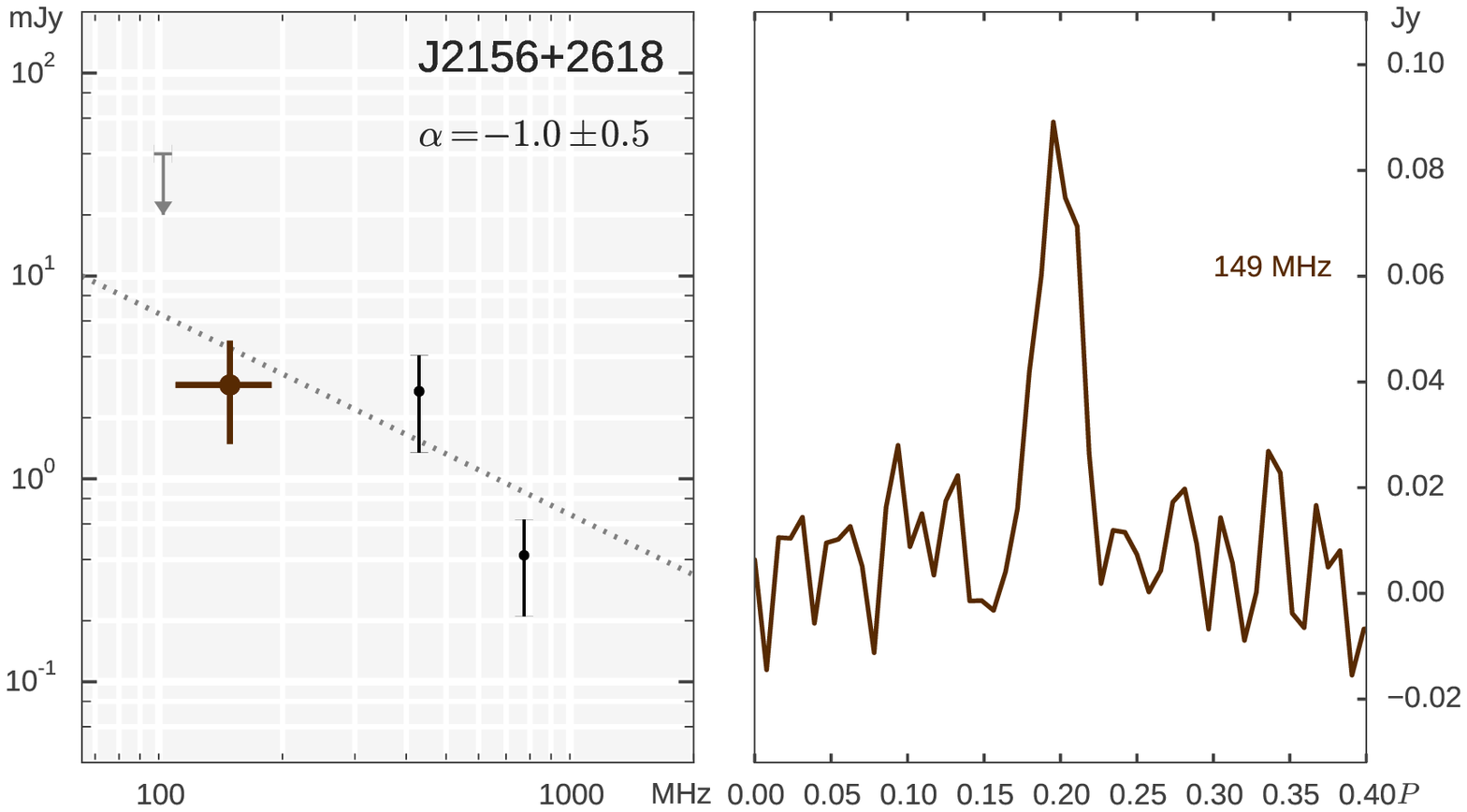}\includegraphics[scale=0.48]{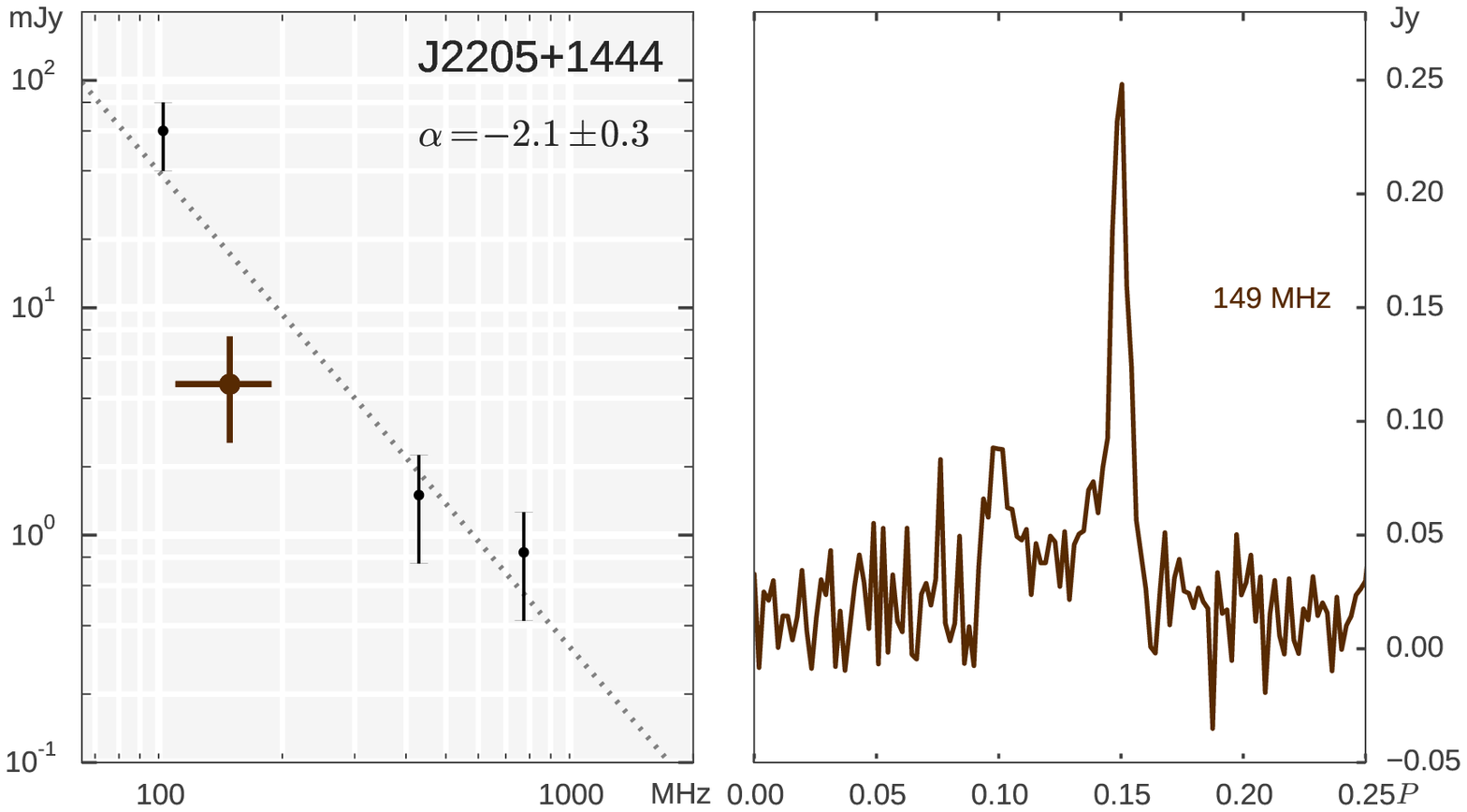}
\includegraphics[scale=0.48]{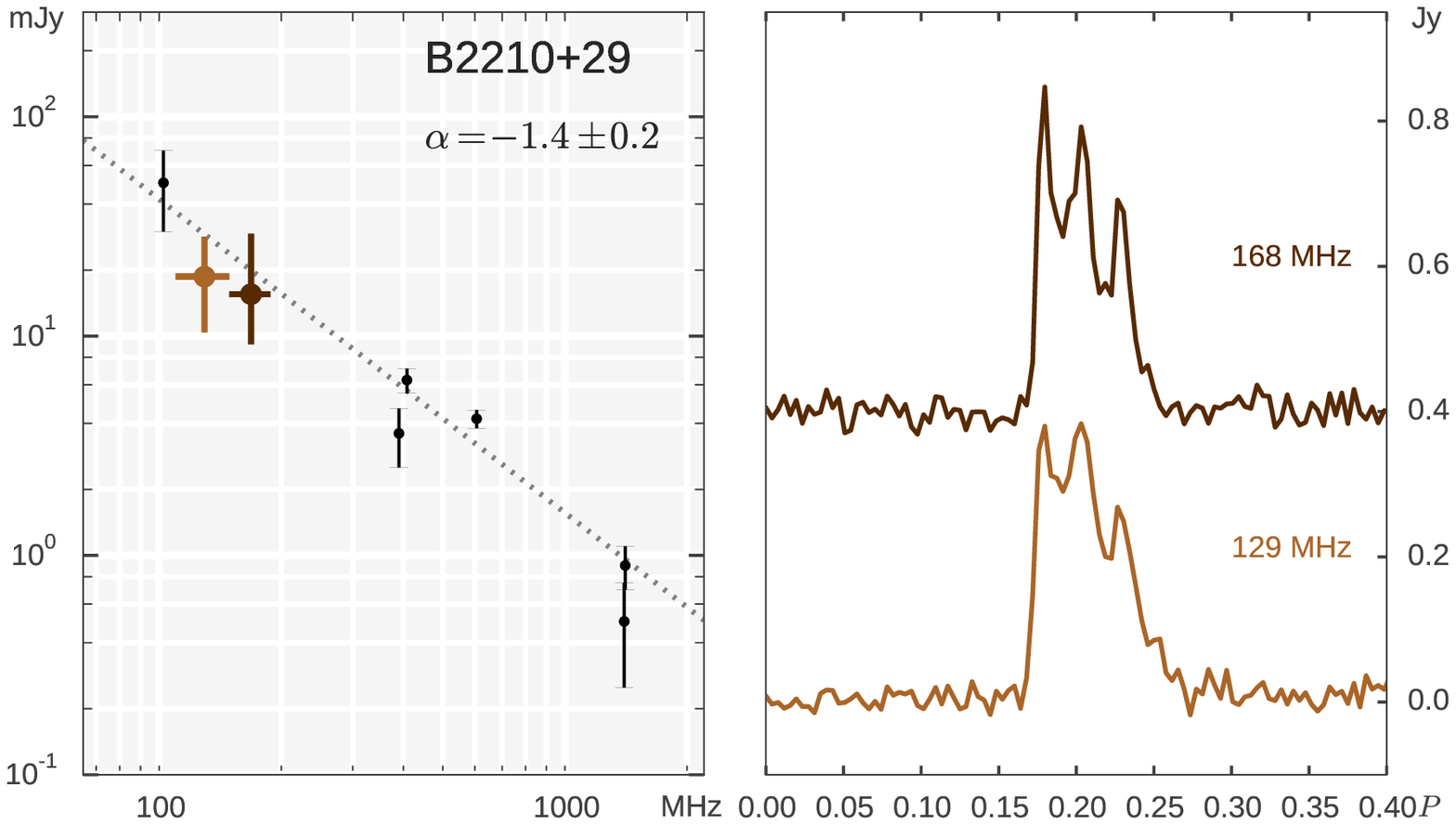}\includegraphics[scale=0.48]{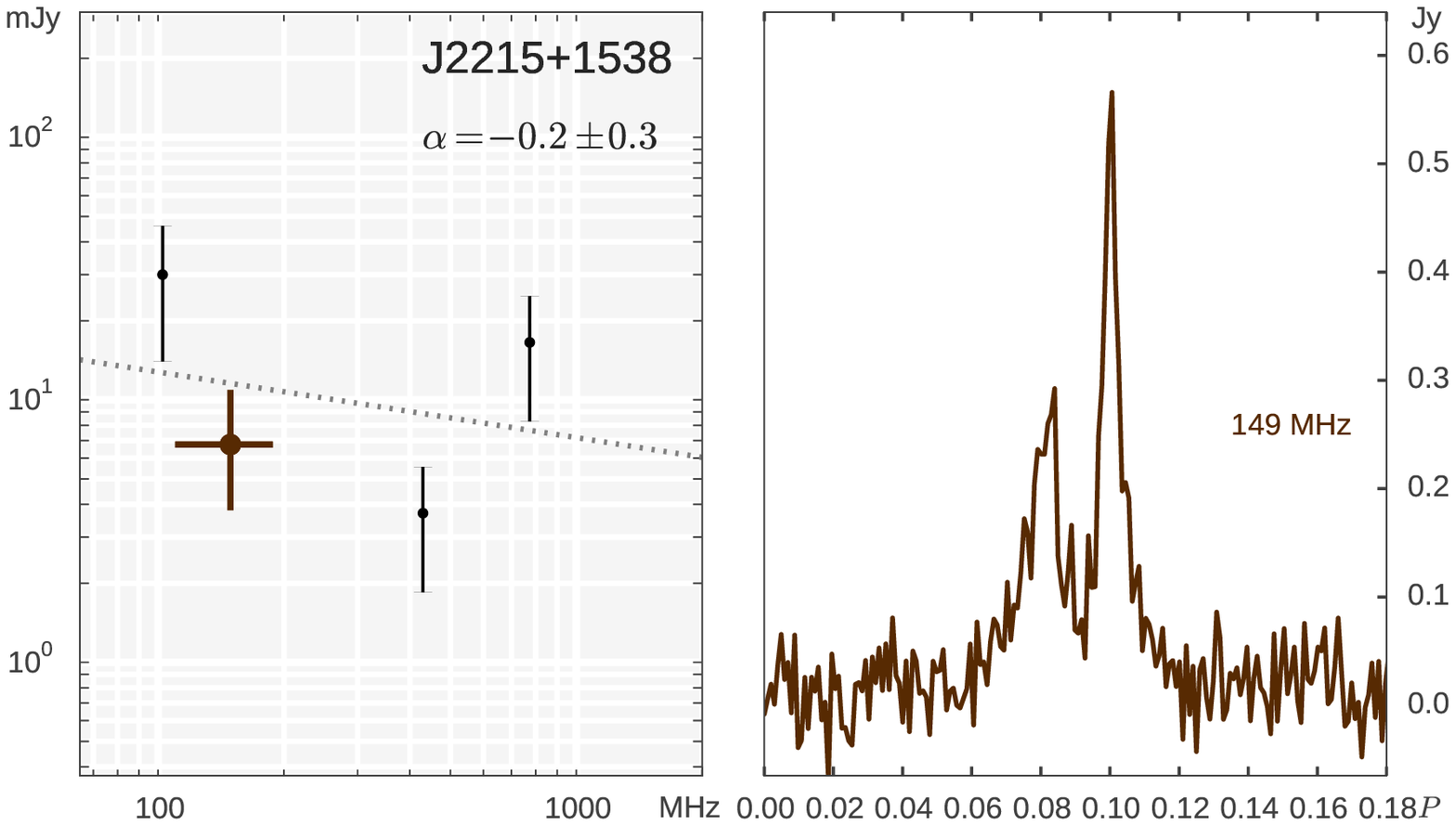}
\includegraphics[scale=0.48]{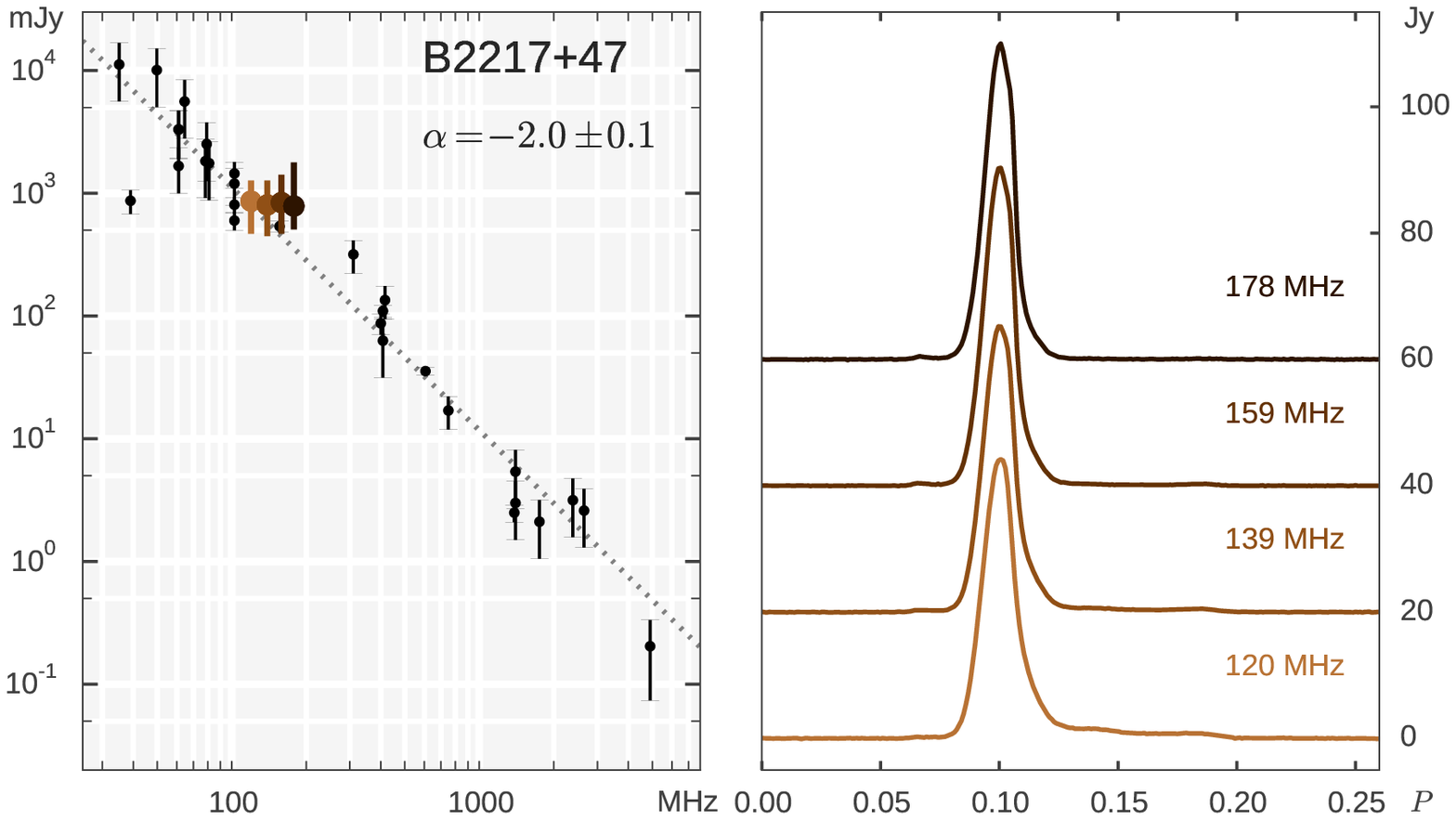}\includegraphics[scale=0.48]{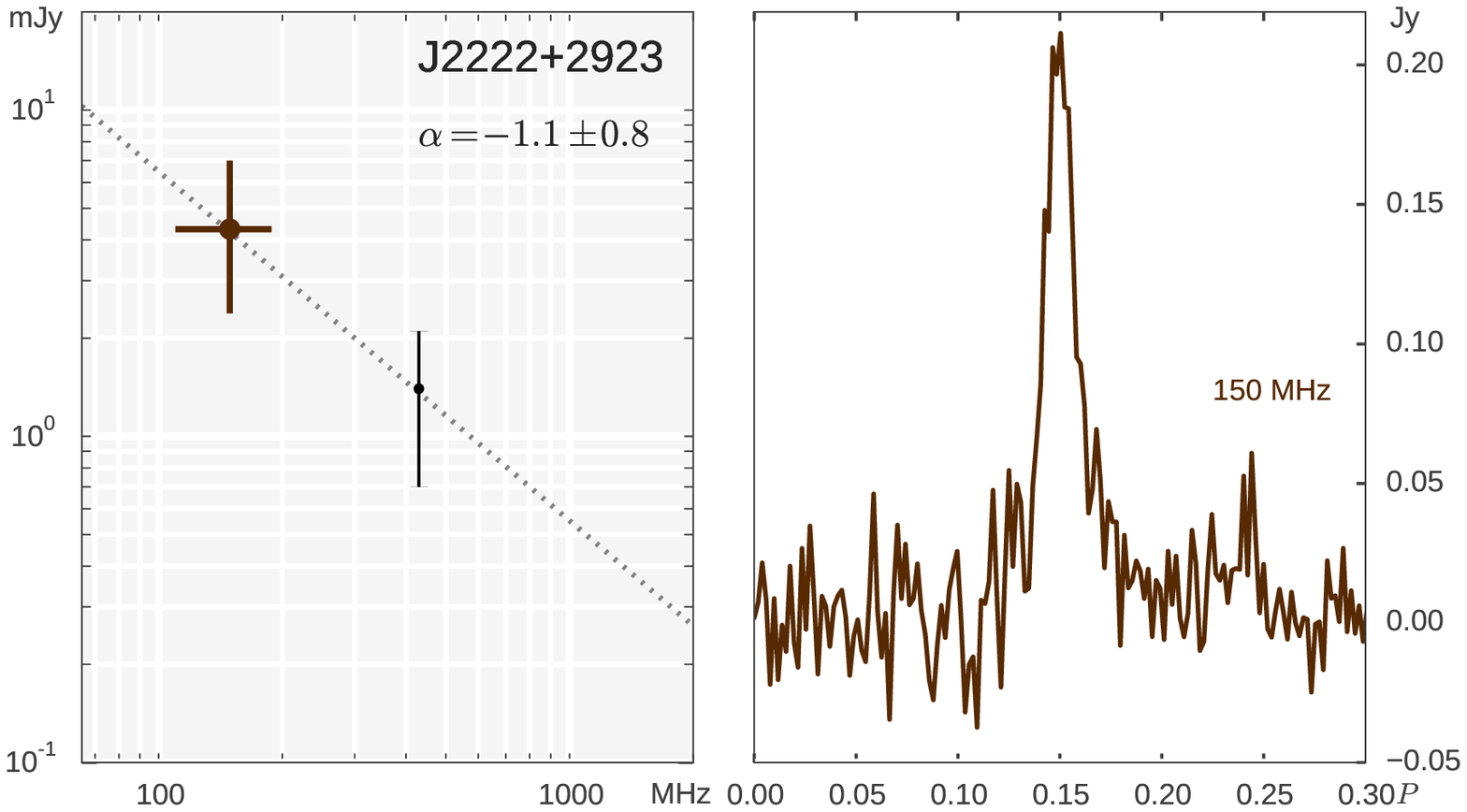}
\includegraphics[scale=0.48]{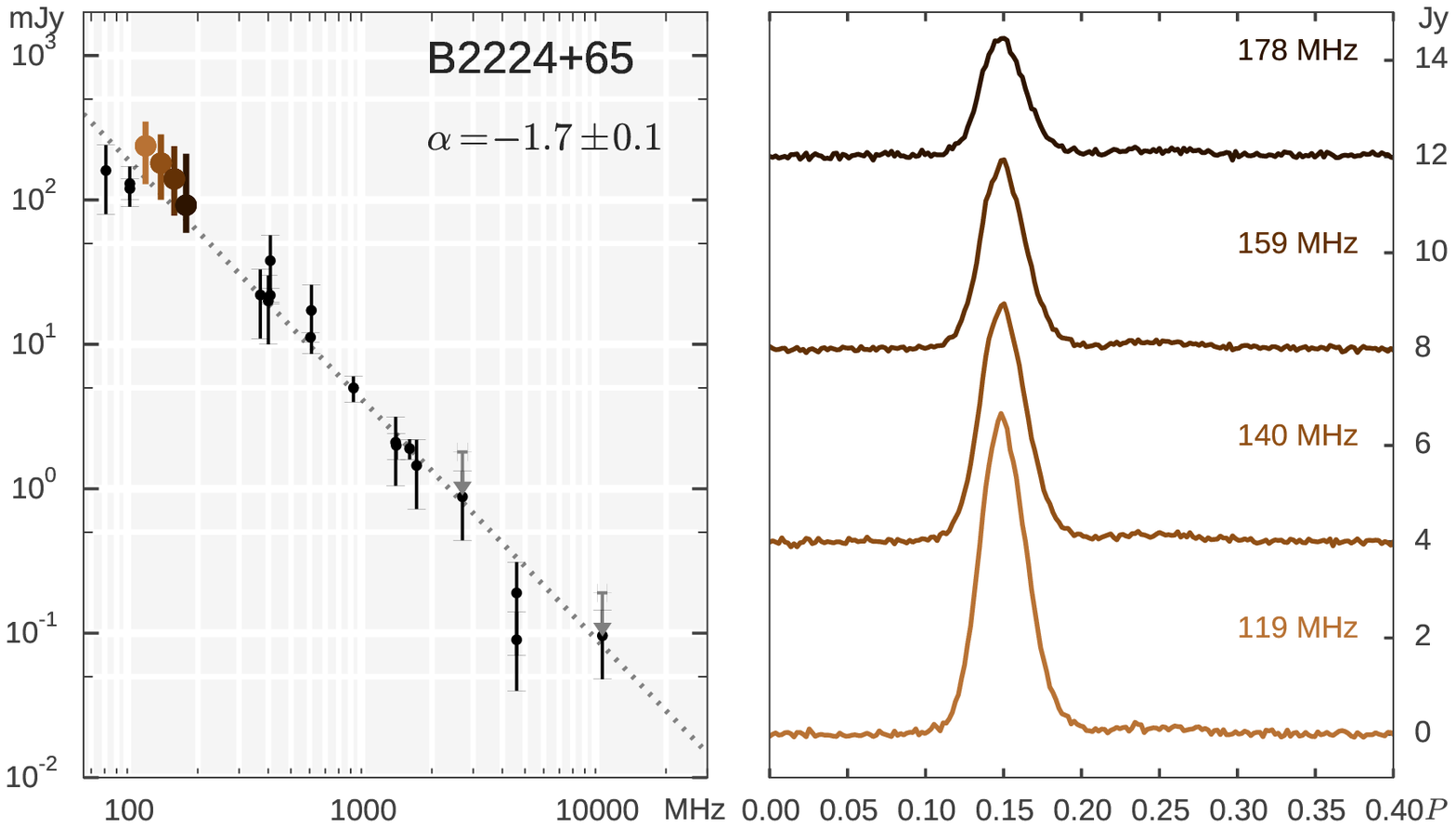}\includegraphics[scale=0.48]{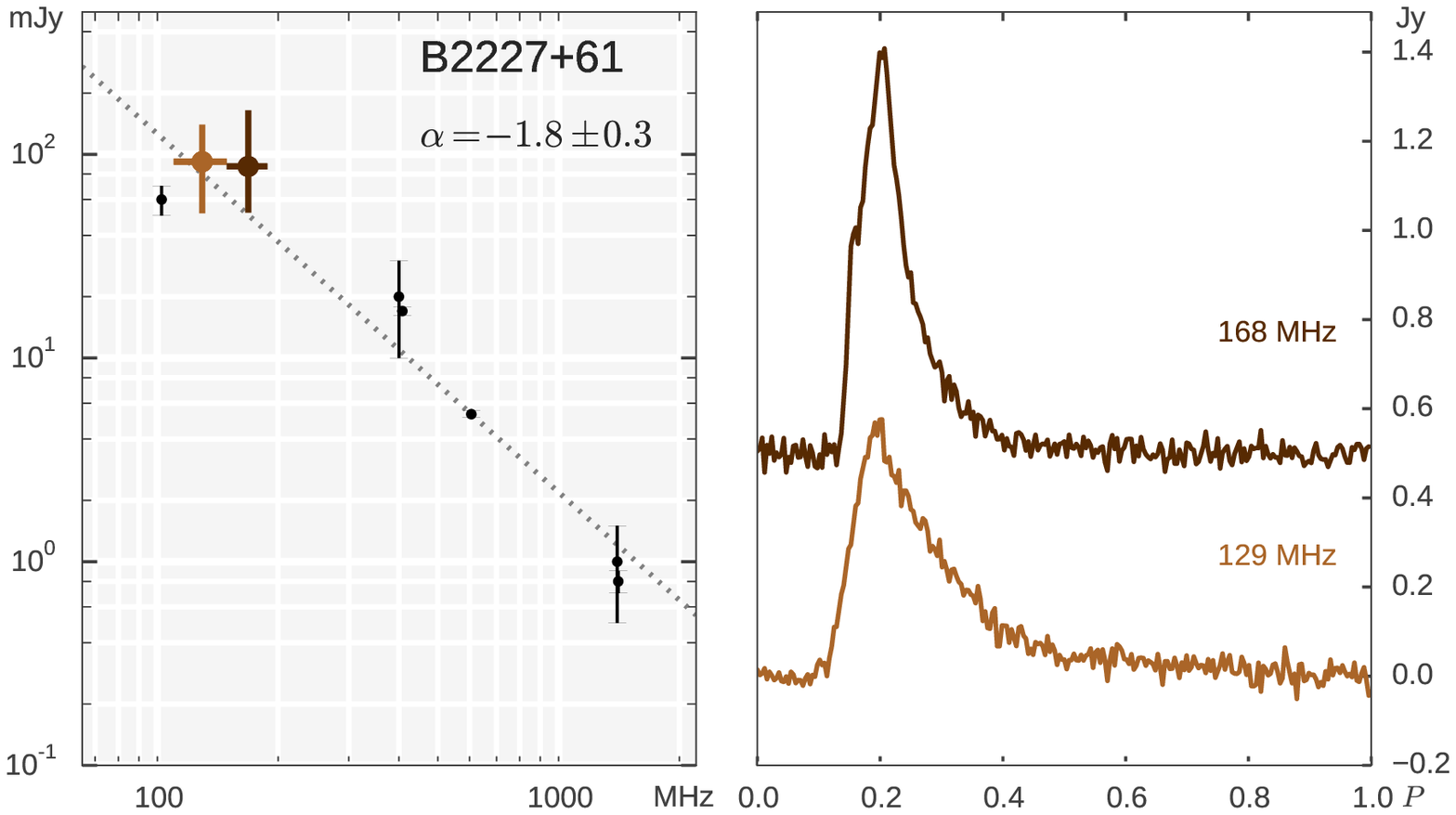}
\includegraphics[scale=0.48]{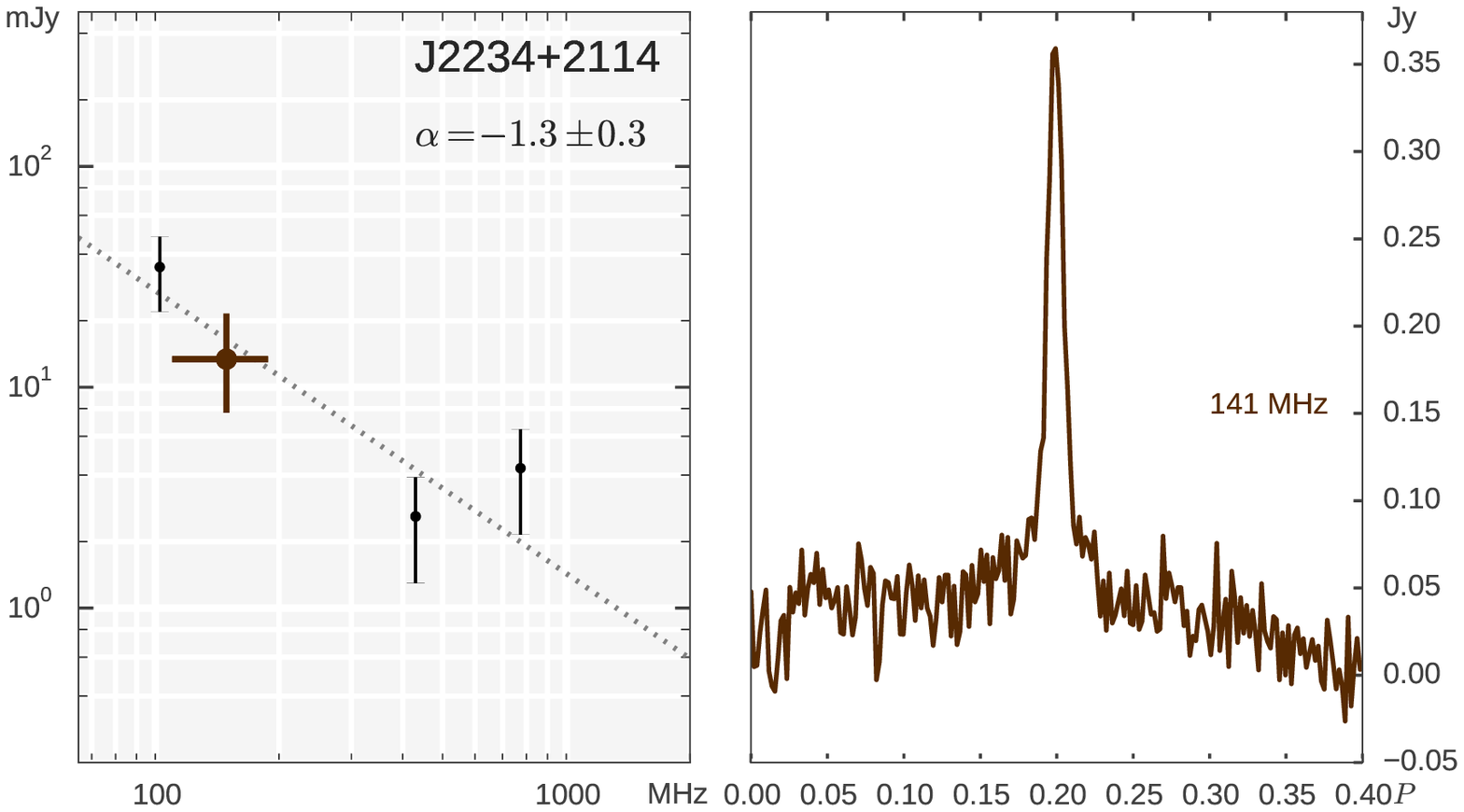}\includegraphics[scale=0.48]{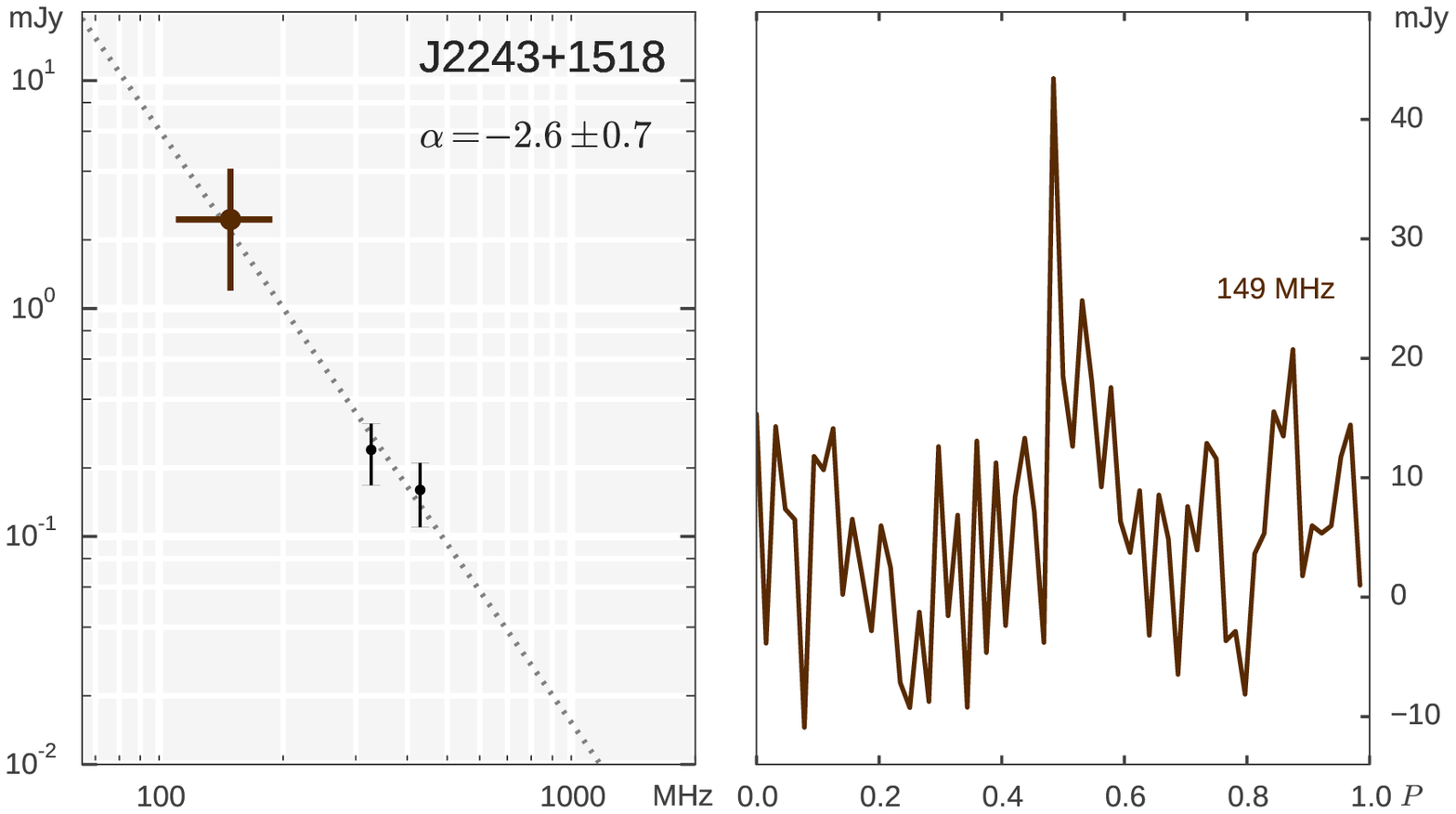}
\caption{See Figure~\ref{fig:prof_sp_1}.}
\label{fig:prof_sp_15}
\end{figure*}

\begin{figure*}
\includegraphics[scale=0.48]{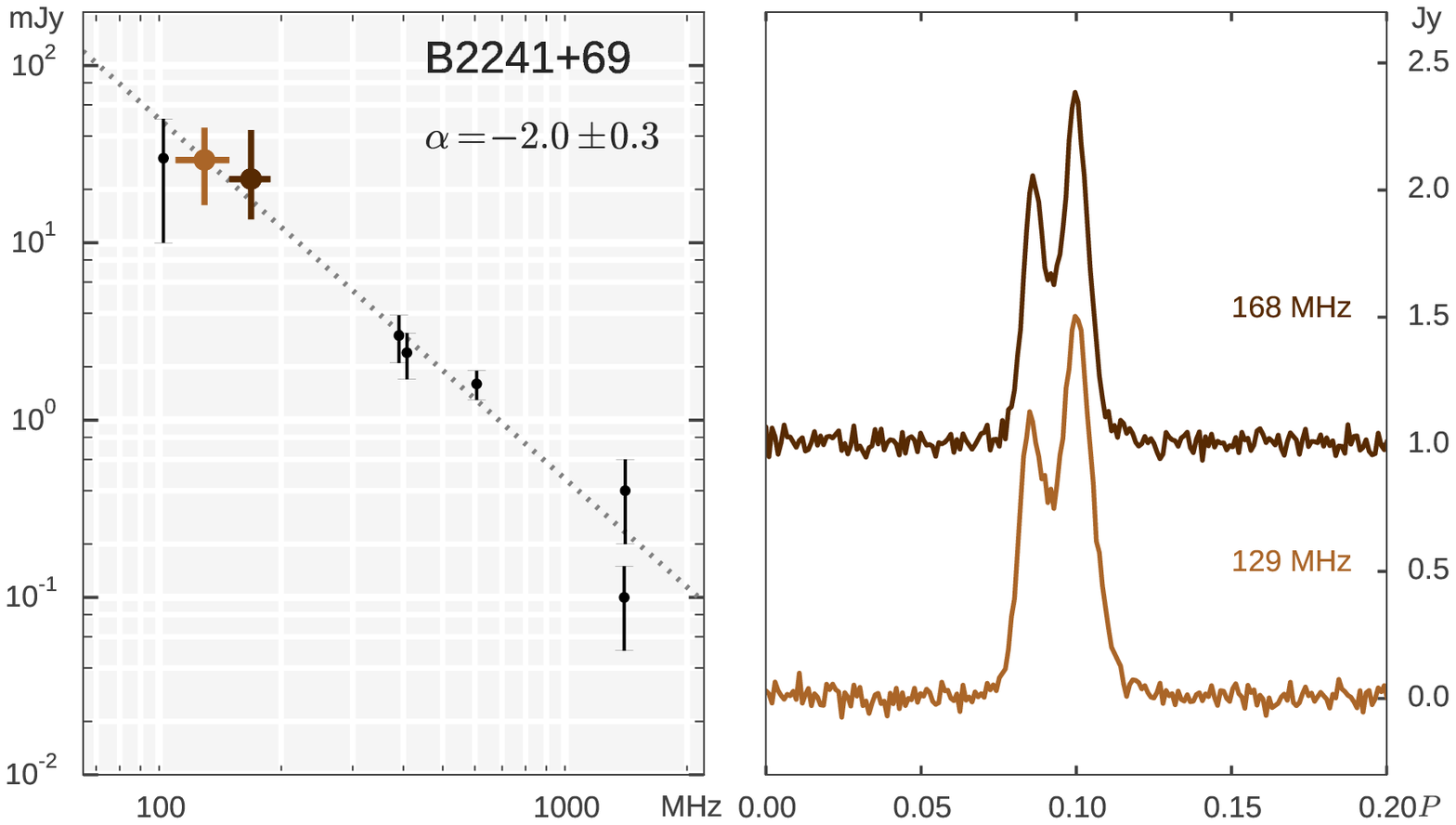}\includegraphics[scale=0.48]{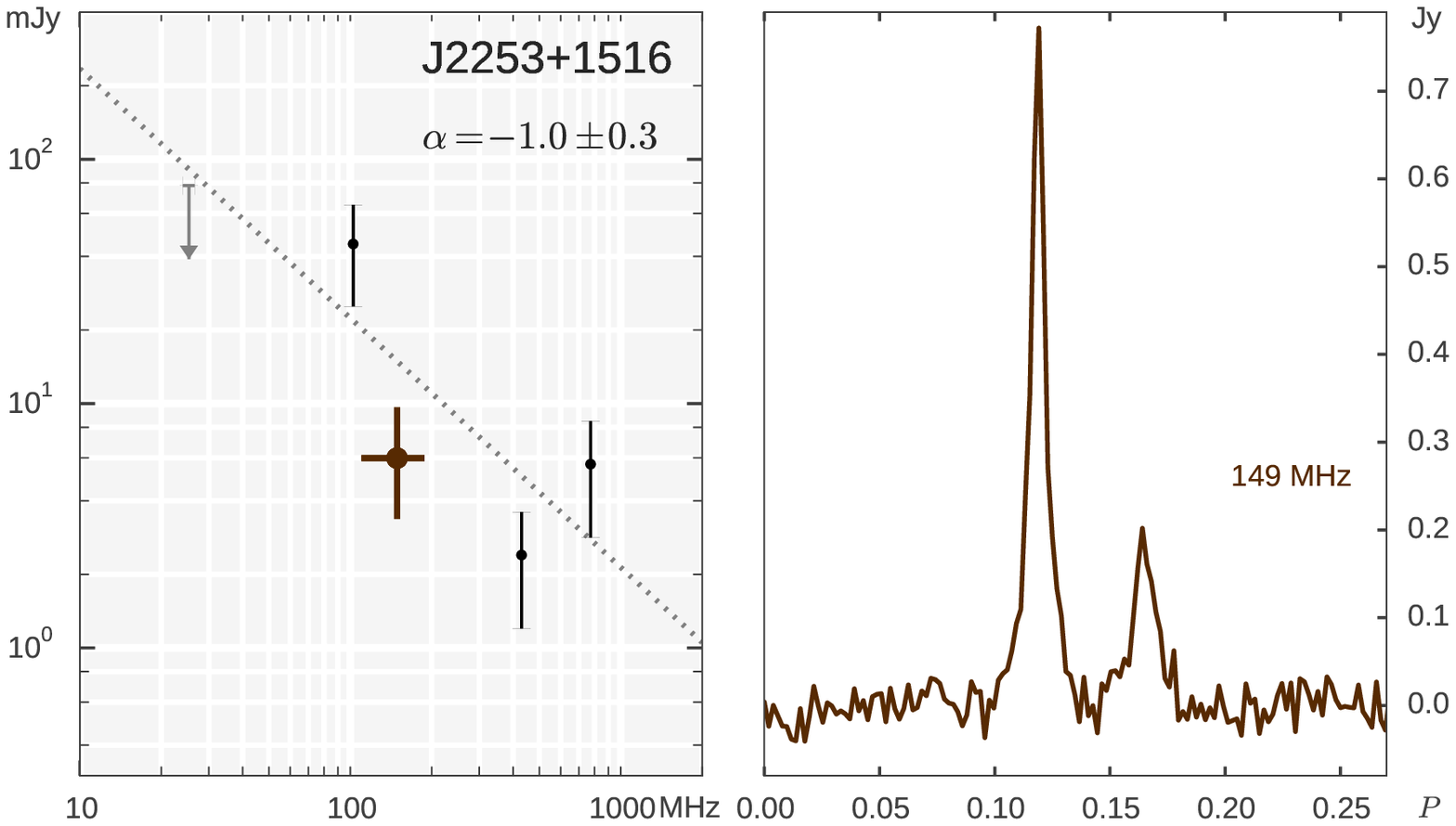}
\includegraphics[scale=0.48]{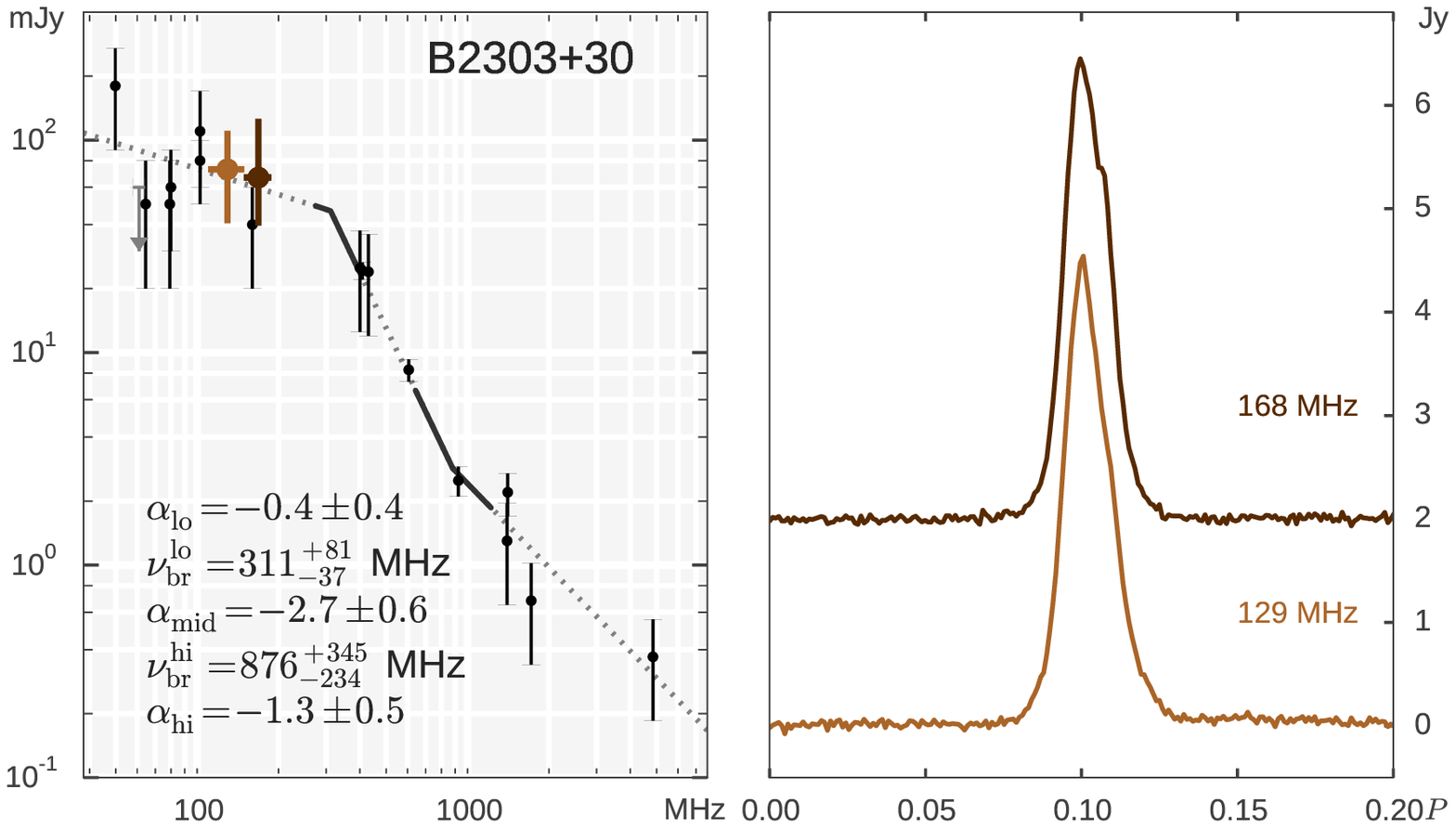}\includegraphics[scale=0.48]{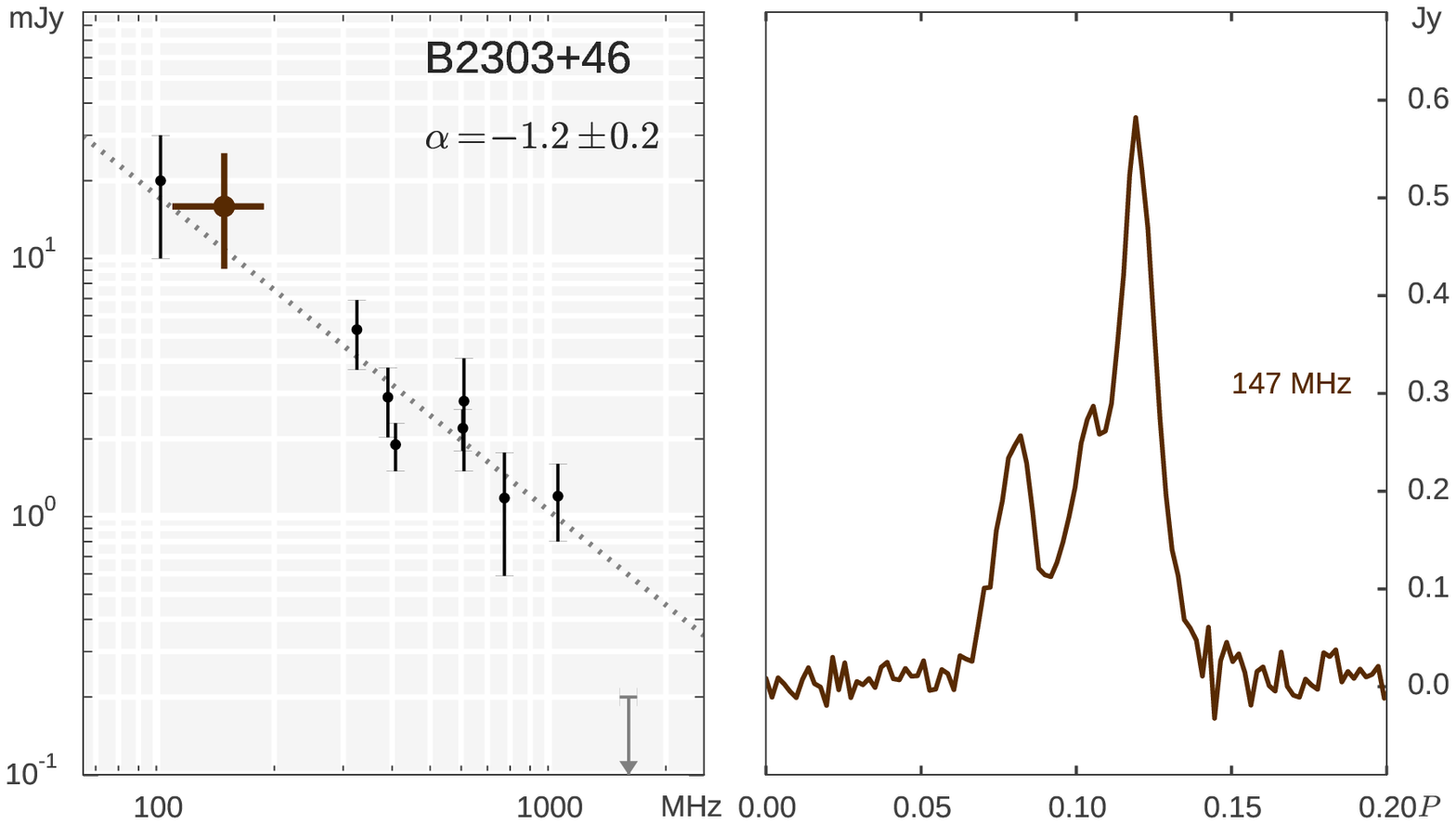}
\includegraphics[scale=0.48]{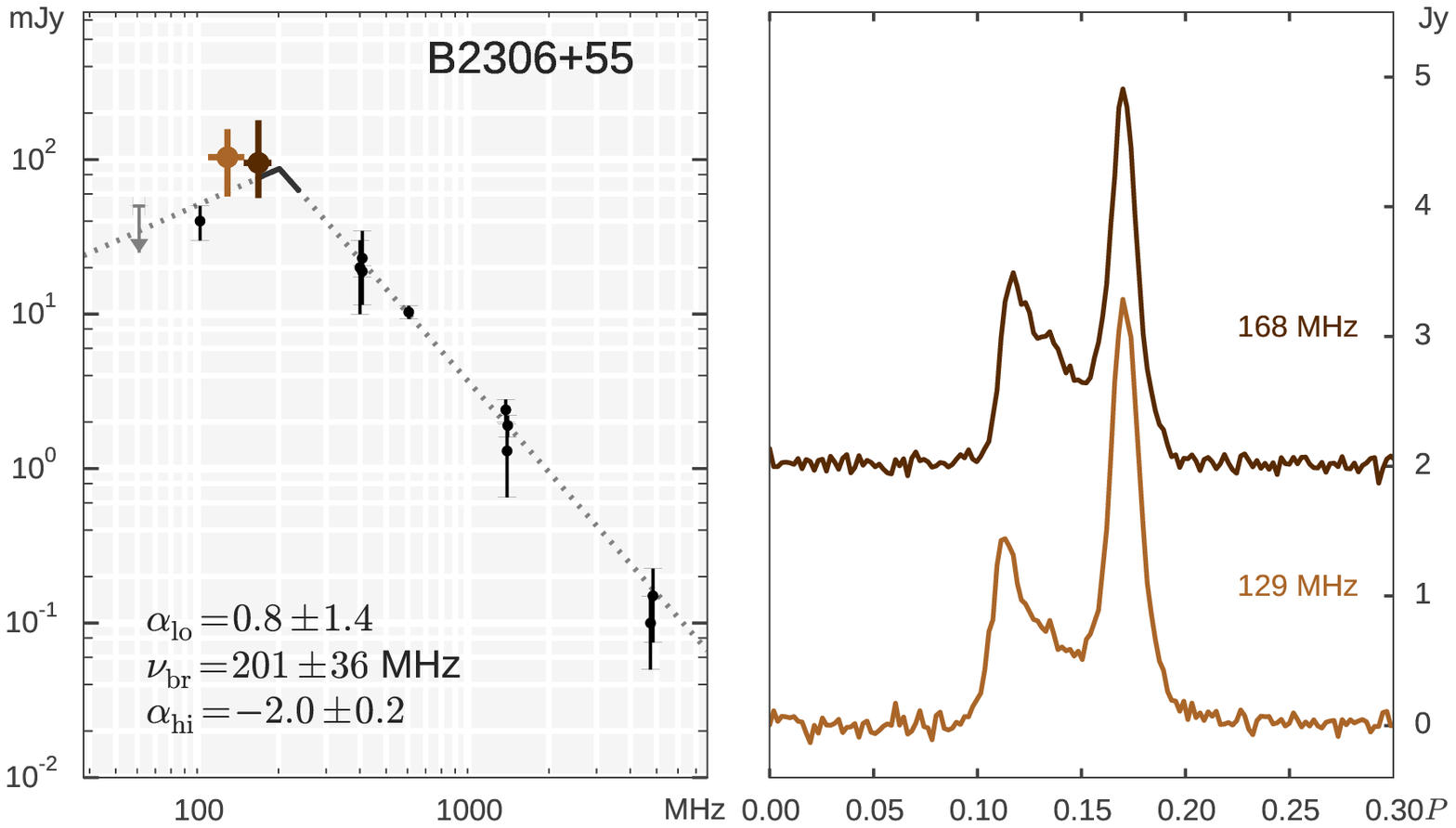}\includegraphics[scale=0.48]{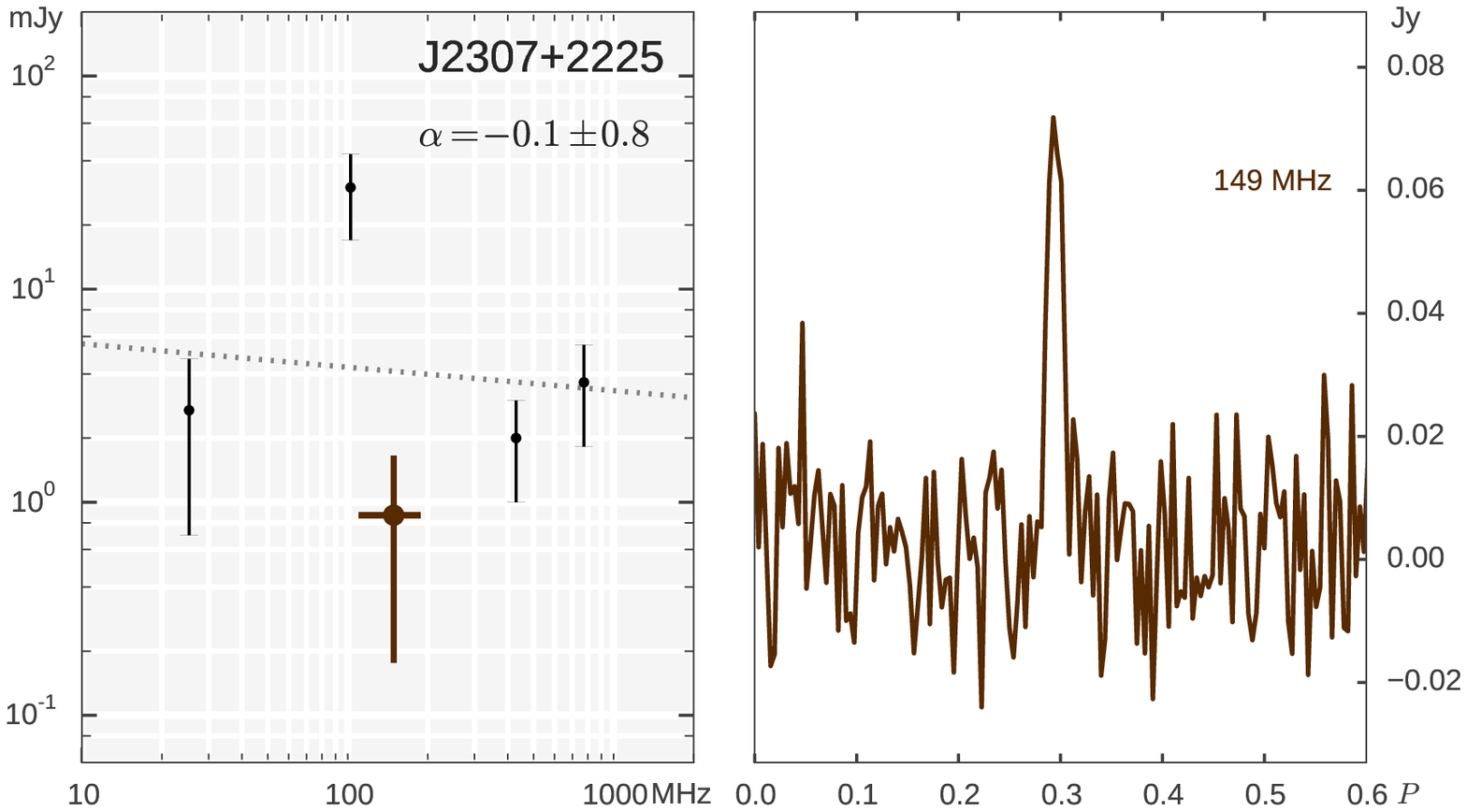}
\includegraphics[scale=0.48]{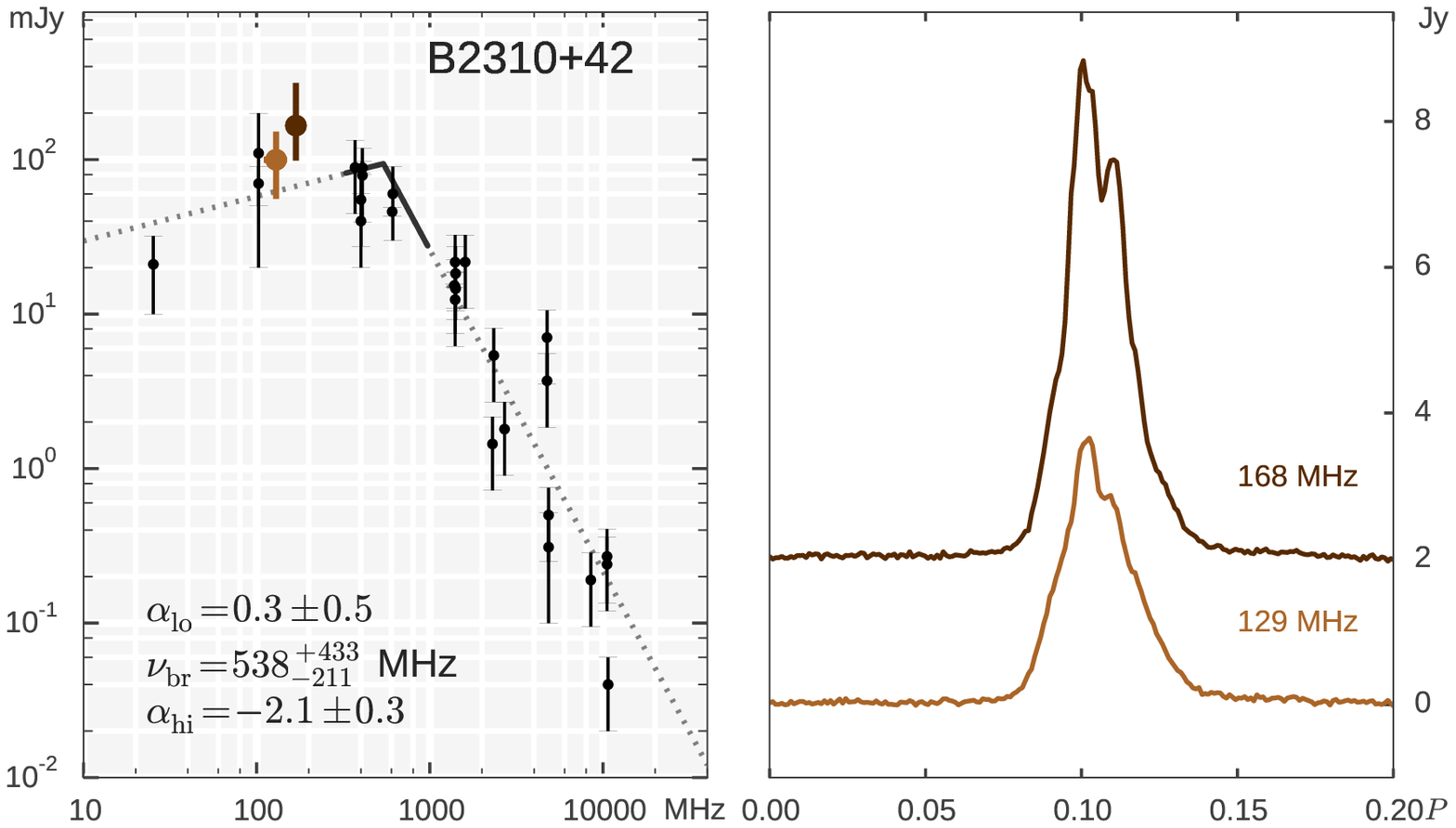}\includegraphics[scale=0.48]{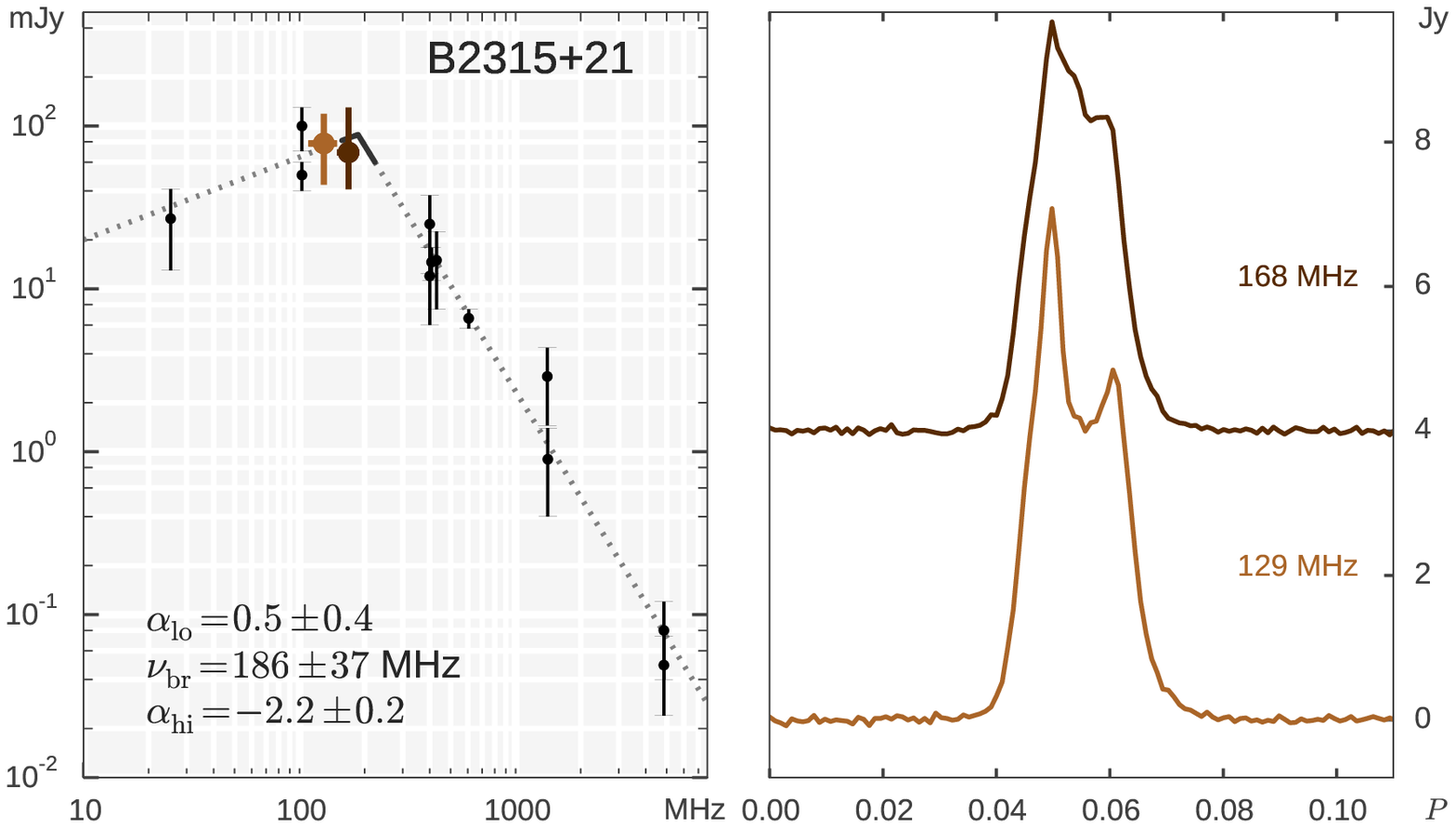}
\caption{See Figure~\ref{fig:prof_sp_1}.}
\label{fig:prof_sp_16}
\end{figure*}

\endgroup

\end{document}